\documentclass[preprint]{aastex}

\usepackage{graphicx}
\usepackage{color,natbib,lscape}

\slugcomment{accepted to ApJ}

\shorttitle{NICI Moving Groups}
\shortauthors{Biller et al.}

\begin{document}


\title{The Gemini NICI Planet-Finding Campaign: \\
The Frequency of Planets around Young Moving Group Stars\footnotemark[0]}


\author{Beth A. Biller\altaffilmark{1}, Michael C. Liu\altaffilmark{2},
  Zahed Wahhaj\altaffilmark{2}, Eric L. Nielsen\altaffilmark{2}, 
  Thomas L. Hayward\altaffilmark{3},   Jared R. Males\altaffilmark{4},
Andrew Skemer\altaffilmark{4}, Laird M. Close\altaffilmark{4}, 
  Mark Chun\altaffilmark{5}, 
  Christ Ftaclas\altaffilmark{1}, 
Fraser Clarke\altaffilmark{6},  Niranjan Thatte\altaffilmark{6}, 
 Evgenya L. Shkolnik\altaffilmark{7},   I. Neill Reid\altaffilmark{8}, 
Markus Hartung\altaffilmark{3}, 
 Alan Boss\altaffilmark{9}, 
 Douglas Lin\altaffilmark{10},
Silvia H.P. Alencar\altaffilmark{11}, 
  Elisabete de Gouveia Dal Pino\altaffilmark{12}, 
  Jane Gregorio-Hetem\altaffilmark{12}, Douglas Toomey\altaffilmark{13}}


\altaffiltext{1}{Max-Planck-Institut f\"ur Astronomie, K\"onigstuhl
  17, 69115 Heidelberg, Germany}
\altaffiltext{2}{Institute for Astronomy, University of Hawaii, 2680
  Woodlawn Drive, Honolulu, HI 96822}
\altaffiltext{3}{Gemini Observatory, Southern Operations Center, c/o AURA, Casilla 603, La Serena, Chile}
\altaffiltext{4}{Steward Observatory, University of Arizona, 933 North Cherry Avenue, Tucson, AZ 85721}
\altaffiltext{5}{Institute for Astronomy, 640 North Aohoku Place, \#209, Hilo,
Hawaii 96720-2700 USA}
\altaffiltext{6}{Department of Astronomy, University of Oxford, DWB,
 Keble Road, Oxford OX1 3RH, U.K.}
\altaffiltext{7}{Lowell Observatory, 1400 West Mars Hill Road  Flagstaff, AZ 86001}
\altaffiltext{8}{Space Telescope Science Institute, 3700 San Martin Drive, Baltimore, MD 21218}
\altaffiltext{9}{Department of Terrestrial Magnetism, Carnegie
 Institution of Washington, 5241 Broad Branch Road, NW, Washington, DC 20015}
\altaffiltext{10}{Department of Astronomy and Astrophysics,
University of California, Santa Cruz, CA 95064}
\altaffiltext{11}{Departamento de Fisica - ICEx - Universidade Federal
 de Minas Gerais, Av. Antonio Carlos, 6627, 30270-901, Belo Horizonte, MG, Brazil}
\altaffiltext{12}{Universidade de Sao Paulo, IAG/USP, Departamento de 
Astronomia, Rua do Matao, 1226, 05508-900, Sao Paulo, SP, Brazil}
\altaffiltext{13}{Mauna Kea Infrared, LLC, 21 Pookela St., Hilo, HI 96720}


\footnotetext[0]{Based
   on observations obtained at the Gemini Observatory, 
   which is operated by the Association of
    Universities for Research in Astronomy, Inc., under a cooperative
    agreement with the NSF on behalf of the Gemini partnership: the
    National Science Foundation (United States), the Science and
    Technology Facilities Council (United Kingdom), the National
    Research Council (Canada), CONICYT (Chile), the Australian
    Research Council (Australia), Minist\'{e}rio da Ci\^{e}ncia e
    Tecnologia (Brazil) and Ministerio de Ciencia, Tecnolog\'{i}a e
    Innovaci\'{o}n Productiva (Argentina).}







\begin{abstract}
We report results of a direct imaging survey for giant planets
around 80 members of the $\beta$ Pic, TW Hya, 
Tucana-Horologium, AB Dor, and 
Hercules-Lyra moving groups, observed as part of the
Gemini NICI Planet-Finding Campaign.   For this sample, we obtained
median contrasts of $\Delta$$H$=13.9 mag at 1'' in combined
CH$_{4}$ narrowband ADI+SDI mode and median contrasts of 
$\Delta$$H$=15.1 mag at 2'' in $H$-band ADI mode.  We found numerous ($>$70) 
candidate companions in our survey images.
Some of these candidates were rejected as common-proper motion
companions using archival data; we reobserved with NICI all other candidates 
that lay within 400 AU of the star and were not in dense stellar fields. 
The vast majority of candidate companions were confirmed as background objects 
from archival observations and/or dedicated NICI campaign followup.  
Four co-moving companions of brown
dwarf or stellar mass were discovered in this moving group
sample: PZ Tel B (36$\pm$6 M$_{Jup}$, 16.4$\pm$1.0 AU, Biller et al. 2010) , 
CD -35 2722B (31$\pm$8 M$_{Jup}$, 67$\pm$4 AU, Wahhaj et al. 2011), 
HD 12894B (0.46$\pm$0.08 M$_{\odot}$, 15.7$\pm$1.0 AU), and 
BD+07 1919C (0.20$\pm$0.03 M$_{\odot}$, 12.5$\pm$1.4 AU).  
From a Bayesian analysis of the achieved H band ADI and ASDI contrasts, 
using power-law models of planet distributions and hot-start evolutionary 
models, we restrict the frequency of 1--20 M$_{Jup}$ companions at semi-major
axes from 10--150 AU to $<$18\% at a 95.4$\%$ confidence level
using DUSTY models and to $<$6\% at a 95.4$\%$ using COND models. 
Our results strongly constrain the frequency of planets within
semi-major axes of 50 AU as well.
We restrict the frequency of 1--20 M$_{Jup}$ companions at semi-major
axes from 10--50 AU to $<$21\% at a 95.4$\%$ confidence level
using DUSTY models and to $<$7\% at a 95.4$\%$ using COND models.
This survey is the deepest search to date for giant planets around young 
moving group stars.
\end{abstract}





\section{Introduction}

In the last decade, $\sim$10 planets and planet candidates
with estimated masses $<$13 M$_{Jup}$ have been imaged in orbit around
young stars and brown dwarfs \citep[e.g.][]{Cha05a, Mar08, Kal08, Laf08,
Lag09, Lag10, Mar10, Tod10, Ire11, Luh11, Kra12, Ram13b, Qua13,
Kuz13, Bow13}.  In total, $\sim$30 companions 
with estimated masses $<$25 M$_{Jup}$ have been imaged. (See
http:$//$exoplanet.eu for a compilation of these objects.)
These discoveries have provided a wealth
of new information about young giant planets, as well as some surprises. 
Prior to these detections, models predicted that young gas giant
planets at moving group ages (10-300 Myr) would likely have cool photospheres with 
prominent methane absorption features \citep[][]{Bar03, Bur03},
i.e. that these objects would be spectral analogs to T-type brown dwarfs.  
However, all known directly-imaged planets at these ages 
\citep[specifically~2MASS~1207b~and~the~HR~8799~planets,][]{Cha05a,Mar08,
Mar10}  have lacked methane absorption and 
show extremely red colors, likely due to dust clouds 
and/or non-equilibrium chemistry in their 
atmospheres \citep[]{Bow10, Ske11, Bar11a,Bar11b, Cur11}.

Additionally, all of these companions except for $\beta$ Pic b
\citep[][]{Lag09, Lag10}, HR 8799e \citep[][]{Mar10}, 
and LkCa 15b \citep[][]{Kra12} lie at
projected separations greater than 20 AU, considerably
wider than giant planets in our own solar system.
Such widely separated companions pose a
challenge for the accepted model of core-accretion formation, 
which likely formed the closer-in ($<$10 AU)  population of planets detected to
date via radial velocity studies \citep[e.g.,~][]{Mor09, Jan12, 
Dod09}.  However, given that only $\sim$10 such companions 
have been imaged to date,   
it is perhaps premature to make statements based on such a small sample.
Thus, it is a priority to discover additional companions as well as to 
constrain on the distributions of their
semimajor axes, eccentricities, masses, etc.

In the last decade, a number of deep, adaptive-optics aided surveys
with sample sizes $>$20 stars   
have been completed at 8-m telescopes to search for 
additional planetary companions.  Many of these have been 
conducted in the 1.6 $\mu$m $H$-band or 2.2 $\mu$m $K$-band
\citep[][]{Mas05, Bil07, Laf07b, Apa08, Cha10}, 
while others have focused further into the infrared (3.5-5 $\mu$m) in
the $L$, $L'$, or $M$ bands \citep[][]{Kas07, Hei10a, Hei10b,Ram13a}.
A number of very large scale surveys ($>$100 stars) are ongoing 
or recently completed, including the NICI Campaign at Gemini-South
(Liu et al. 2010, this publication, Wahhaj et al. 2013a, 
Nielsen et al. 2013, Wahhaj et al. 2013b), 
the NACO large program using NACO at the VLT (Buenzli et al. 2010), 
SEEDS (Strategic Exploration of Exoplanets and Disks with Subaru)
using HiCIAO at Subaru \citep[][]{Tha09, Car13}, 
and the International Deep Planet Survey
(IDPS) using primarily Gemini and Keck \citep[][]{Vig12}.


The host stars of currently known directly-imaged
planets fall into three categories: (1) members of young ($\lesssim$10
Myr) star-forming clusters or OB associations
\citep[e.g.~Taurus~and~Upper~Sco;][]{Laf08, Tod10, Kra12}
, (2) members of nearby  young moving groups 
\citep[ages~of~10--300~Myr,~e.g.][]{Lag10, Cha05a,Mar08}, and (3) unassociated nearby 
young stars \citep[e.g.][]{Kal08}.  Of these three categories of
targets, moving group objects are particularly compelling
targets for direct imaging searches.
The extremely young ages of star-forming clusters
translates into considerably brighter planets, but at distances
$\geq$140 pc the inner working angles of current instruments
generally only allow detection of companions at projected separations
$\gtrsim$50 AU.  Unassociated nearby young stars 
often do not have well constrained ages, a limitation for estimating the
mass of any companion detected and for 
deriving statistics for the survey sensitivities.  Moving group stars
provide a unique nearby young sample with well-constrained ages and distances.

We have observed 80 young moving group stars as a part of a dedicated science
campaign using the Near-Infrared Coronagraphic Imager (NICI) at the 
8.1 m Gemini South Telescope \citep[][]{Chu08}.
NICI is a dedicated adaptive optics (AO) instrument tailored expressly for direct
imaging of exoplanet companions, combining several techniques to attenuate starlight and suppress
speckles for direct detection of faint companions to bright stars:
(1) Lyot coronagraphy, (2) dual-channel imaging for Spectral
Differential Imaging \citep[SDI;][]{Rac99, Mar05, Bil07}, and (3) operation in
a fixed Cassegrain rotator mode for Angular Differential Imaging
\citep[ADI;][]{Liu04, Mar06, Laf07a,Bil08}.
While each of these techniques has been
used individually in large planet-finding surveys 
\citep[e.g.~][]{Bil07, Laf07b}, the NICI
Campaign is the first time all three have been employed
simultaneously in a large survey.

From 2008 December to 2012 September, the NICI Planet-Finding Campaign
\citep[][]{Liu10} obtained deep, high-contrast AO imaging of a
carefully selected sample of over 200 young, nearby stars.  
Over the course of the Campaign, we discovered 4 new brown dwarf
companions to young stars: PZ Tel B \citep{Bil10}, CD -35 2722B \citep{Wah11},
HD 1160C \citep{Nie12}, and HIP 79797Bab \citep{Nie13}.  Here we report 
results from the subsample of 80 young stars that are members of
the $\beta$ Pic, TW Hya, AB Dor, Tucana-Horologium, and 
Hercules-Lyra moving groups.

\section{Moving Group Sample}

Moving groups are associations of young stars (10--300 Myr) 
that are unconnected to regions of ongoing star-formation.
These associations were not discovered until the late 1990's, as moving 
group members are often dispersed across a wide part of the sky
\citep[e.g.~][]{Zuc04, Tor08}.
Moving group members are identified by a combination of 
youth indicators (Li absorption, high X-ray luminosity,
etc.) and space motion coincident with other cluster members.
We have focused our survey on 5 young moving groups with members 
generally within $\sim$60 pc of the Earth.  

\subsection{TW Hya Association}

The star TW Hya was the first pre-main sequence (henceforth PMS) 
star identified outside of a star-forming region, initially identified
by \citet[][]{Ruc83} as an isolated T Tauri star.
\citet[][]{dlR89} and \citet[][]{Gre92} identified
4 additional T Tauri stars within 10 degrees of TW Hya.
\citet[][]{Kas97} were the first to label these stars as the TW Hya 
Association, based on their strong lithium absorption features,
X-ray fluxes, and similar Hipparcos parallaxes.
Since then, $>$20 TW Hya members have been identified 
\citep[see~][]{Web99, Ste99, Jay99, Zuc01a, Giz02, Son02, Son03,
  Kas08,Zuc04b, Sch05, Mam05, Loo07, Tor08, Fer08, DSi09, Loo10a,
  Loo10b, Rod11, Shk11}.  
Based on lithium absorption and X-ray flux strength, the association
is assigned an age of $\approx$10 Myr \citep[][]{Kas97, Web99}. 
The mean distance of the TW Hya association is 48$\pm$13 pc
\citep[][]{Tor08, Wei13}.

\subsection{$\beta$ Pic Moving Group}

The circumstellar disk around the nearby A star 
$\beta$ Pic was first imaged by 
\citet[][]{Smi84}, leading to the identification of $\beta$ Pic as a
young star with planet formation having occurred in the 
recent past.  \citet[][]{Bar99} found two
additional M type stars (GJ 799 and GJ 803) with matching 
Galactic space motions to $\beta$ Pic.  \citet[][]{Zuc01b}
cemented the existence of the $\beta$ Pic moving group with the 
confirmation of 18 additional members on the basis of their
Galactic space motions.  Over 60 members of the $\beta$ Pic moving
group have been identified to date 
\citep[][]{Son03, Zuc04, Tor06, Tor08, Fer08, DSi09, 
Lep09, Ric10, Sch10, Kis11,Sch12, Shk12}.
From the color-magnitude
diagram placement of these stars as well as their
lithium absorption ages, an age of $\approx$12 Myr is estimated for this
moving group \citep[][]{Zuc04}, with a mean distance of 
31$\pm$21 pc \citep[][]{Tor08}.

\subsection{Tucana-Horologium Association}

Young stars are often far-IR excess sources.  Based on this fact, 
\citet[]{Zuc00} searched the Hipparcos catalog 
for stars with similar space motions and within a 6 degree radius of
24 stars detected at 60 $\mu$m with IRAS.  
From this search, they identified $\sim$10 stars with distances of $\sim$45 pc
and ages of $\sim$30 yr, which they named the Tucana Association.
\citet[]{Tor00} found a group of $\sim$10 stars through X-ray
emission and ground-based spectroscopy that showed youth 
indicators and are associated with the previously identified isolated
T Tauri star EP Eri, which they titled the Horologium Association.
As the stars in these two associations share the same space
motions, ages, distances, and volume density, they are now
considered to be part of the same association \citep[][]{Zuc01c}.
Over 60 stars have been identified to date in the Tucana-Horologium
association \citep[][]{Son03, Zuc04, Tor08, Fer08, DSi09, Kis11,
  Zuc11}, with a mean distance of 48$\pm$7 pc \citep[][]{Tor08}.

\subsection{AB Dor Moving Group}

The star AB Dor is notable as an ultrafast rotator which is also extremely
X-ray active, at a distance of only 15 pc and an age of $\sim$100 Myr
\citep[][]{Luh05}.  AB Dor itself, in fact, is a quadruple system with a close M-dwarf
companion and a wider separation M-dwarf binary 
\citep[][]{Gui97, Clo05, Nie05, Clo07}.
\citet{Zuc04b} identified $\sim$30 nearby star systems with 
similar space motions to AB Dor as well characteristics of youth,
which they designated the AB Dor moving group.
Over 50 stars have been identified to date in the AB Dor moving group
\citep[][]{Zuc04, Tor08, Fer08, DSi09, Sch10, Zuc11, Sch12,Shk12}, 
with a mean distance of 34$\pm$26 pc \citep[][]{Tor08}. 

\subsection{Hercules-Lyra Association}

\citet[][]{Gai98} first identified 4 young solar analogues with
similar space motions towards Hercules.  \citet[][]{Fuh04}
identified a further 15 late-type stars with similar space motions
and gave the whole complex the name Hercules-Lyra.  
The existence of the Hercules-Lyra association was initially disputed, as 
the candidate members possessed a wide age spread inconsistent
with a single moving group, with 
some of the initially identified stars 
possessing ages (derived from lithium absorption and chromospheric activity)
much younger or older than the average association age of
$\sim$200 Myr.  \citet[][]{Lop06} confirmed 
the existence of the Hercules-Lyra association and 
winnowed down the 27 initial candidate members 
to 10 confirmed members with an average distance of 20$\pm$10 pc
and age of $\sim$200 Myr.

\section{Observations}

We observed 80 stars in nearby young moving groups as part of the NICI
Campaign -- 14 stars from the TW Hya association,  
30 stars from the $\beta$ Pic moving group, 12 stars from the 
Tucana-Horologium association, 19 stars from the AB Dor moving group,
4 stars from the Hercules-Lyra association, and 1 star (BD +1 2447) 
which is either a Hercules-Lyra or AB Dor moving group member.  
The survey sample was selected from a larger sample of moving group
stars compiled from the literature.  
Observations were prioritized according to the probability of 
detecting a planet around a given survey star
\citep[][]{Liu10}, as predicted
 by Monte Carlo simulations similar to those described in Section 5.1.

\begin{figure}
\includegraphics[width=6in]{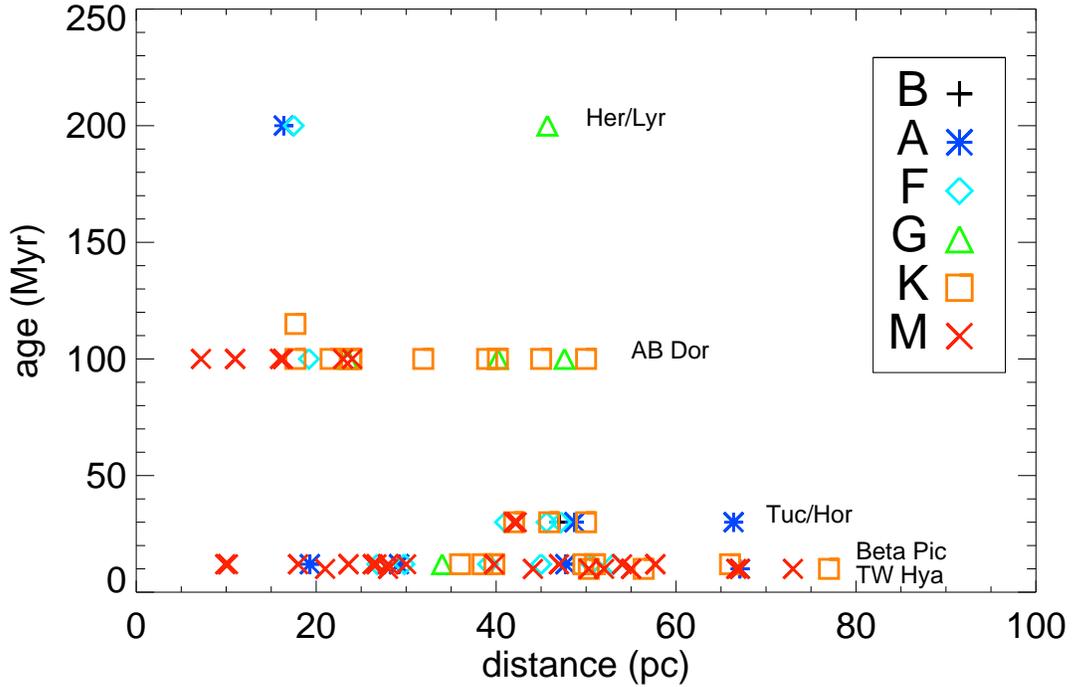}
\caption{ NICI Campaign moving group sample plotted as a function of 
distance and age.  Plot symbols give object spectral type and each
moving group is labeled at the appropriate age.  The majority (85$\%$)
of stars in our sample have ages less than 100 Myr and distances less than 60 pc.
We observed 14 stars from the TW Hya association, 
30 stars from the $\beta$ Pic moving group, 12 stars from the 
Tucana-Horologium association, 19 stars from the AB Dor moving group.
4 stars from the Hercules-Lyra association, and 1 star (BD +1 2447) 
which is either a Hercules-Lyra or AB Dor moving group member.  
\label{fig:MG_sample}
}
\end{figure}

\begin{figure}
\includegraphics[width=3.2in]{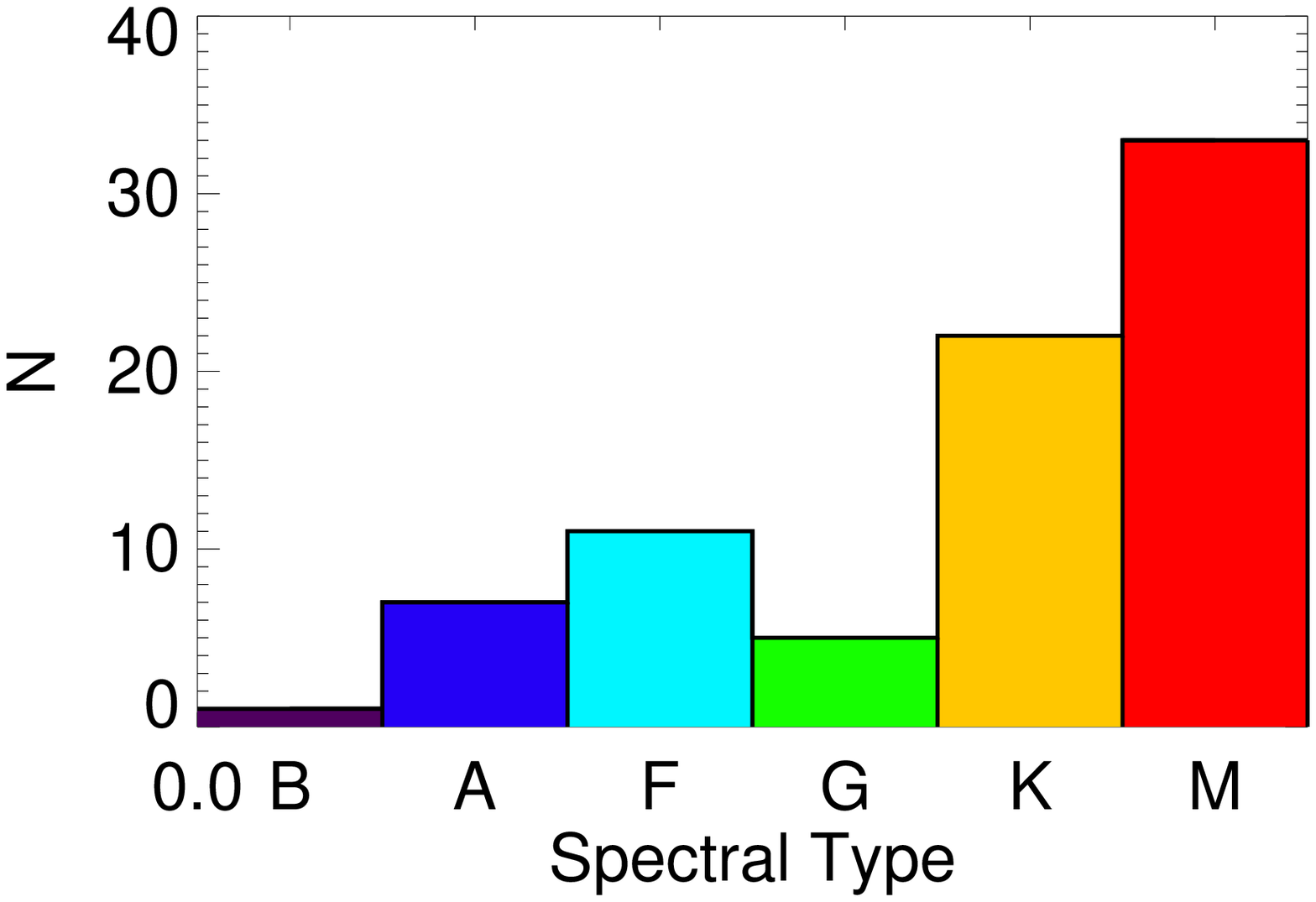}
\includegraphics[width=3.2in]{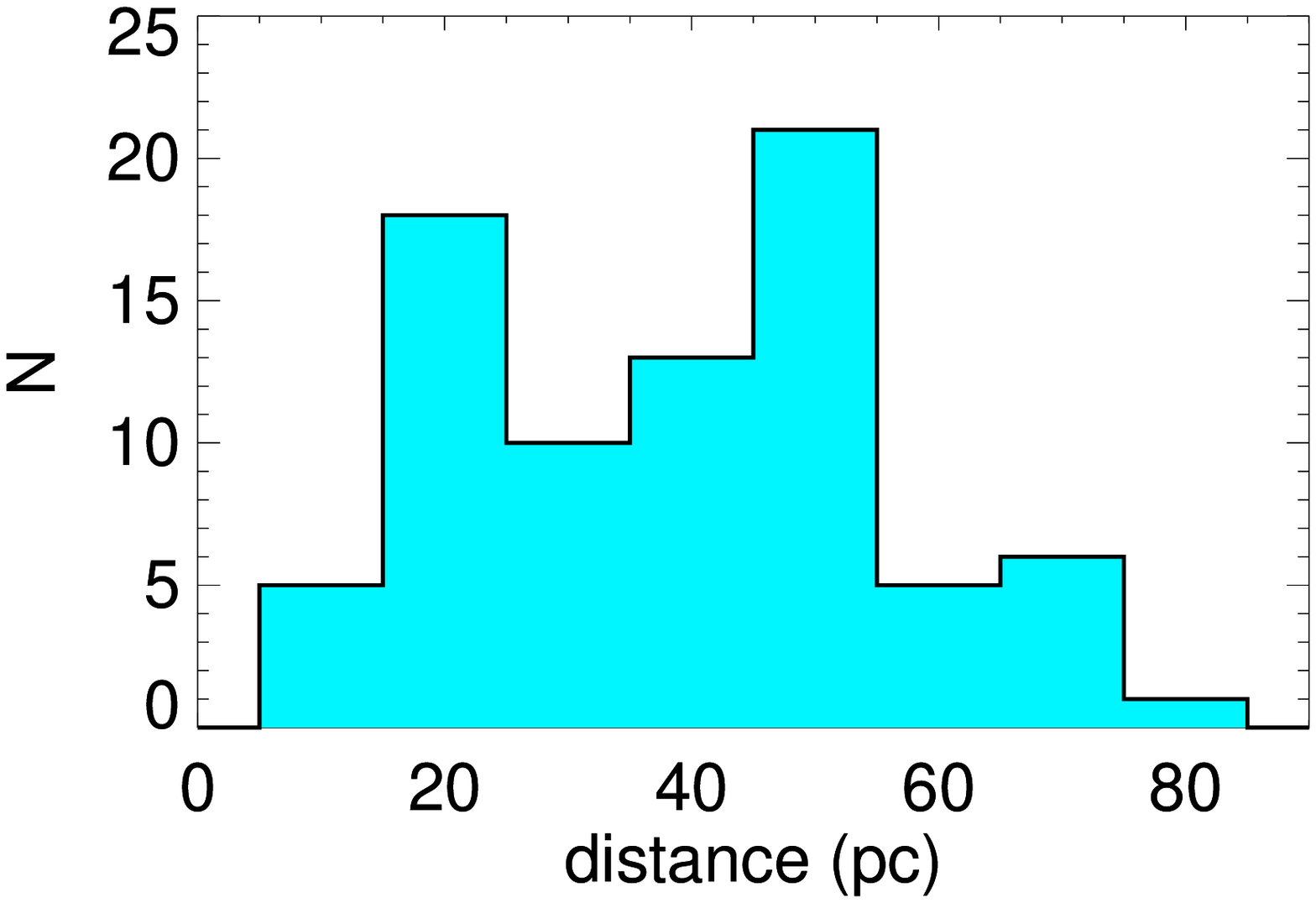}
\caption{ Histograms of the spectral types and distances 
  of our 80 MG sample objects.  Column width is 1 spectral type 
and 10 pc respectively.  The median distance for this sample is
  39.8 pc.  We observed 1 B star, 7 A stars, 11 F stars, 5 G stars, 23 K
stars, and 33 M stars.  About 70$\%$ of our sample have spectral types of 
K or later; thus, our sample is comprised largely of lower mass stars.
\label{fig:MG_hists}}
\end{figure}

The survey sample is listed in 
Table~\ref{tab:properties} and is plotted as a function 
of age, distance, and spectral type in Figure~\ref{fig:MG_sample}.
Histograms of the spectral type and distance distributions are presented in Figure~\ref{fig:MG_hists}.  
The majority (85$\%$) of sample stars have ages less than 100 Myr and distances
less than 60 pc.  The median distance is 39.8 pc.  
We observed 1 B star, 7 A stars, 11 F stars, 5 G stars, 23 K stars,
and 33 M stars.  Thus, our moving
group sample is primarily composed of lower mass stars.
Observations of our survey sample are listed in
Table~\ref{tab:observations}.  We only report observations which
contain at least 10 individual images in the ADI or ASDI sequences, in
order to achieve the field rotation needed by our ADI processing
pipeline to obtain reliable detections (see Section~\ref{sec:obs_strategy}
for details). 

\subsection{The Near-Infrared Coronagraphic Imager at Gemini South}

NICI was specifically designed to provide the high contrasts 
necessary to directly image young extrasolar giant planet.  
NICI's 85-element curvature AO system provides AO correction
of $\sim$30-45$\%$ Strehl in $H$ band \citep[][]{Chu08}.  The AO beam is then reflected
into the science camera, where it passes through a partially transparent
focal plane mask.  The focal plane mask is a flat-topped Gaussian,
which suppresses $>$99.5$\%$ of the incoming starlight \citep[$\Delta$CH$_{4S}$=6.39$\pm$0.03~mag,
$\Delta$$H$=5.94$\pm$0.05~mag;][]{Wah11}, 
thus reducing scattered light from the central star and increasing
the attained contrast.  A variety of these semi-transparent masks are
available for use with NICI; we utilized the 0.32$\arcsec$ radius 
mask for NICI Campaign observations, thus providing an effective inner working 
angle of 0.32$\arcsec$ for faint companions, although tight stellar
companions can still be detected in the innermost regions.
The partially-transparent mask also 
allows us to attain very precise photometry and astrometry, 
as we can simultaneously obtain unsaturated images of both the primary
and faint companions.  The beam then passes through a hard-edged
pupil stop, which reduces diffracted light from 
PSF artifacts associated with 
the Gemini-South secondary mirror.   For observations in dual-channel
mode, the beam is split using a dichroic
and passes into two separate science cameras.  For the majority of
the Campaign, a 50/50 beamsplitter was utilized, resulting in the loss
of half of the incoming light to each channel, but from the beginning of
2012, this beamsplitter was replaced by an H/K dichroic, boosting
throughput when imaging simultaneously in these two filters.  
Different filters may be chosen for each science camera; thus NICI's 2-camera 
capability can provide simultaneous color information.
Both cameras have fields of view of 18$\times$18$\arcsec$, with a
platescale of 17.96 mas
for the science camera using the 1.578 $\mu$m $CH_{4} S$ filter 
(henceforth "blue channel''  or "off-methane channel") and a platescale of 17.94
for the science camera using the 1.652 $\mu$m $CH_{4} L$ 4\% filter
(henceforth "red channel" or "on-methane channel")
for the science camera using the 1.578 $\mu$m.

\subsection{Observing Strategy~\label{sec:obs_strategy}}

NICI Campaign observations were conducted in two separate modes: (1)  single channel H-band ADI (Angular Differential
Imaging) mode and (2) dual-channel methane band combined ADI+SDI (Spectral Differential Imaging) mode.
Both SDI and ADI techniques seek to distinguish real objects from speckles. SDI achieves this by exploiting a
spectral feature in the desired target
\citep[e.g.~the~1.6~$\mu$m~methane~absorption~feature~observed~in~substellar~objects~with~a~T~spectral~type][]{Geb02,
  Cus05}.  Images are taken simultaneously both within and
outside the chosen absorption feature.  Due to the simultaneity
of the observations, the stellar point-spread functions in the two
NICI channels, including the coherent speckle patterns, are nearly identical.
In contrast, any faint companion with the chosen absorption feature is
bright in one filter and faint in the other. Subtracting the two images thus removes the starlight and speckle
patterns while a real companion with the chosen absorption feature remains in the image. In other words, the
absorption band image acts as an ideal reference point spread function (henceforth PSF) for the off-absorption
band image. Utilizing a signature spectral feature of substellar
objects can help distinguish between true methanated companions and likely background objects,
e.g. a background object will be subtracted out by the SDI subtraction since it
will not have methane absorption.   However, this mode is sensitive even to
companions without this absorption feature, as 
a real companion will appear fixed in separation relative to
the star in both filters, while a speckle will modulate with the Airy
pattern and appear further from the star
in the red filter relative to the blue filter.

ADI employs a different strategy in order to decorrelate real companions from speckles.
For ADI observations, the rotator is left off at the Cassegrain focus
or set to follow the elevation angle at the Nasmyth focus, allowing the telescope optics
rotate relative to the sky.  In a sequence of images taken at
different parallactic angles, a real companion will move relative to
the detector along with the sky, while the speckles will remain fixed.
 From a series of images, a reference PSF
can thus be constructed for and subtracted from each individual image, attenuating quasi-static speckle structure.
Combining both SDI and ADI techniques (henceforth ASDI) 
thus allows an even greater degree of speckle supression.

In order to take advantage of both the higher contrast available within 1.5'' using
the ASDI mode (due to improved speckle suppression from the SDI subtraction)
and the improved sensitivity available outside of 1.5'' 
with the ADI mode (due to the wider bandpass used during our ADI
observations), most NICI Campaign stars were observed in both modes.
For ASDI mode, we observed 
simultaneously in the off-methane (central $\lambda$=1.578 $\mu$m;
width=0.062 $\mu$m; $CH_{4}  S$ 4\%) and on-methane
(central $\lambda$=1.652 $\mu$m; width=0.066 $\mu$m;
$CH_{4} L$ 4\%) bands using NICI's dual-channel imaging
capability.  ADI data were taken with the broadband $H$ filter in the
blue channel (central $\lambda$=1.65 $\mu$m, width=0.29 $\mu$m)
Stars fainter than $H$=8 mag were observed only in single-channel ADI
mode, as the contrast within 1.5'' was similar to that achievable
in the ASDI mode.  Stars close to the Galactic Bulge were only observed in
ASDI mode, as ADI mode often yielded enormous numbers ($>$50 per field) of background 
field objects.  

Typically, we obtained 20 minutes on-sky data in ADI mode and 40
minutes on-sky data in ASDI mode for each star. 
Observations were carefully scheduled in order to maximize field rotation while avoiding
too much blurring during single exposures.  We aimed to obtain at least
5$^{\circ}$ field rotation in ADI mode observations and at least 15$^{\circ}$ field
rotation in ASDI mode observations.  This ensures on-sky rotations of
at least 3$\times$FWHM of the PSF at 5'' 
separation from the primary in ADI mode and at least
3$\times$FWHM of the PSF at 1'' separation in ASDI mode.  Typical FWHMs of the
PSF ranged between 3-4 pixels.  Out of 68 stars with ADI mode
observations, all but 4 have at least one dataset with sky rotation
$>$5 degrees.  Out of 56 stars with ASDI 
observations, all but 7 have at least one dataset with sky rotation
$>$15 degrees, and only one ASDI observation has 
sky rotation $<$10 degrees. 

For ASDI, individual exposure times were chosen to produce
high S/N in the speckle halo while avoiding saturation in this region.
In ADI mode, exposure times of 4 to 60 s were used, allowing the halo
to saturate if needed.  For bright stars that saturate 
in the ADI exposures, short exposures were interleaved
with deep exposures in order to provide unsaturated images
of the star behind the partially transparent mask (henceforth the 
``starspot'') for accurate photometry.

\subsection{Data Reduction}

All observations are processed using a custom pipeline described in 
\citet[][]{Wah13b}.  
Here we briefly summarize procedures for both ADI and ASDI datasets;
some data processing steps pertain only to the ASDI mode and are noted
as such below.  For all data, the pipeline first
applies dark, flatfield, and distortion corrections.
For ADI data, all images are centroided and aligned to the first
exposure in the sequence.
For ASDI data, images from the two science cameras
are then centroided and aligned.
Datasets where the starspot is unsaturated are aligned 
using the starspot centroid position in 
each science exposure.
For saturated images, the structure of the saturated PSF 
is used to align the images \citep[][]{Wah13b}. 
Specifically, the peak of the primary is still discernible as a 
negative image and can be used to centroid.  
We have estimated that the centroiding accuracy of the saturated
images is 9 mas by comparing these to the centroids of
unsaturated short-exposure images obtained right before and after the
long exposures.
Image filters (i.e. unsharp masking or catch filtering) are applied 
frame-by-frame.  In the ASDI case, the red-channel image is subtracted
from the blue-channel image for each science exposure.  A
high-fidelity PSF is built for the entire observation by median
combining the stack of reduced images and then subtracted
from each individual science exposure.  Finally, the reduced
PSF-subtracted images are registered, rotated to a common 
sky orientation, and stacked to produce a final image.
In the ASDI case, 3 final output images are produced: the full
subtracted reduction as well as single-channel ADI reductions for the blue and 
red channel images respectively, which can be added to achieve deeper 
sensitivity.  This ensures that no planet candidates are missed 
due to spectral self-subtraction in the ASDI mode.

\section{Results}

\subsection{Contrast Curves and Minimum Detectable Masses}

In order to robustly measure the contrast achieved by our pipeline
reductions, we generate 95\%-completeness contrast curves following
the method described in Wahhaj et al. (2013b).  The
95\%-completeness technique accounts for self-subtraction losses 
endemic to ADI and SDI data, unlike simple measurements based solely 
on the noise level of the data.  Briefly, the data are first
pipeline-processed, rotationally misaligned (derotated in the 
opposite direction of the actual parallactic angle rotation), 
and stacked to create a companion-free reduction.  The 1$\sigma$ contrast curve is 
calculated from the standard deviation found in 3 pixel annuli
as a function of separation from the primary star.
Next, a set of 20$\sigma$ simulated
companions (1340 total simulated companions, at separations of 
0.36$\arcsec$ to 6.3$\arcsec$ and uniformly distributed in azimuth
in 67 concentric rings), produced by scaling the image of the primary star 
behind the partially-transparent mask, is inserted into the individual 
raw images, and the new data are re-reduced as before.  
The 20$\sigma$ simulated companions are recovered in the reduced 
data and used to evaluate the flux losses and artifacts in input
contrasts due to the pipeline.  Finally, the recovered companions 
(now with flux loss effects and other pipeline artifacts incorporated)
are reinserted into the original reduction and scaled in intensity
until they meet our detection criteria.  The contrast at which 95\% of
the simulated companions are detected is presented as the 
95\%-completeness contrast curve.  

\begin{figure}
\includegraphics[width=6in]{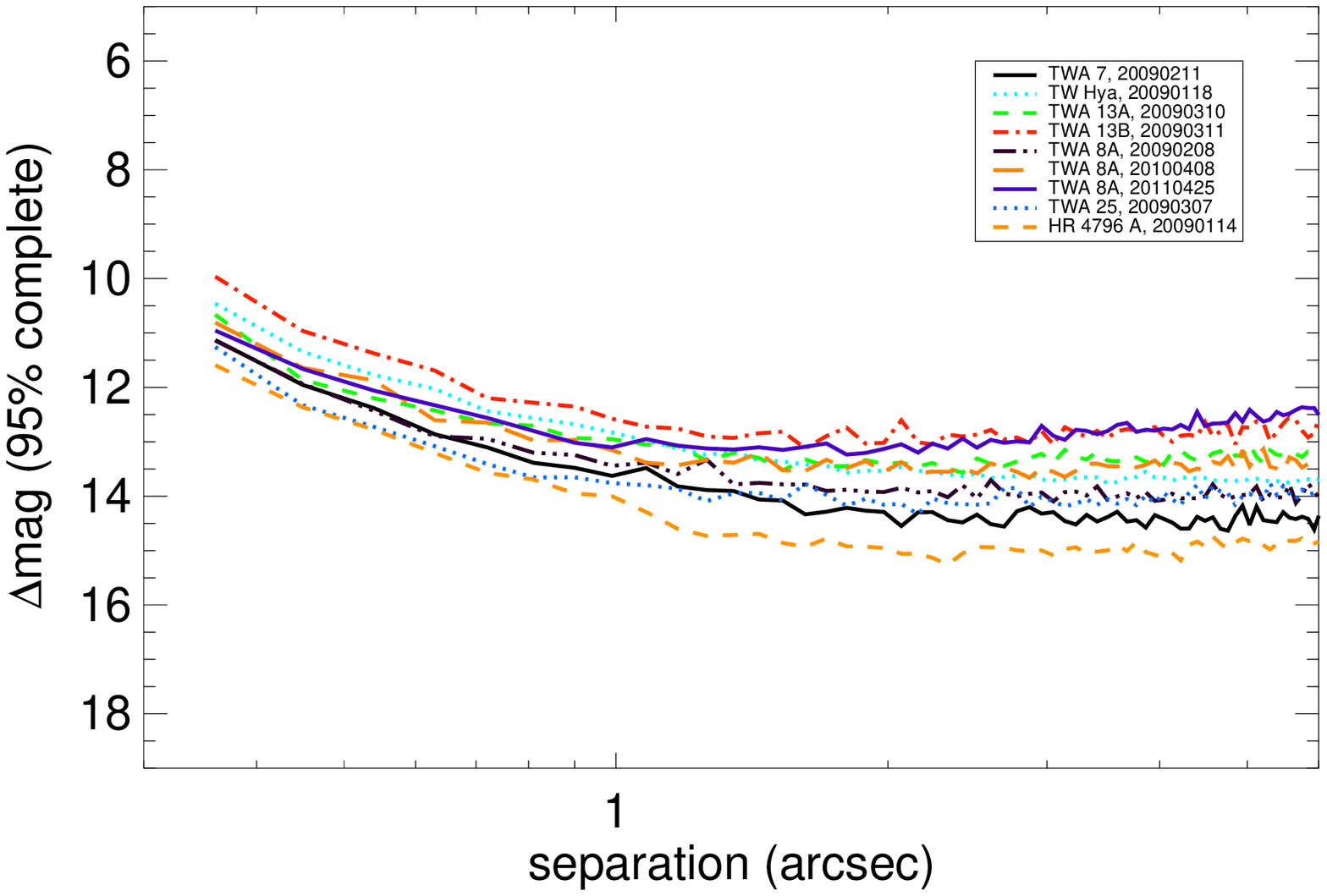}
\includegraphics[width=6in]{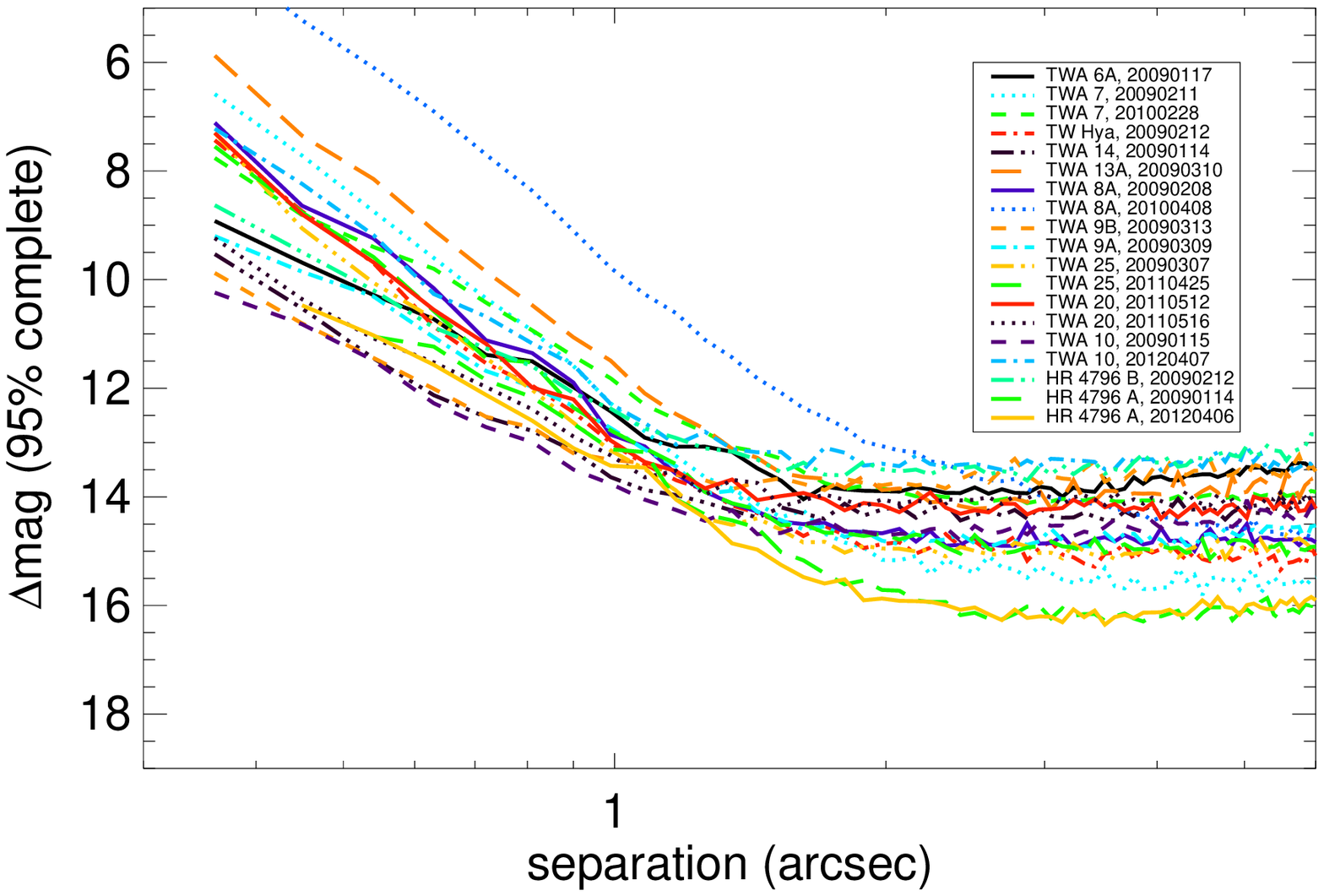}
\caption{95\% completeness contrast curves for TW Hya association
  stars, with plot range from 0.3'' (coronagraph inner working angle)
  to 6''.  A contrast of 15 magnitudes represents a flux ratio of
  10$^{6}$.  Top: ASDI contrasts.  Bottom: ADI contrasts.}
\label{fig:TWHya_contrasts}
\end{figure}

\begin{figure}
\includegraphics[width=6in]{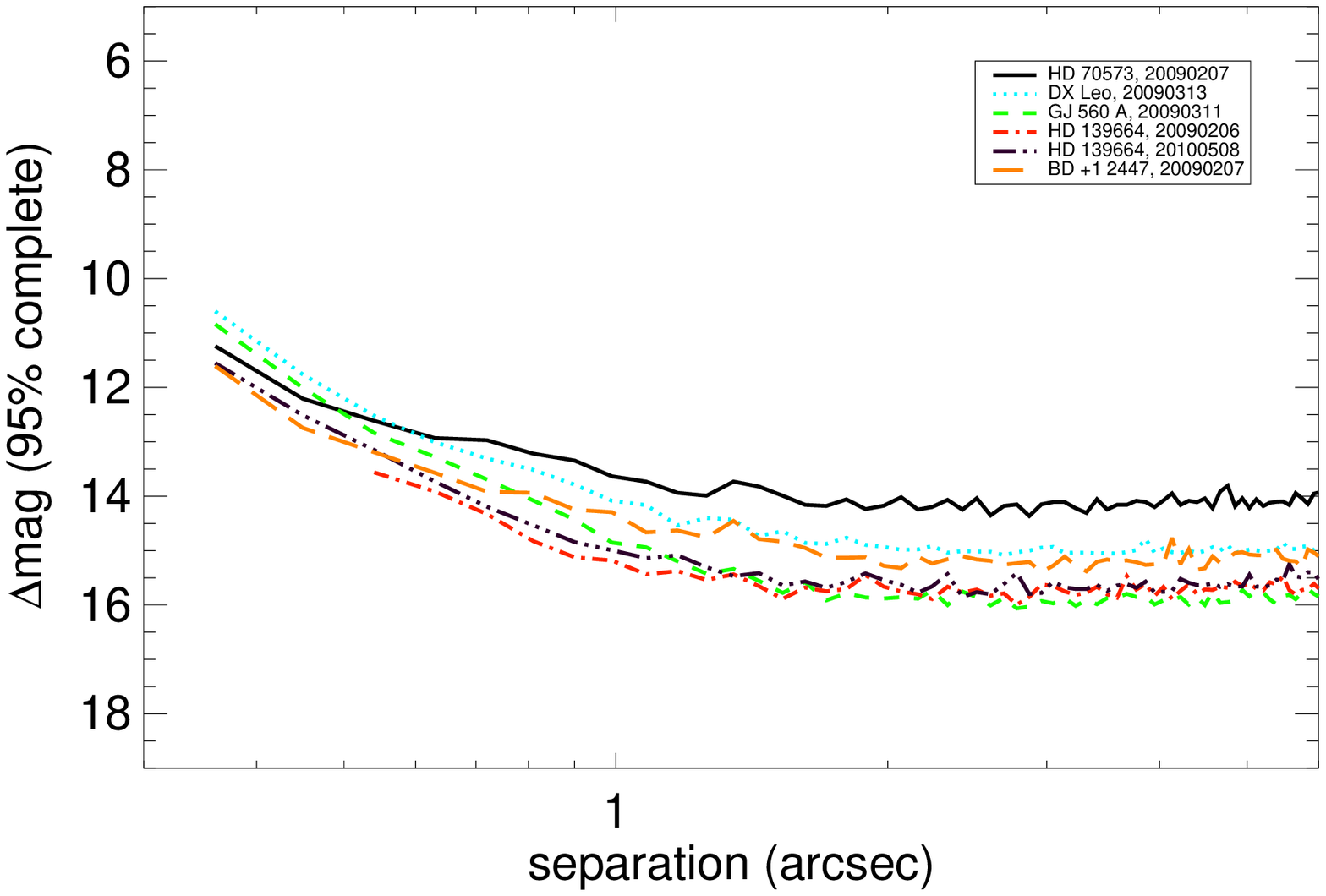}
\includegraphics[width=6in]{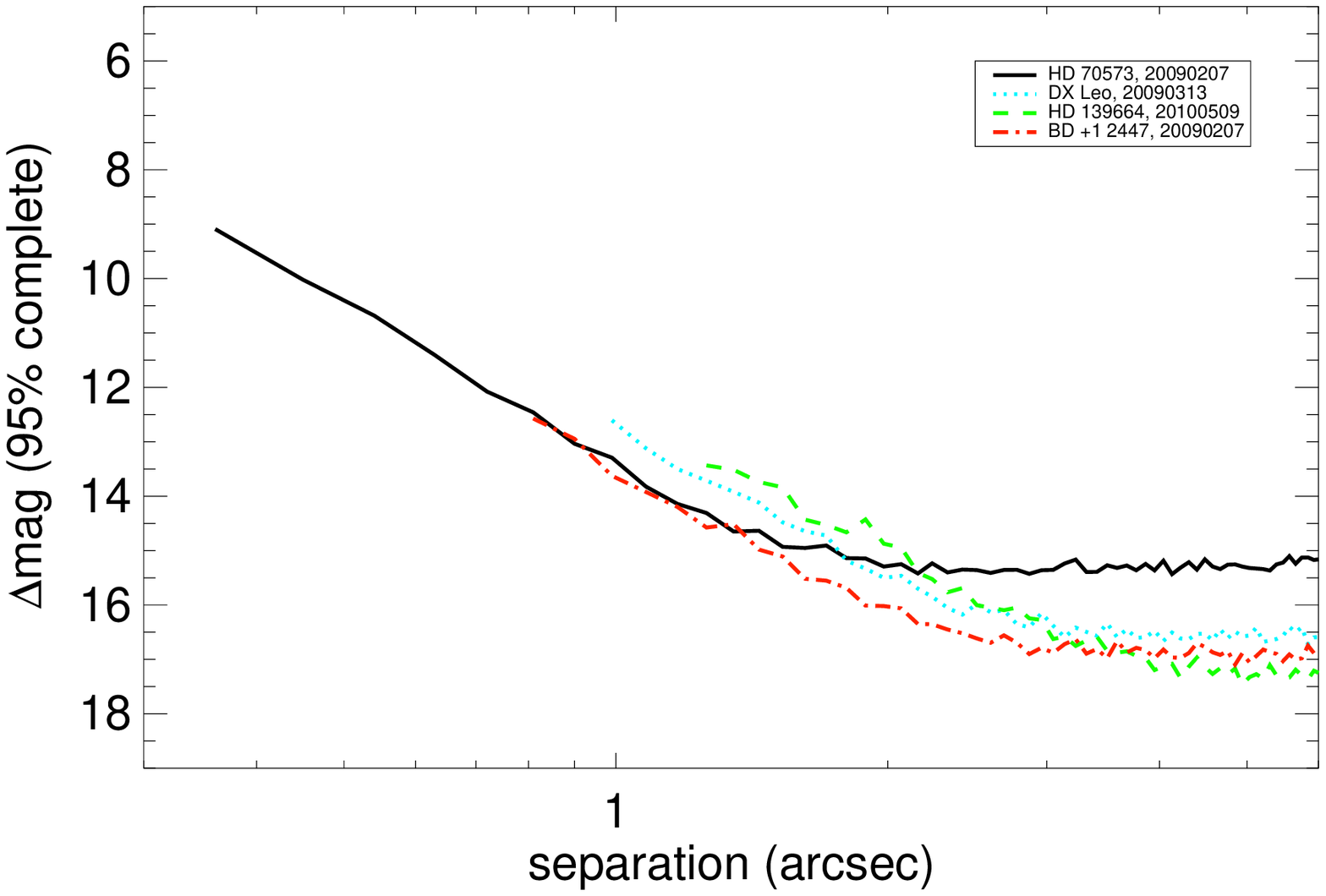}
\caption{95\% completeness contrast curves for Hercules-Lyra
  association stars, with plot range from 0.3'' (coronagraph inner working angle) to 6''.  Top:
   ASDI contrasts.  Bottom:  ADI contrasts.}
\label{fig:HerLyr_contrasts}
\end{figure}

The 95\% completeness
contrast curves for the moving group sample are presented in Figures~\ref{fig:TWHya_contrasts}
to \ref{fig:HerLyr_contrasts}.  Tables of measured contrast
 are presented for the ASDI subtracted 
reductions in Table~\ref{tab:contrastsASDI} and 
for the ADI reductions in Table~\ref{tab:contrastsADI}.

\begin{figure}
\includegraphics[width=3.2in]{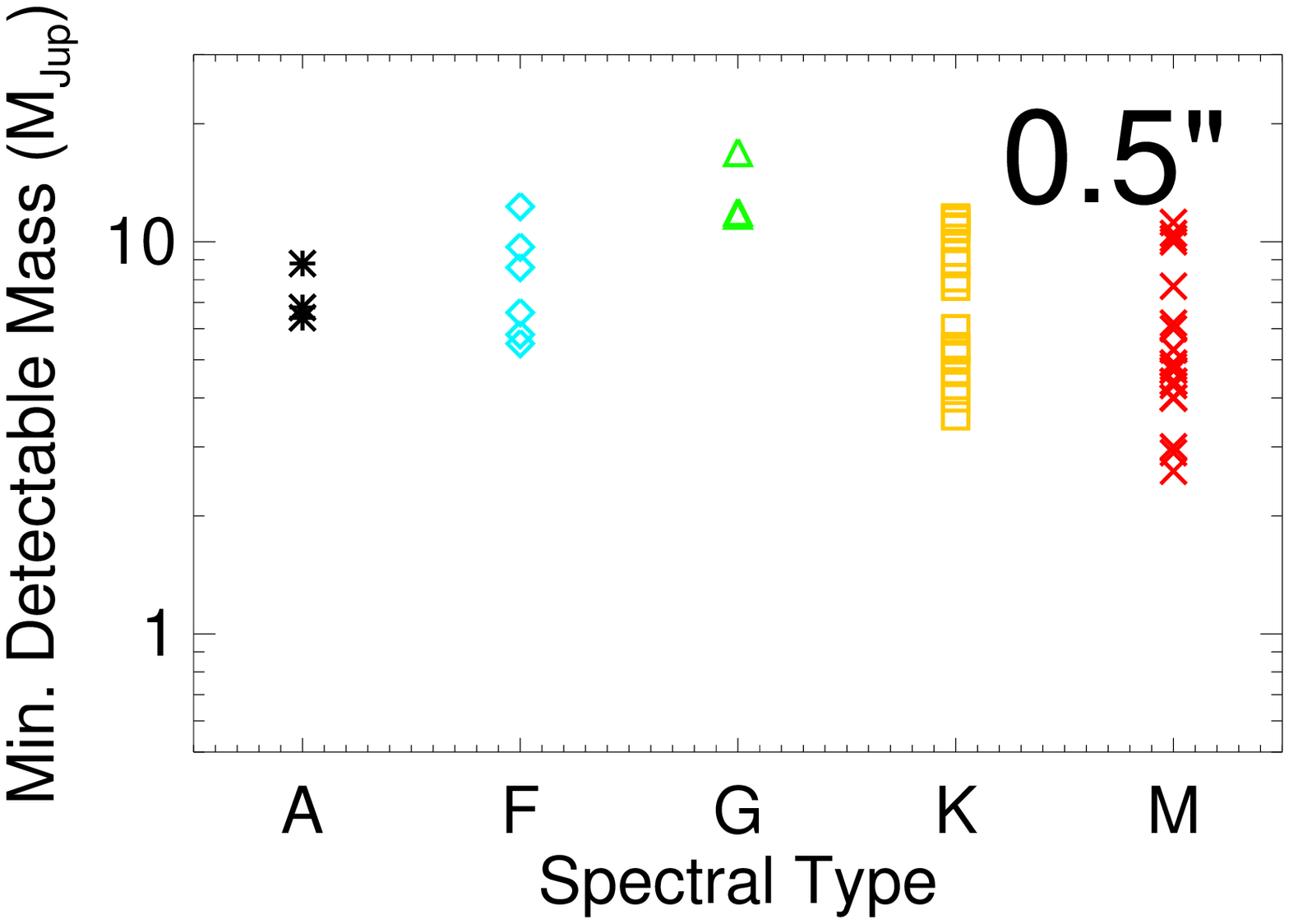}
\includegraphics[width=3.2in]{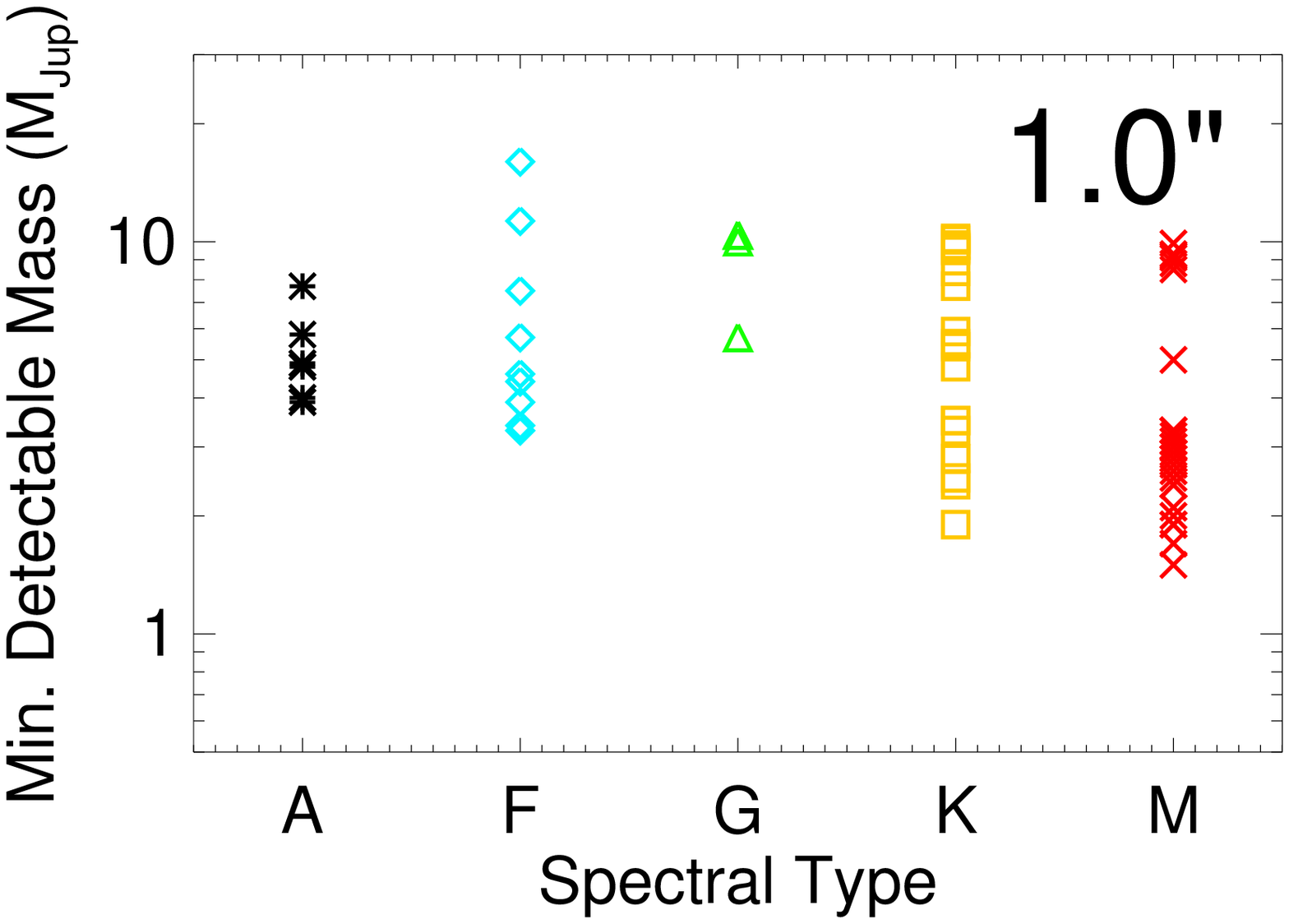}
\includegraphics[width=3.2in]{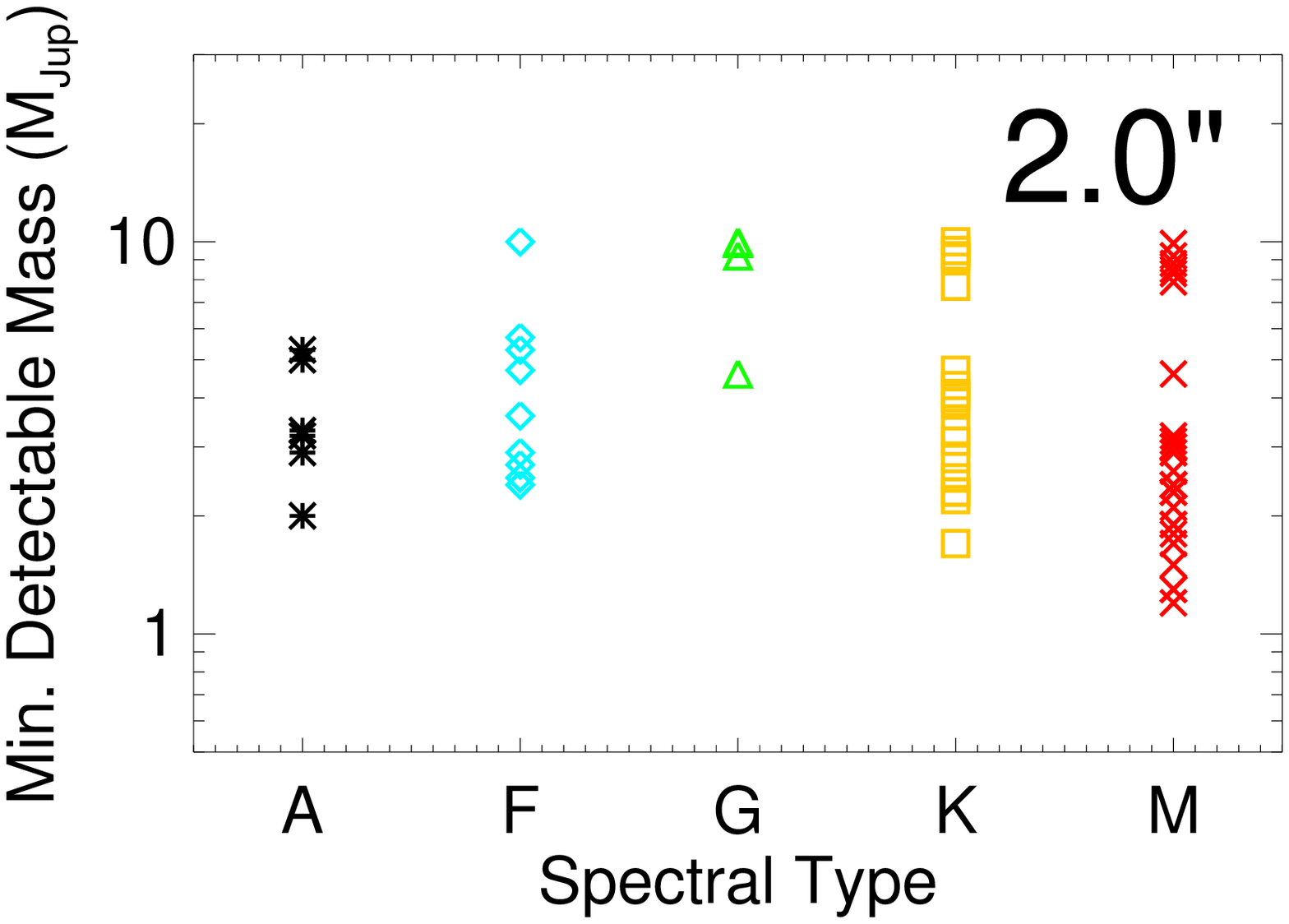}
\includegraphics[width=3.2in]{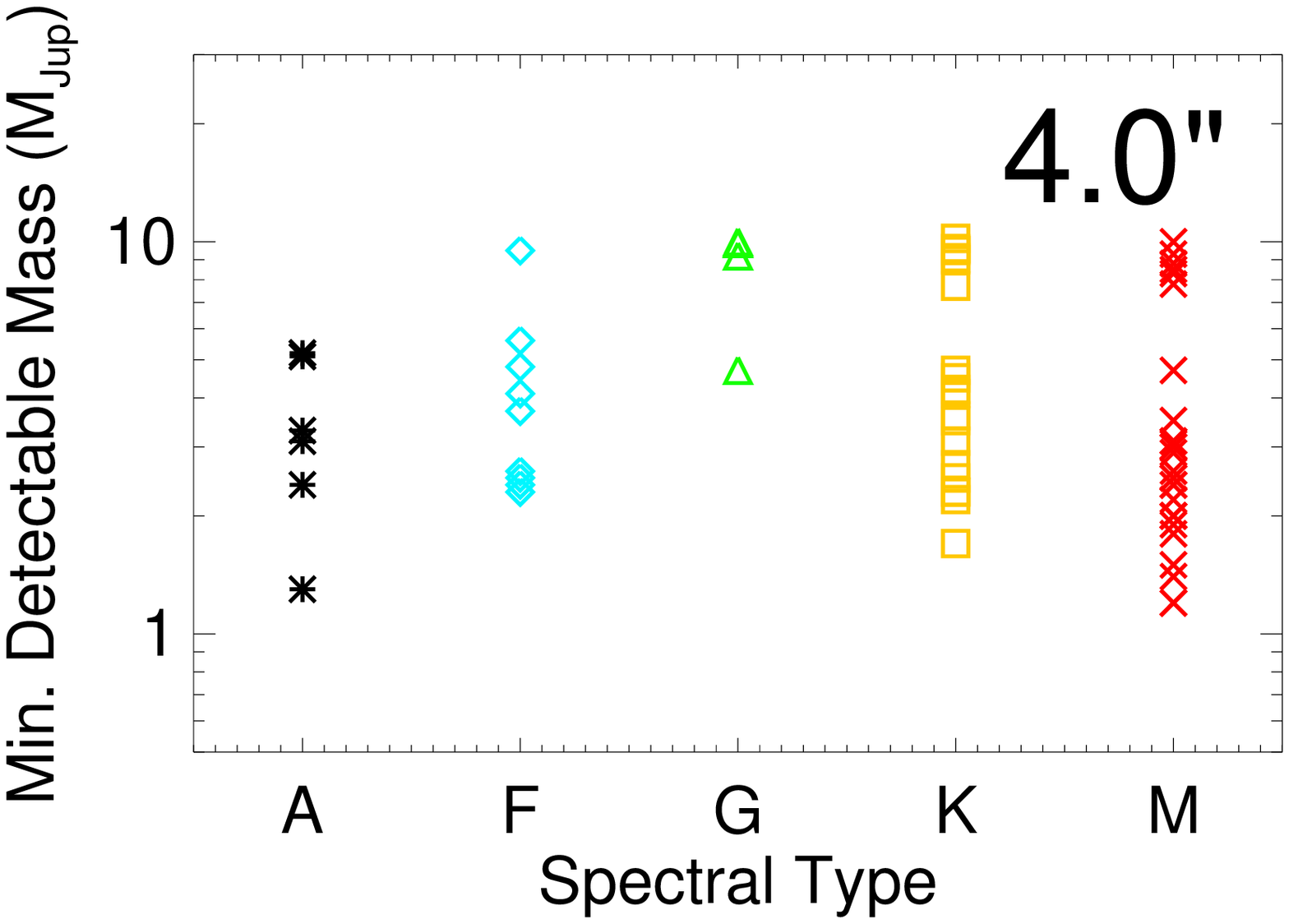}
\caption{Minimum detectable masses as a function of spectral
type at 0.5$\arcsec$, 1$\arcsec$, 2$\arcsec$, and 4$\arcsec$, 
using the DUSTY models of \citet{Bar02} to convert from contrasts
to masses.  At 0.5$\arcsec$ we are sensitive to companions of $\leq$13 M$_{Jup}$
for all but one star.   At 2$\arcsec$ we are sensitive to 
companions with masses $<$10 M$_{Jup}$ for all stars.
The minimum detectable mass varies by star (according to spectral type,
magnitude, distance, etc.) but we are generally sensitive to
$\geq$5 M$_{Jup}$ companions at 2$\arcsec$.
around all sample stars.
\label{fig:minmass_plots_DUSTY}}
\end{figure}

\begin{figure}
\includegraphics[width=3.2in]{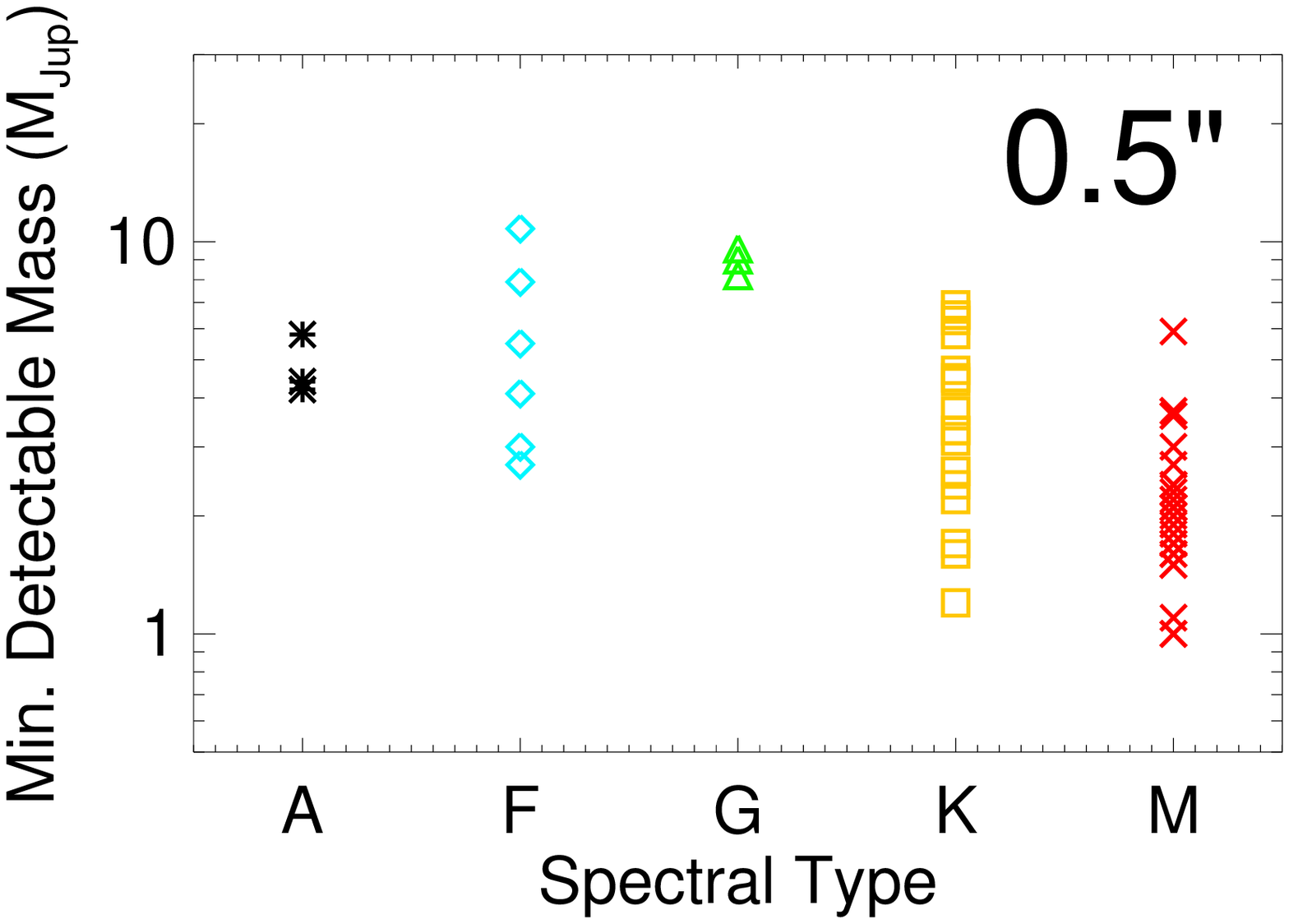}
\includegraphics[width=3.2in]{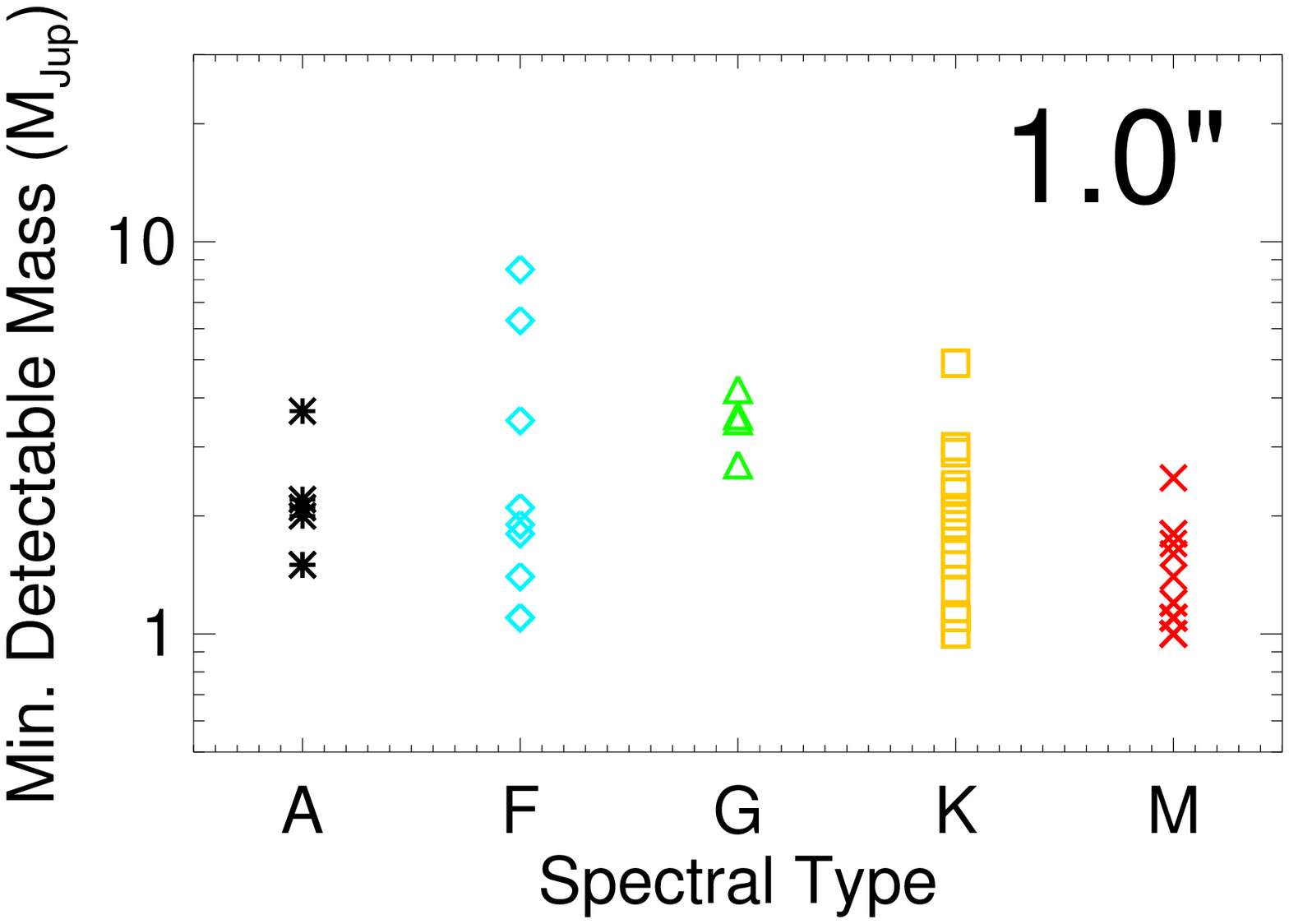}
\includegraphics[width=3.2in]{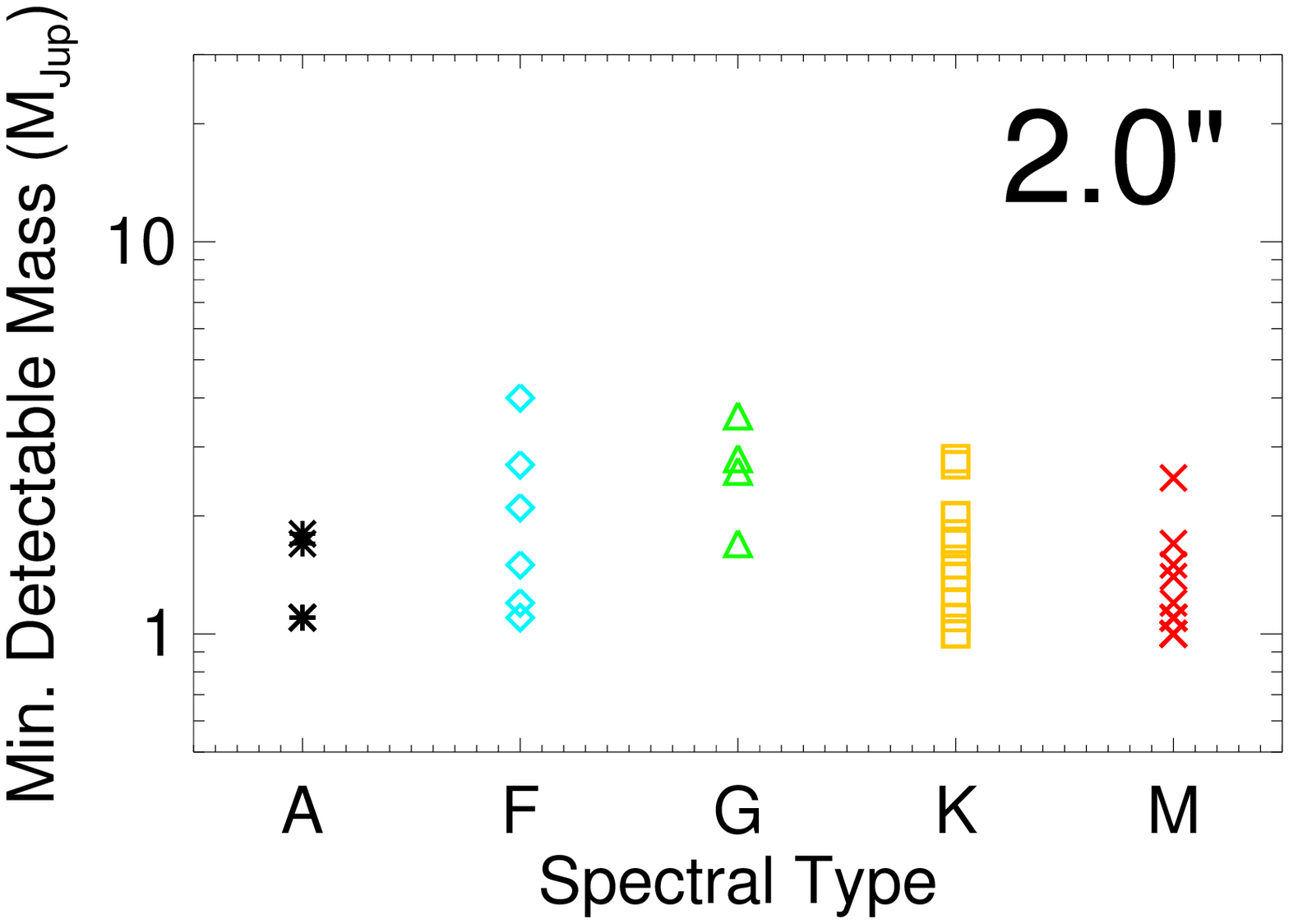}
\includegraphics[width=3.2in]{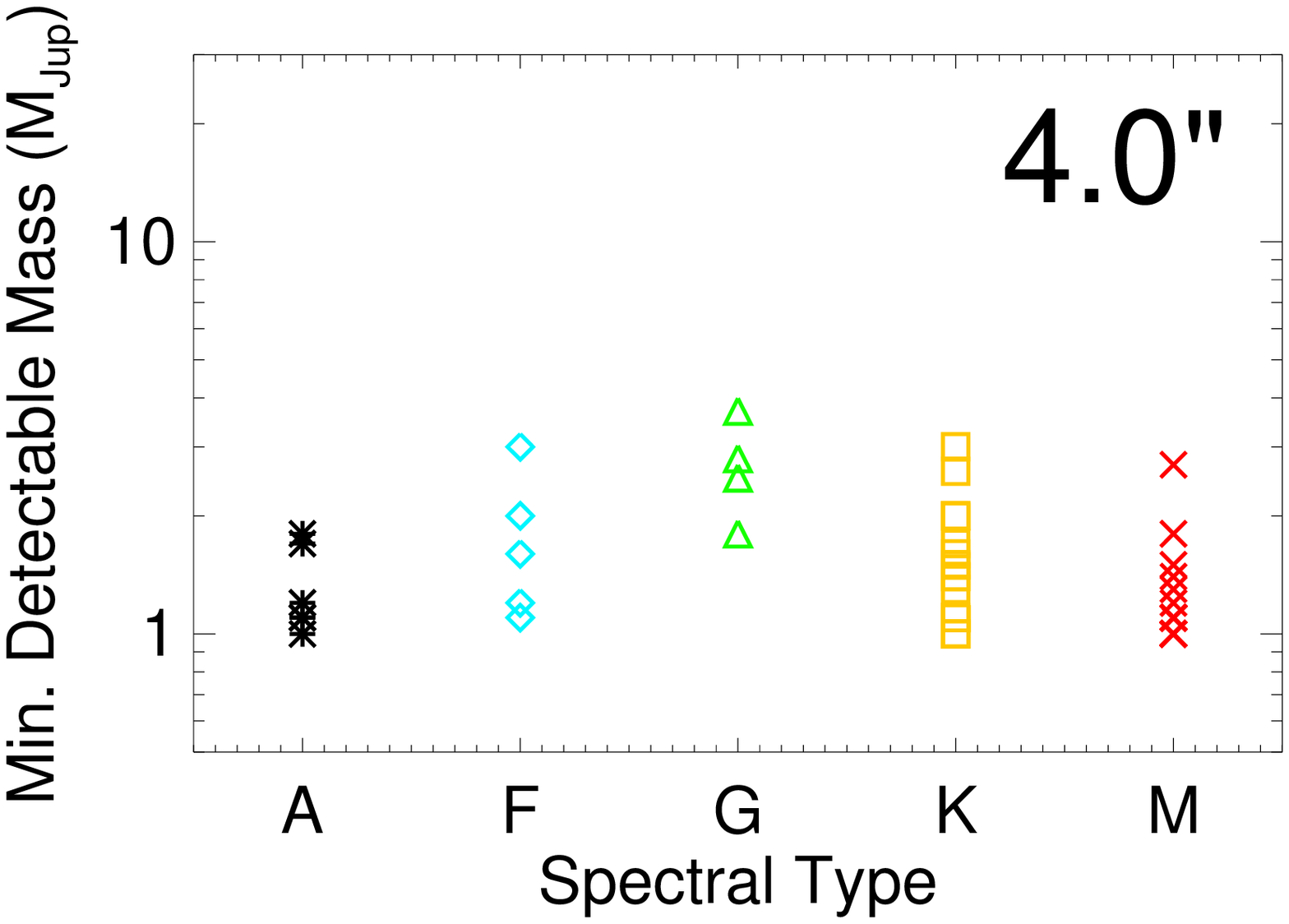}
\caption{Minimum detectable masses as a function of spectral
type at 0.5$\arcsec$, 1$\arcsec$, 2$\arcsec$, and 4$\arcsec$, 
using the COND models of \citet{Bar03} to convert from contrasts
to masses.  At 0.5$\arcsec$ we are sensitive to companions of $\leq$10 M$_{Jup}$
for all but one star.   At 2$\arcsec$ we are sensitive to 
companions with masses $<$5 M$_{Jup}$ for all stars.
The minimum detectable mass varies by star (according to spectral type,
magnitude, distance, etc.) but we are generally sensitive to
$\geq$5 M$_{Jup}$ companions at 2$\arcsec$ around all sample stars.
\label{fig:minmass_plots_COND}}
\end{figure}

For our ADI contrast curves, we convert measured contrast to 
maximum detectable apparent magnitude in
Table~\ref{tab:appmagADI} and minimum detectable mass in 
Tables~\ref{tab:minmassADI_DUSTY} and Tables~\ref{tab:minmassADI_COND}.  
We interpolated from both the DUSTY and COND models of \citet{Cha00}
and \citet{Bar02,Bar03} using the maximum detectable apparent magnitudes, distance, and age of each
stars to estimate the minimum detectable mass curves.
At some point as they cool and dust condenses from their atmospheres, directly imaged
exoplanets are predicted to transition from red, dusty
L dwarf spectra (DUSTY) to T dwarf spectra with methane absorption features (COND). 
However, no directly imaged planet to date has yet to show strong methane
absorption in the near-IR, with only weak methane absorption
observed at longer wavelengths \citep{Ske12}.
Thus as this transition has not been observed, we choose here to
present minimum detectable masses according to both of these models.
Minimum detectable masses as a function of spectral
type at 0.5$\arcsec$, 1$\arcsec$, 2$\arcsec$, and 4$\arcsec$ 
are presented in Figure~\ref{fig:minmass_plots_DUSTY} 
using the DUSTY models \citep{Bar02} and in 
Figure~\ref{fig:minmass_plots_COND} using the COND models
\citep{Bar03}.  For the more conservative DUSTY 
model case, at 0.5$\arcsec$ we are sensitive to companions of $\leq$13 M$_{Jup}$
for all but one star.   At 2$\arcsec$ we are sensitive to 
companions with masses $\leq$10 M$_{Jup}$ for all stars.
The minimum detectable mass varies by star (according to spectral
type, magnitude, distance, etc.) but we are generally sensitive
to $\geq$5 M$_{Jup}$ companions at 2$\arcsec$ around all sample stars.
We do not present minimum detectable masses in ASDI subtracted mode here,
as this requires knowledge 
of a potential companion's $H$-band spectrum.  For an example of such an analysis of 
ADI self-subtraction as a function of radius, see Nielsen et al. (2013).

\subsection{Astrometry of Candidate Companions}

We found numerous candidate companions in our images.  
Candidates were first identified using an automated
finding algorithm and then verified by eye.  For the
entire NICI Campaign sample, candidate companions were found for $\sim$50$\%$
of observed stars.  The vast majority of these objects are
not expected to be true co-moving companions.  To test whether a candidate companion
is co-moving with its parent star requires reobserving after enough
time has elapsed for significant proper motion and/or parallactic
motion of the star in the 
sky, ideally at the $\geq$3 pixel ($\geq$50 mas) or greater level.  

After identifying candidate companions in our reduced images, we first 
checked if any older archival data from VLT, HST or Gemini 
were available.  In this manner, we were able to immediately identify a number of
bright candidates as background objects. 
Astrometry for candidates observed at multiple epochs with NICI 
as well as other telescopes is presented in Tables \ref{table_comp2a}
and \ref{table_comp2b}.

For objects with {\it HST} NICMOS observations, we retrieved
data from the {\it HST} MAST archive and used the
mosaic files.  Images taken at different telescope roll angles
were subtracted to remove the slowly changing speckle pattern
(henceforth roll subtraction).  For datasets with images taken at only 
one roll angle, images were rotated by 180$^{\circ}$ and subtracted
from themselves.  We typically performed roll subtraction without any
subpixel alignment as most of the candidates were well outside the
region where PSF subtraction was important.  \citet[]{Low05} found the
position of the star behind the NICMOS coronagraph using acquistion
images and slew vectors gleaned from HST engineering telemetry, and
claim that the difference image diffraction spikes do not give an
accurate measure of the star's position.  Our candidate companions
followed up with NICMOS archival data are generally at wide separations ($>$2'')
 and with large time baselines (usually $\geq$3 years) relative to the NICI epoch;
thus, we often did not require an extremely accurate knowledge of the 
central star position in order to determine if they were background objects.  To see
if the simpler method of using the diffraction spikes could be used,
we tested this method on 10 stars in the \citet[]{Low05} sample by
measuring the position of the same companions they reported.  We found
a mean difference of 1.2 pixels from their positions.  Taking this to
be entirely due to our centroiding method, we combine it in quadrature
with their reported 1.05 pixel (0.08") uncertainty to calculate a
total uncertainty of 1.6 pixels, or 0.12".  

Data from Gemini-NIRI were reduced using a custom ADI script
\citep[][]{Clo10}.  Due to saturation of the primary stars, we estimate our 
astrometric uncertainty to be $\sim$2 pixels, or 0.044".

Candidates within 400 AU from the star and not in dense stellar fields
that were not confirmed or rejected as common-proper
motion companions using archival data were
reobserved with NICI. 
NICI astrometry was measured relative to the unsaturated starspot 
position in either the science or short exposures.  
The uncertainties in the separation and 
PA are estimated to be 0.009\arcsec (0.5 pixel)
and 0.2$^{\circ}$ respectively, when the primary is
unsaturated, and 0.018\arcsec (1 pixel)  and 0.5$^{\circ}$ when the 
primary is saturated (Wahhaj et al. 2013b).

\begin{figure}
\centerline{
\includegraphics[width=2.0in]{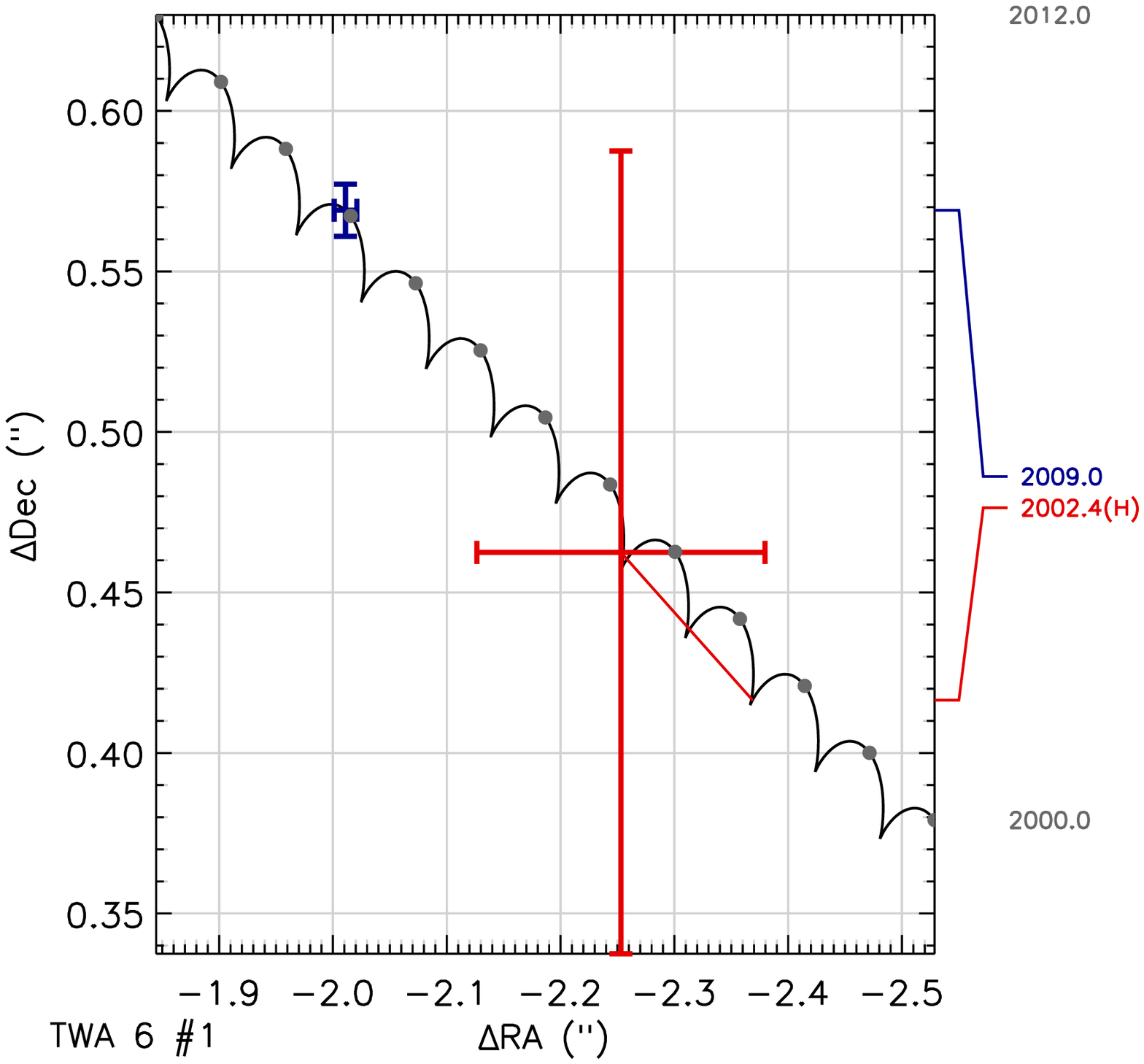}
\hskip -0.3in
\includegraphics[width=2.0in]{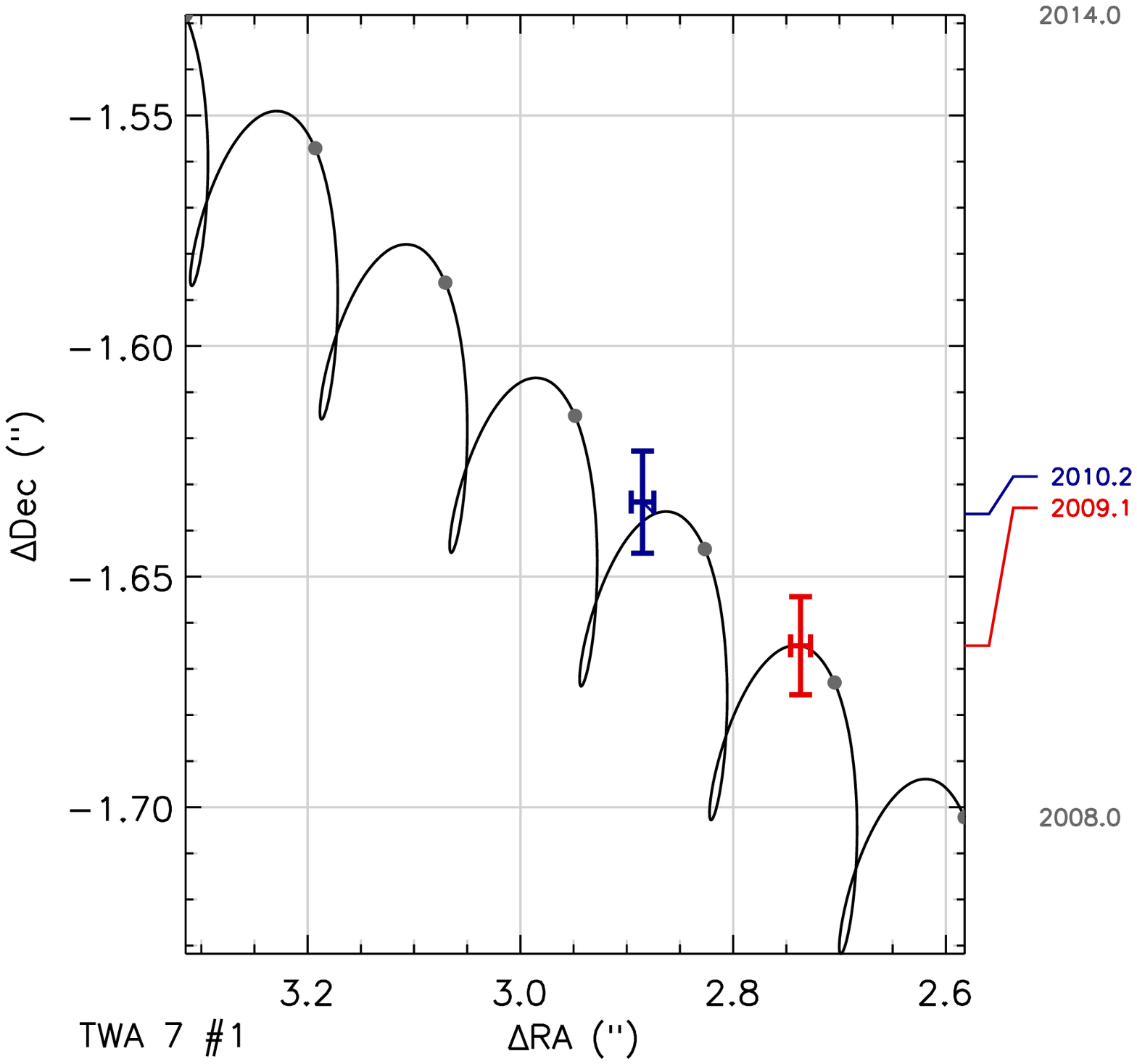}
\hskip -0.3in
\includegraphics[width=2.0in]{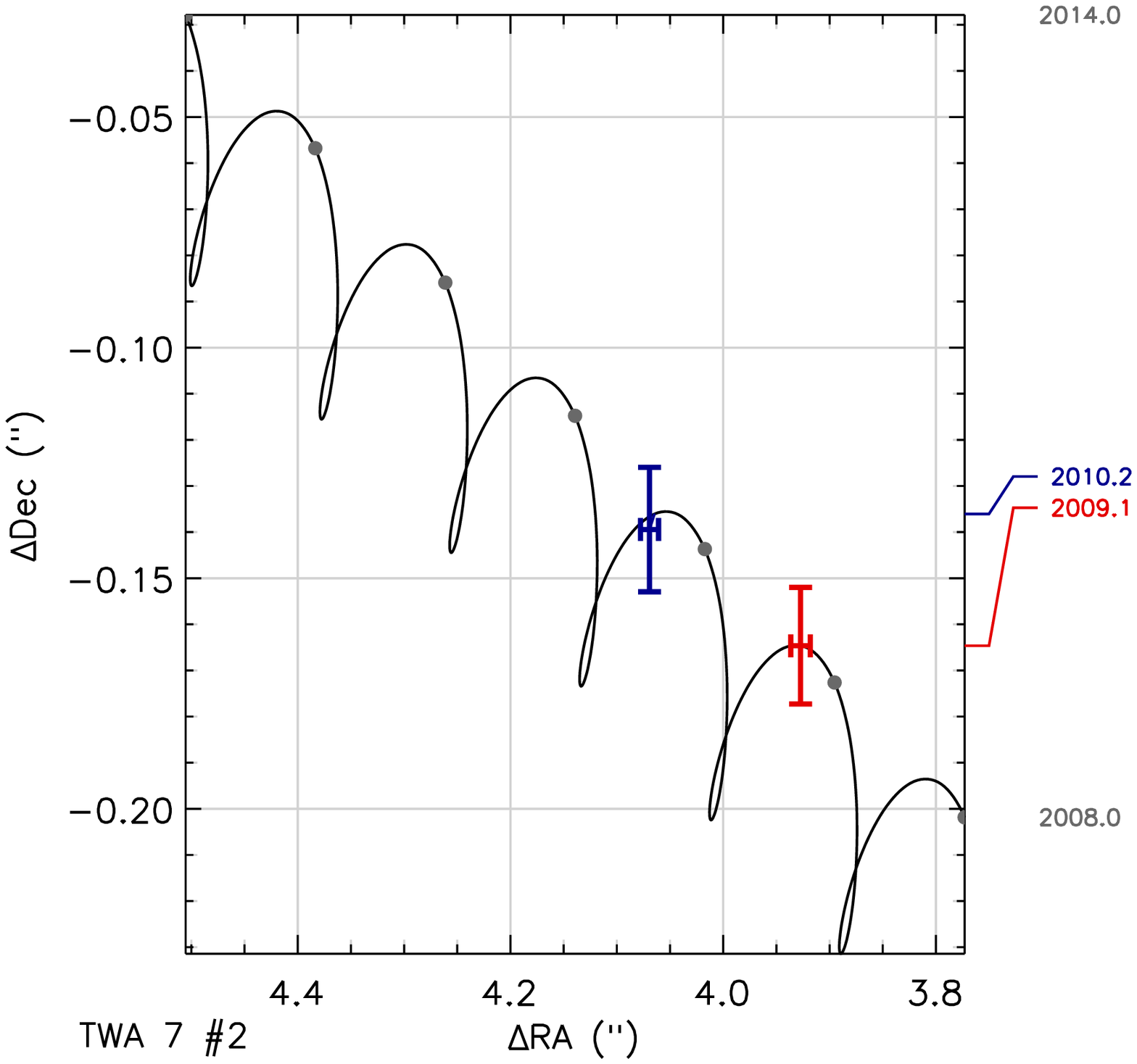}
\hskip -0.3in
\includegraphics[width=2.0in]{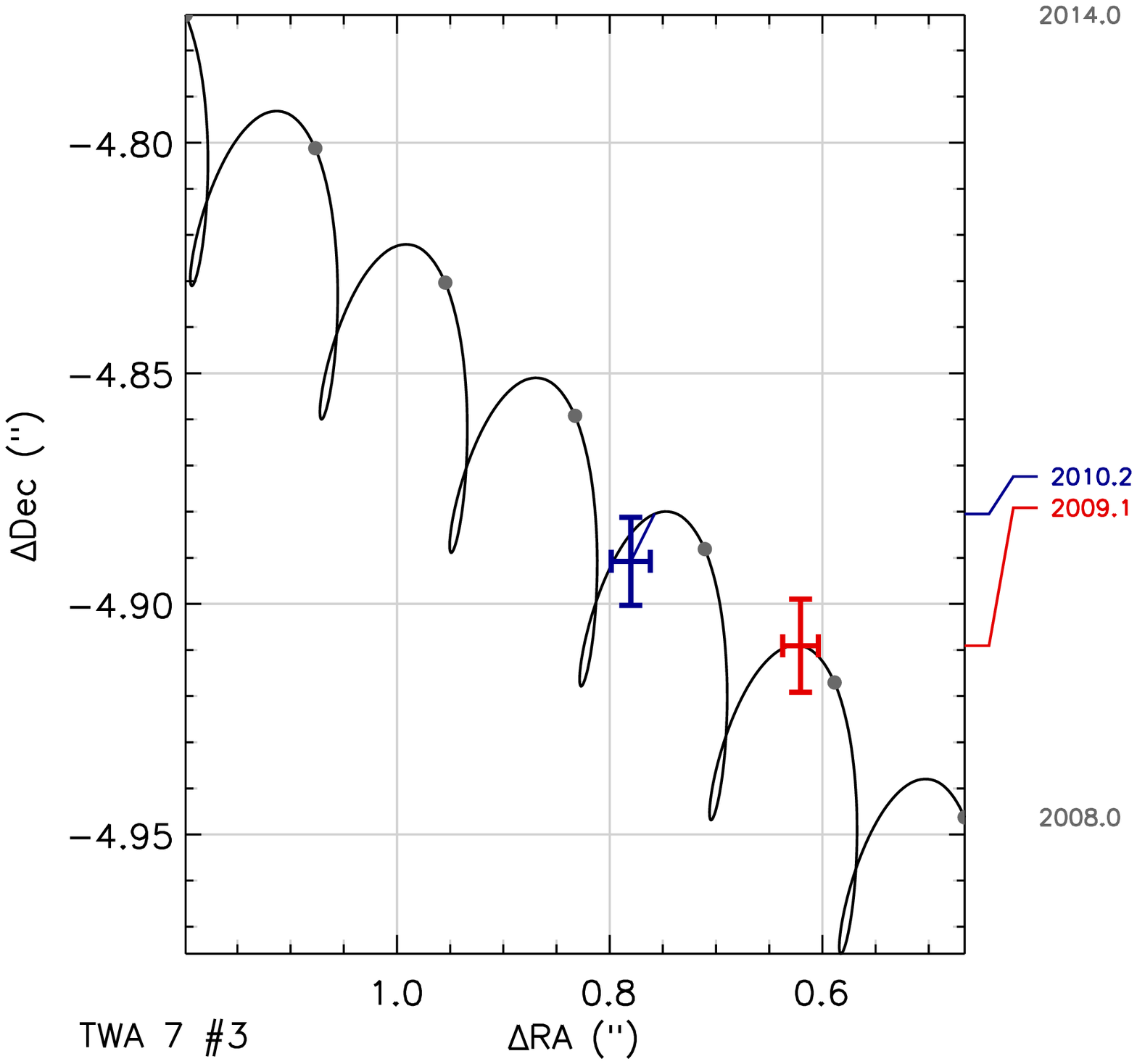}
}
\vskip -0.2in
\centerline{
\includegraphics[width=2.0in]{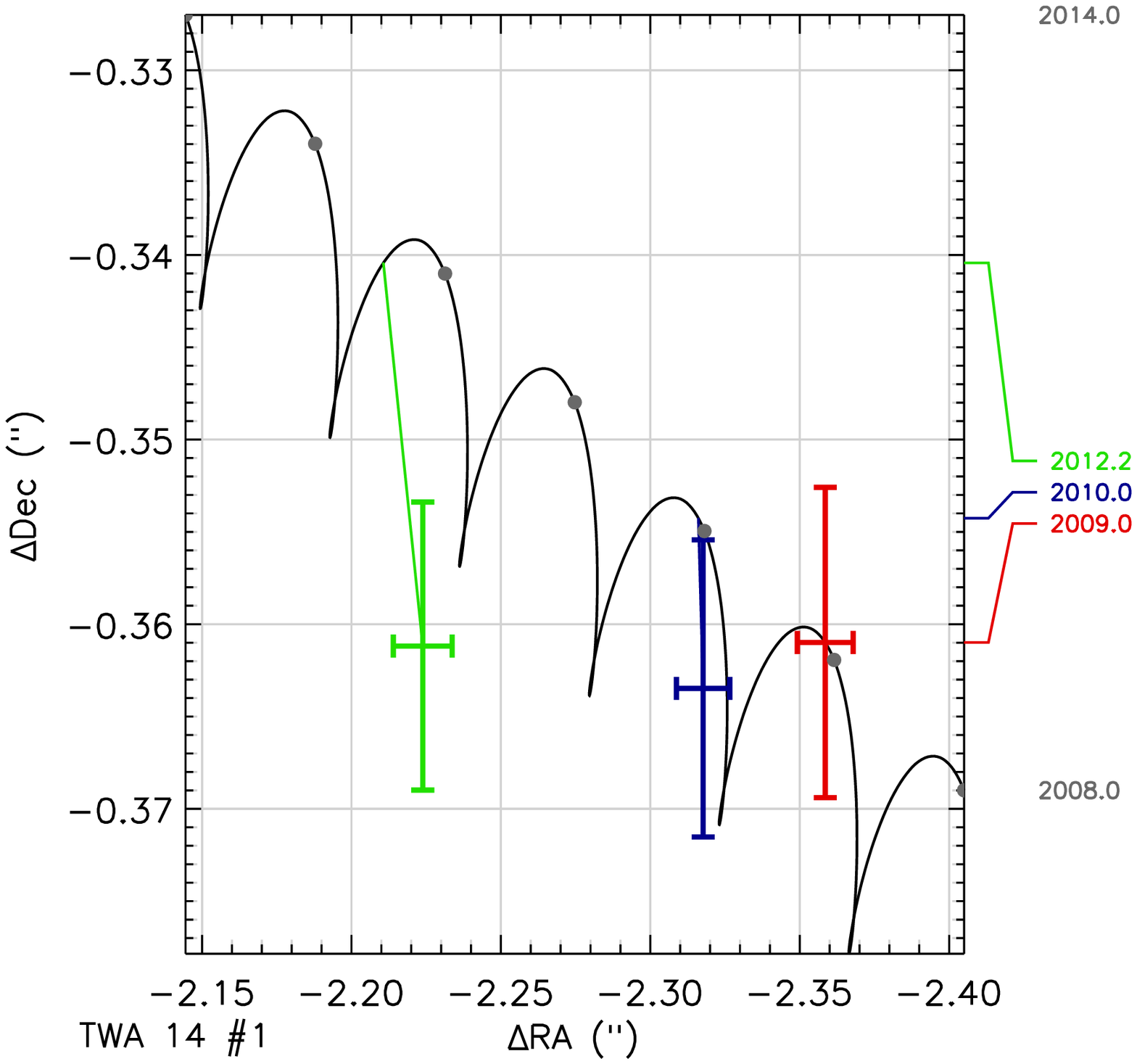}
\hskip -0.3in
\includegraphics[width=2.0in]{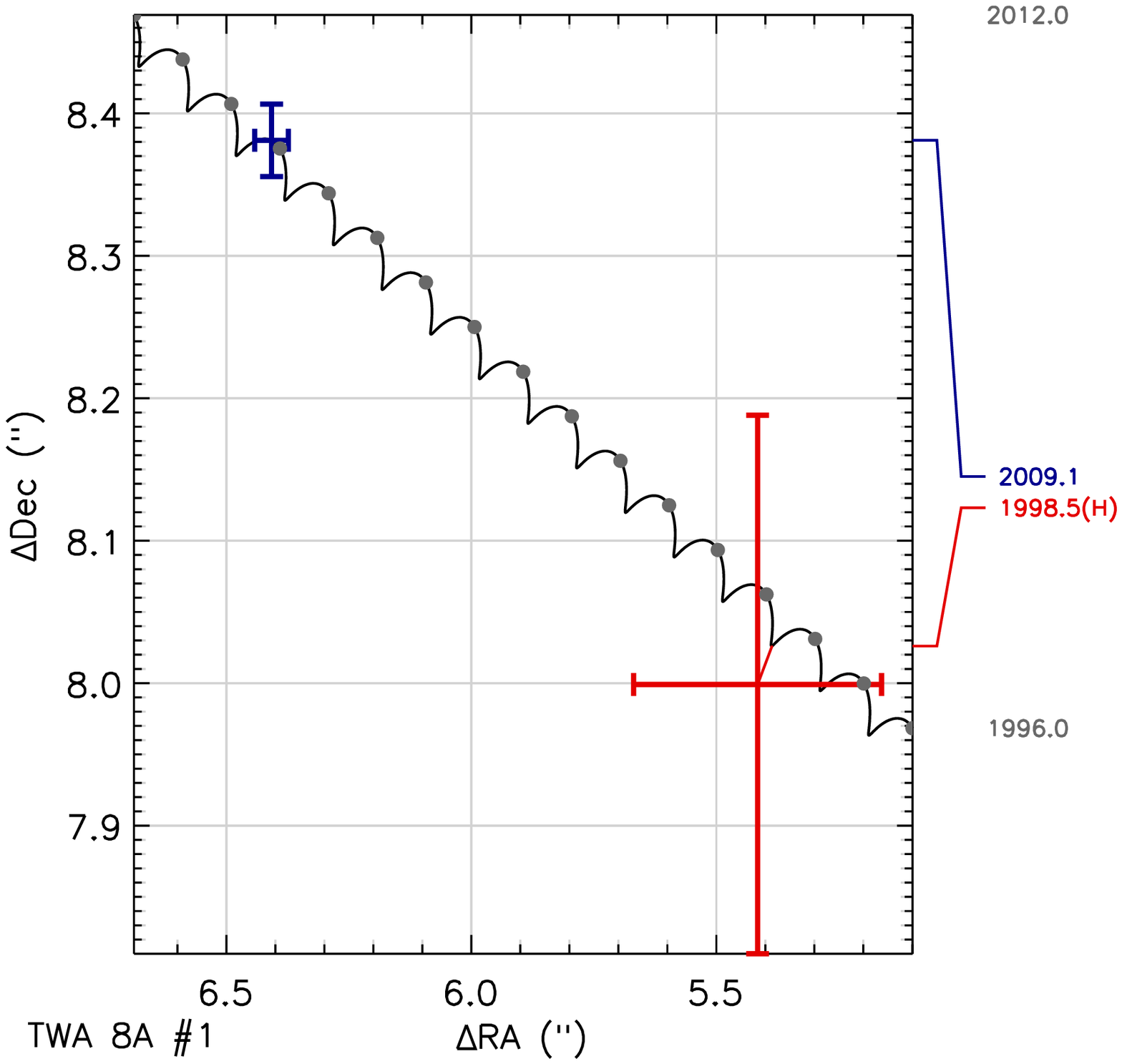}
\hskip -0.3in
\includegraphics[width=2.0in]{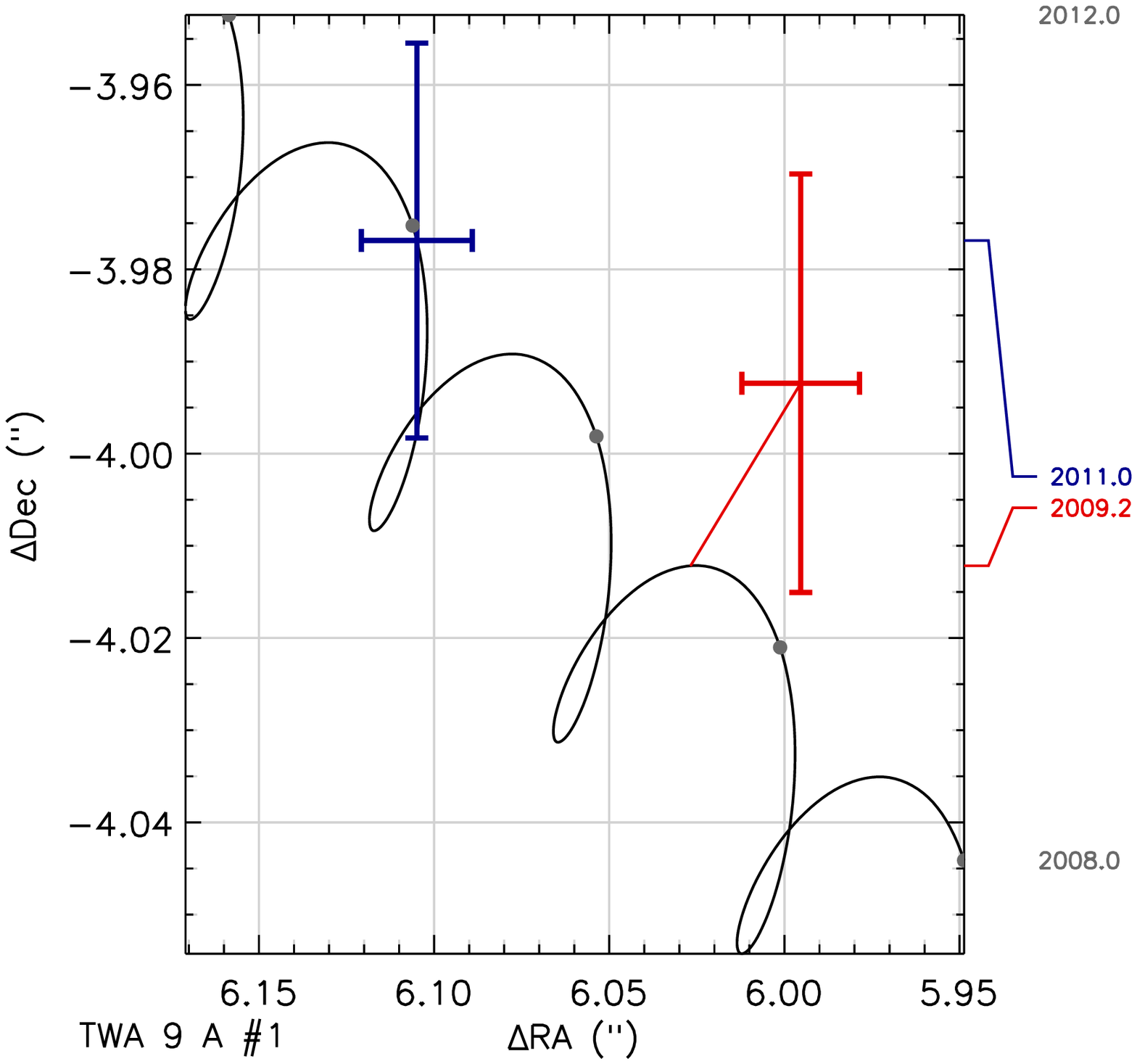}
\hskip -0.3in
\includegraphics[width=2.0in]{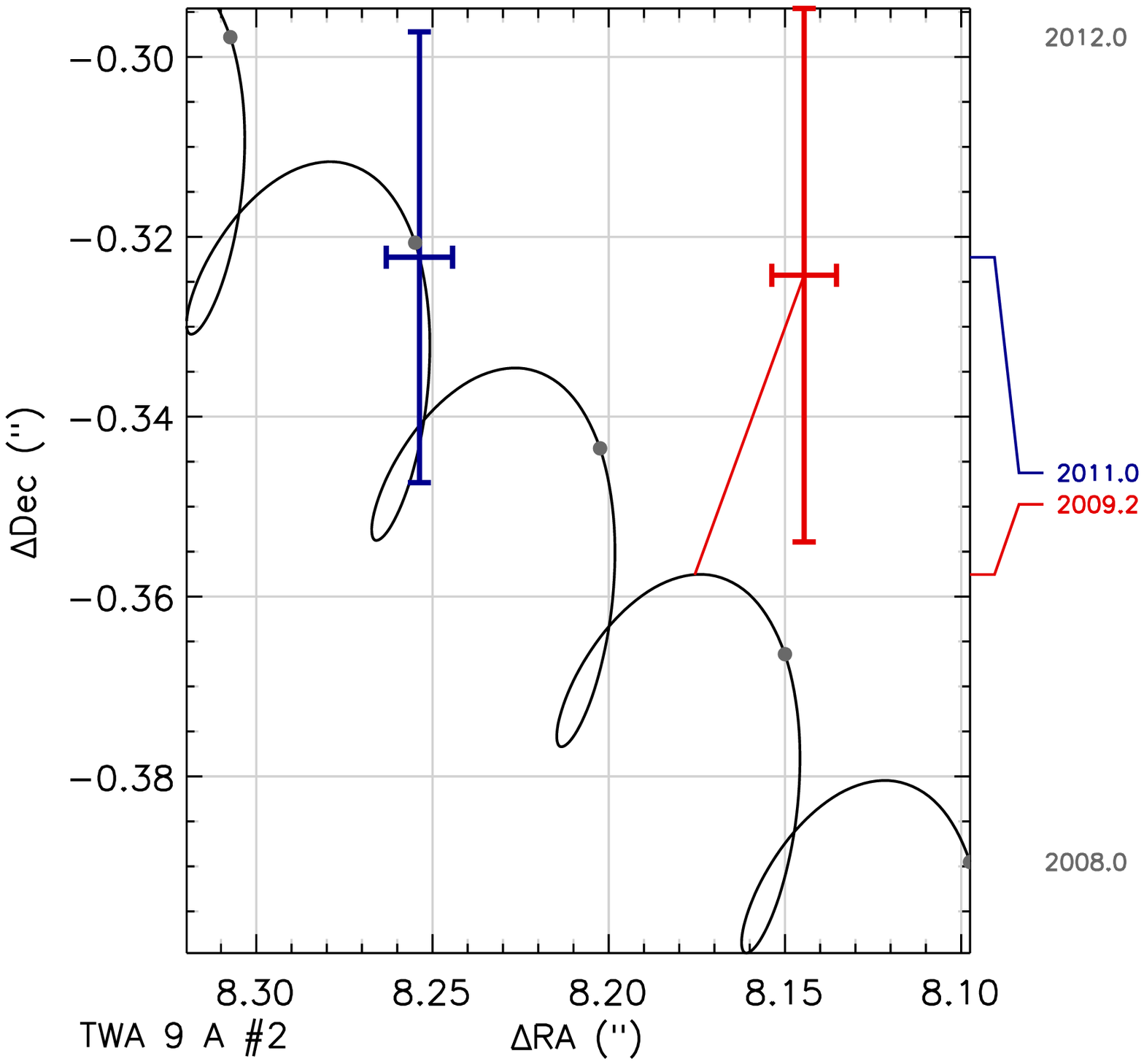}
}
\vskip -0.2in
\centerline{
\includegraphics[width=2.0in]{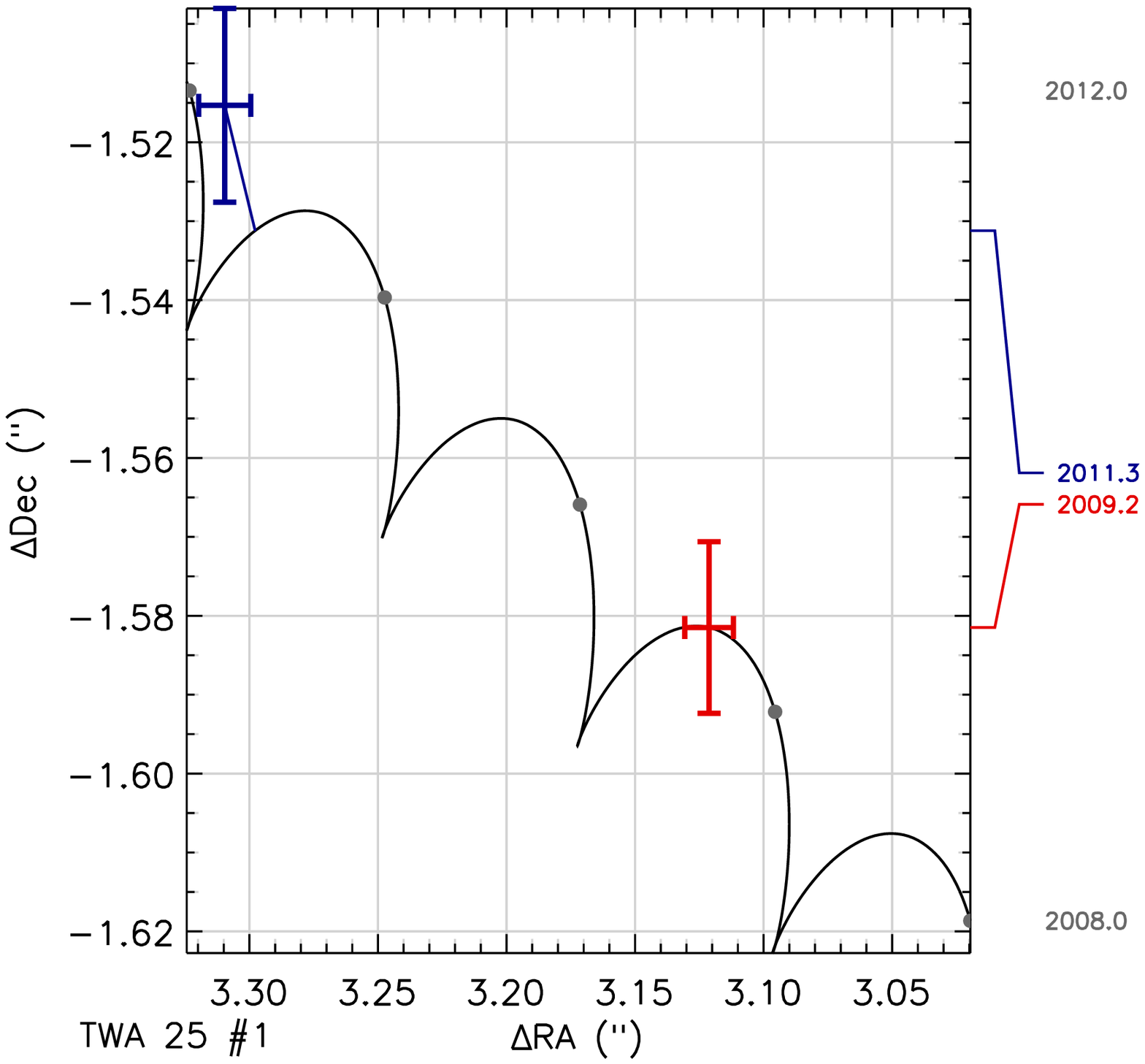}
\hskip -0.3in
\includegraphics[width=2.0in]{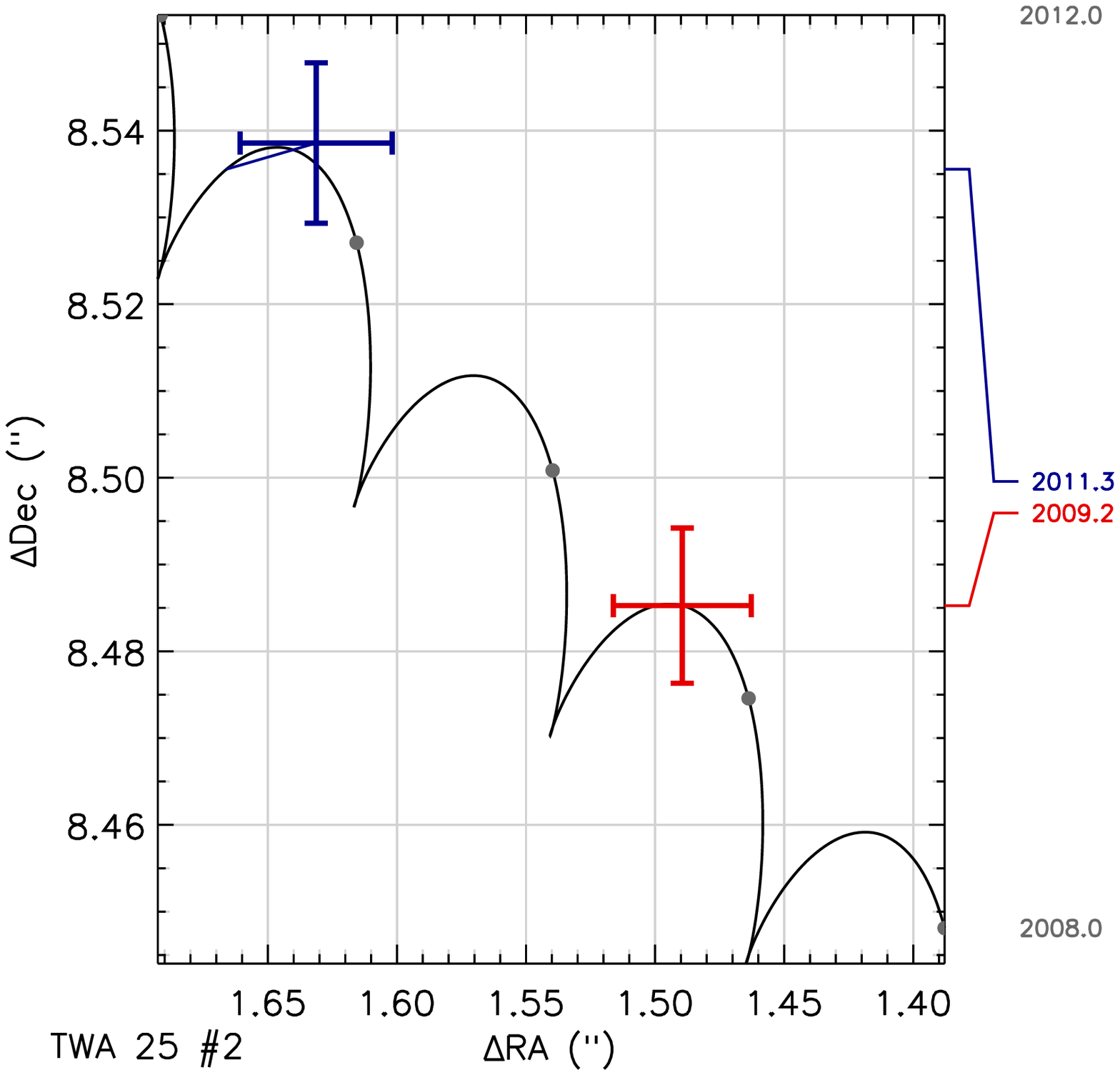}
\hskip -0.3in
\includegraphics[width=2.0in]{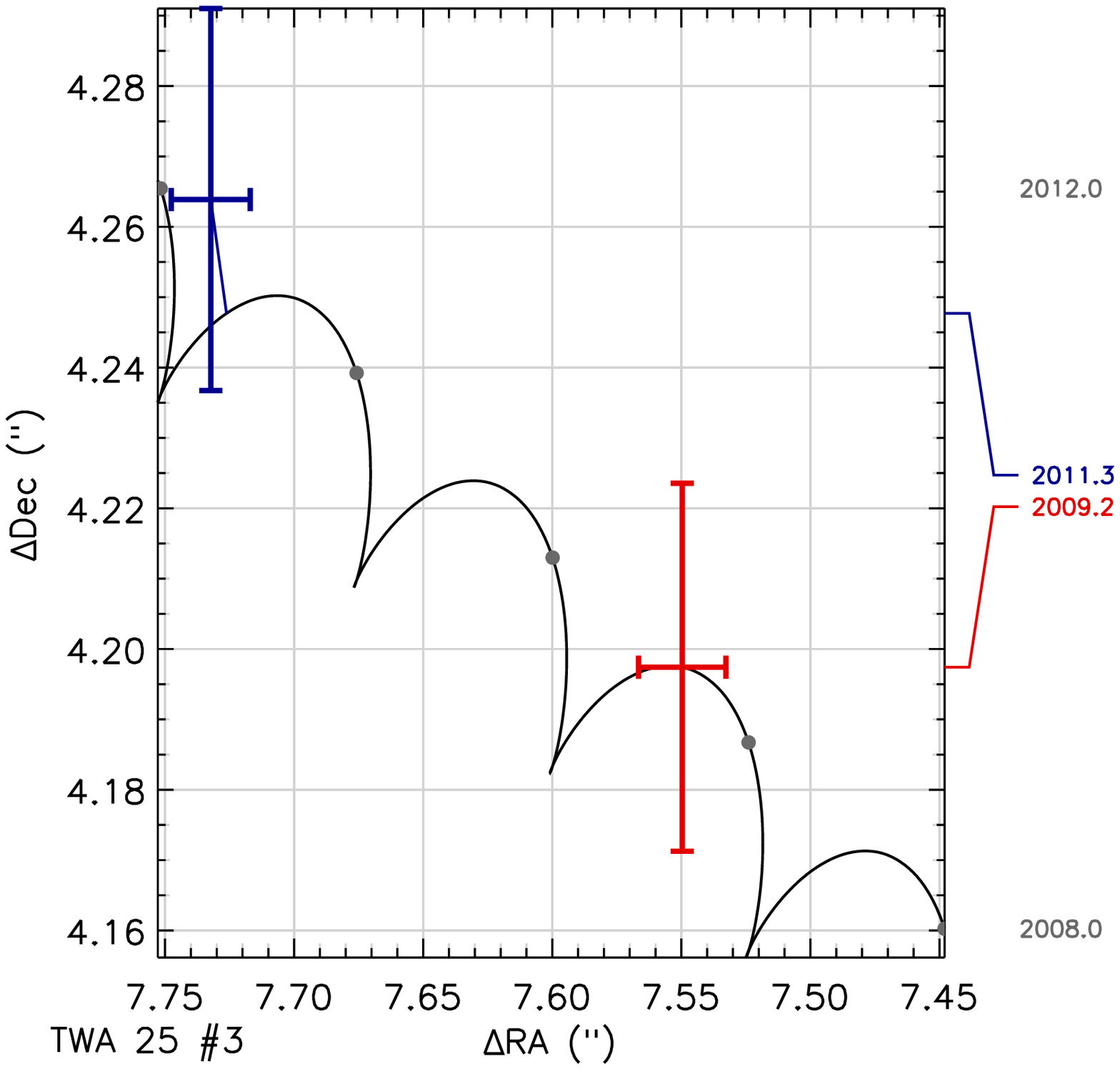}
\hskip -0.3in
\includegraphics[width=2.0in]{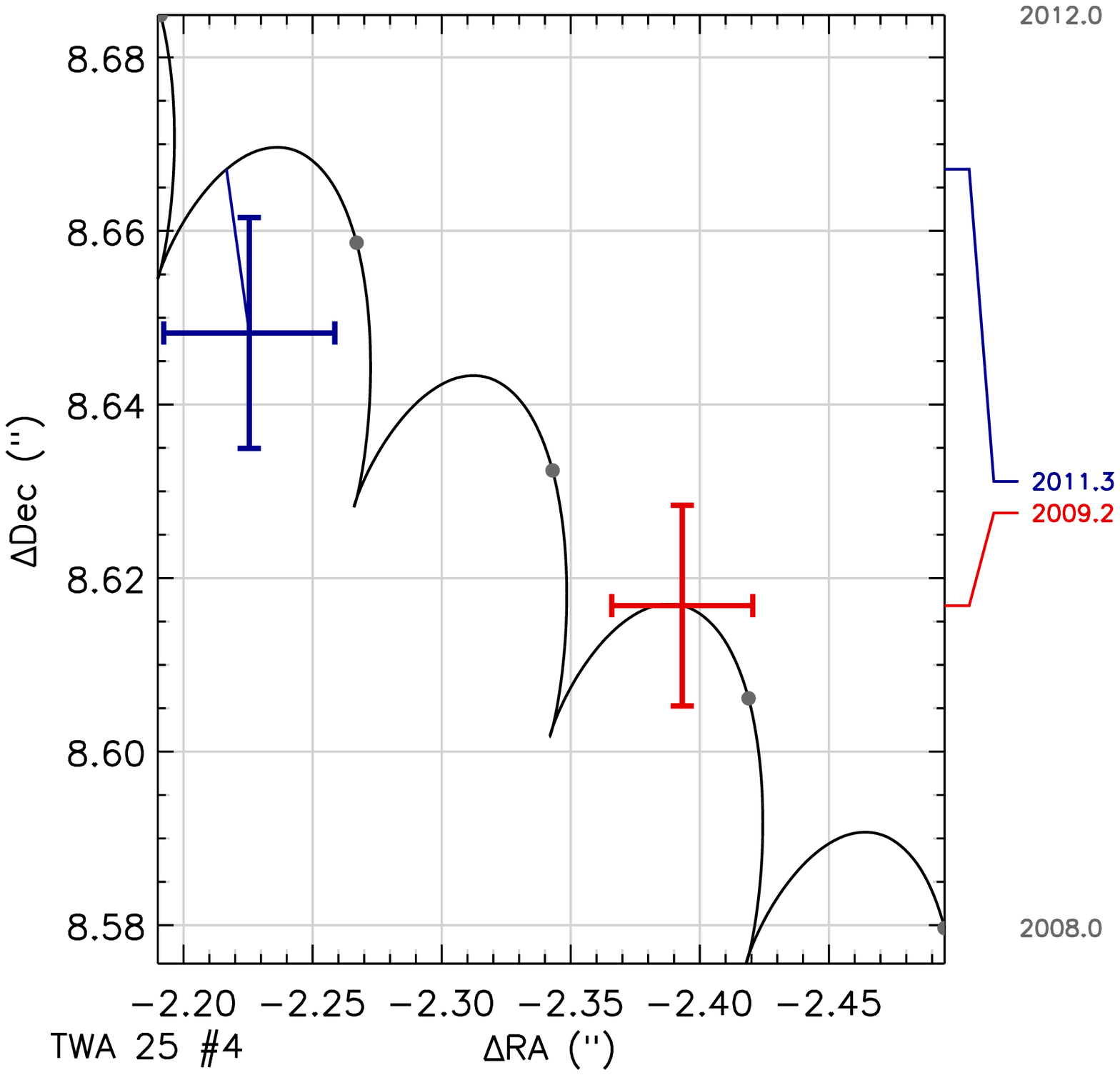}
}
\vskip -0.2in
\centerline{
\includegraphics[width=2.0in]{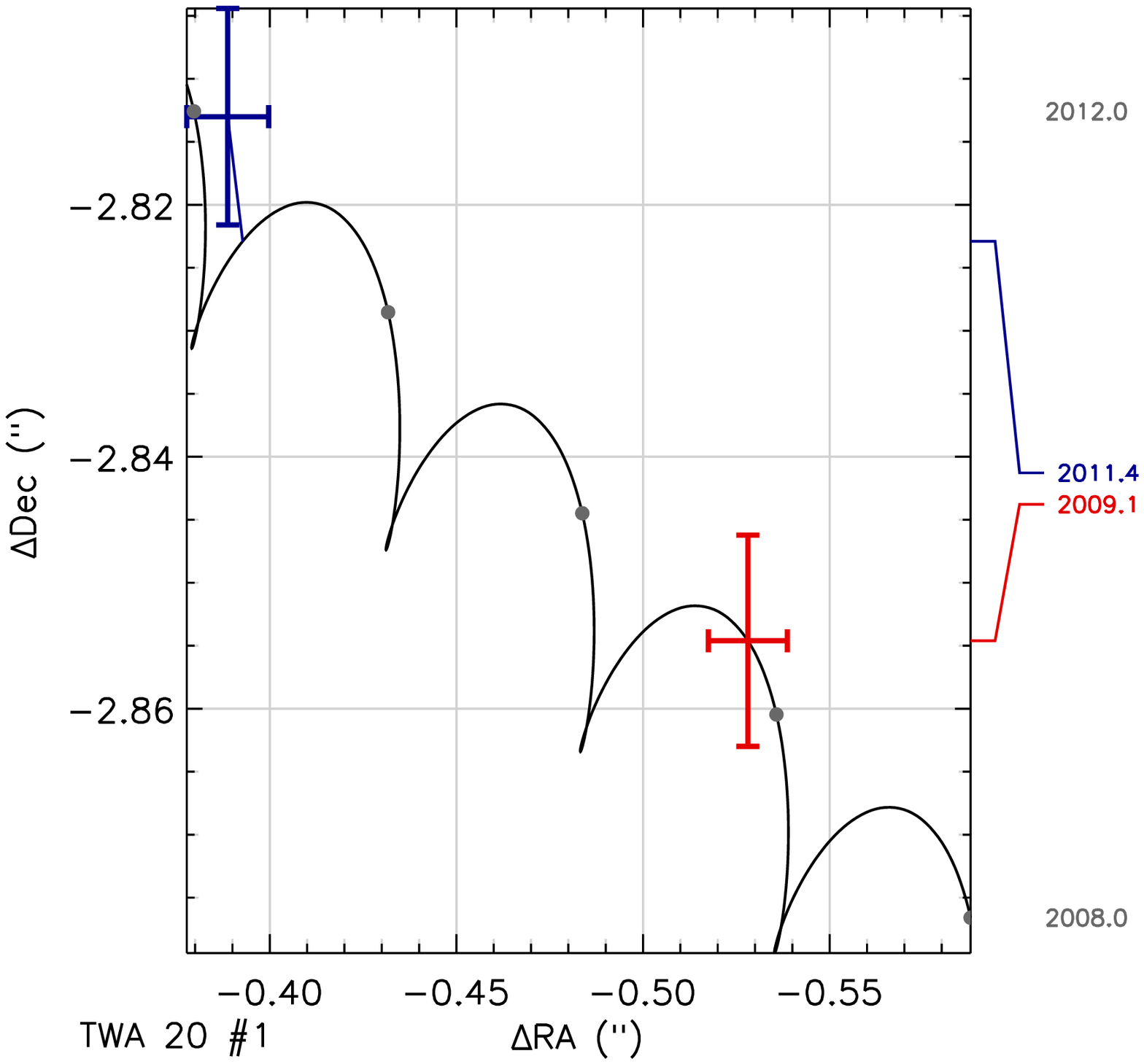}
\hskip -0.3in
\includegraphics[width=2.0in]{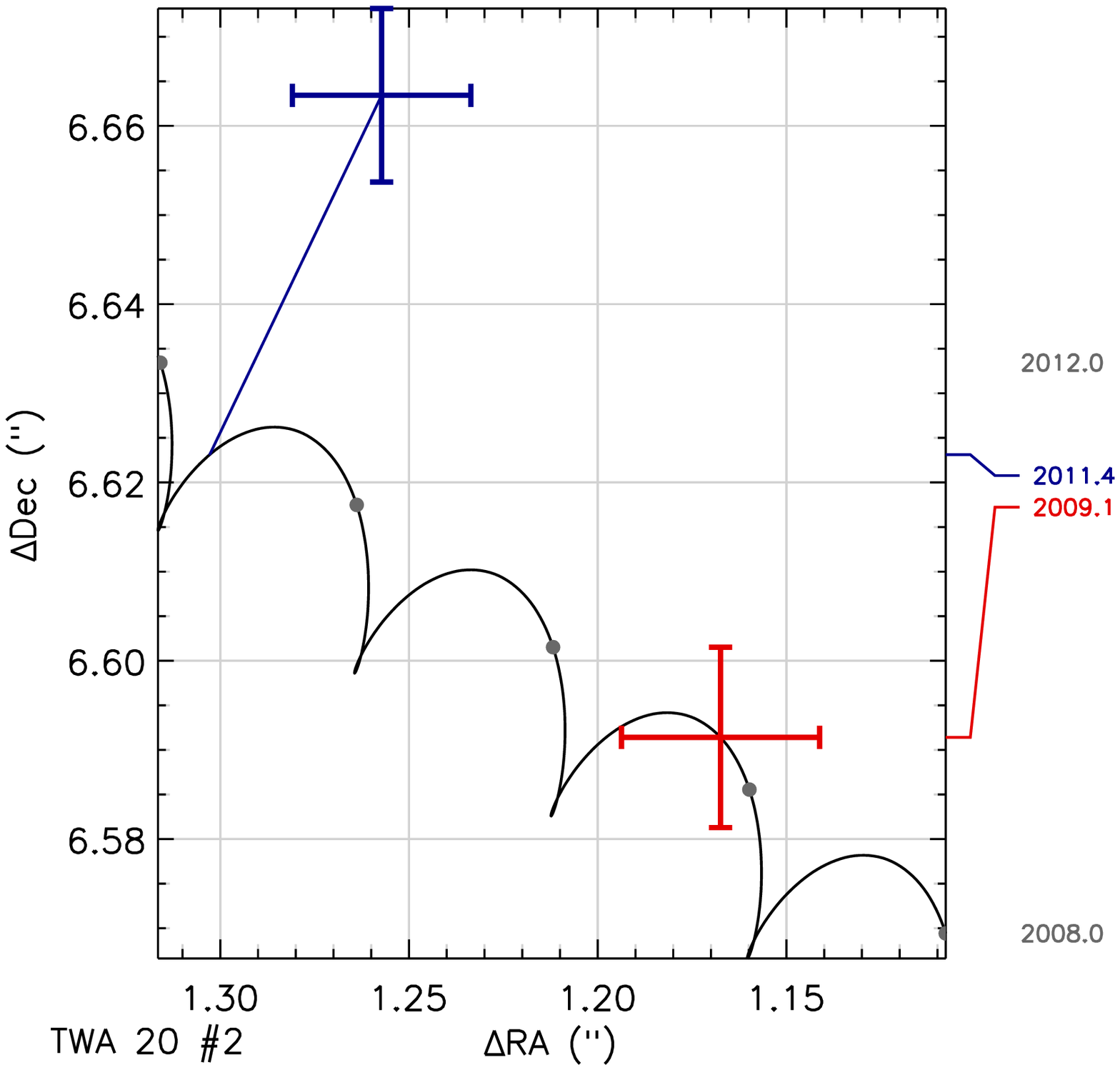}
\hskip -0.3in
\includegraphics[width=2.0in]{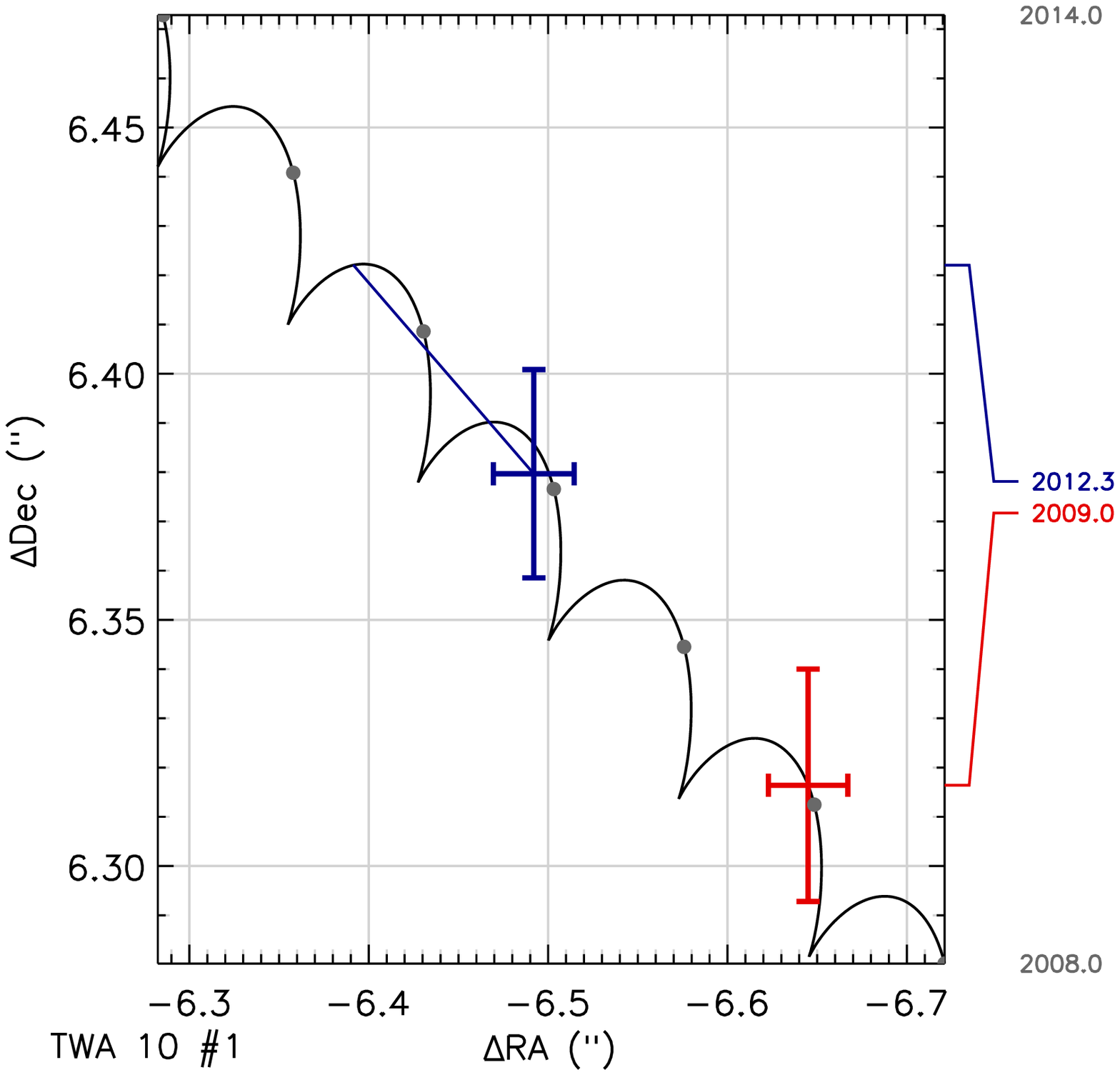}
\hskip -0.3in
\includegraphics[width=2.0in]{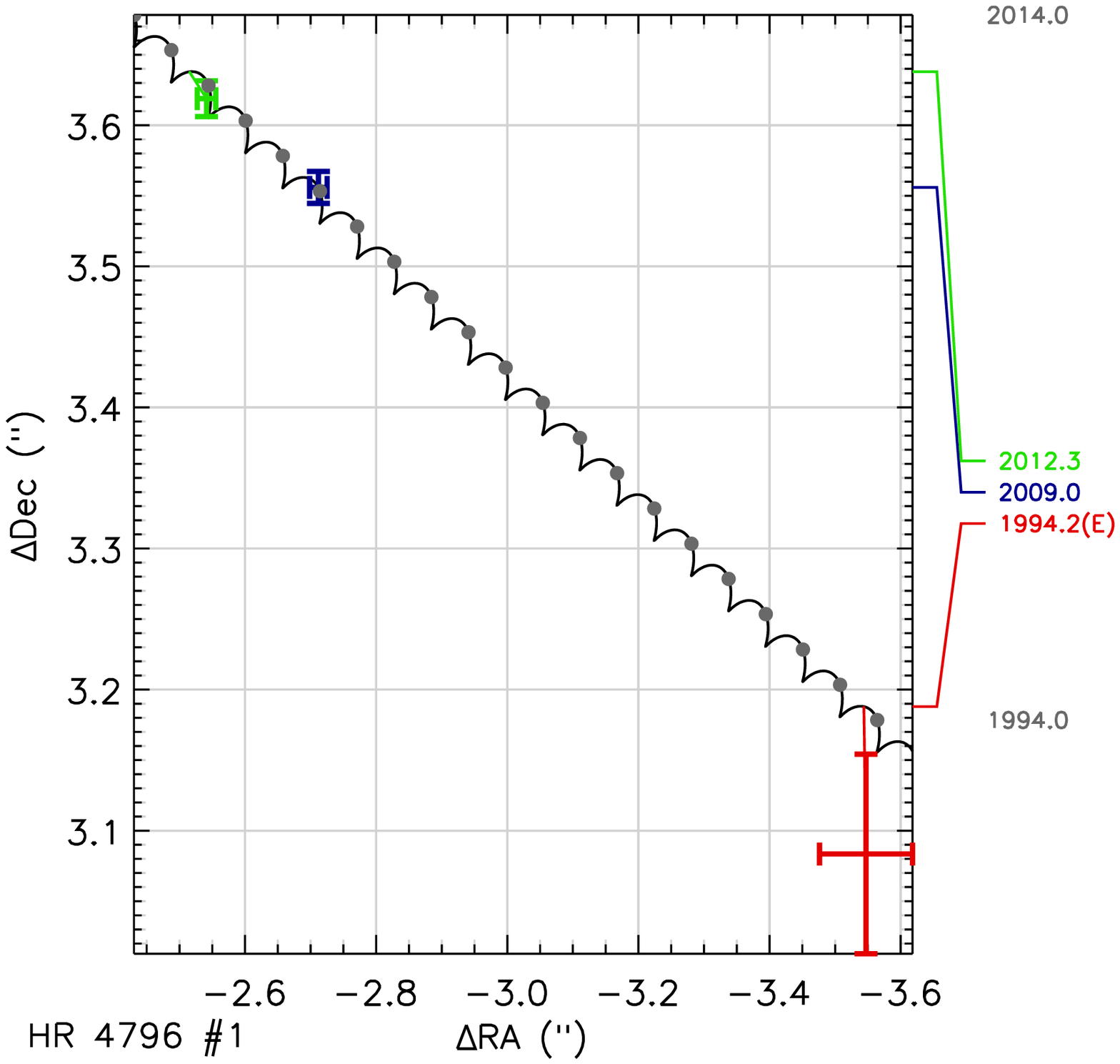}
}
\caption{On-sky plots for TW Hya association objects.  For each candidate, the background
track (black curve) is calculated from the proper motion and parallax of the star and position of the candidate
at the initial reference epoch. Astrometry at the reference epoch and additional epochs are shown as points
with error bars, and a colored line (1st epoch red, 2nd epoch blue, 3rd epoch
green) connects the position at additional epochs to the expected position on
the background track.  The labels at the right of each plot give the epochs of each astrometric data point, at
the vertical position corresponding to the location on the background track for that epoch. When the epoch
is given alone, the observation was conducted with the NICI instrument. Otherwise observational data are
taken from VLT NACO (V), Keck NIRC2 (K), VLT ISAAC (I), ESO 3.6m (E), and Gemini NIRI (G).
\label{fig:TWHya_skyplots}}
\end{figure}

\begin{figure}
\centerline{
\includegraphics[width=2.0in]{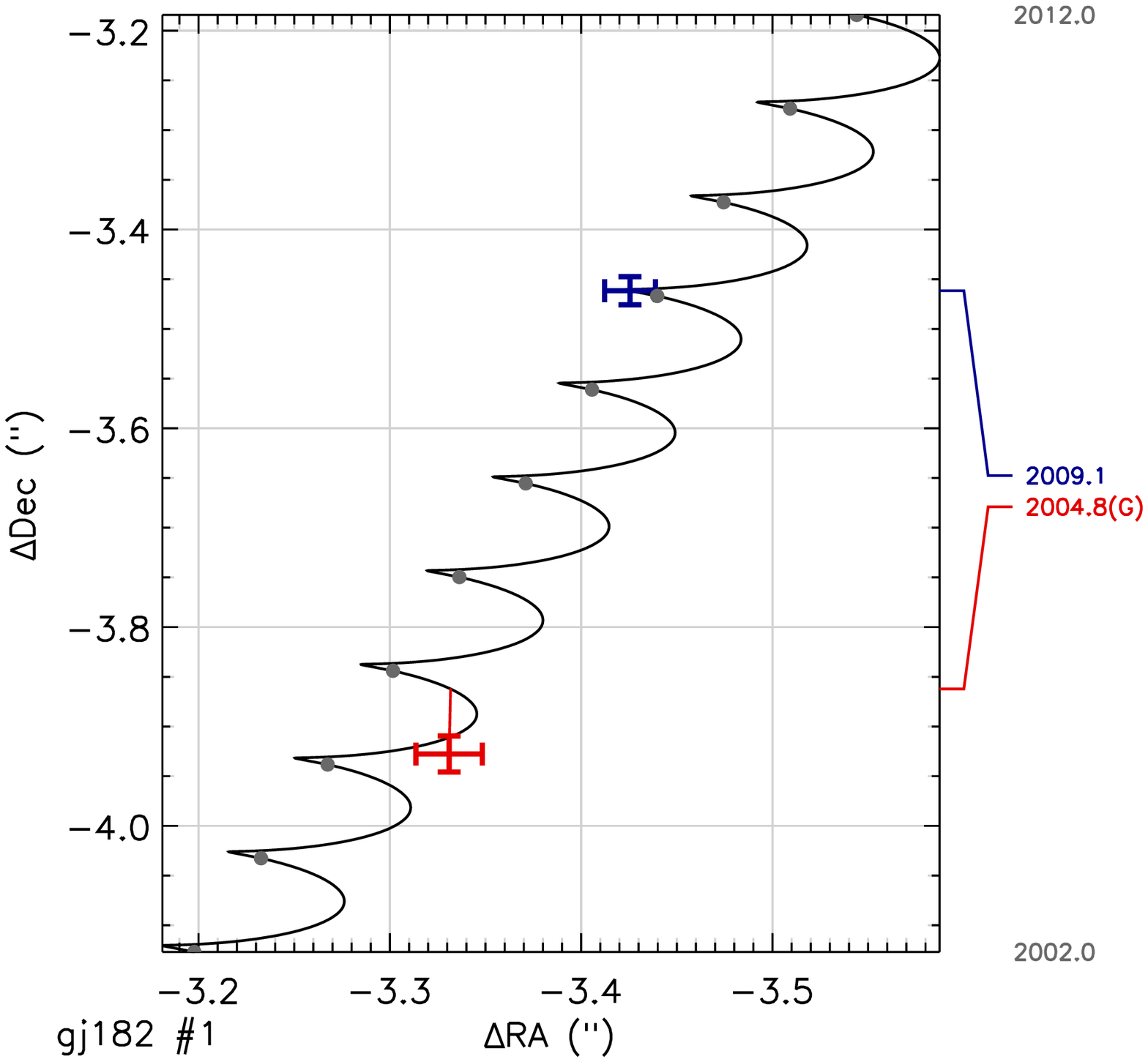}
\hskip -0.3in
\includegraphics[width=2.0in]{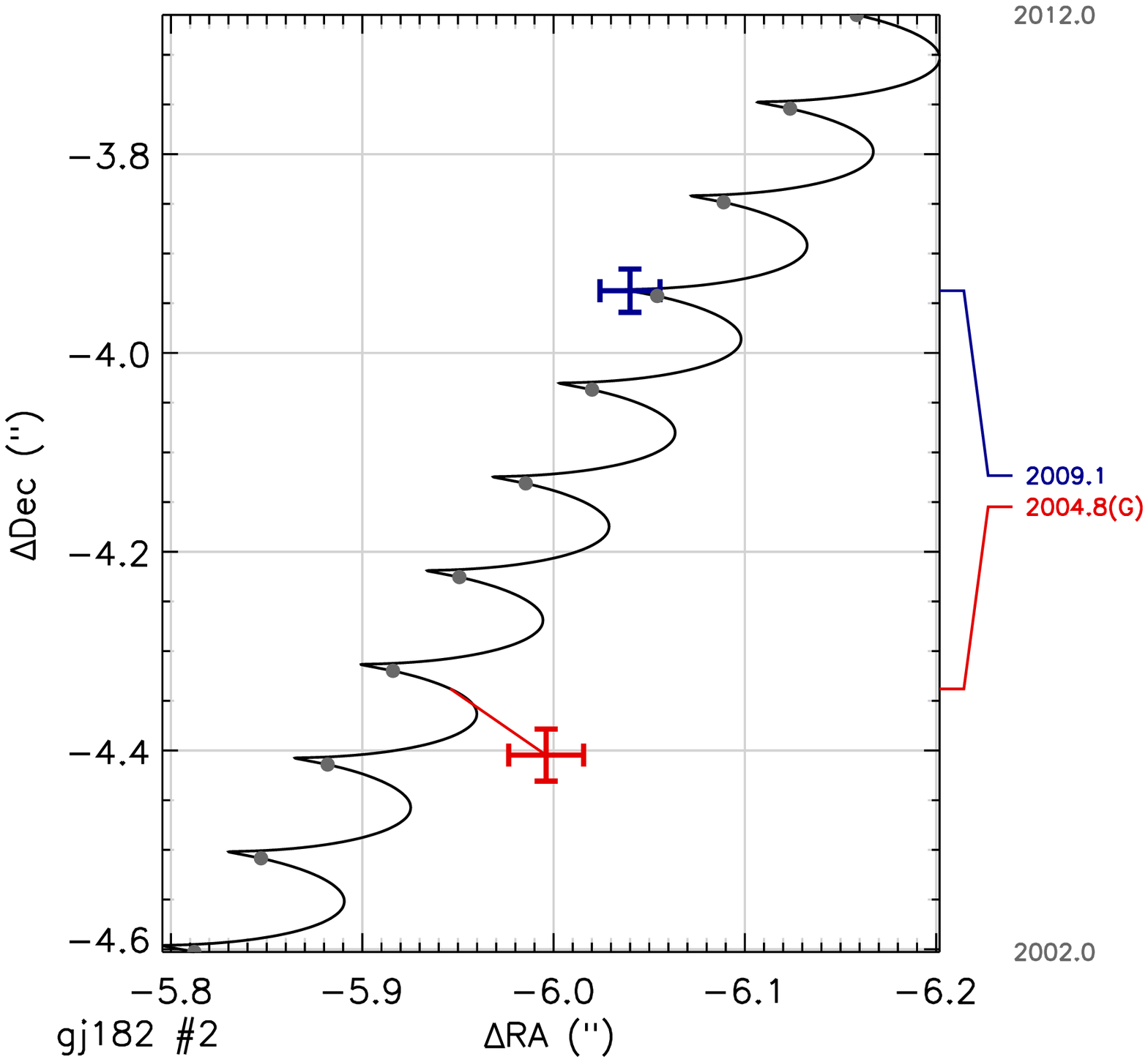}
\hskip -0.3in
\includegraphics[width=2.0in]{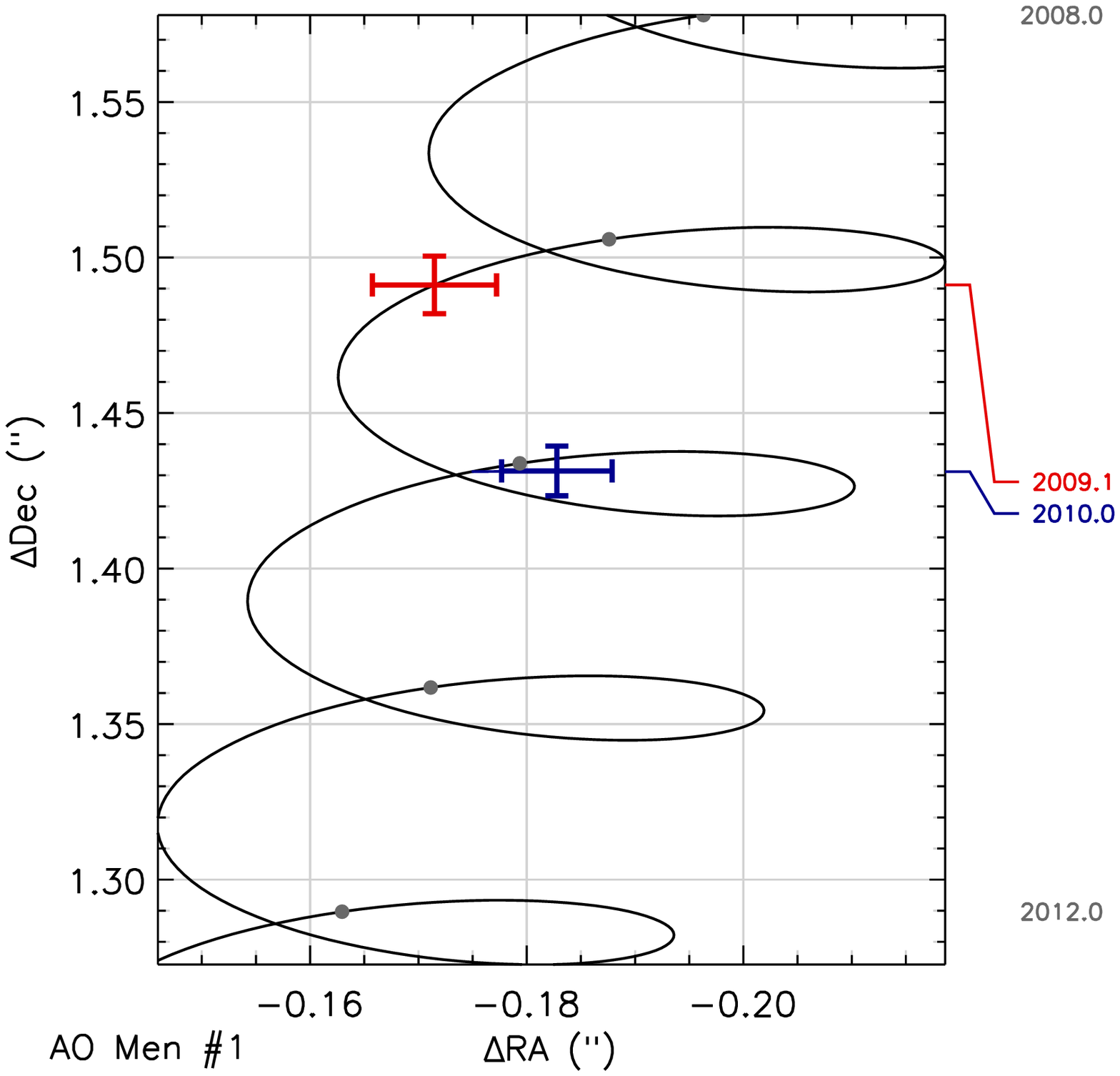}
\hskip -0.3in
\includegraphics[width=2.0in]{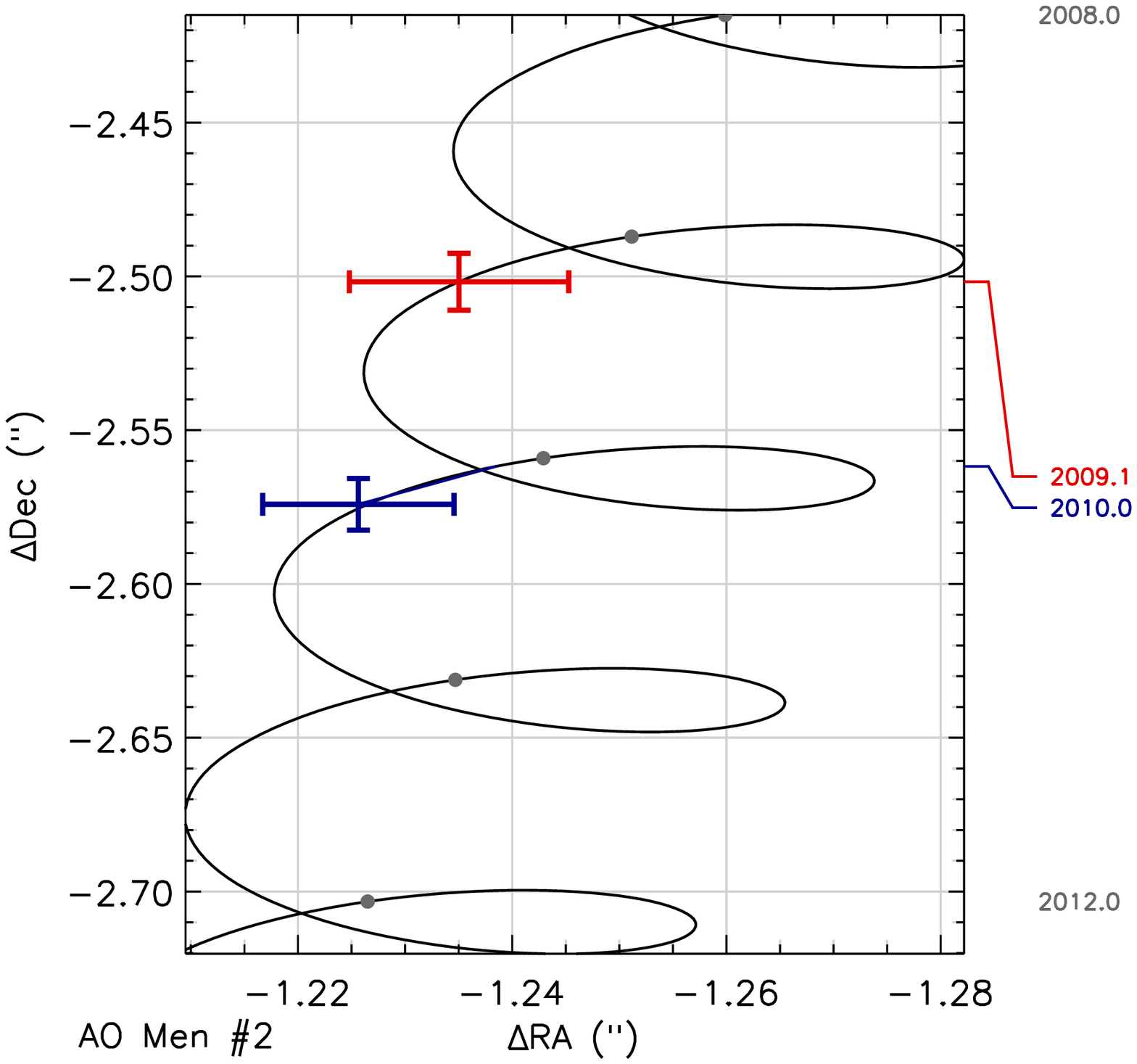}
}
\vskip -0.2in
\centerline{
\includegraphics[width=2.0in]{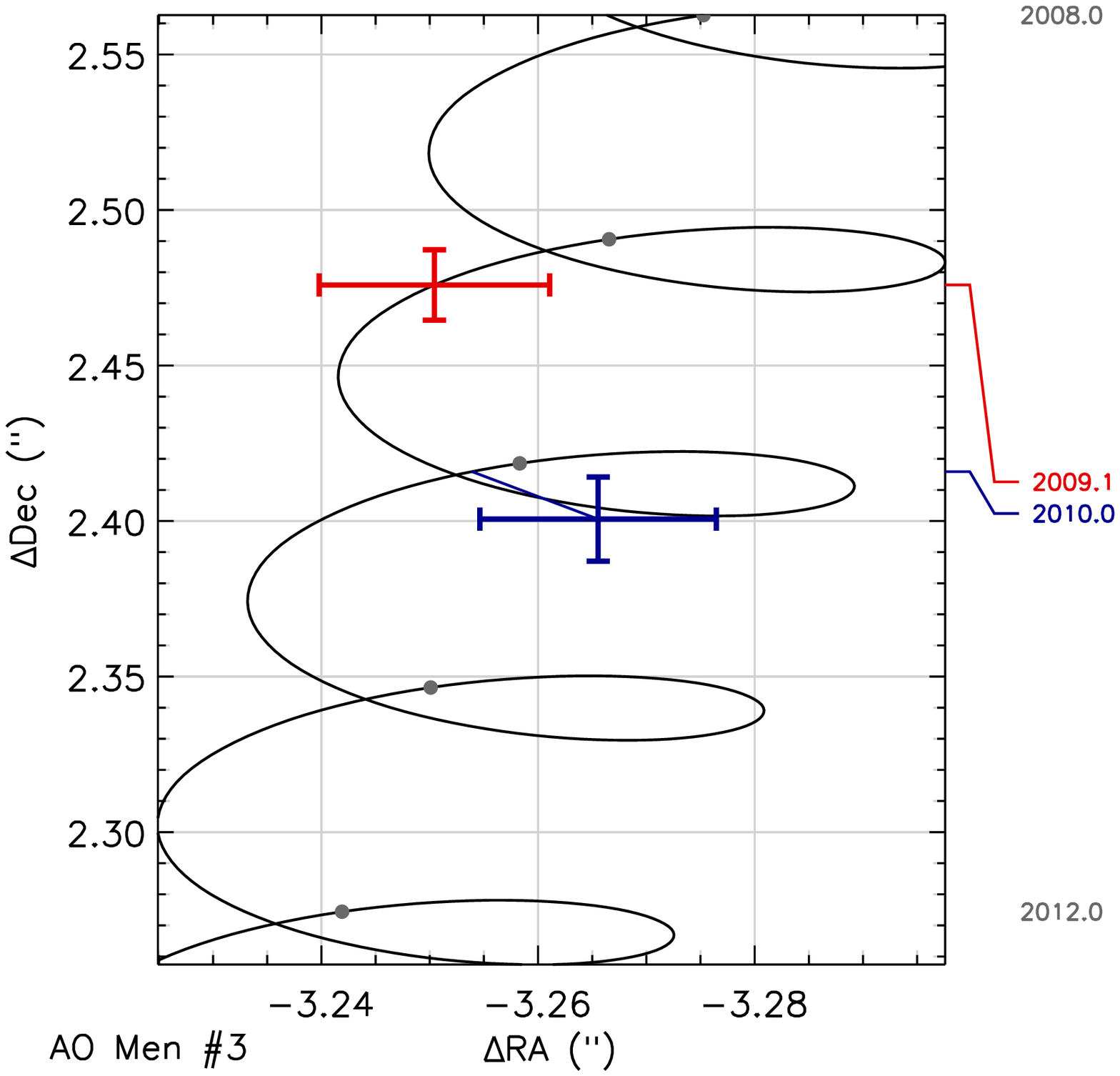}
\hskip -0.3in
\includegraphics[width=2.0in]{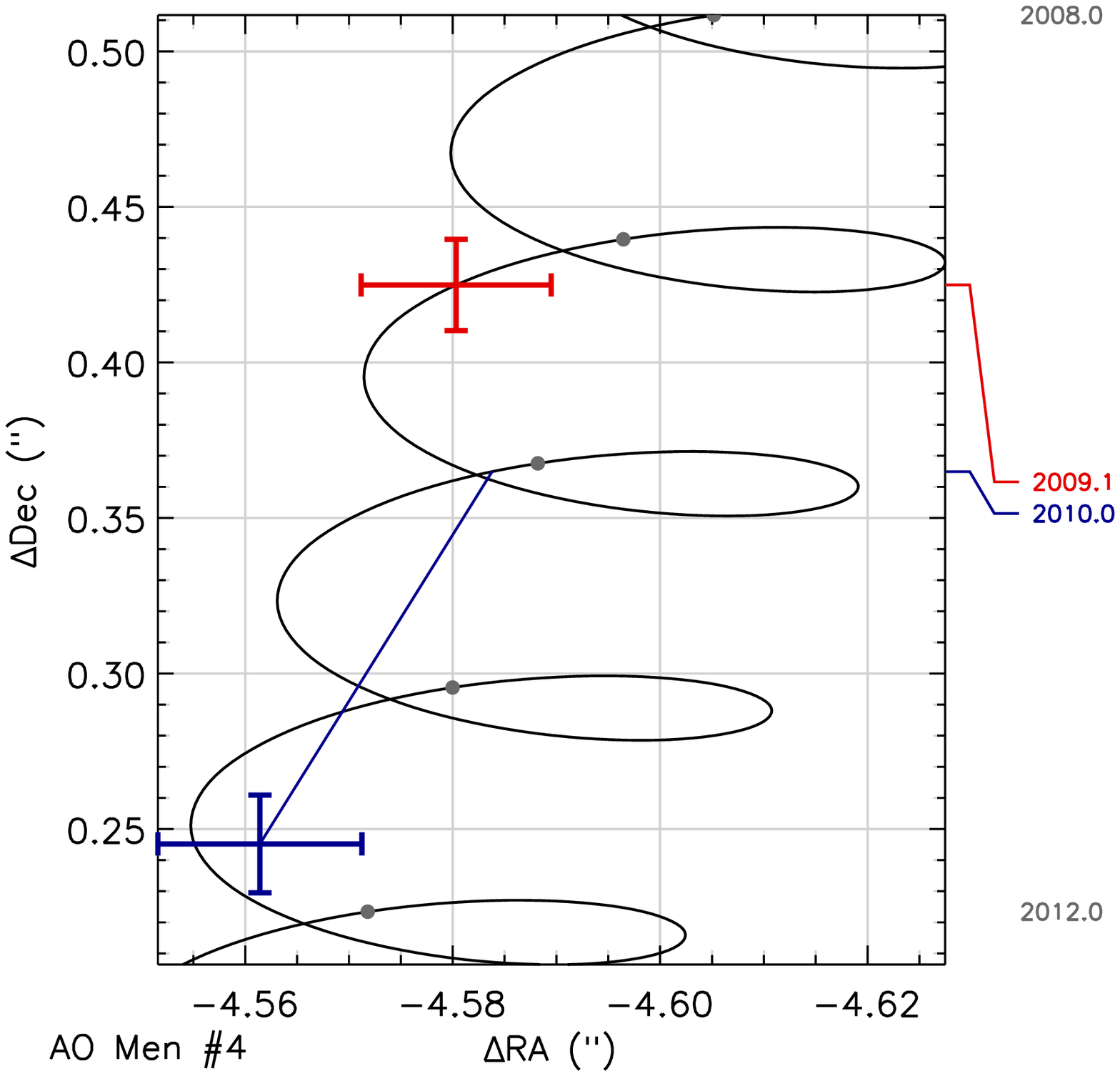}
\hskip -0.3in
\includegraphics[width=2.0in]{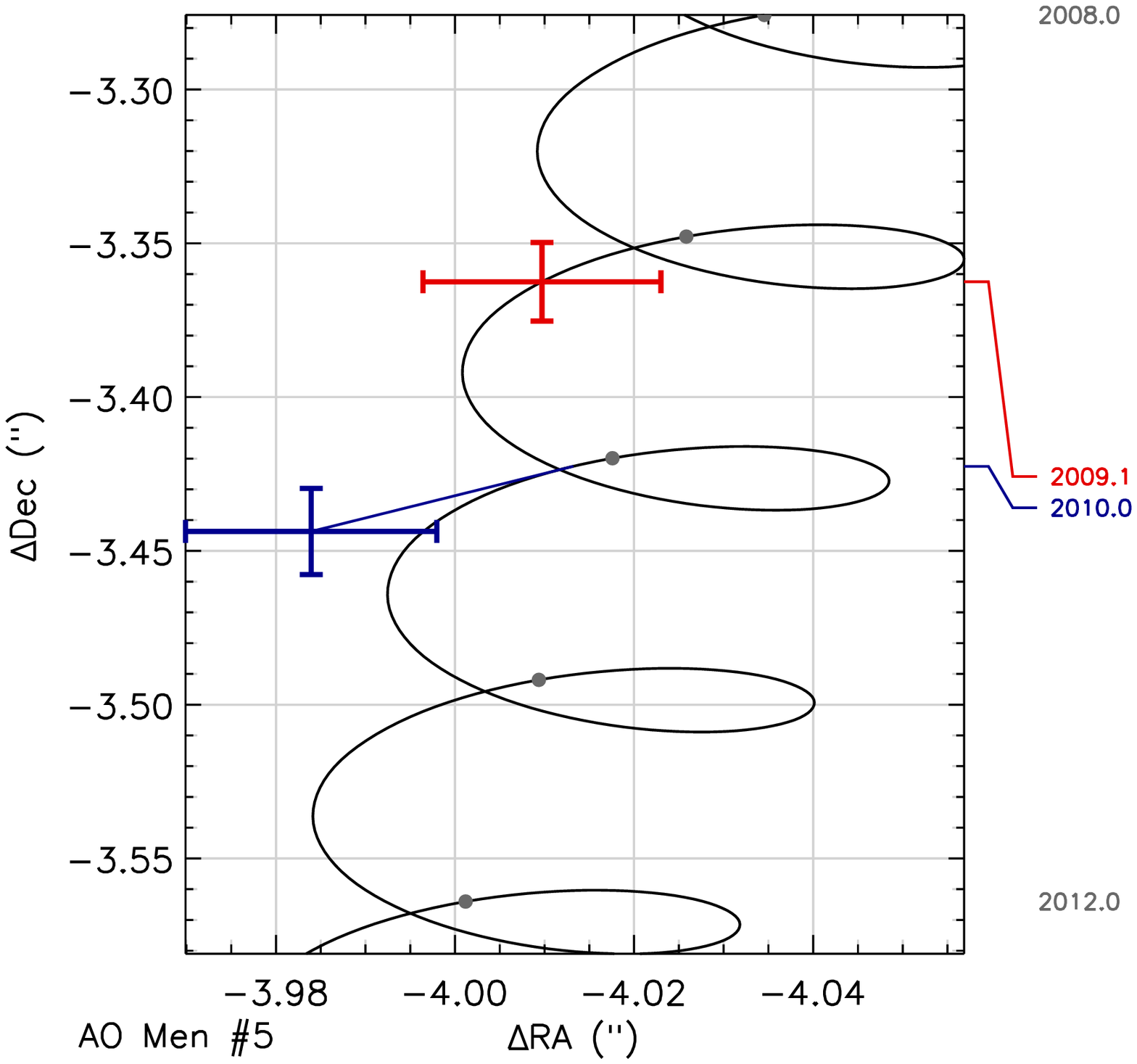}
\hskip -0.3in
\includegraphics[width=2.0in]{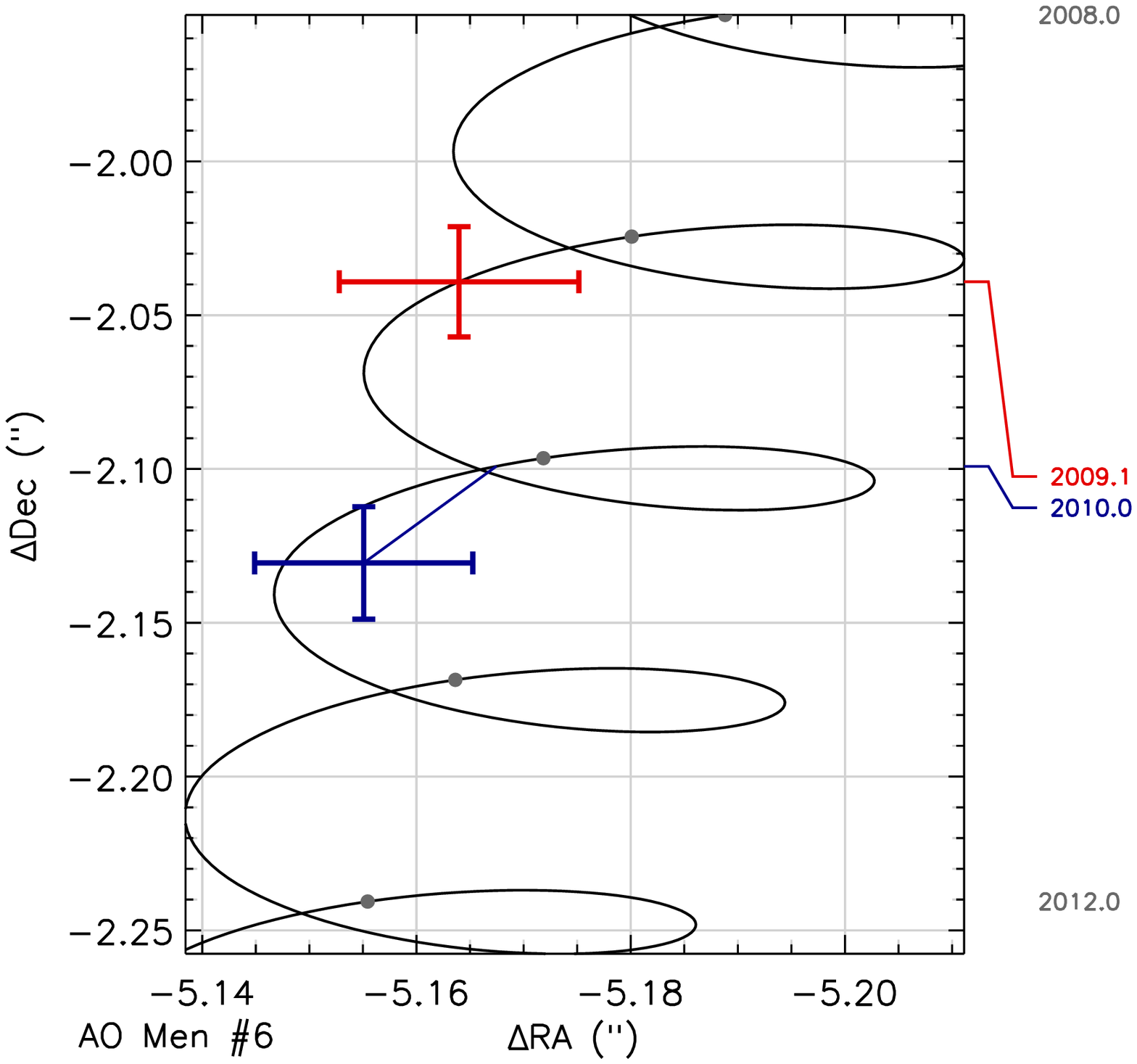}
}
\vskip -0.2in
\centerline{
\includegraphics[width=2.0in]{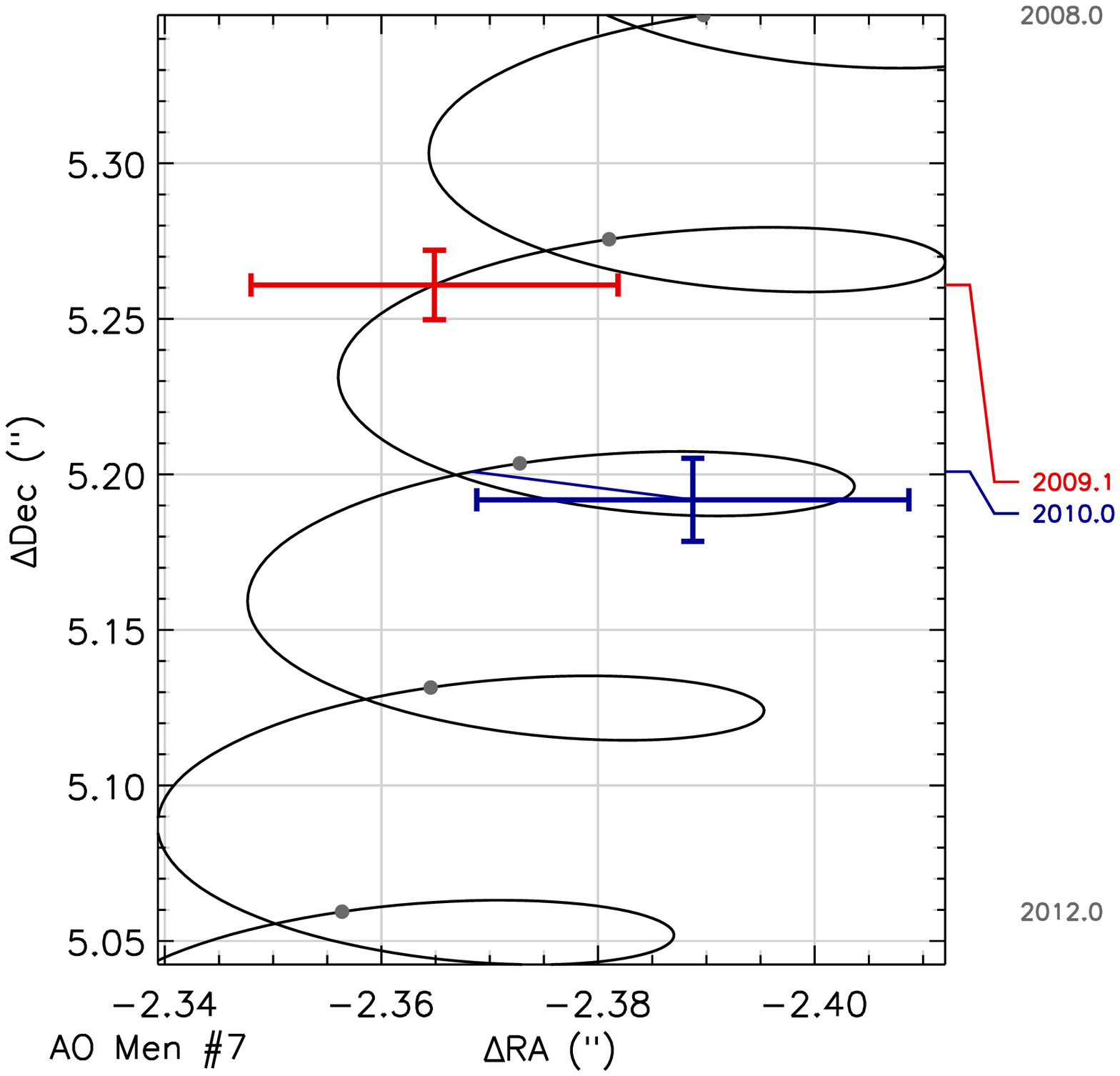}
\hskip -0.3in
\includegraphics[width=2.0in]{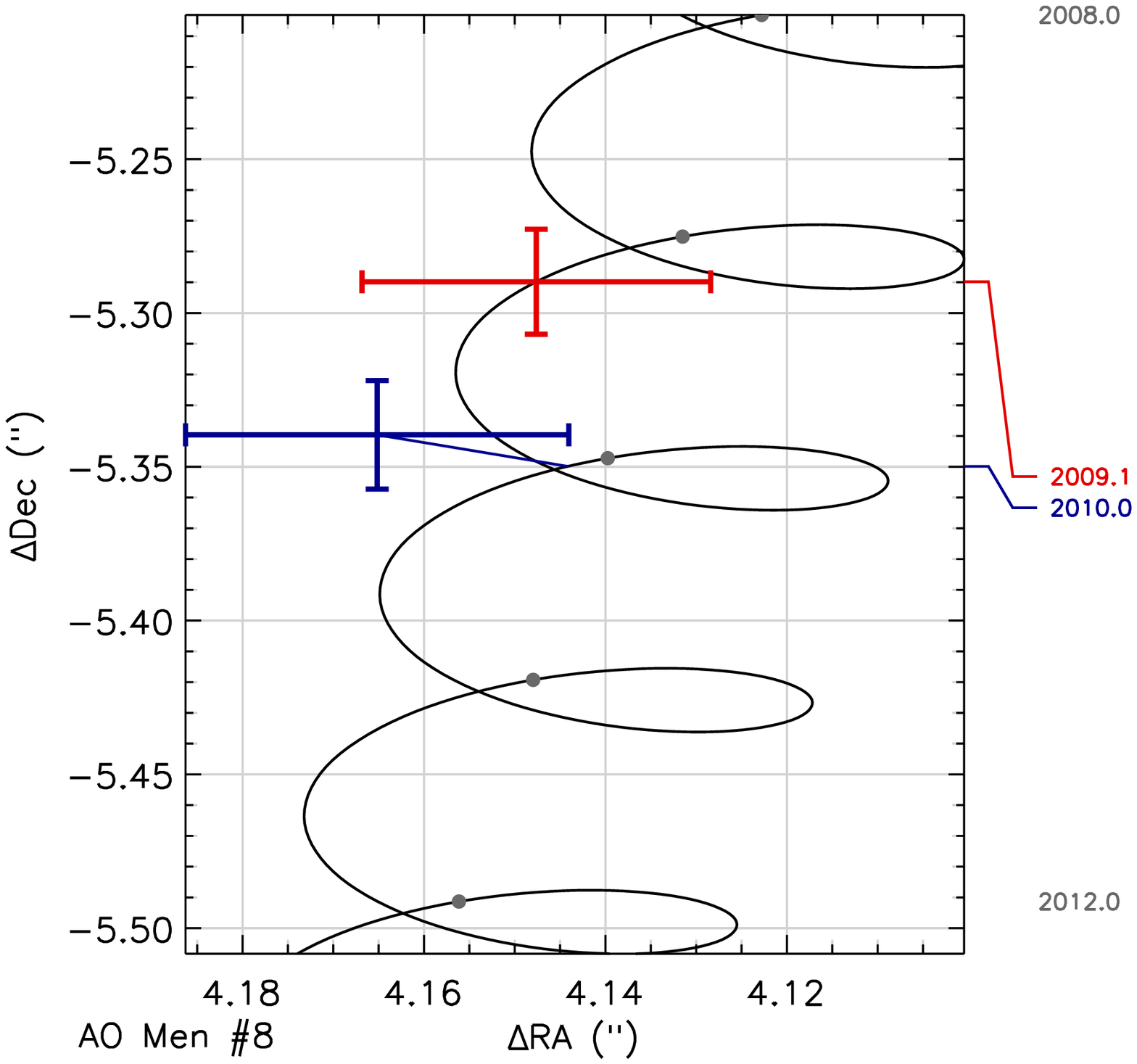}
\hskip -0.3in
\includegraphics[width=2.0in]{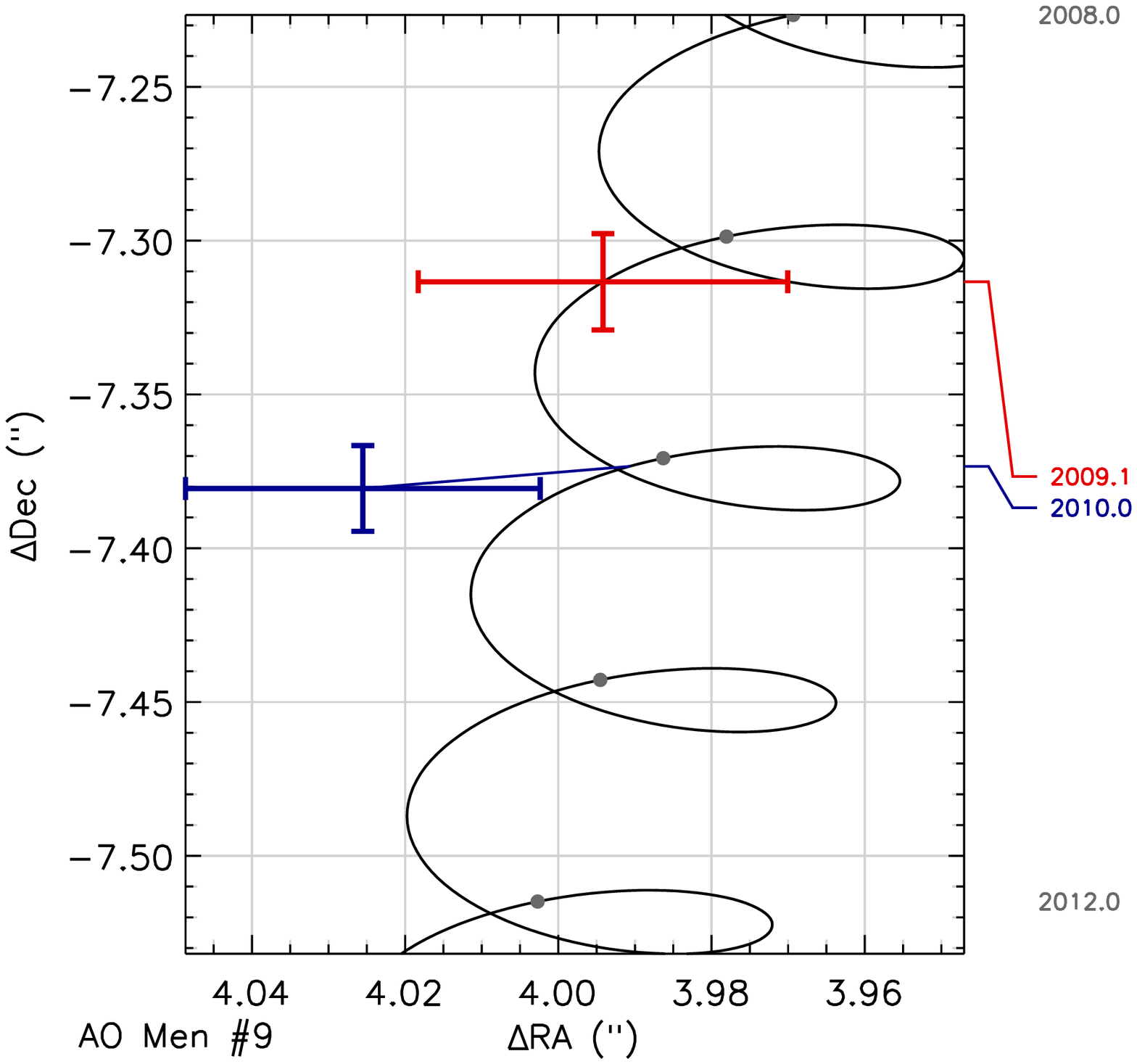}
\hskip -0.3in
\includegraphics[width=2.0in]{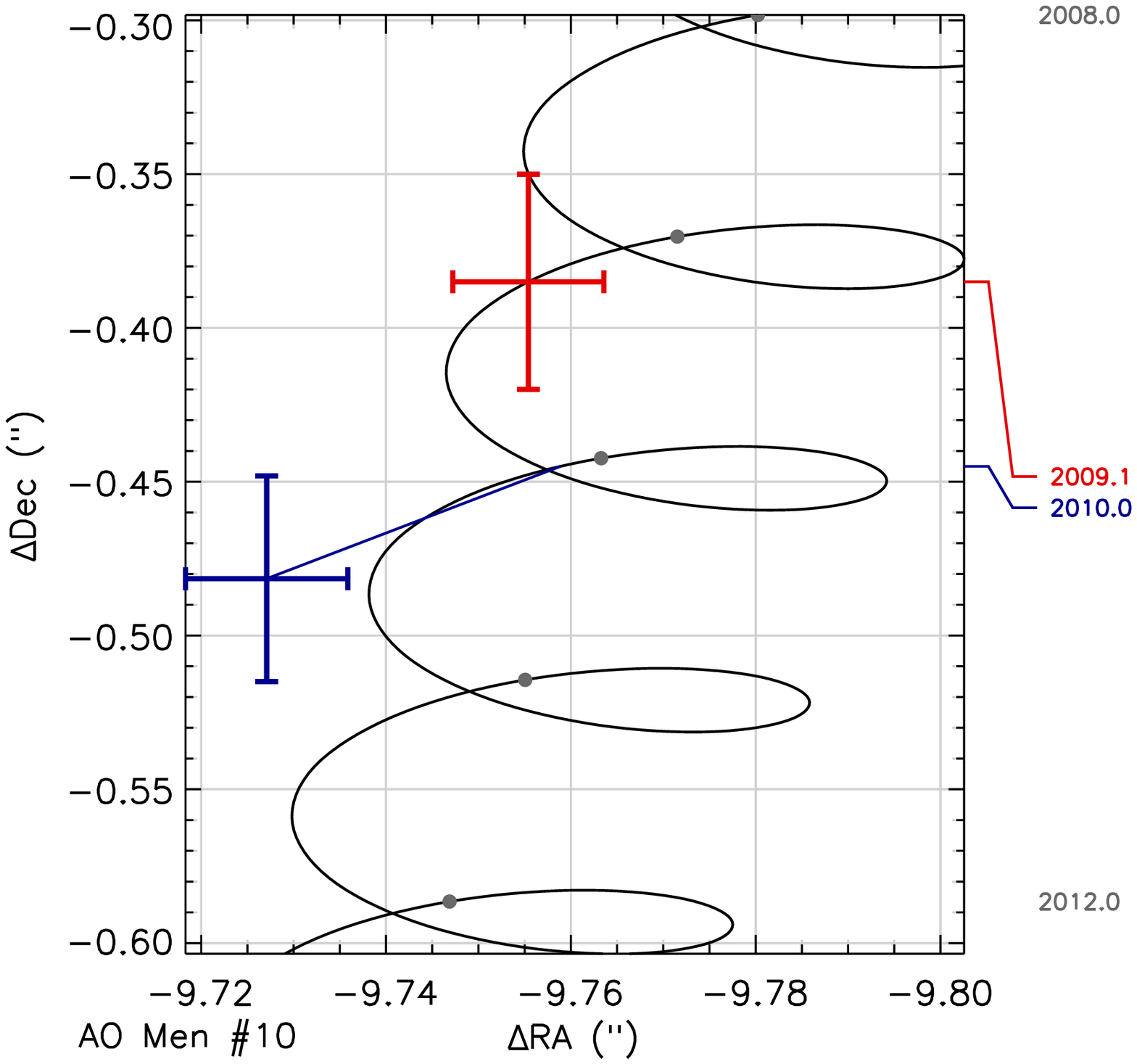}
}
\vskip -0.2in
\centerline{
\includegraphics[width=2.0in]{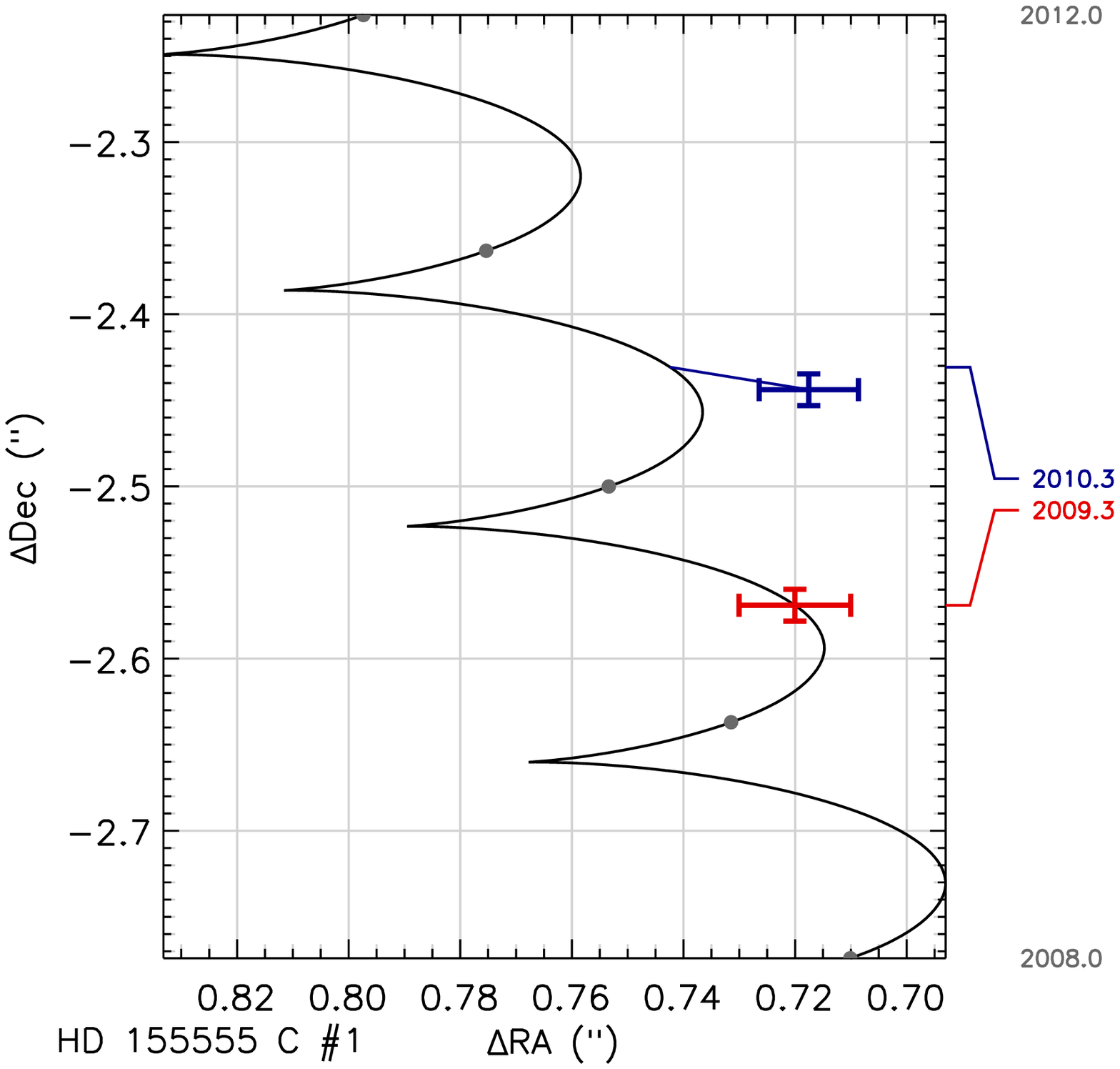}
\hskip -0.3in
\includegraphics[width=2.0in]{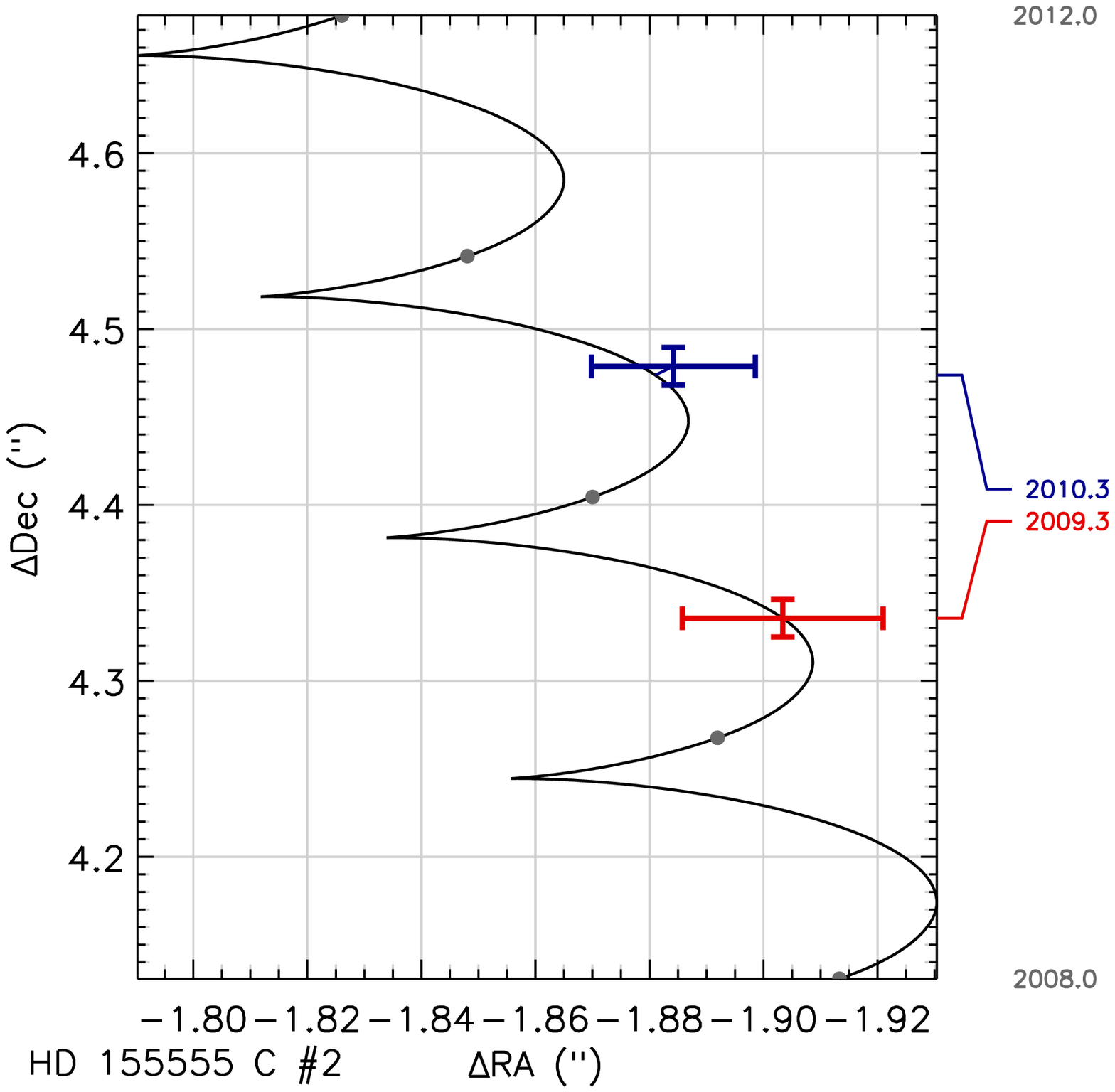}
\hskip -0.3in
\includegraphics[width=2.0in]{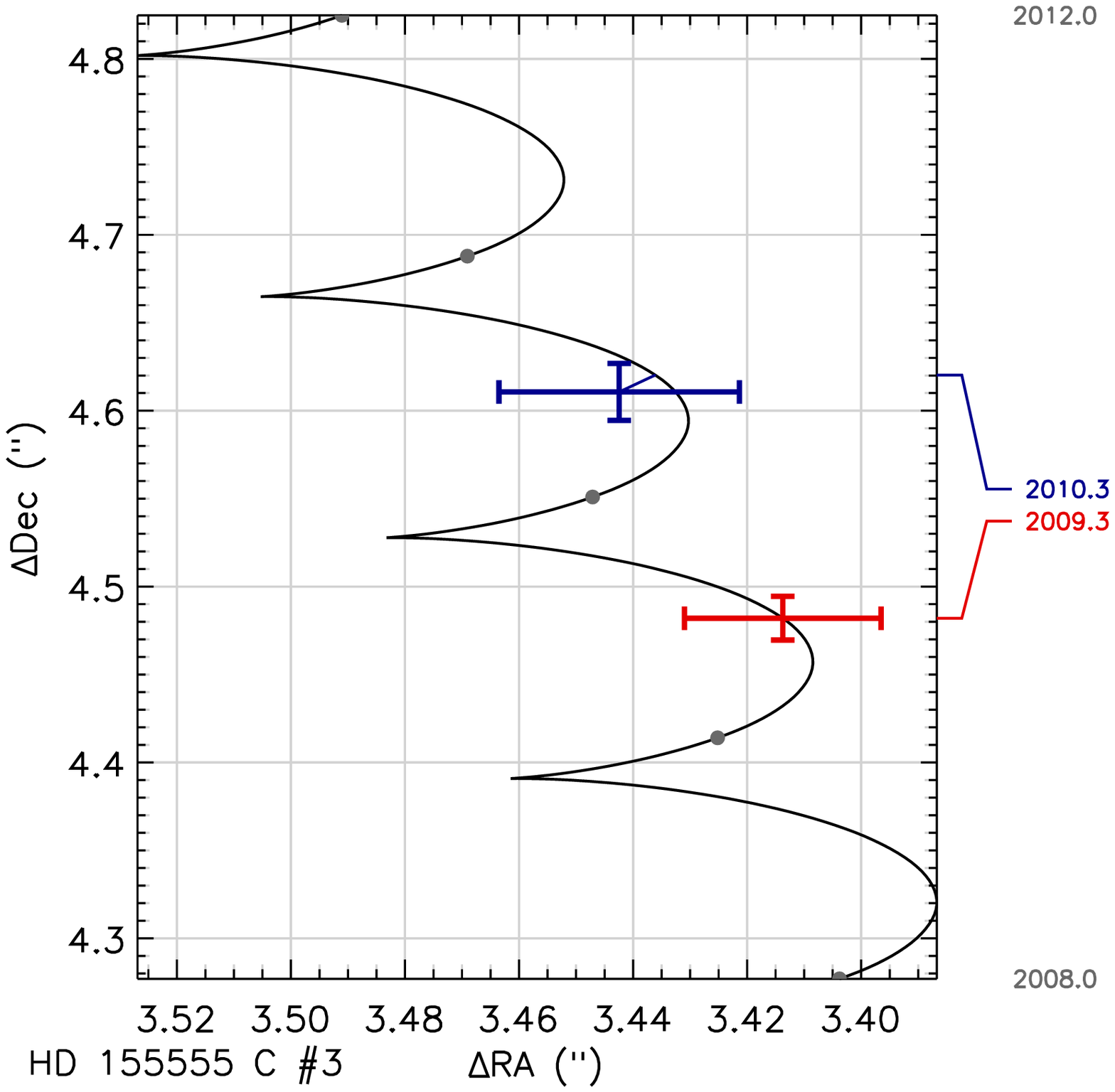}
\hskip -0.3in
\includegraphics[width=2.0in]{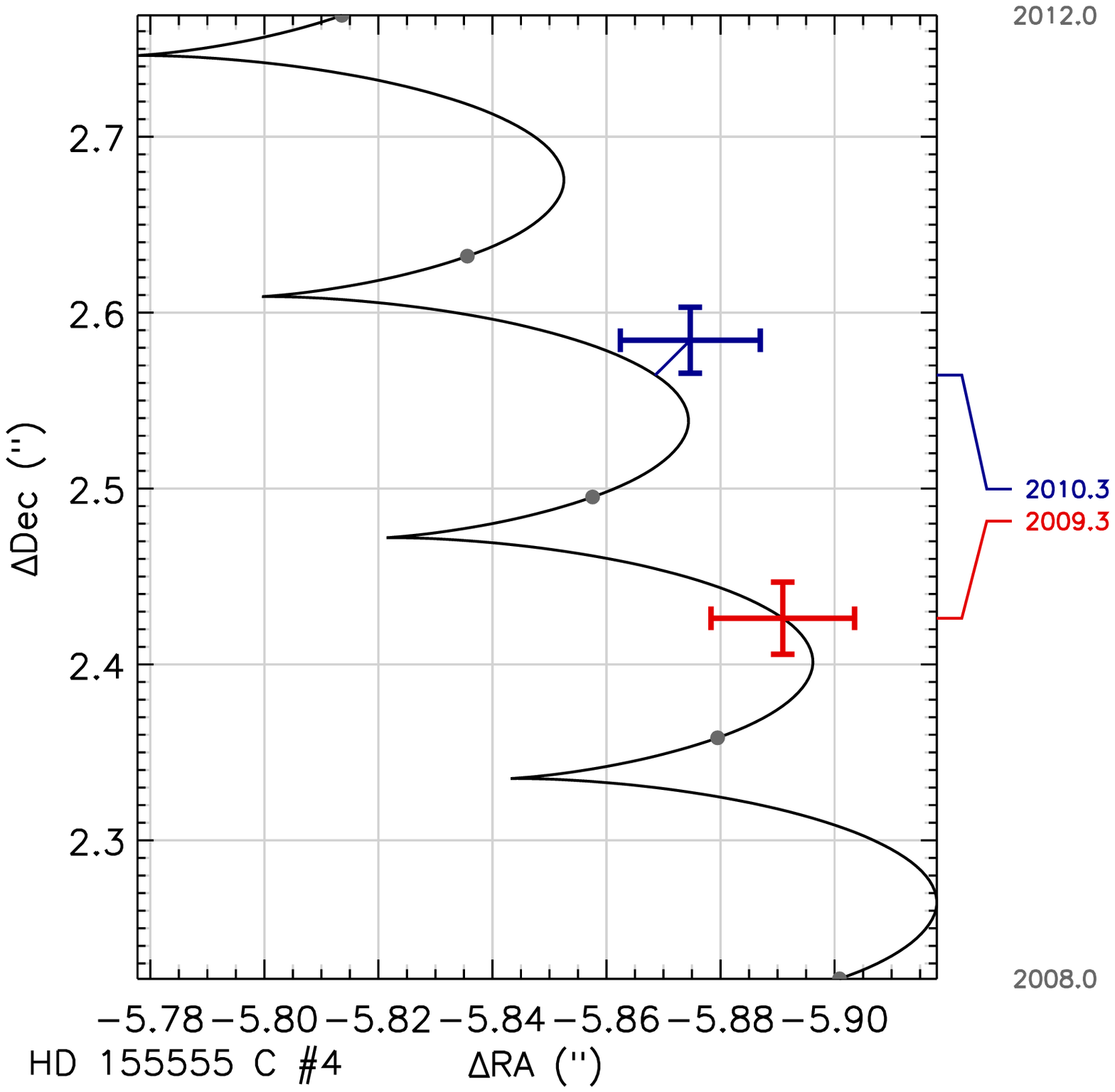}
}
\caption{On-sky plots for $\beta$ Pic MG objects.
\label{fig:BetaPic_skyplots1}}
\end{figure}

\begin{figure}
\vskip -0.2in
\centerline{
\includegraphics[width=2.0in]{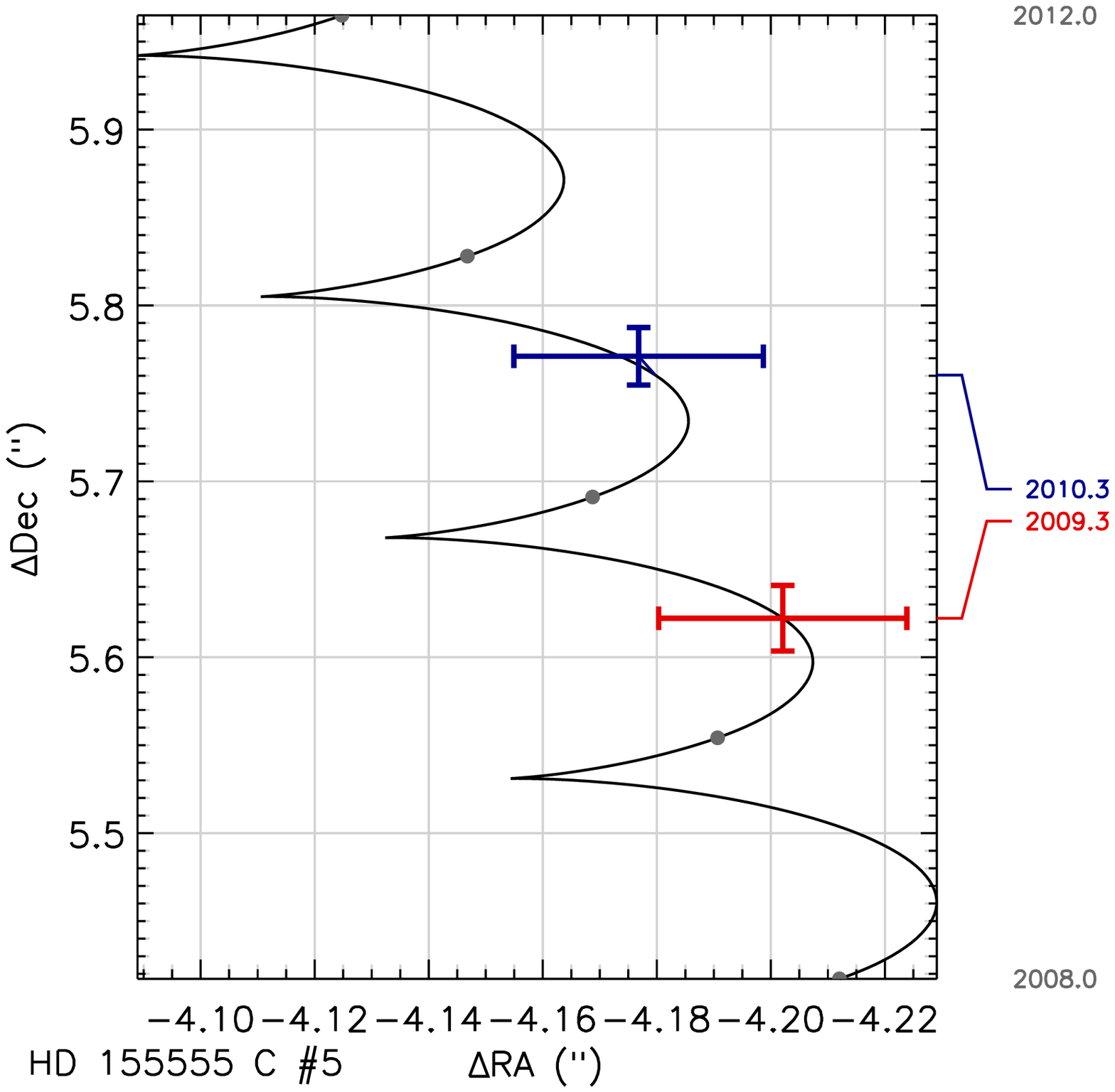}
\hskip -0.3in
\includegraphics[width=2.0in]{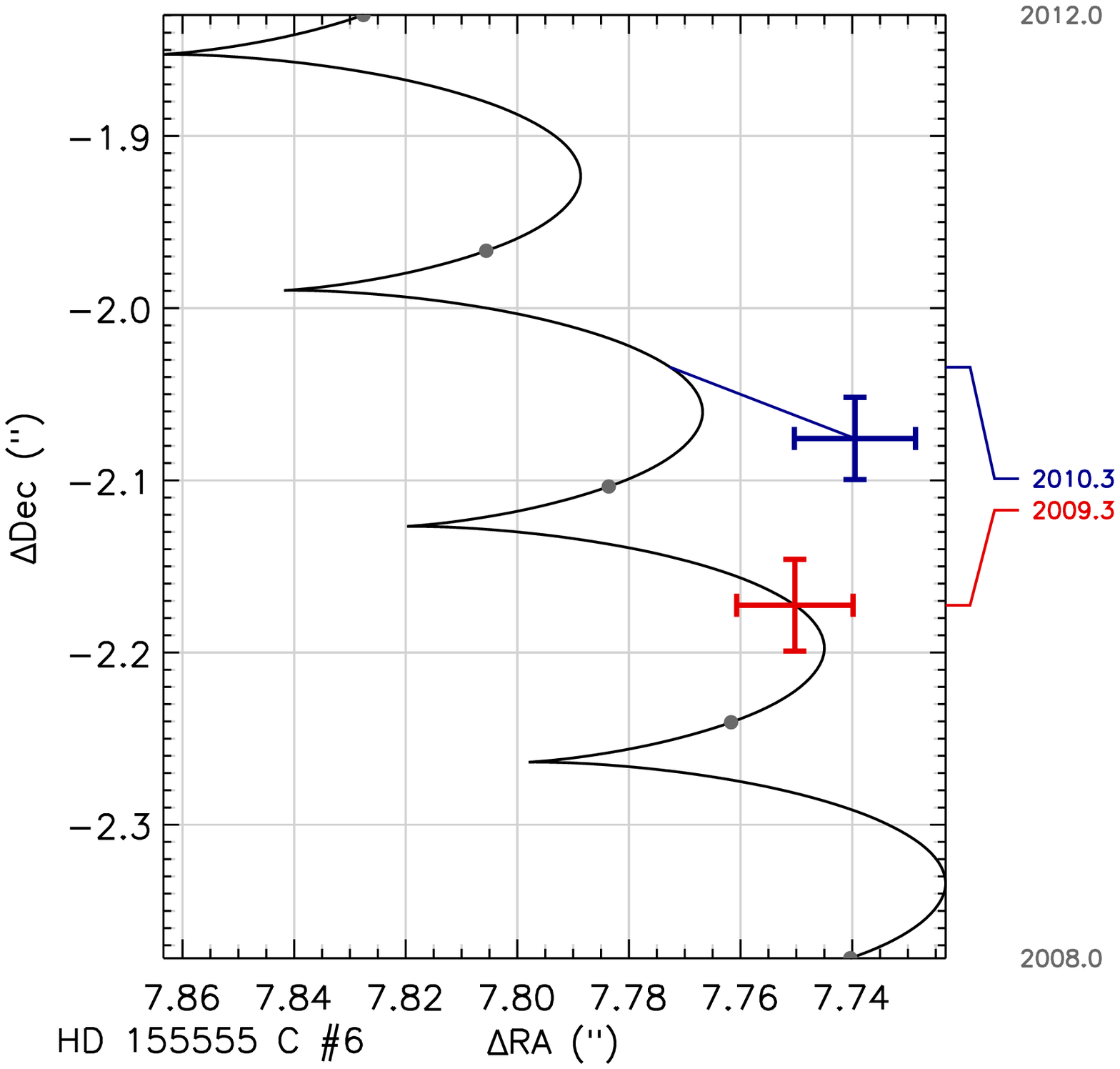}
\hskip -0.3in
\includegraphics[width=2.0in]{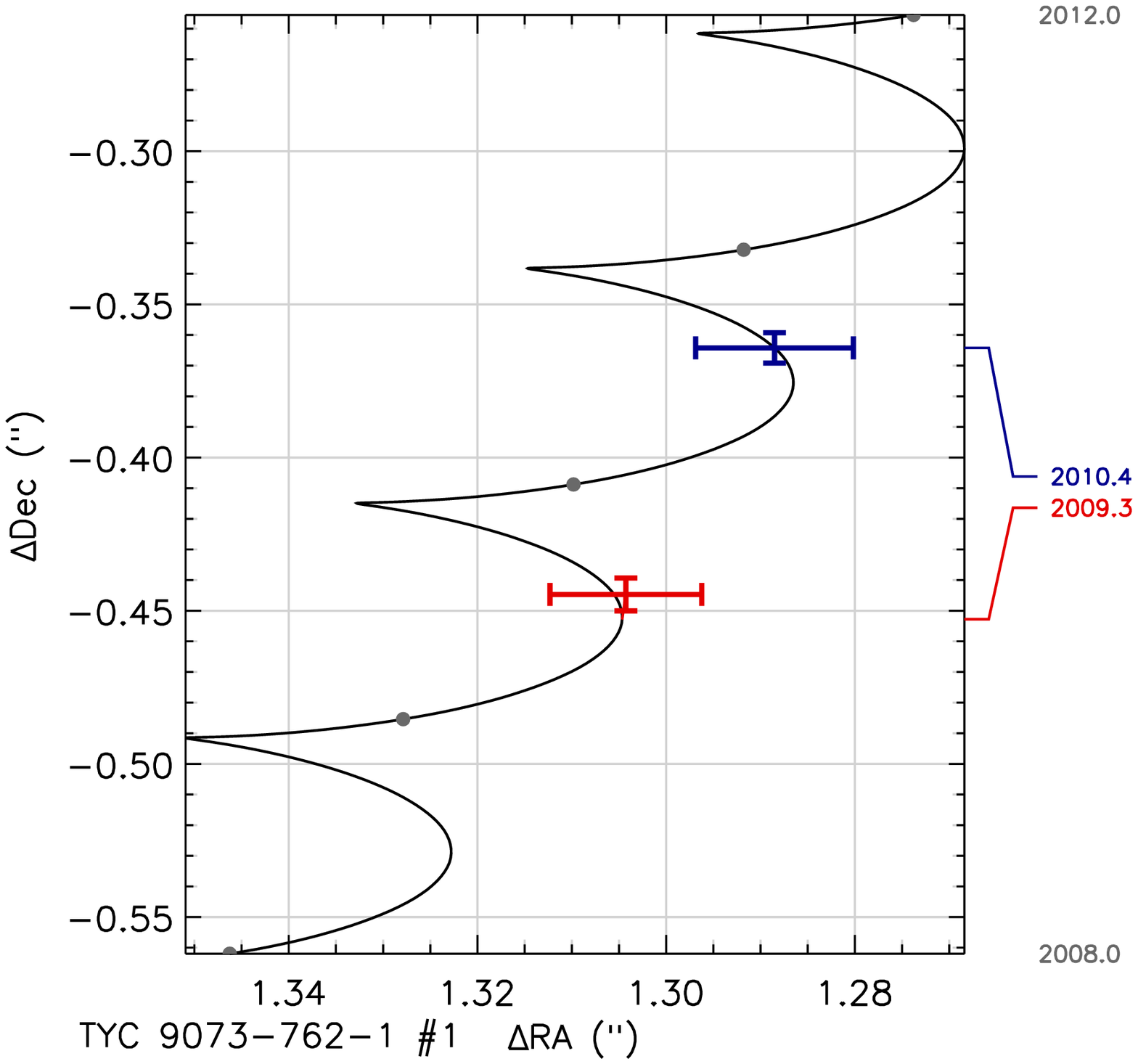}
\hskip -0.3in 
\includegraphics[width=2.0in]{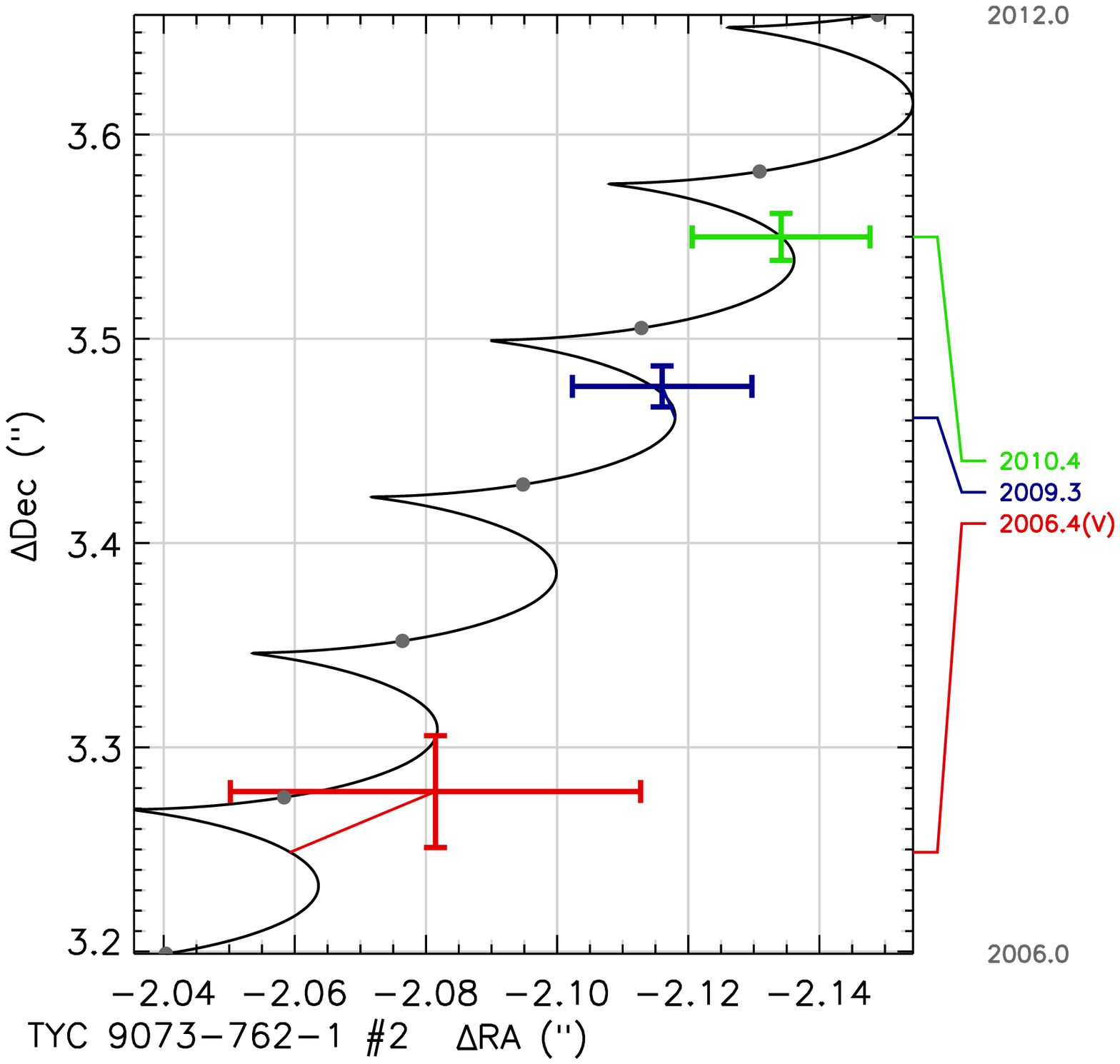}
}
\vskip -0.2in
\centerline{
\includegraphics[width=2.0in]{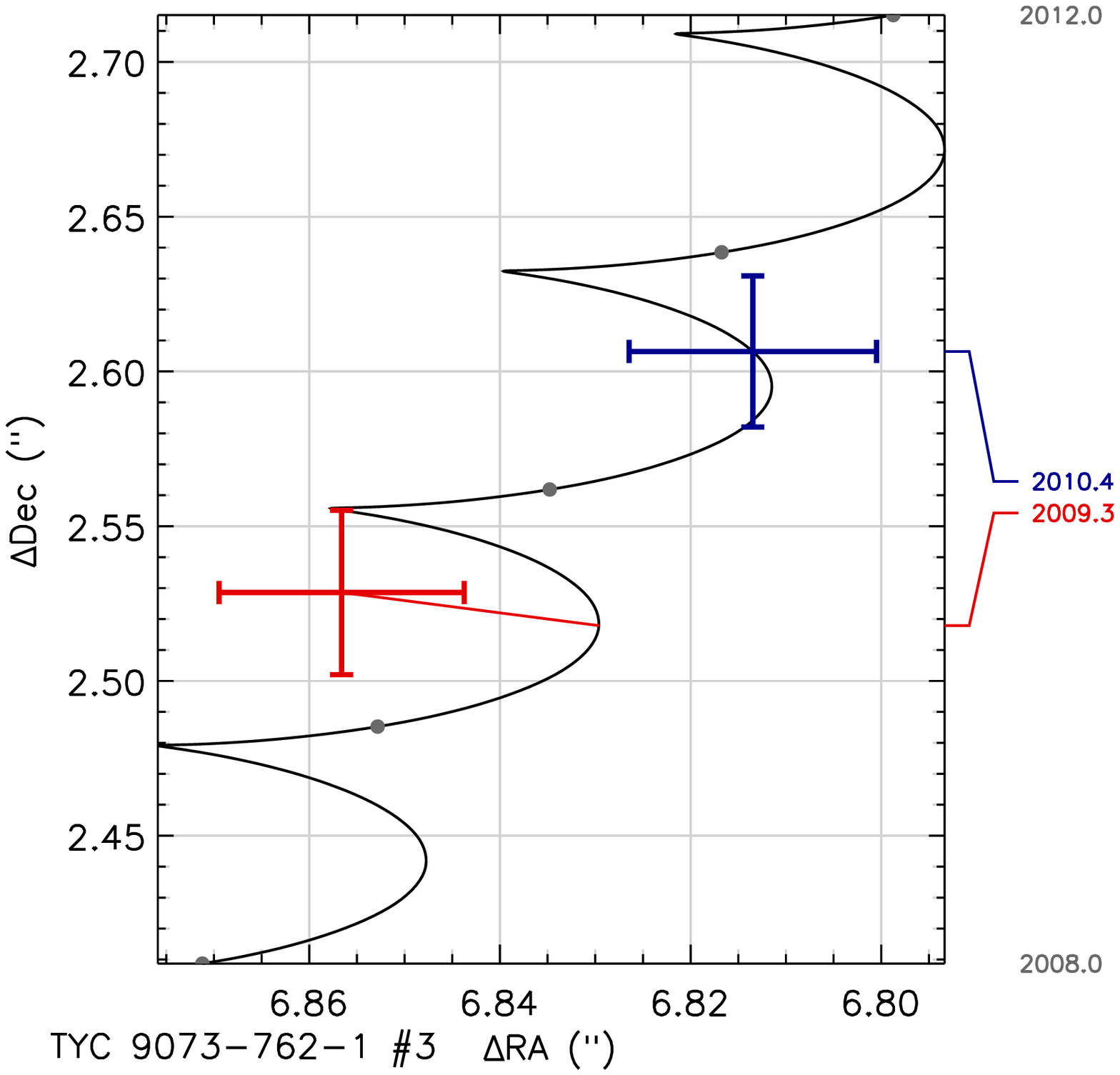}
\hskip -0.3in
\includegraphics[width=2.0in]{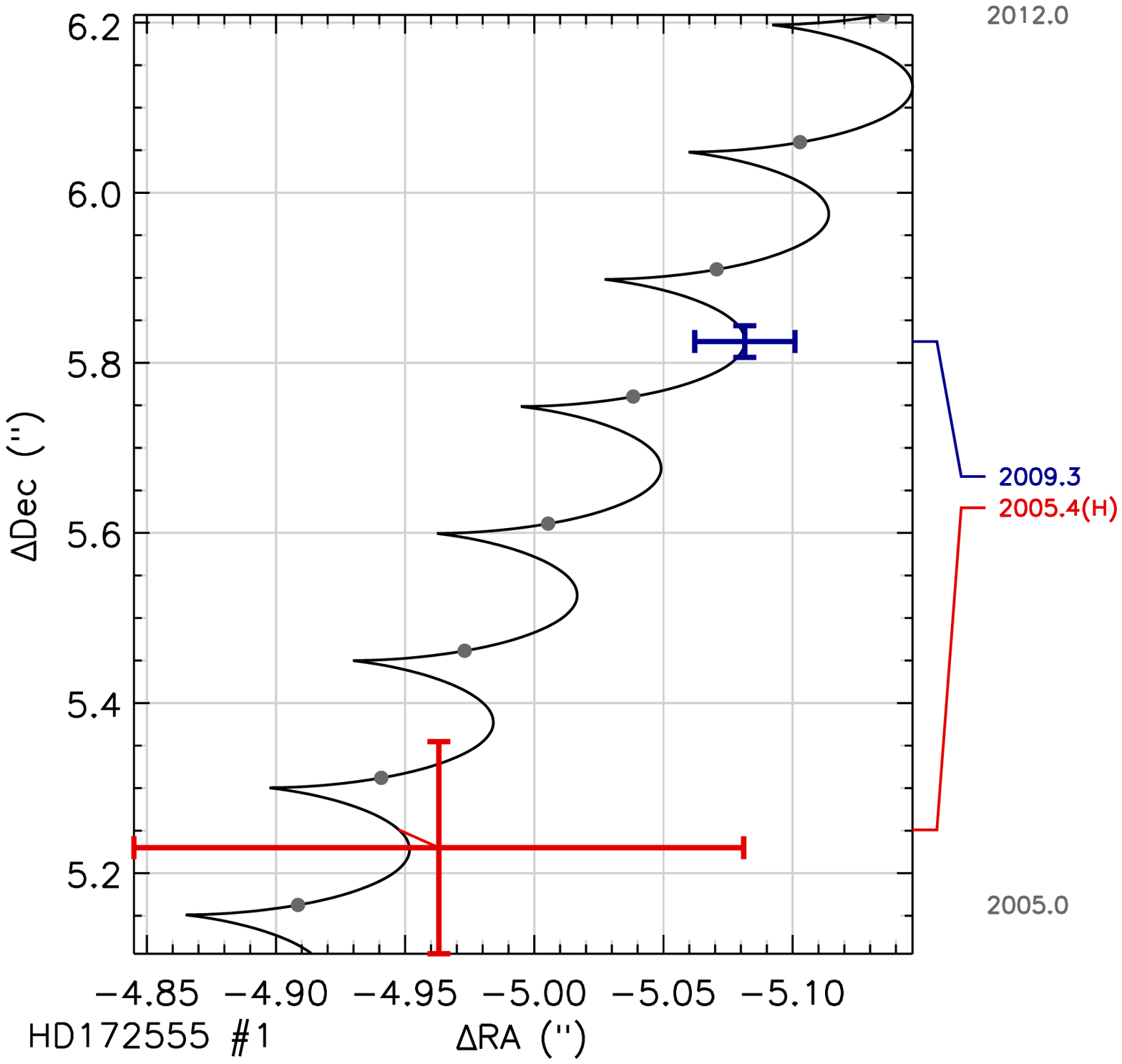}
\hskip -0.3in
\includegraphics[width=2.0in]{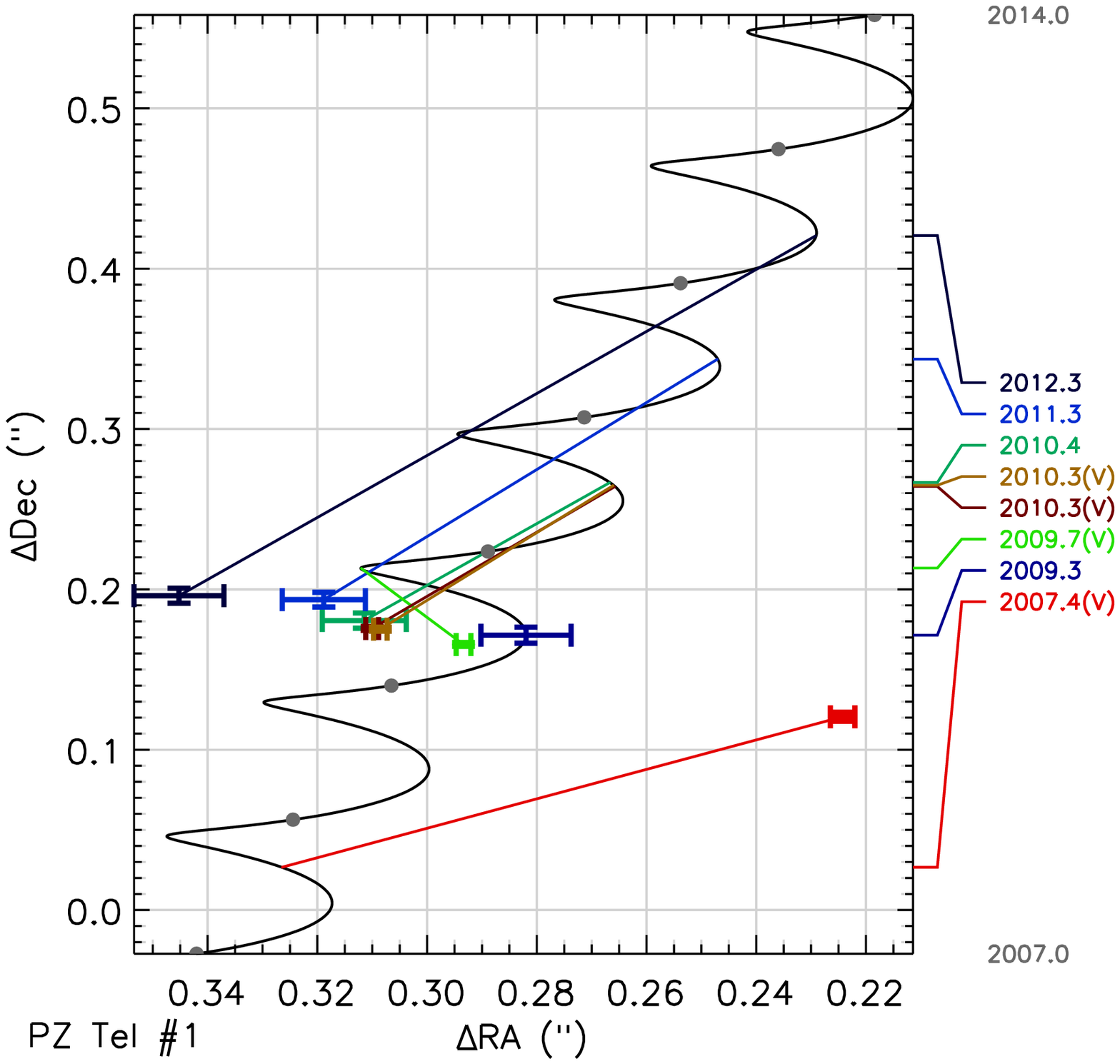}
\hskip -0.3in
\includegraphics[width=2.0in]{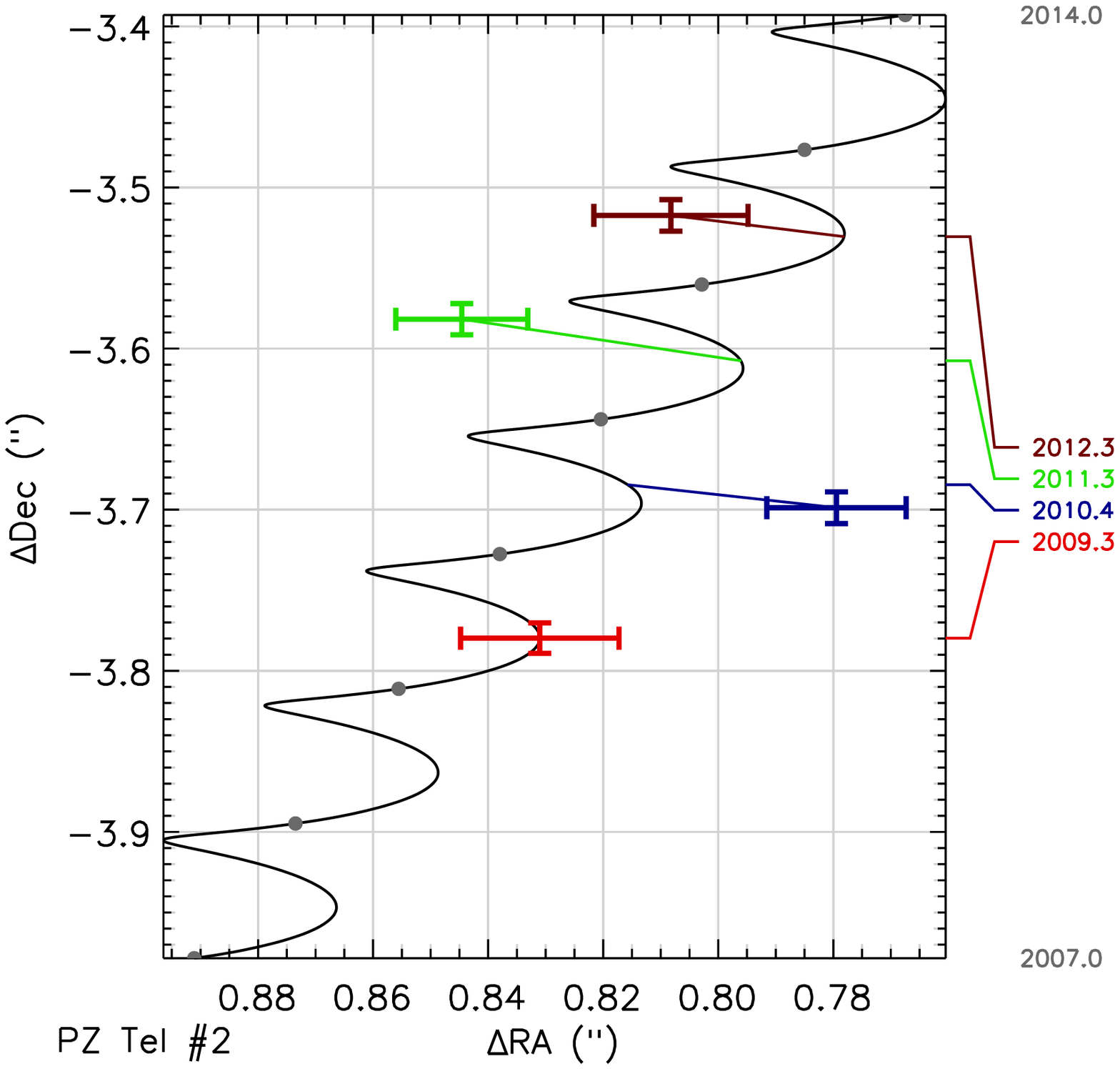}
}
\vskip -0.2in
\centerline{
\includegraphics[width=2.0in]{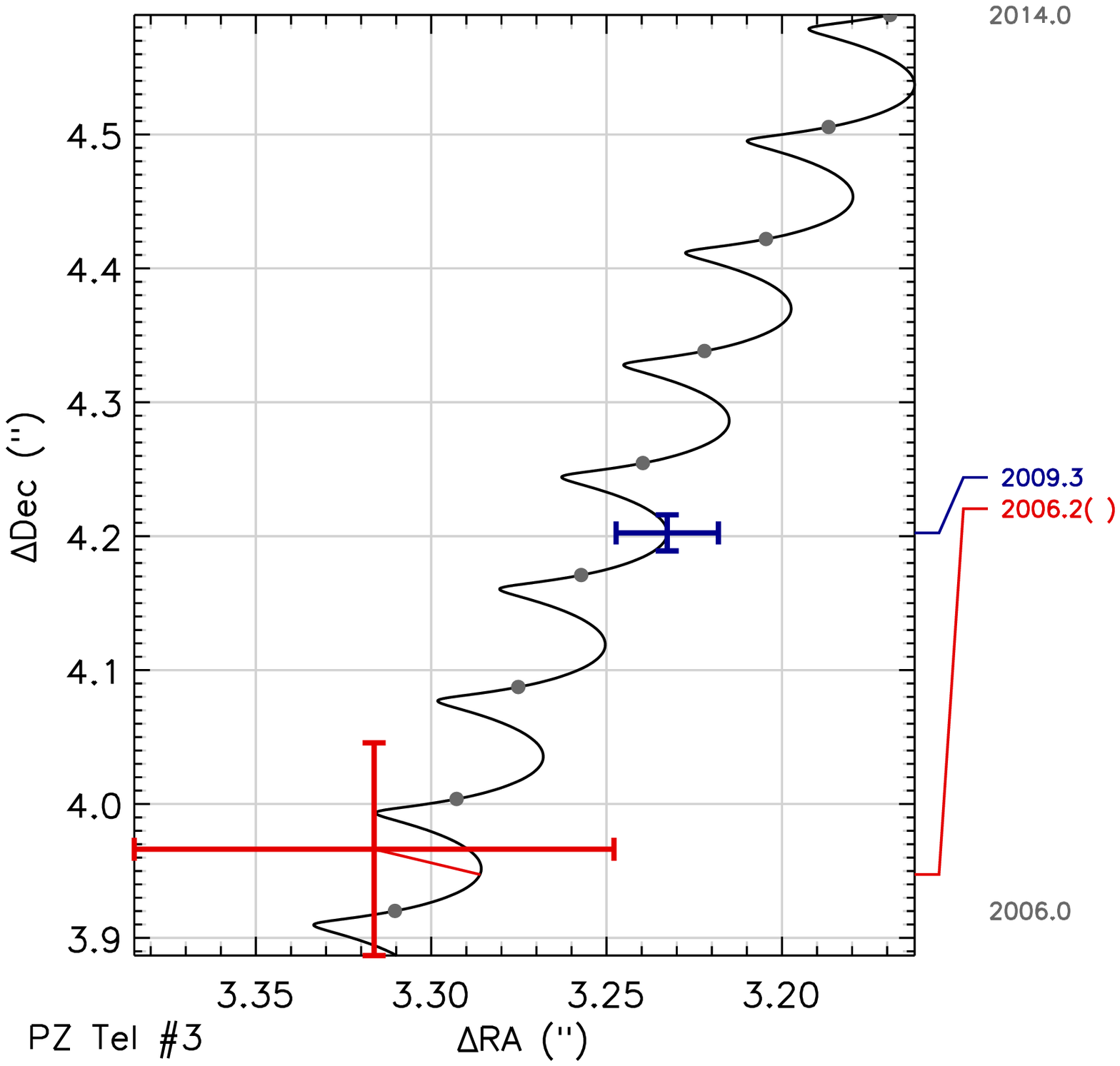}
\hskip -0.3in
\includegraphics[width=2.0in]{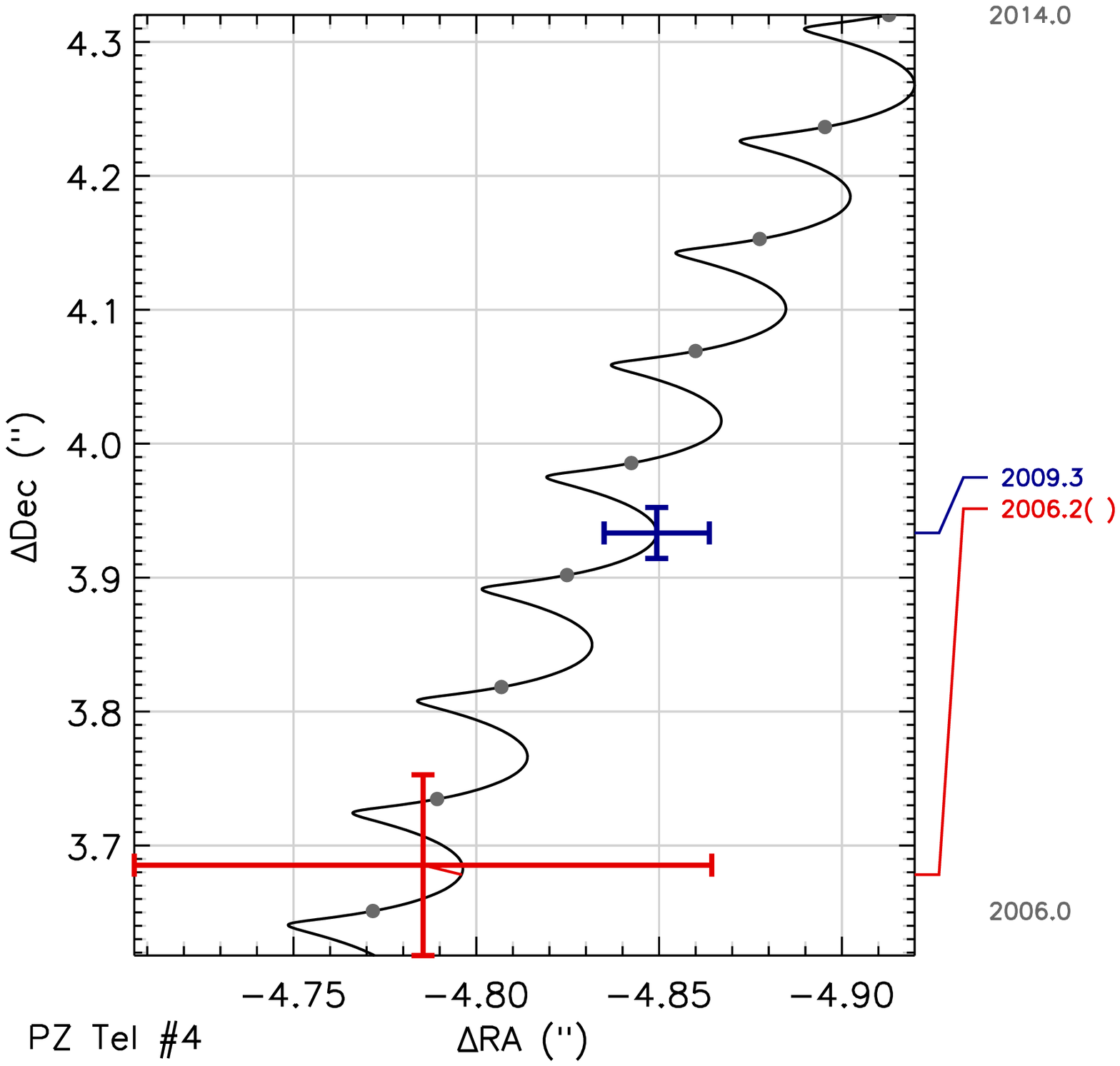}
\hskip -0.3in
\includegraphics[width=2.0in]{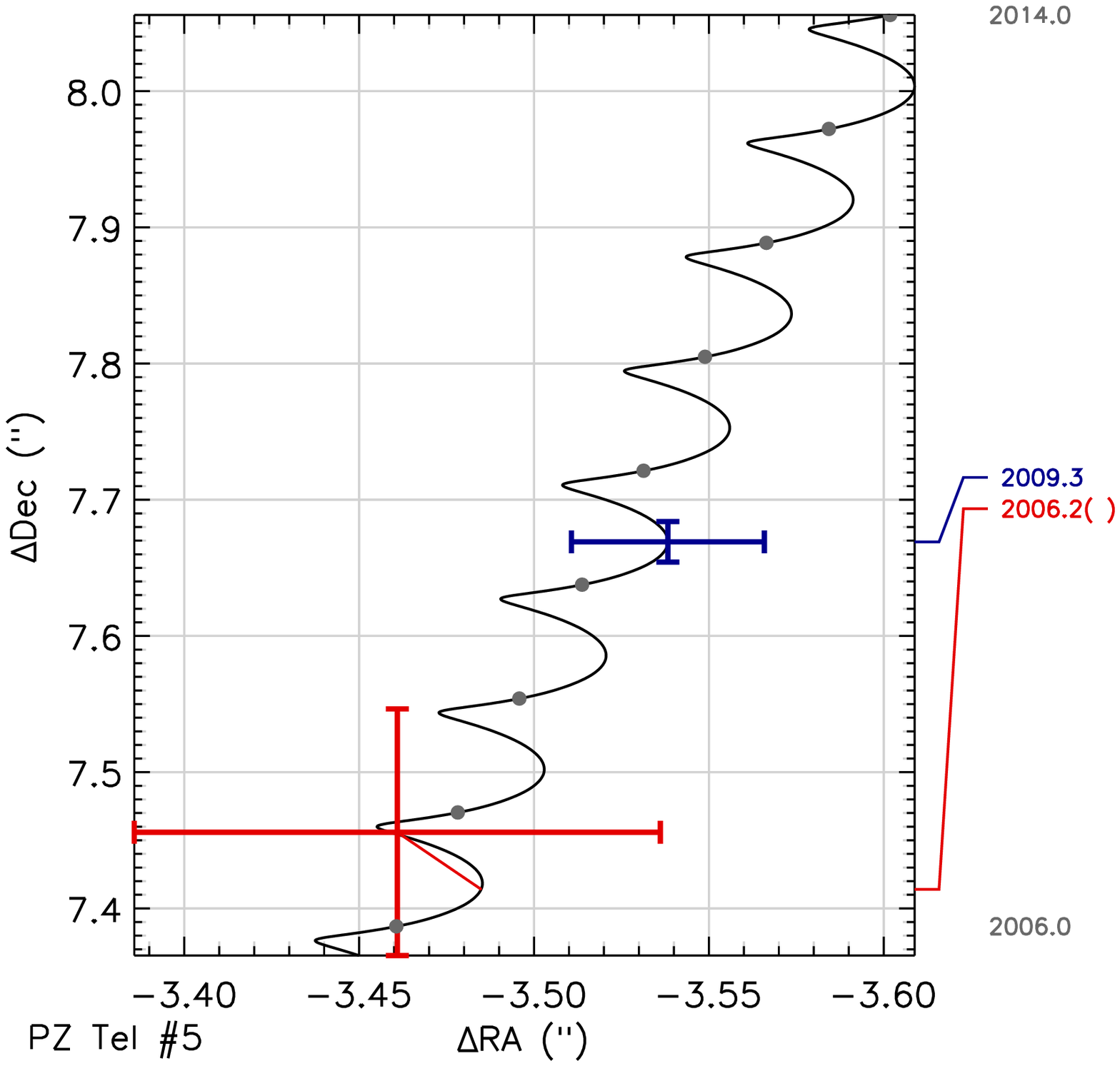}
\hskip -0.3in
\includegraphics[width=2.0in]{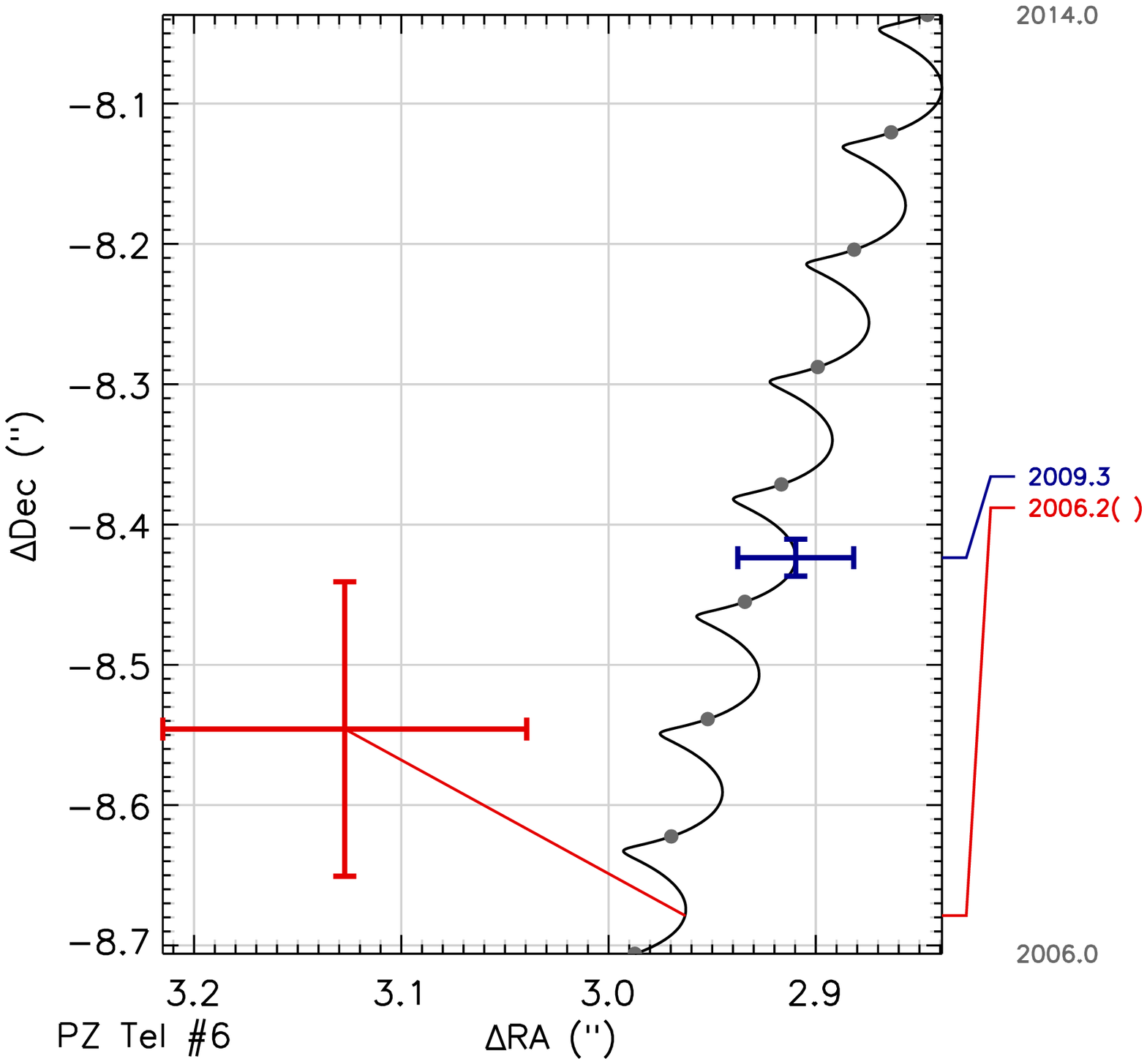}
}
\vskip -0.2in
\centerline{
\includegraphics[width=2.0in]{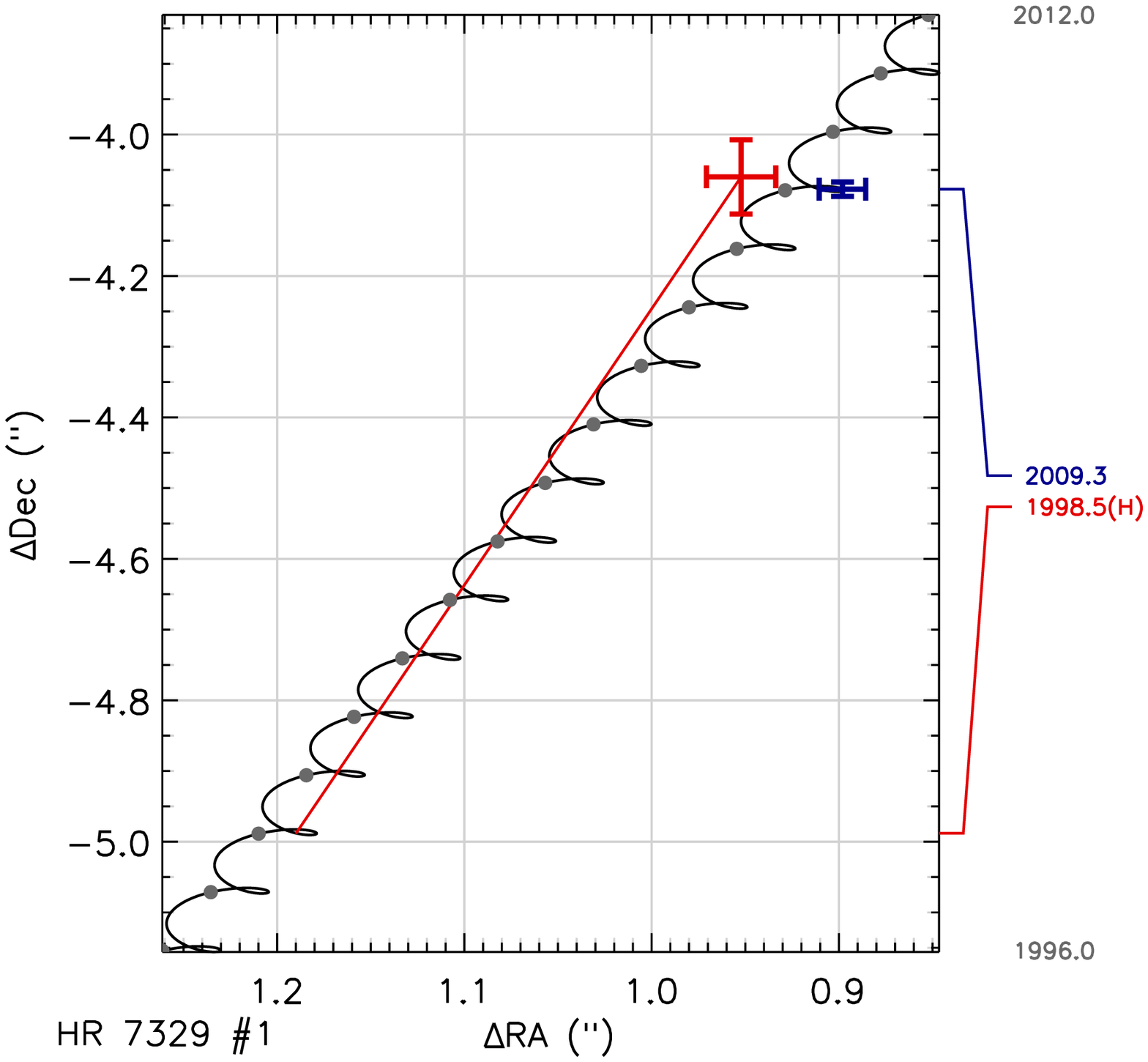}
\hskip -0.3in
\includegraphics[width=2.0in]{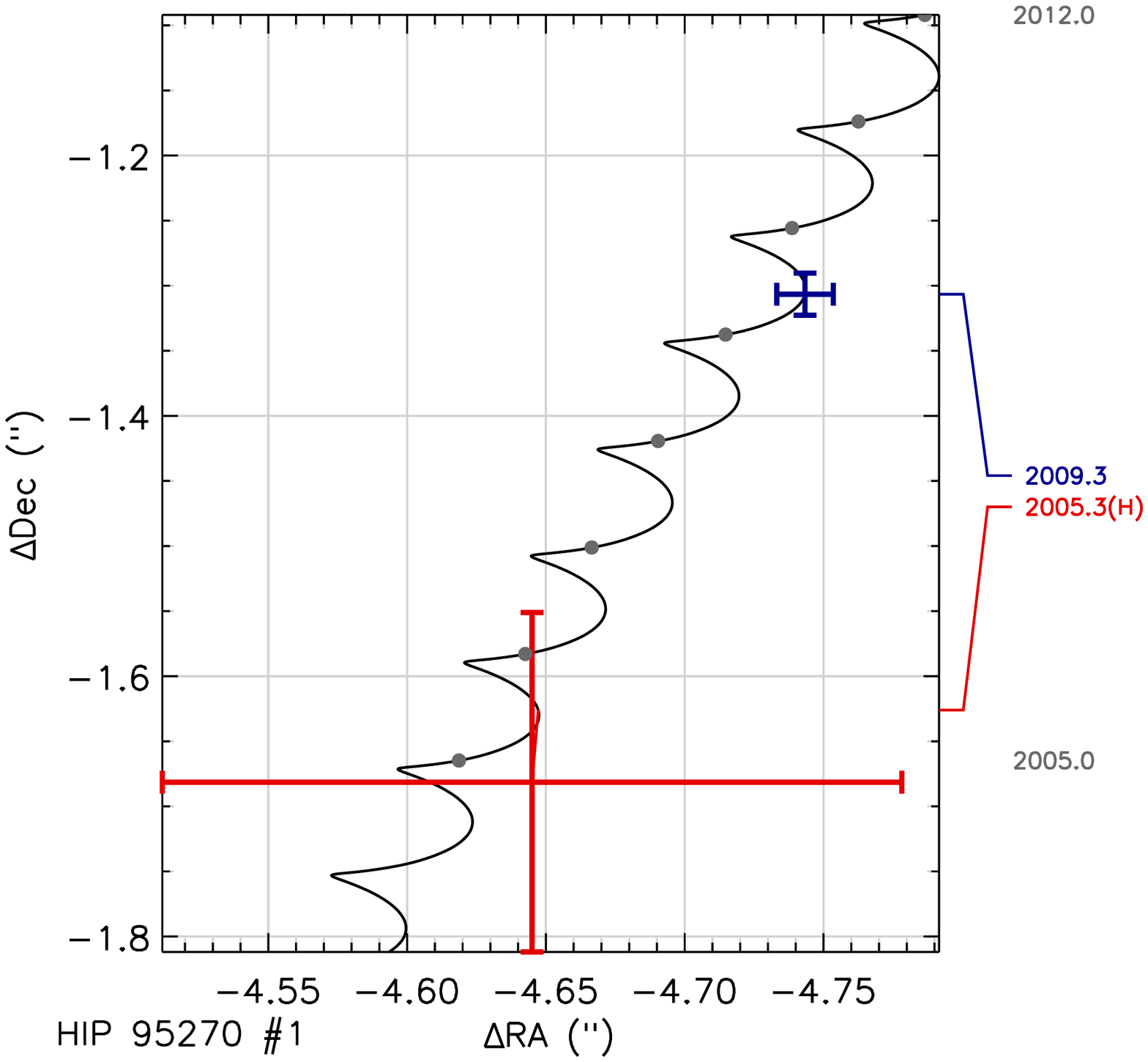}
\hskip -0.3in
\includegraphics[width=2.0in]{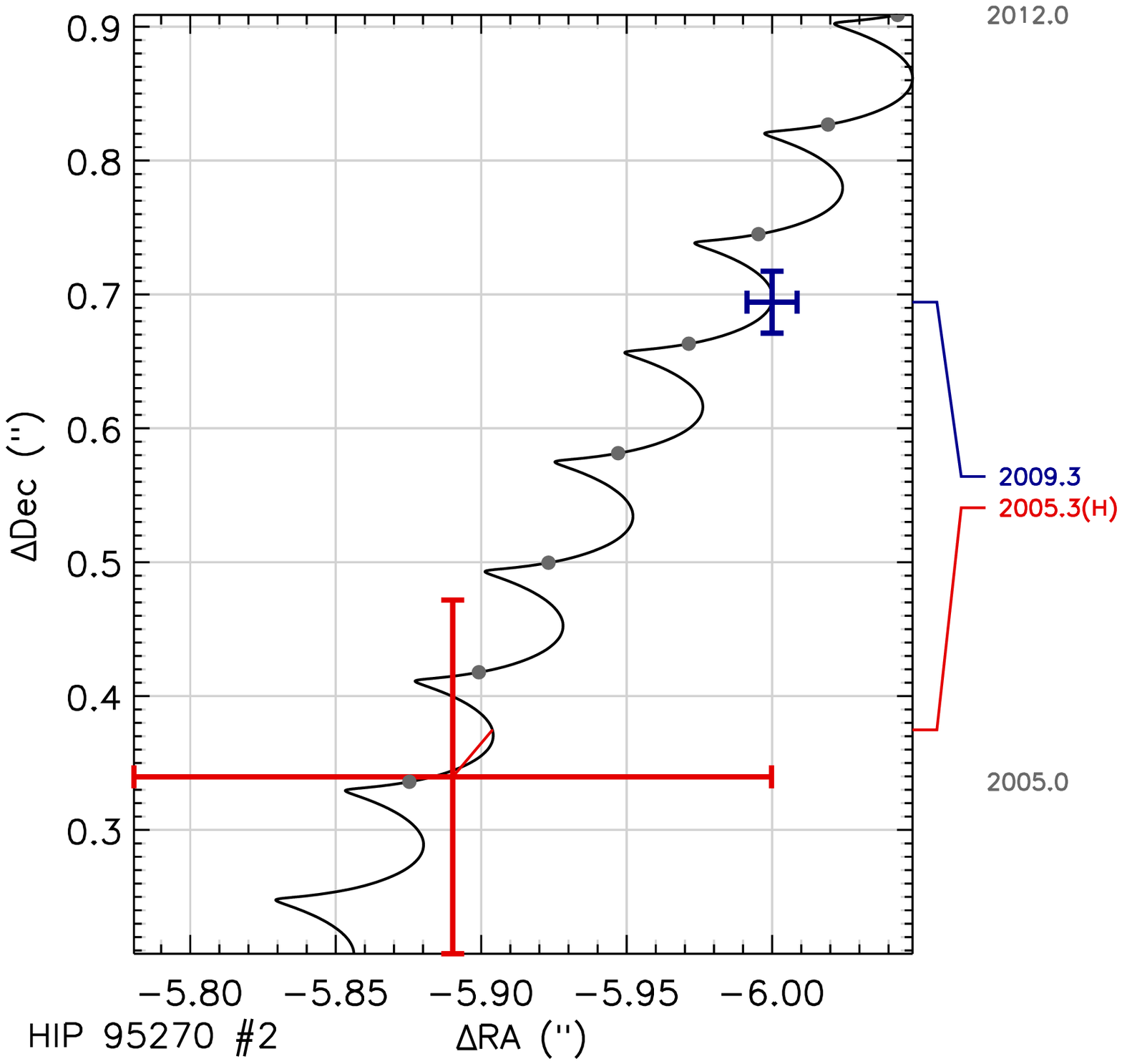}
\hskip -0.3in
\includegraphics[width=2.0in]{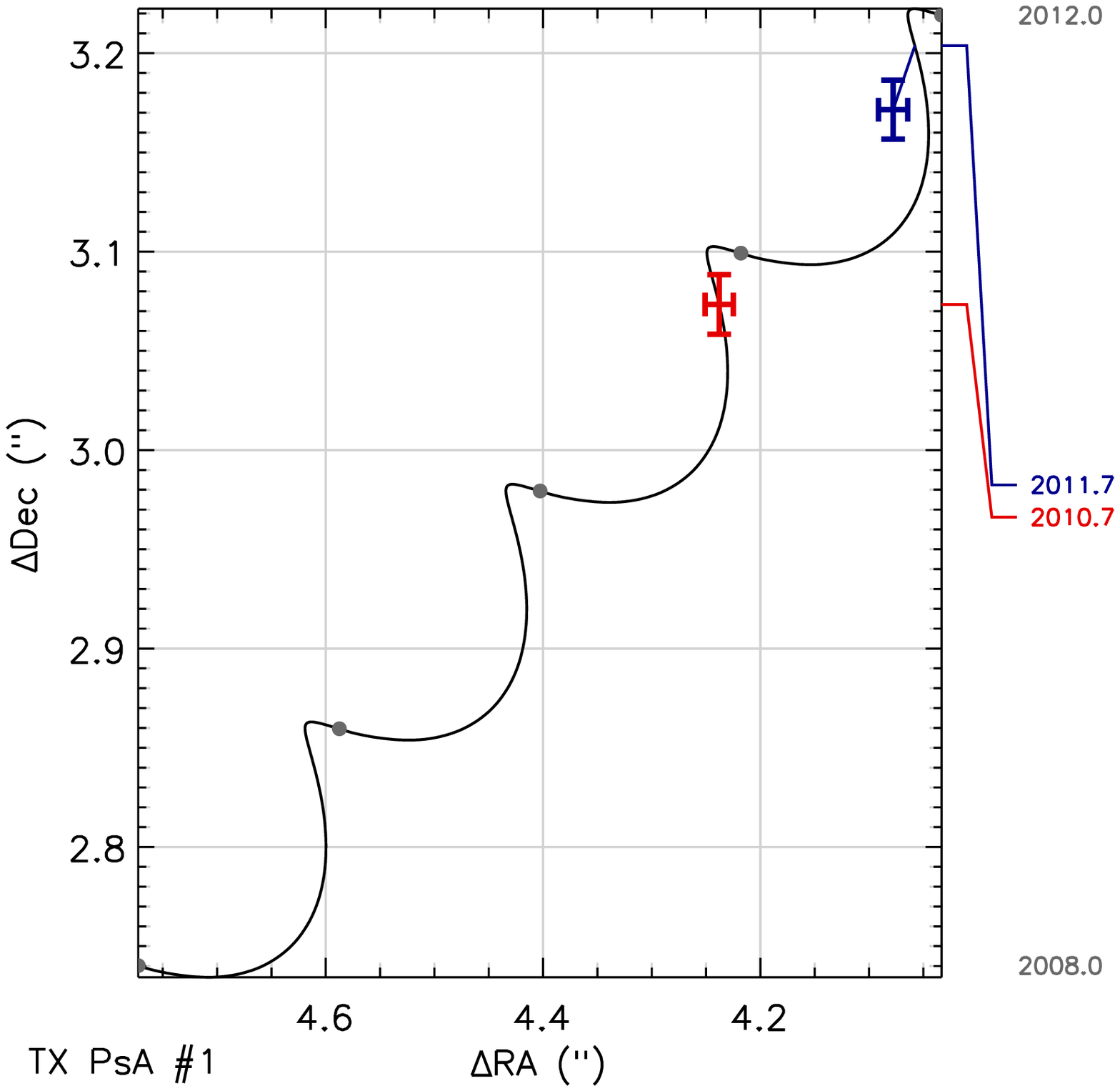}
}
\vskip -0.2in
\centerline{
\includegraphics[width=2.0in]{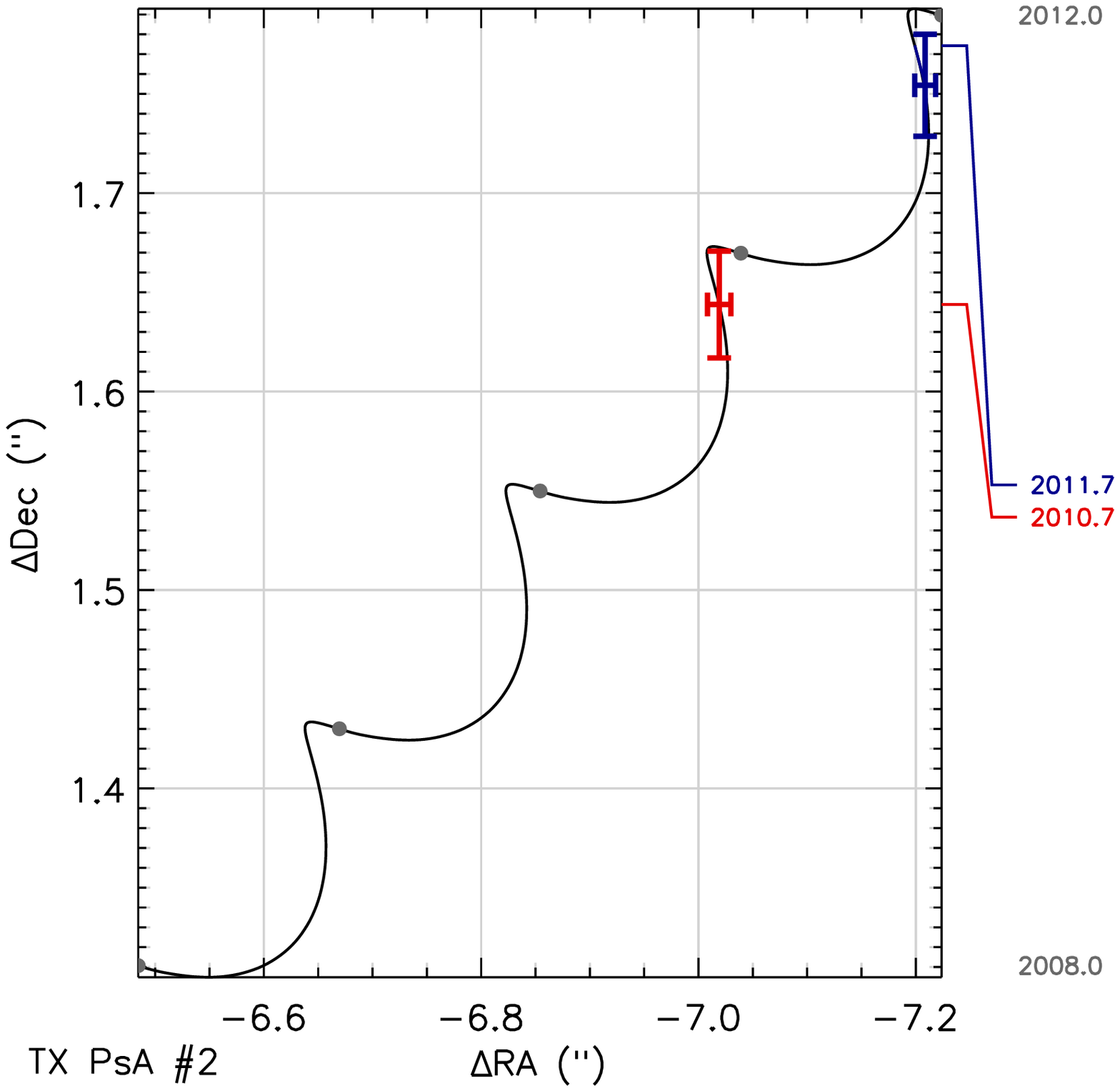}
\hskip -0.3in
}
\caption{\label{fig:BetaPic_skyplots2}
On-sky plots for $\beta$ Pic MG objects (continued).}
\end{figure}

\begin{figure}
\centerline{
\includegraphics[width=2.0in]{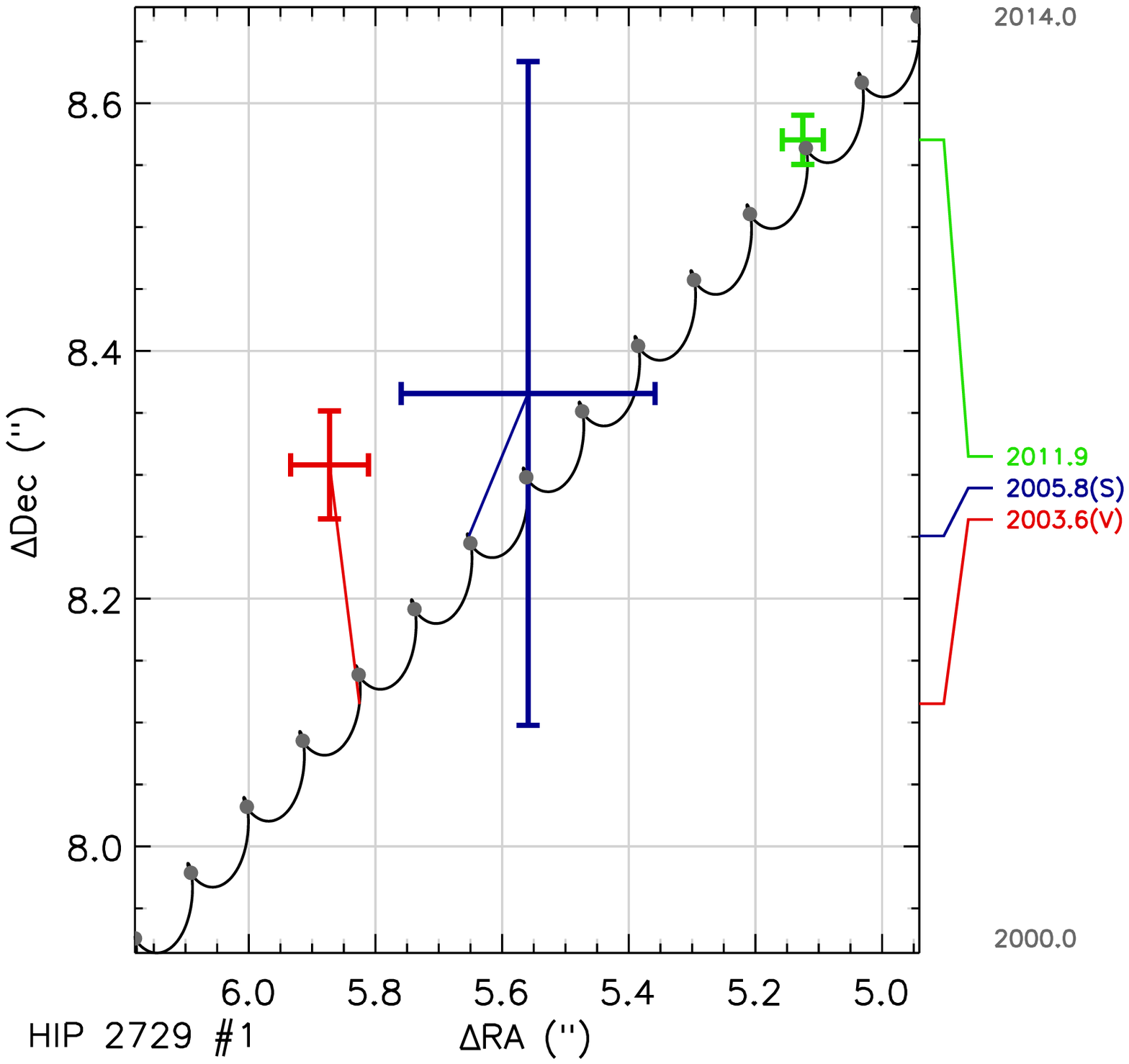}
\hskip -0.3in
\includegraphics[width=2.0in]{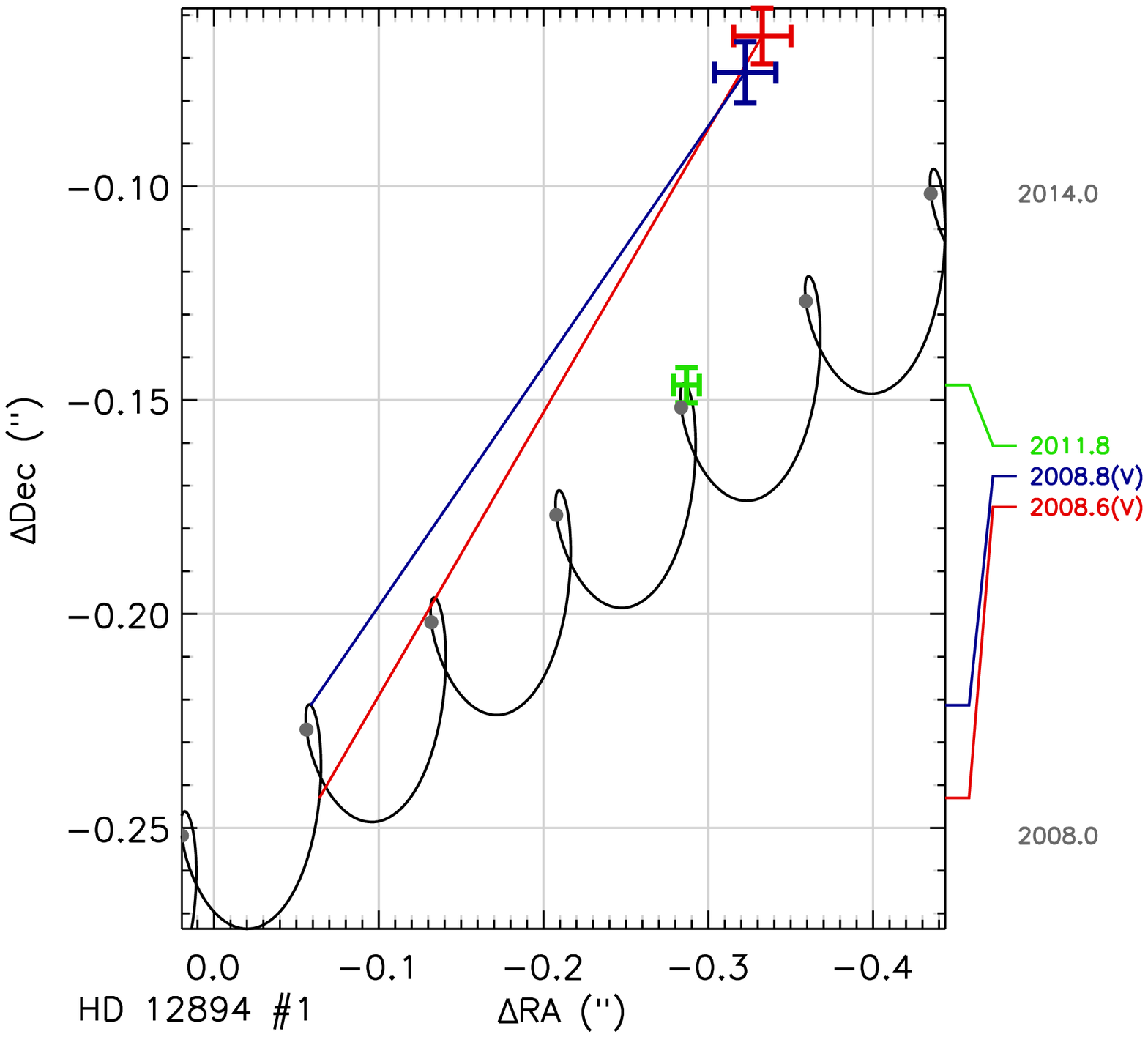}
\hskip -0.3in
\includegraphics[width=2.0in]{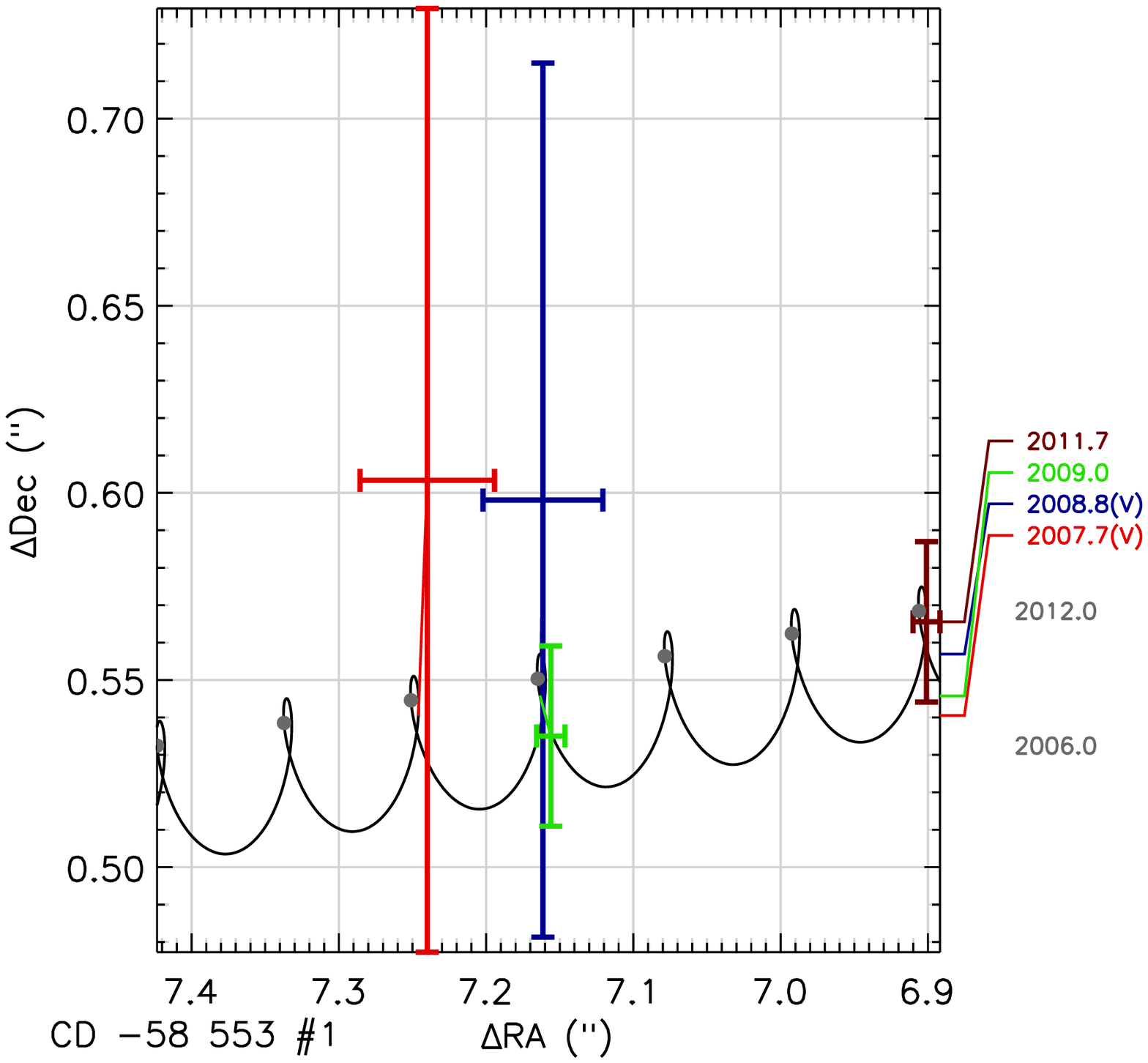}
\hskip -0.3in
\includegraphics[width=2.0in]{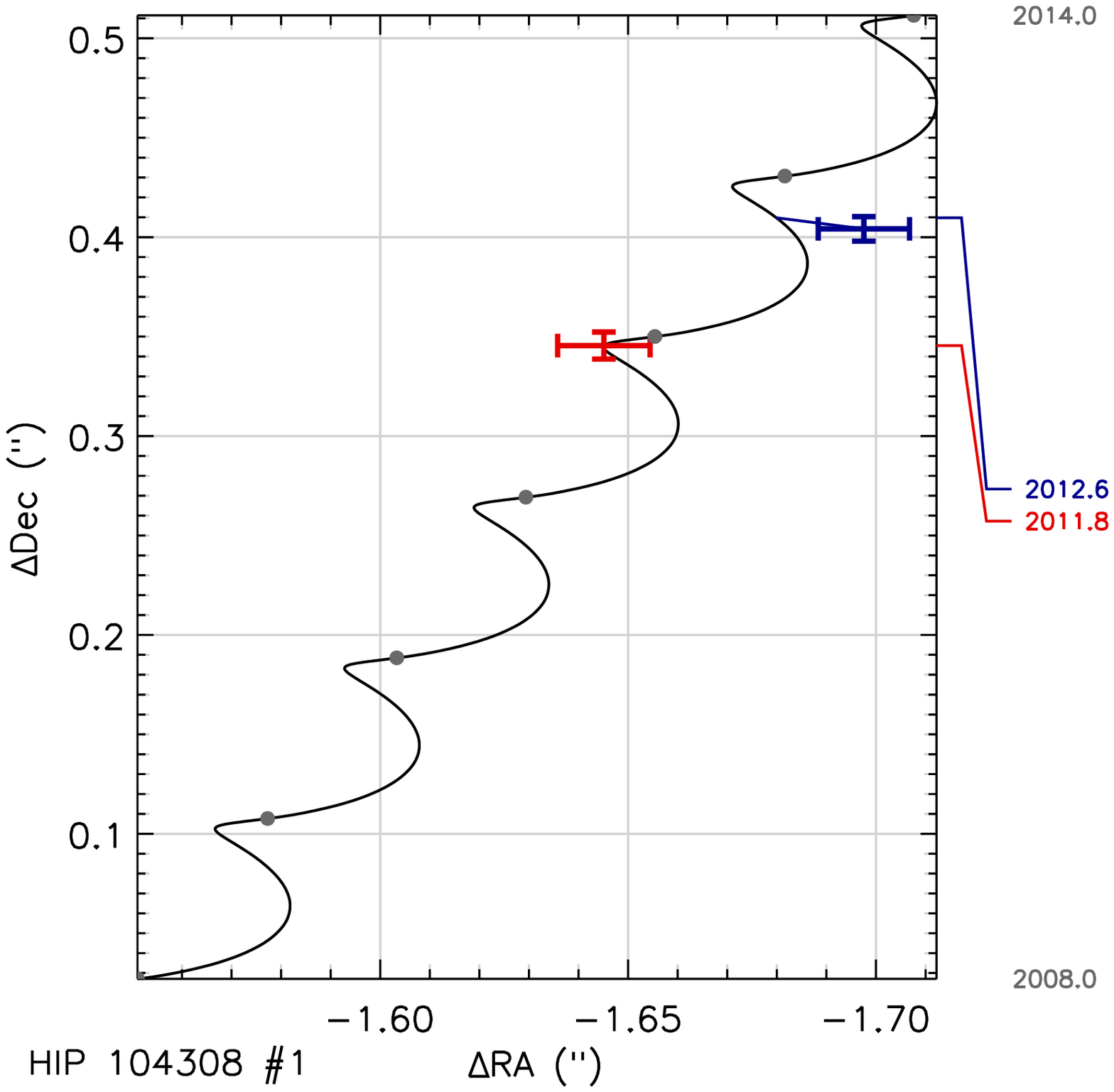}
}
\vskip -0.2in
\centerline{
\includegraphics[width=2.0in]{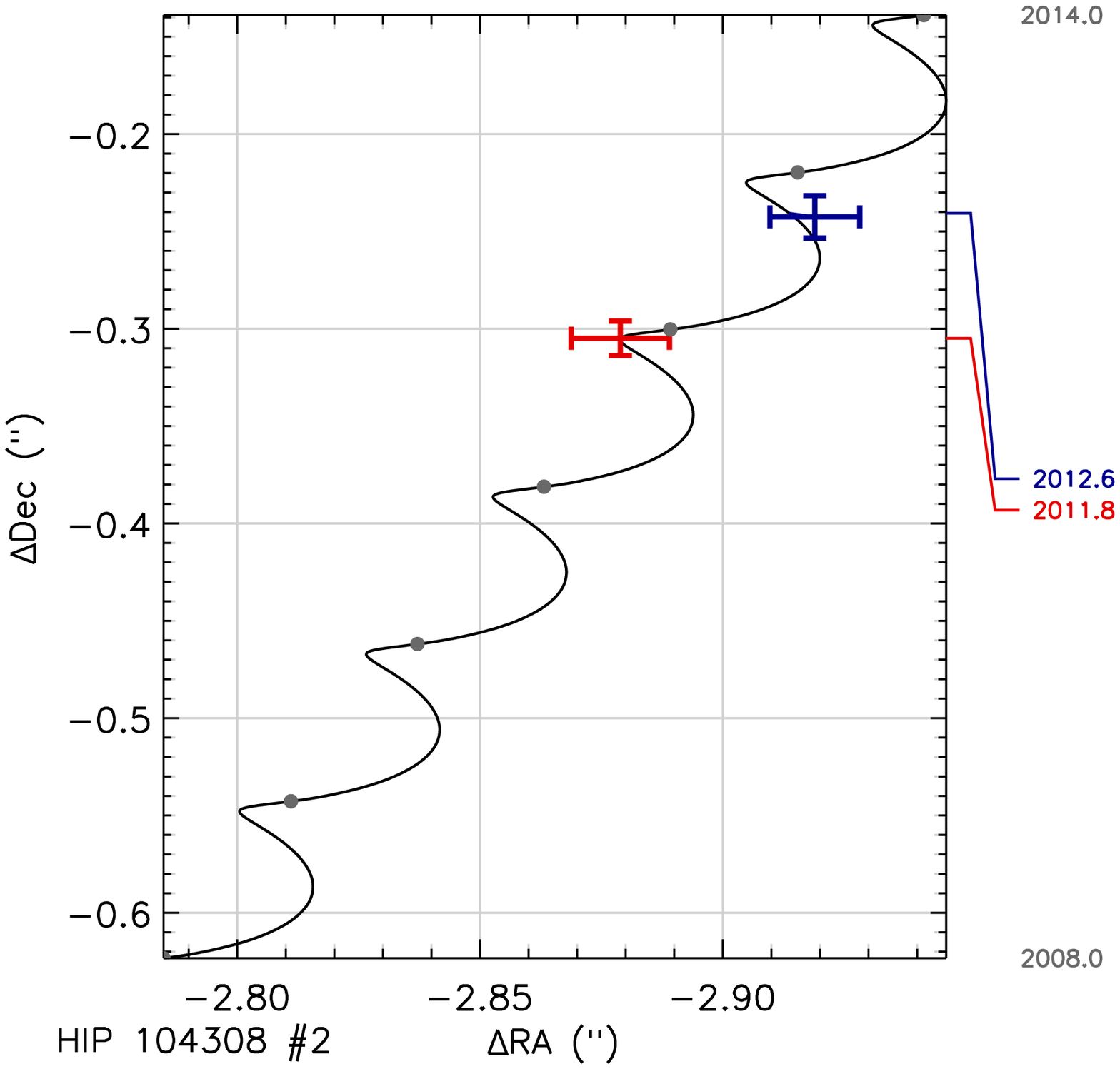}
\hskip -0.3 in
\includegraphics[width=2.0in]{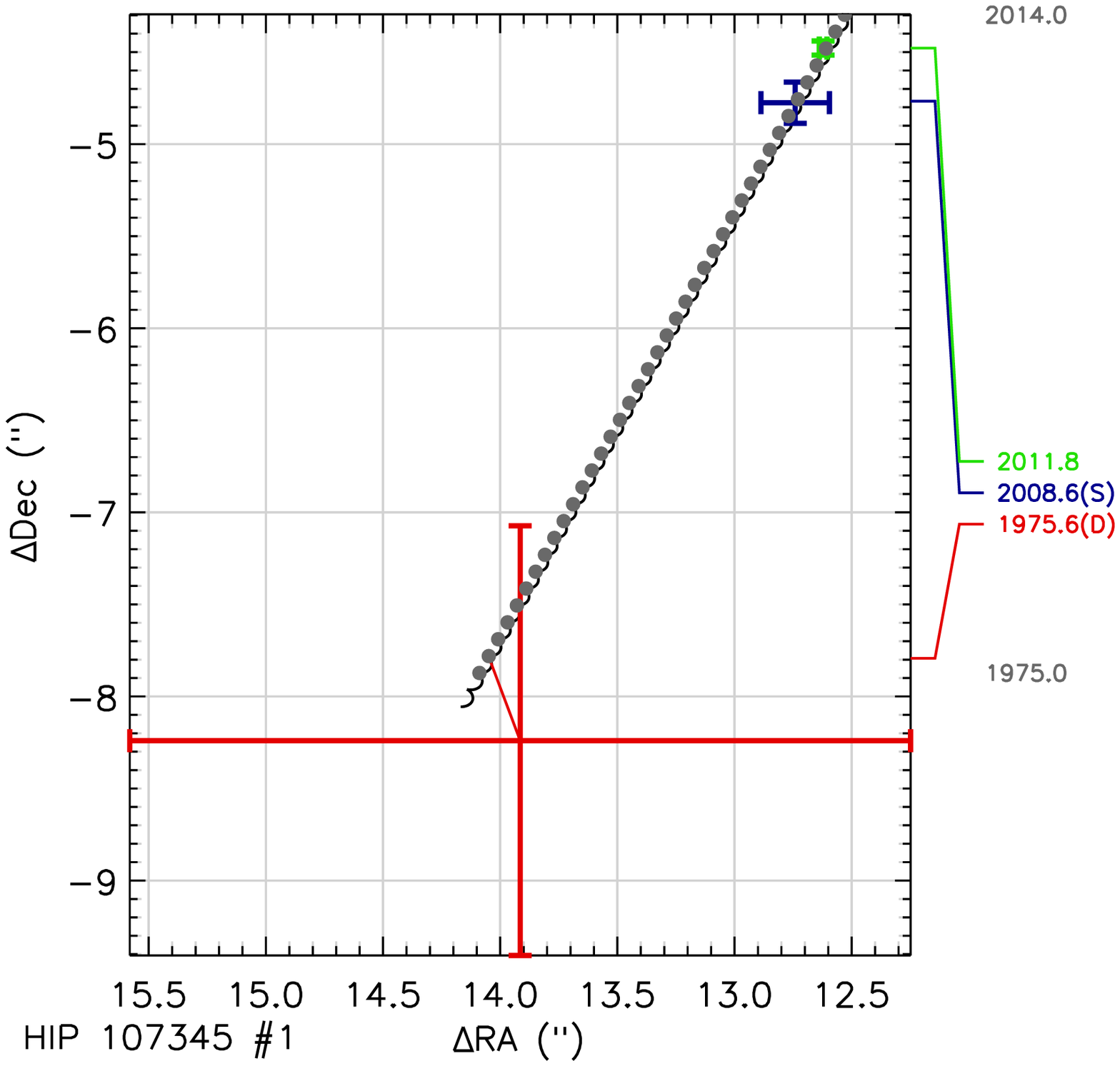}
\hskip -0.3 in
\includegraphics[width=2.0in]{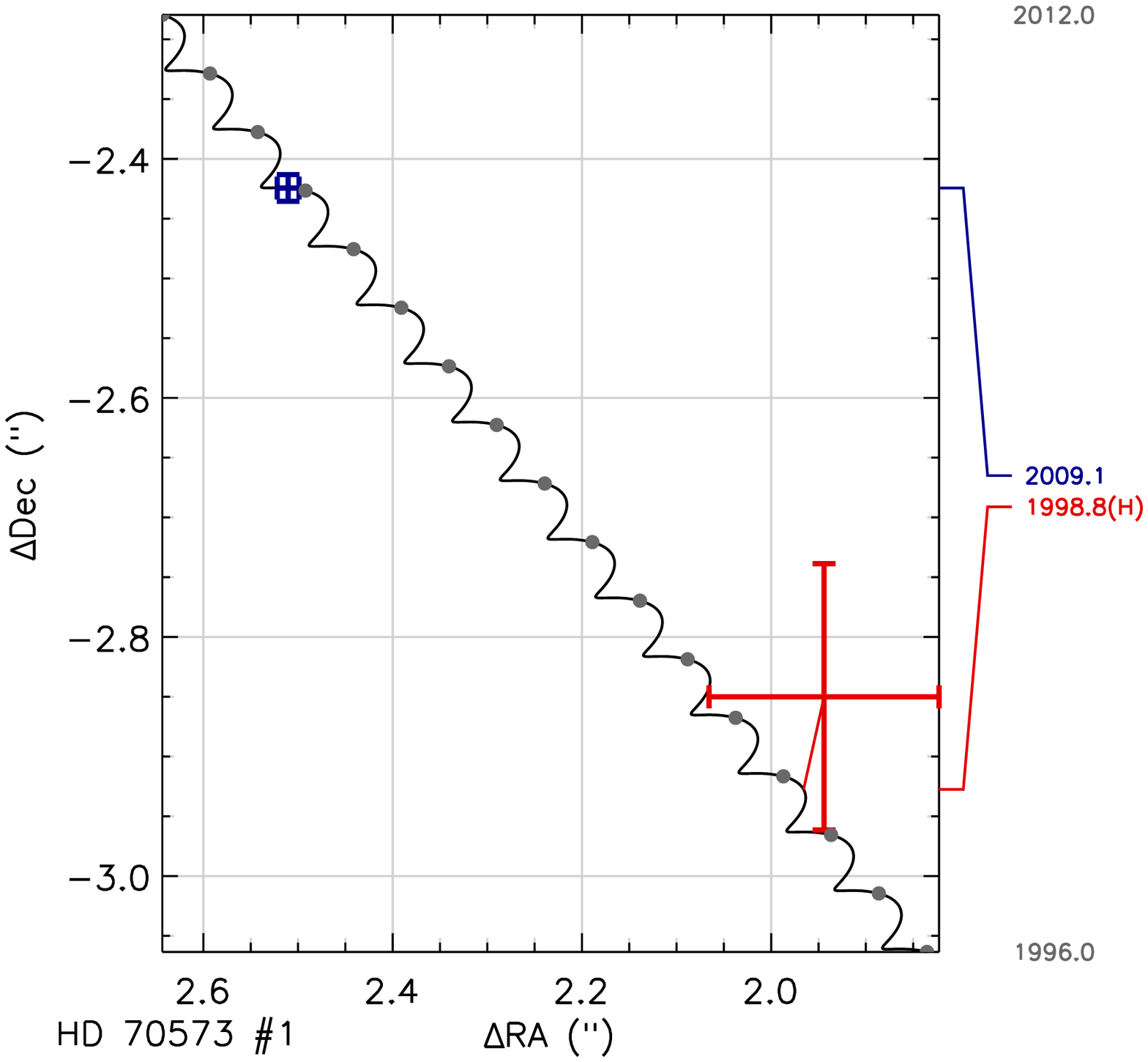}
\hskip -0.3in
\includegraphics[width=2.0in]{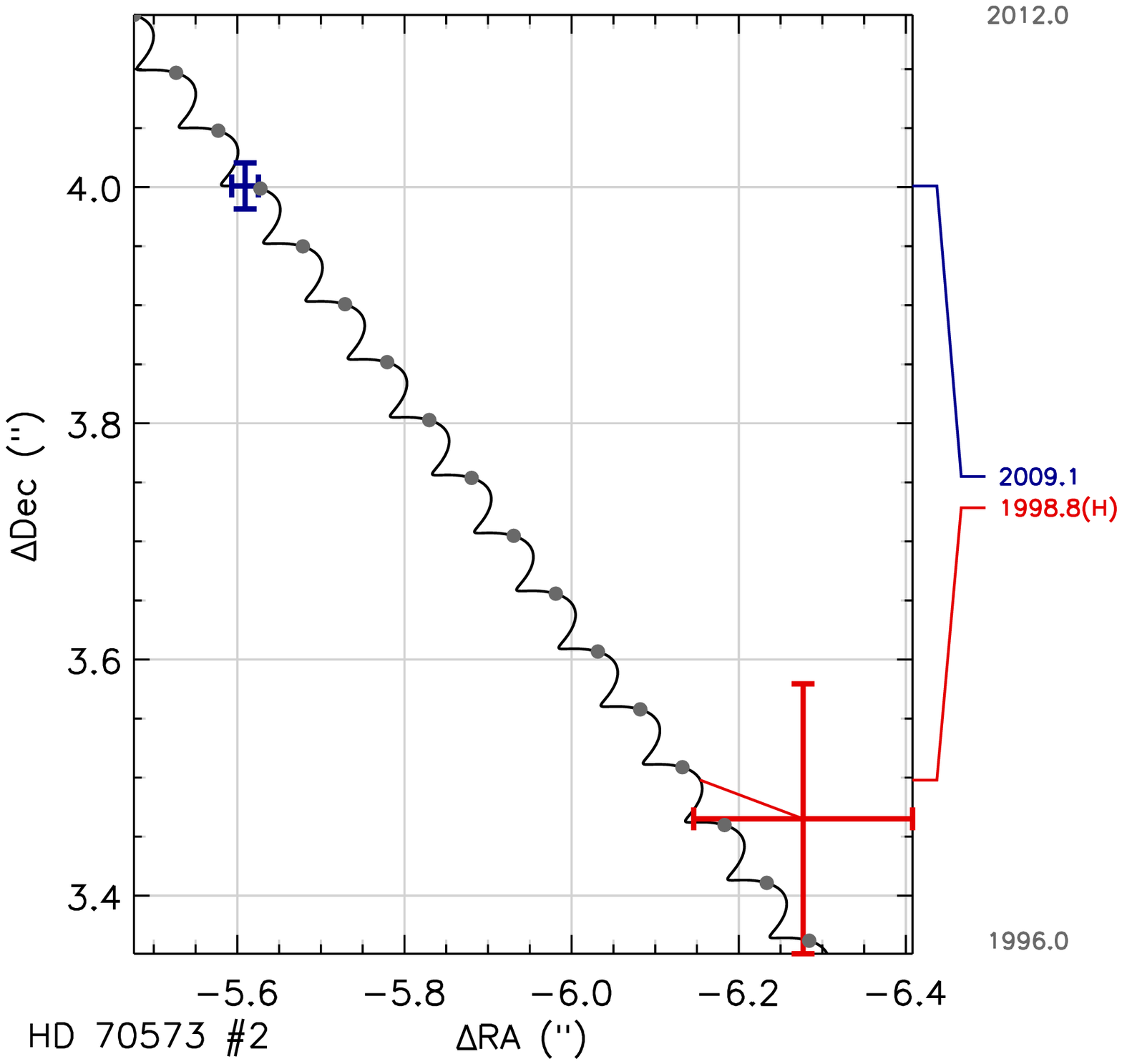}
}
\vskip -0.2in
\centerline{
\includegraphics[width=2.0in]{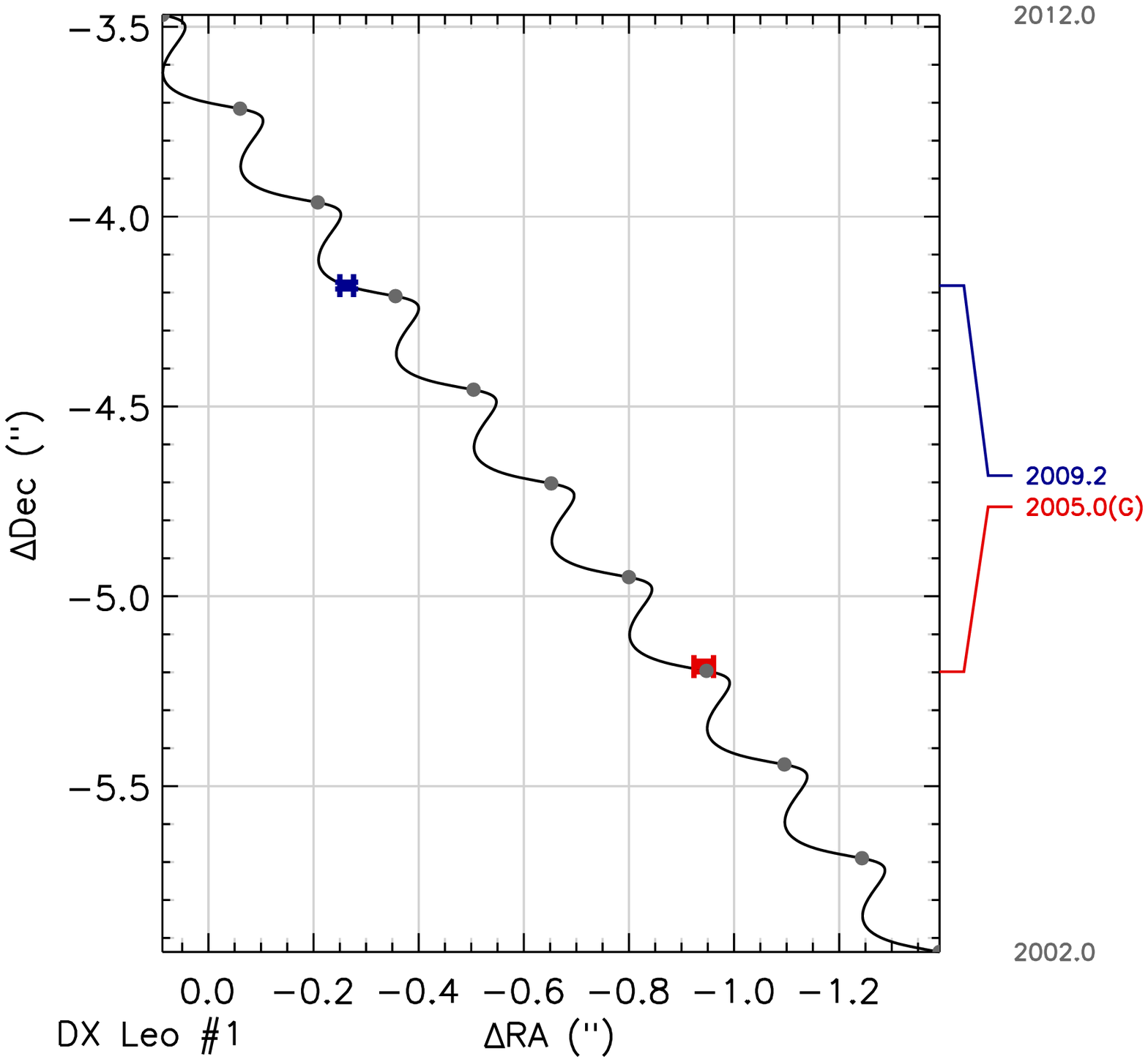}
\hskip -0.3in
\includegraphics[width=2.0in]{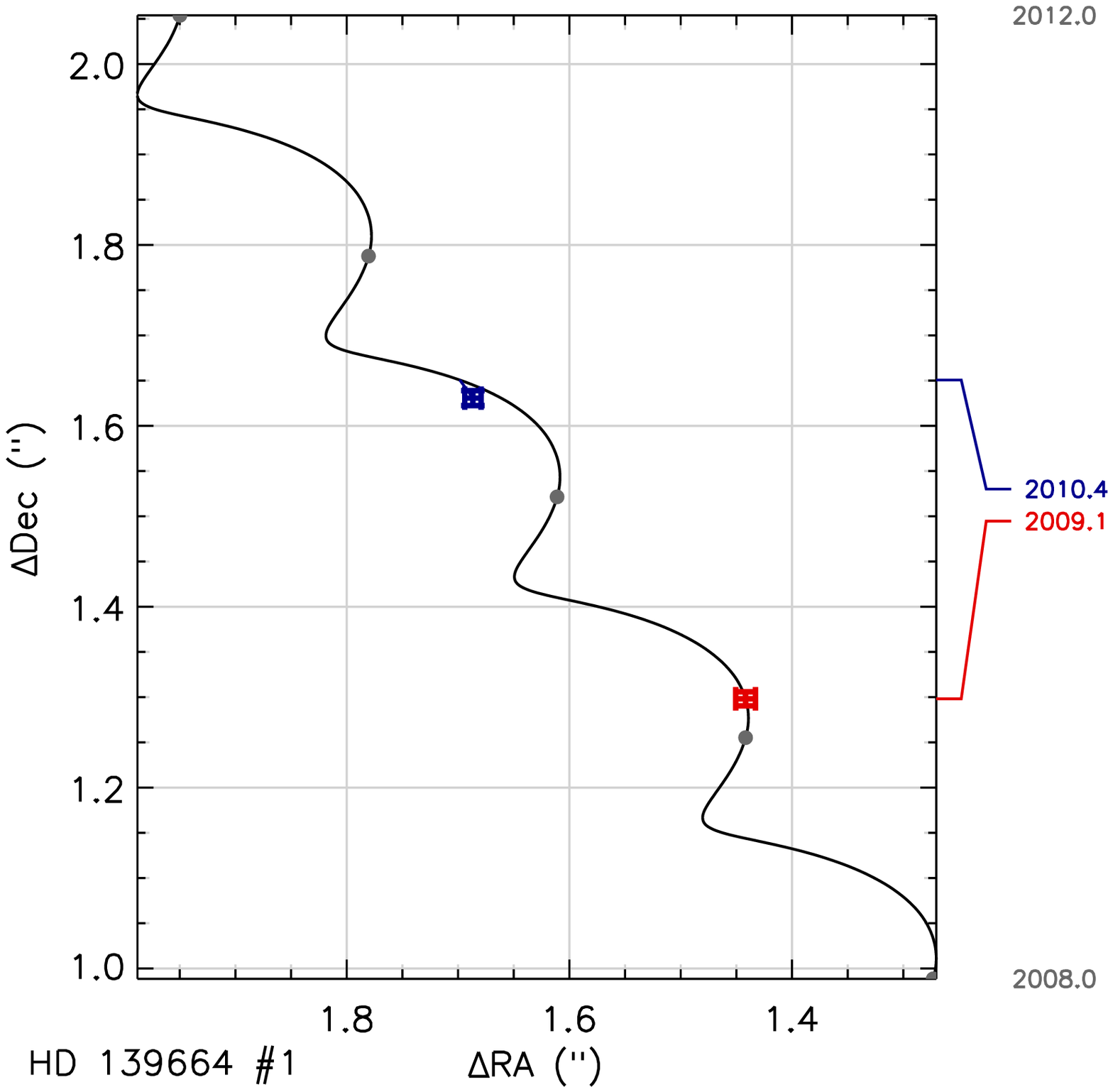}
\hskip -0.3in
\includegraphics[width=2.0in]{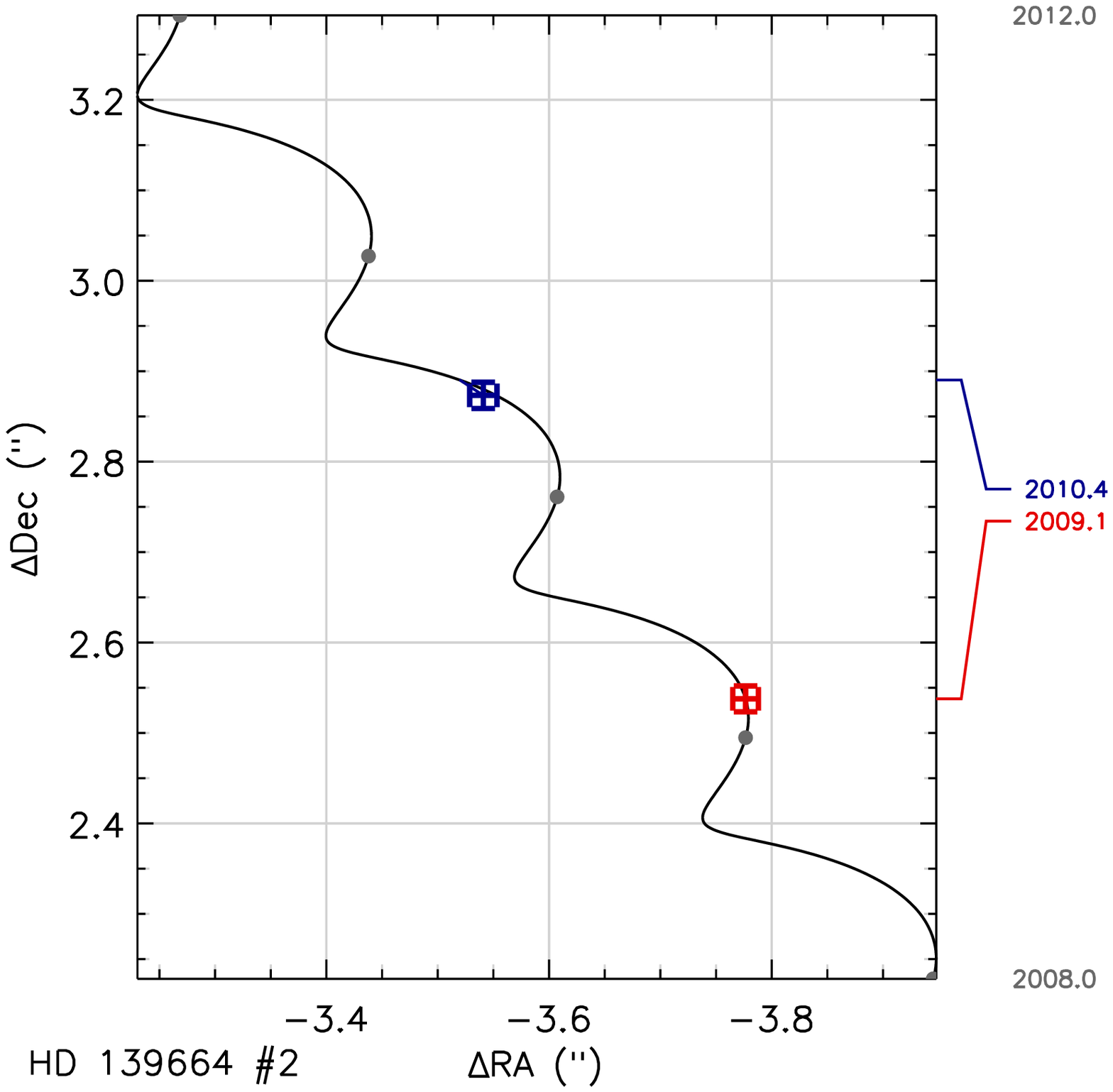}
\hskip -0.3in
\includegraphics[width=2.0in]{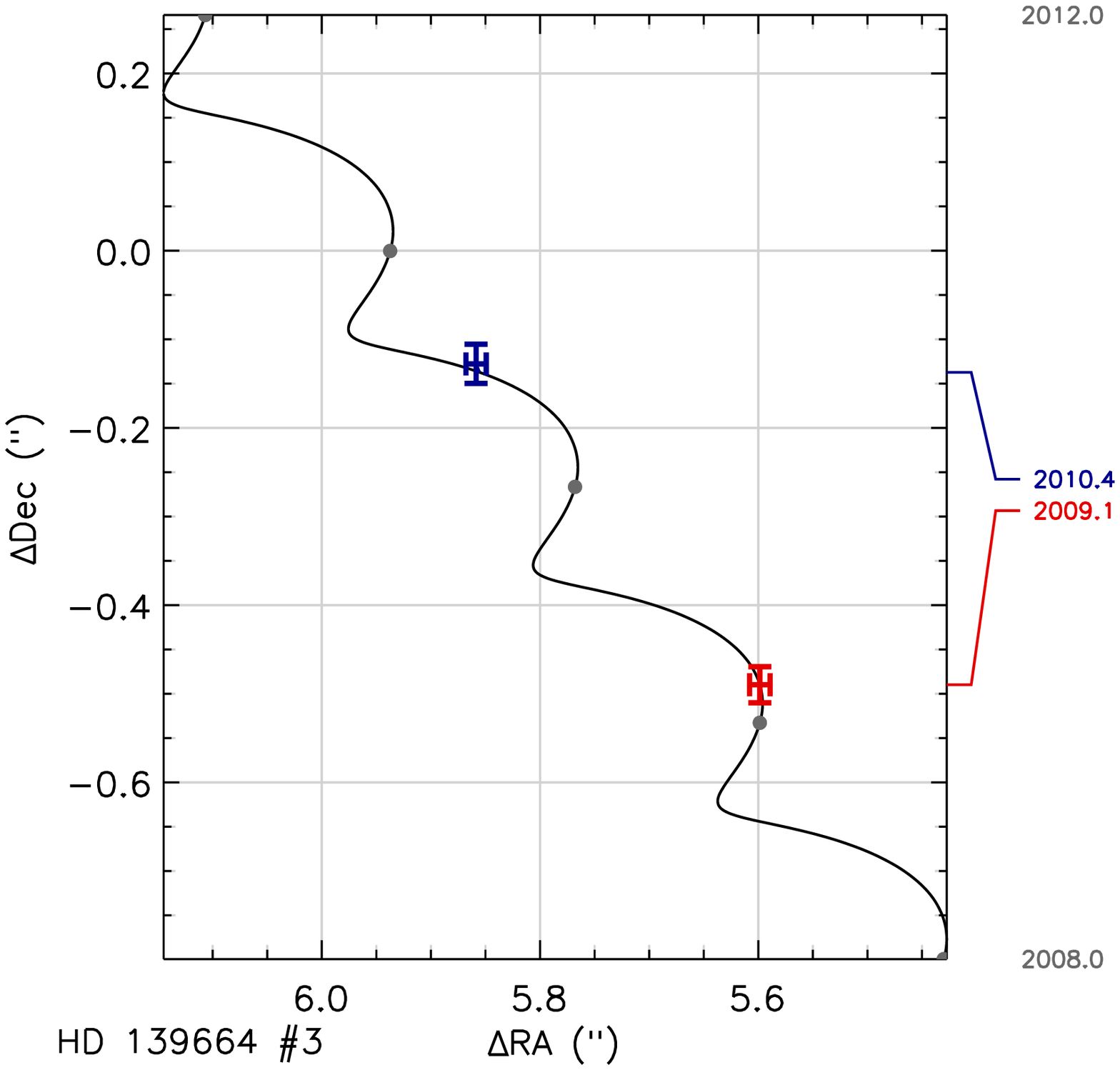}
}
\vskip -0.2in
\centerline{
\includegraphics[width=2.0in]{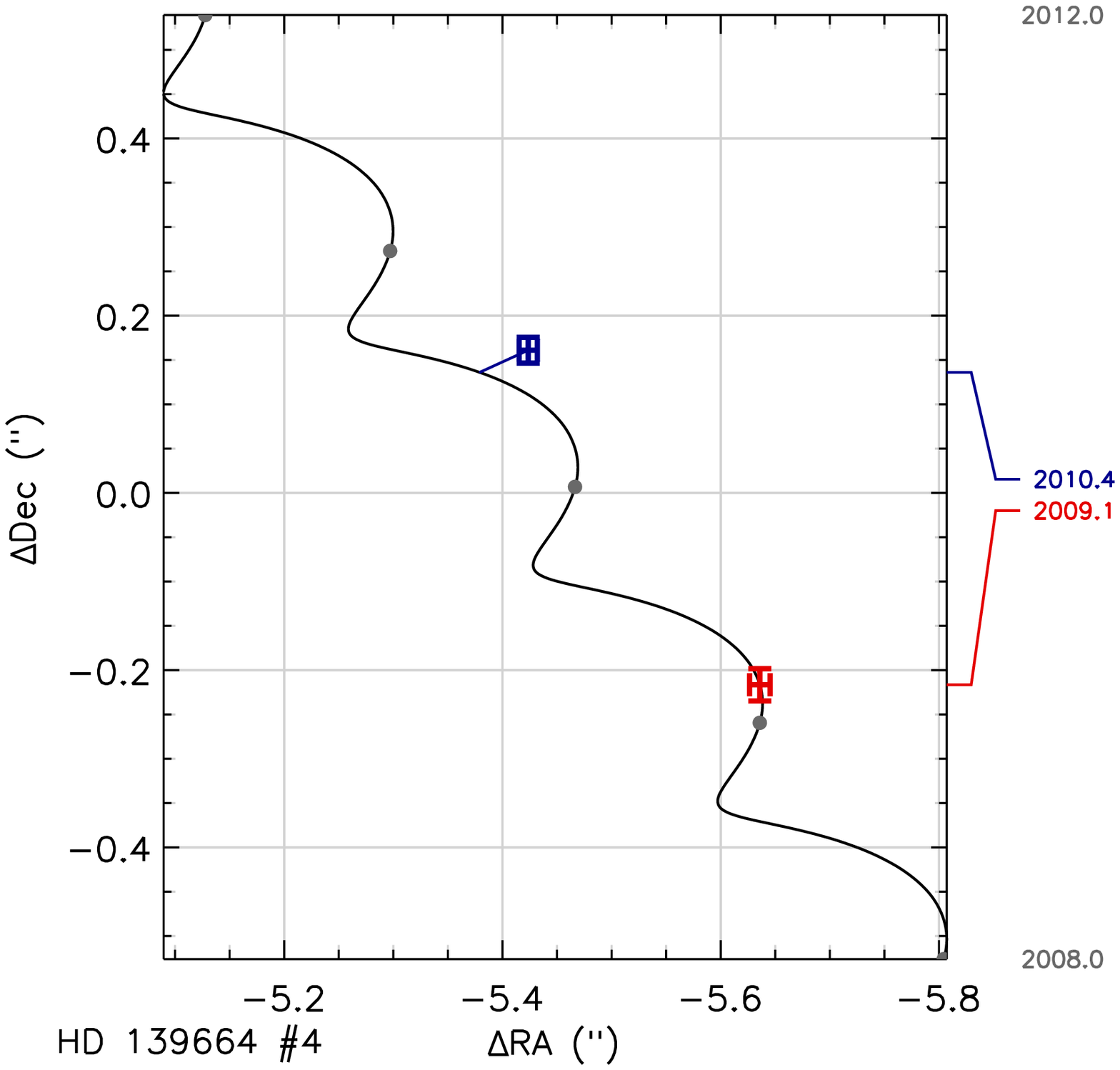}
\hskip -0.3in
\includegraphics[width=2.0in]{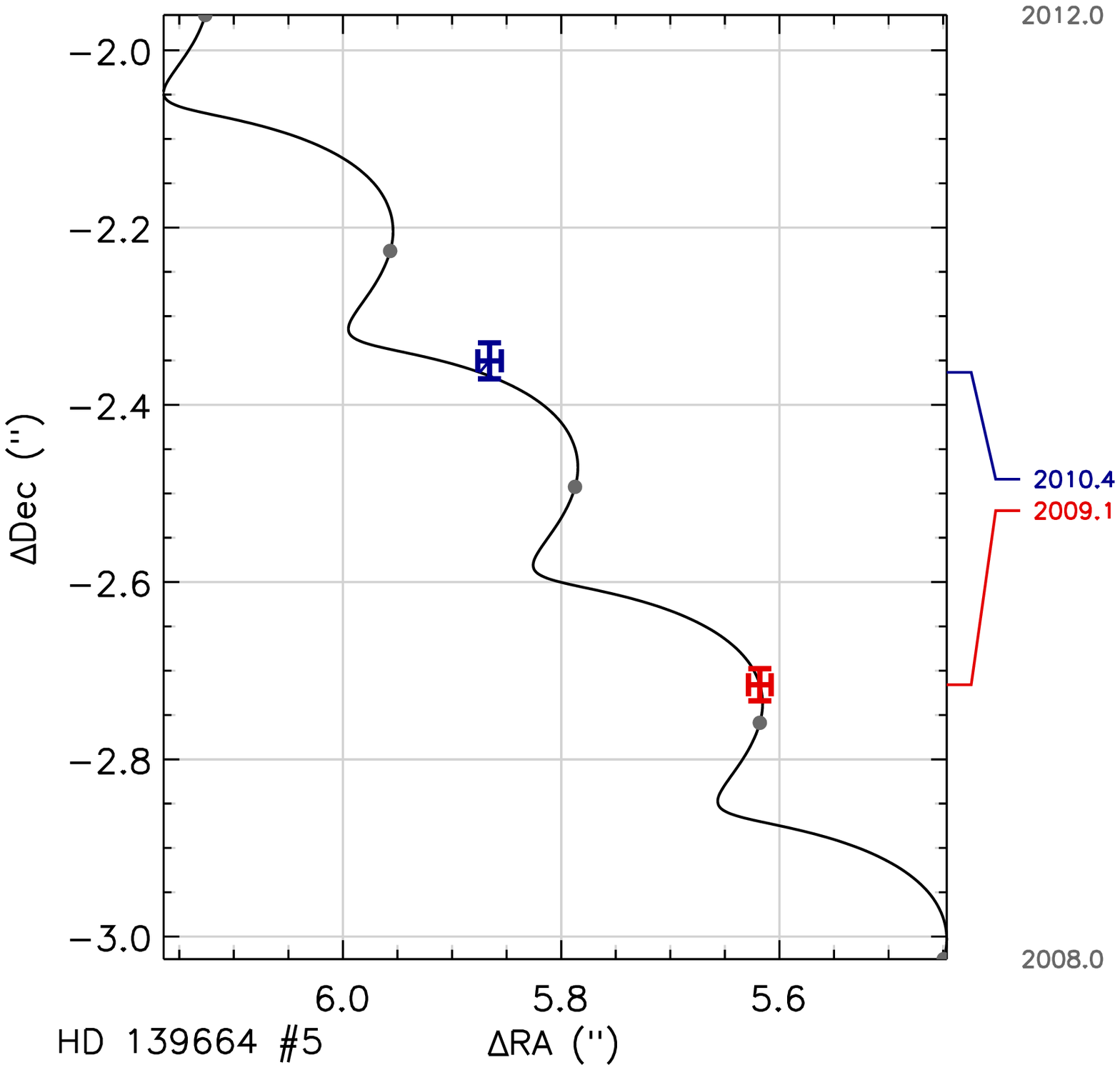}
\hskip -0.3in
\includegraphics[width=2.0in]{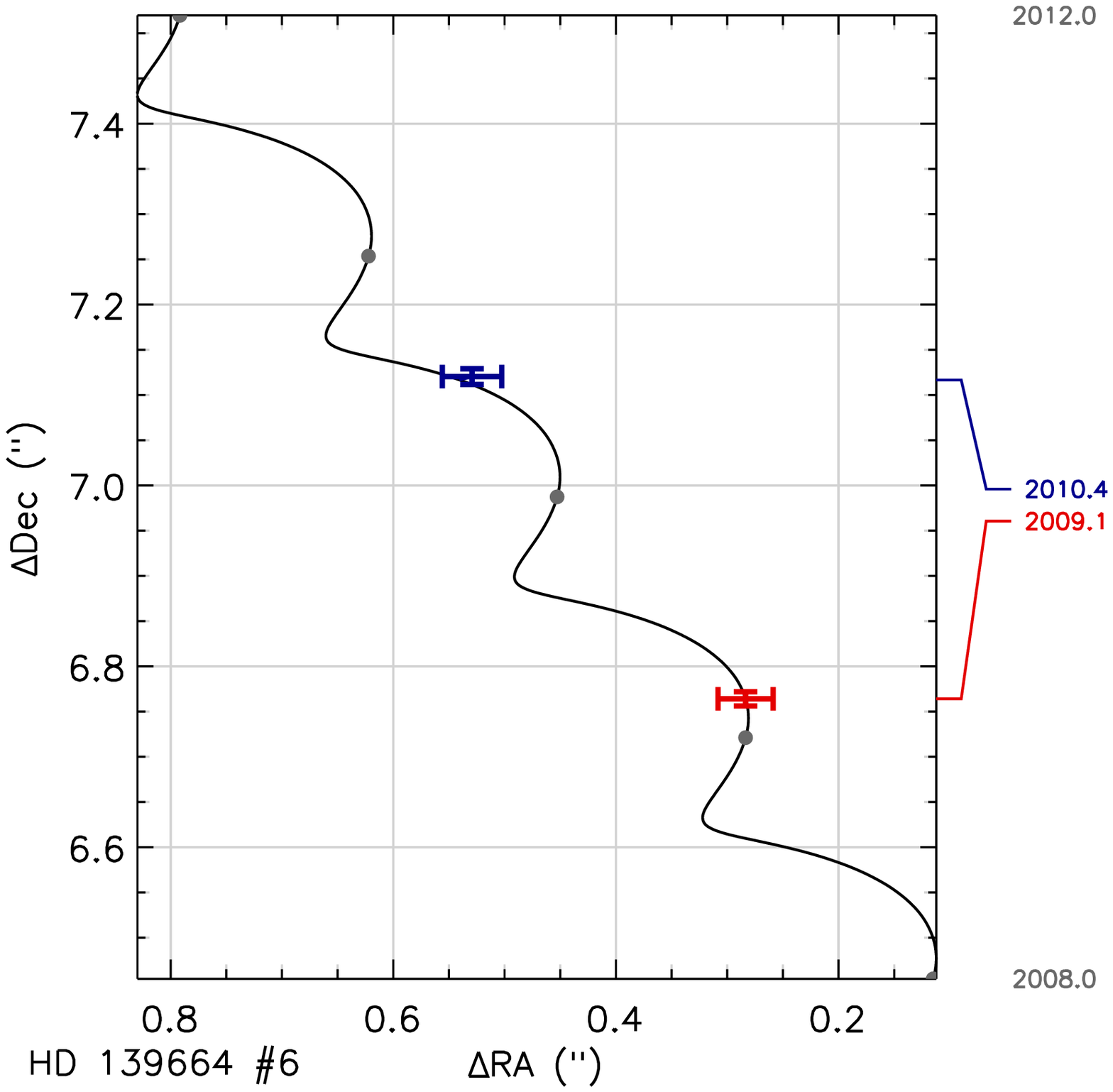}
\hskip -0.3in
}
\vskip -0.2in
\centerline{
}
\caption{\label{fig:TucHor_skyplots} On-sky plots for
  Tucana-Horologium and Hercules-Lyra association objects. }
\end{figure}

\begin{figure}
\centerline{
\includegraphics[width=2.0in]{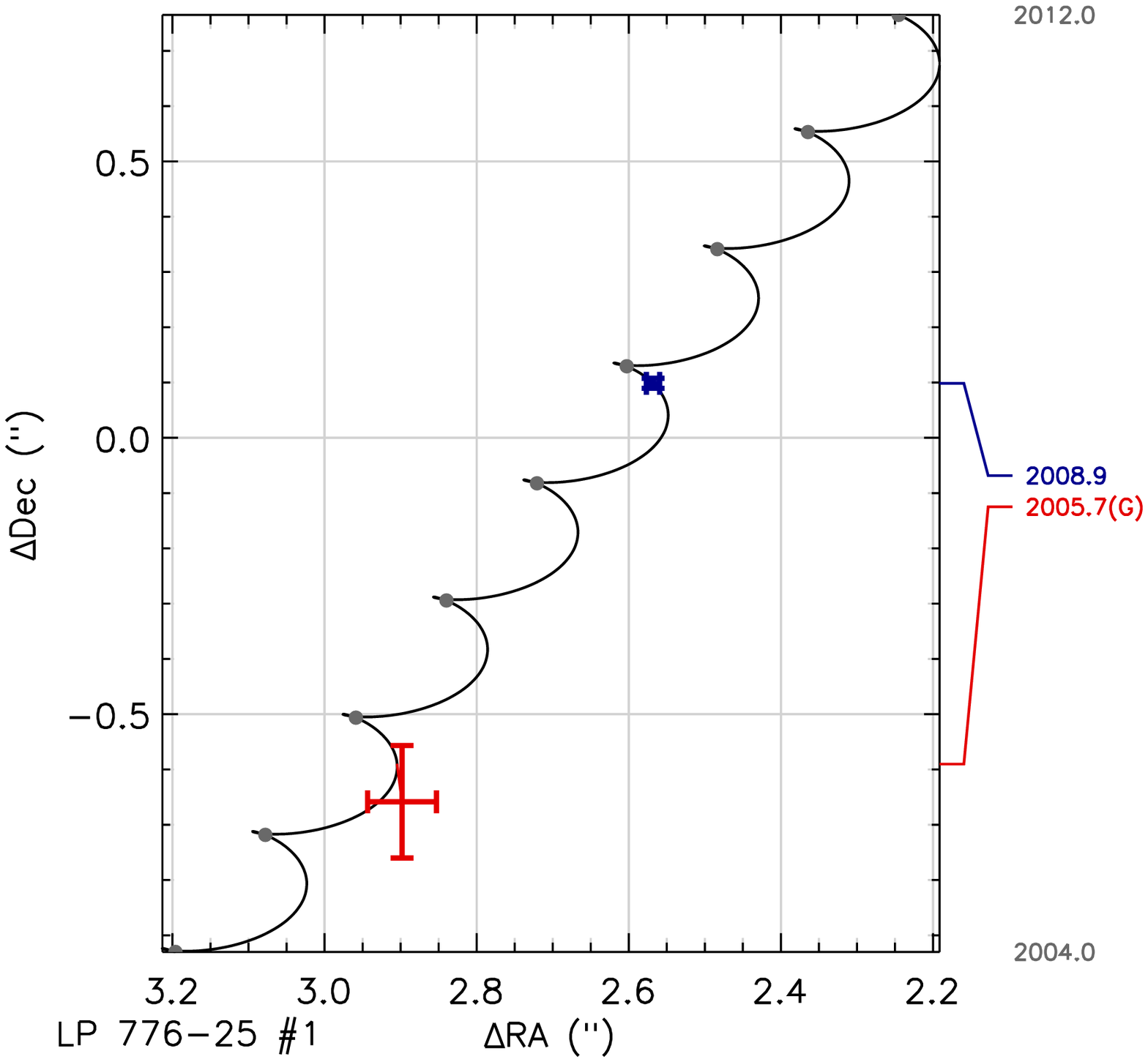}
\hskip -0.3in
\includegraphics[width=2.0in]{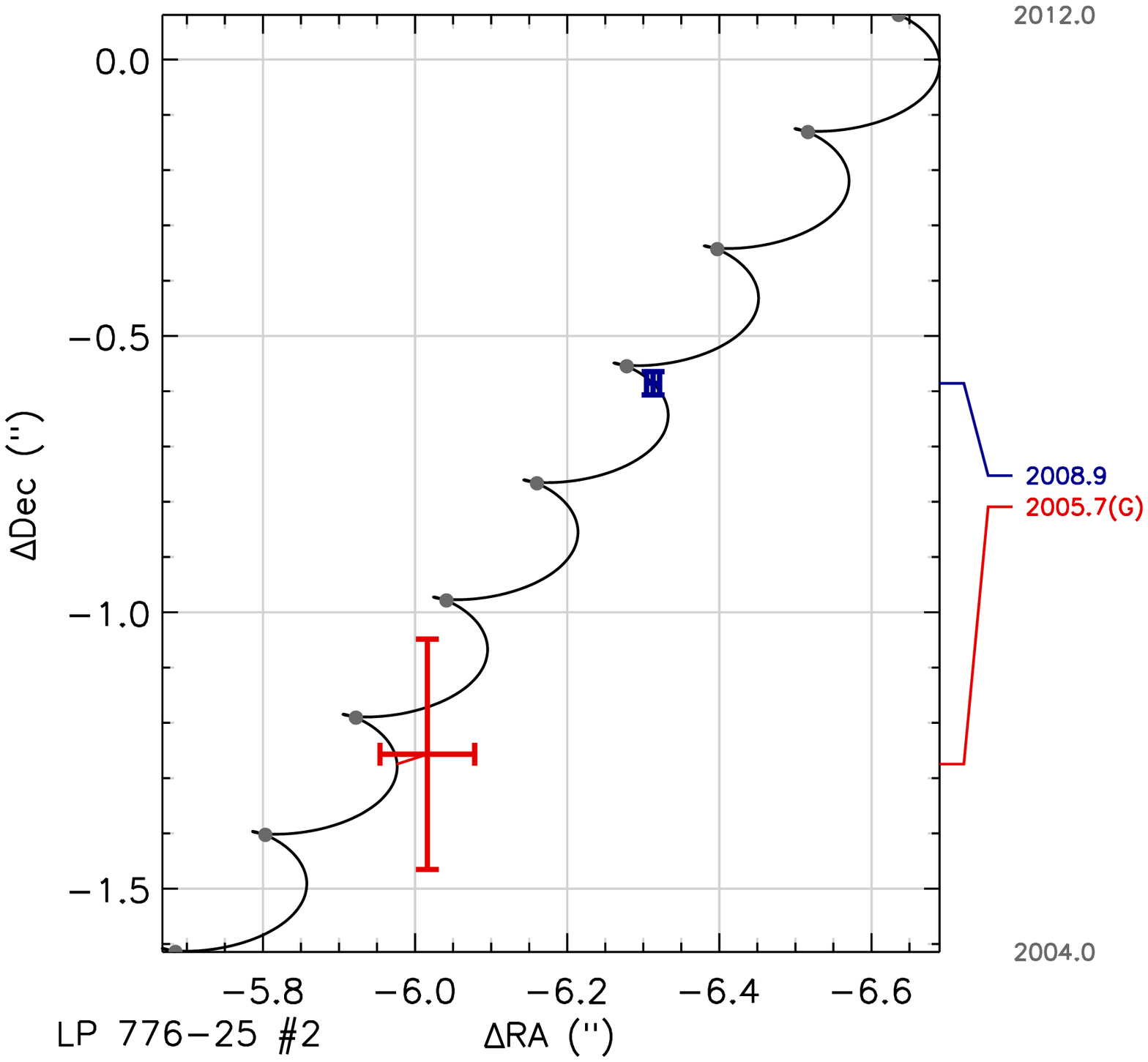}
\hskip -0.3in
\includegraphics[width=2.0in]{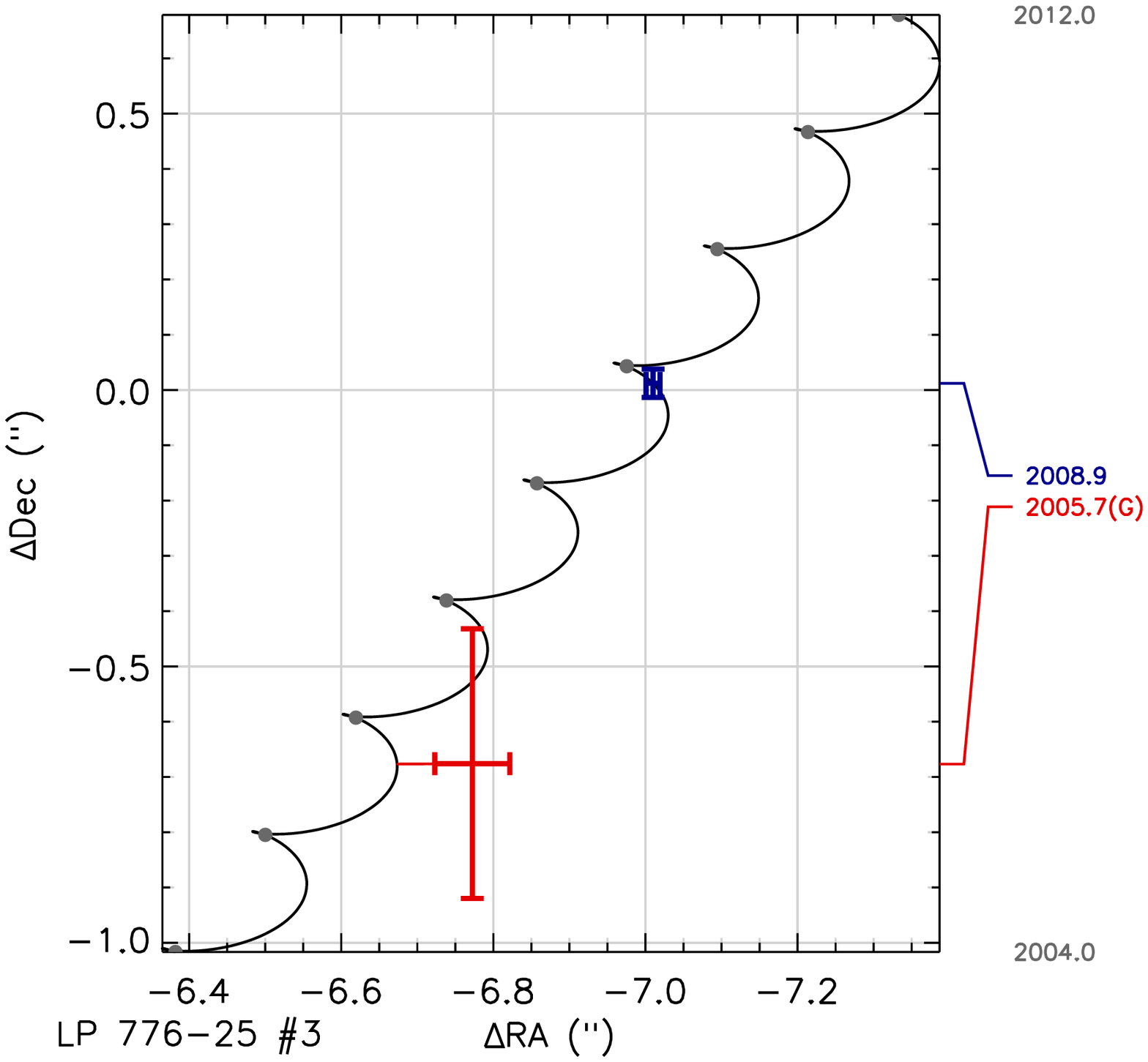}
\hskip -0.3in
\includegraphics[width=2.0in]{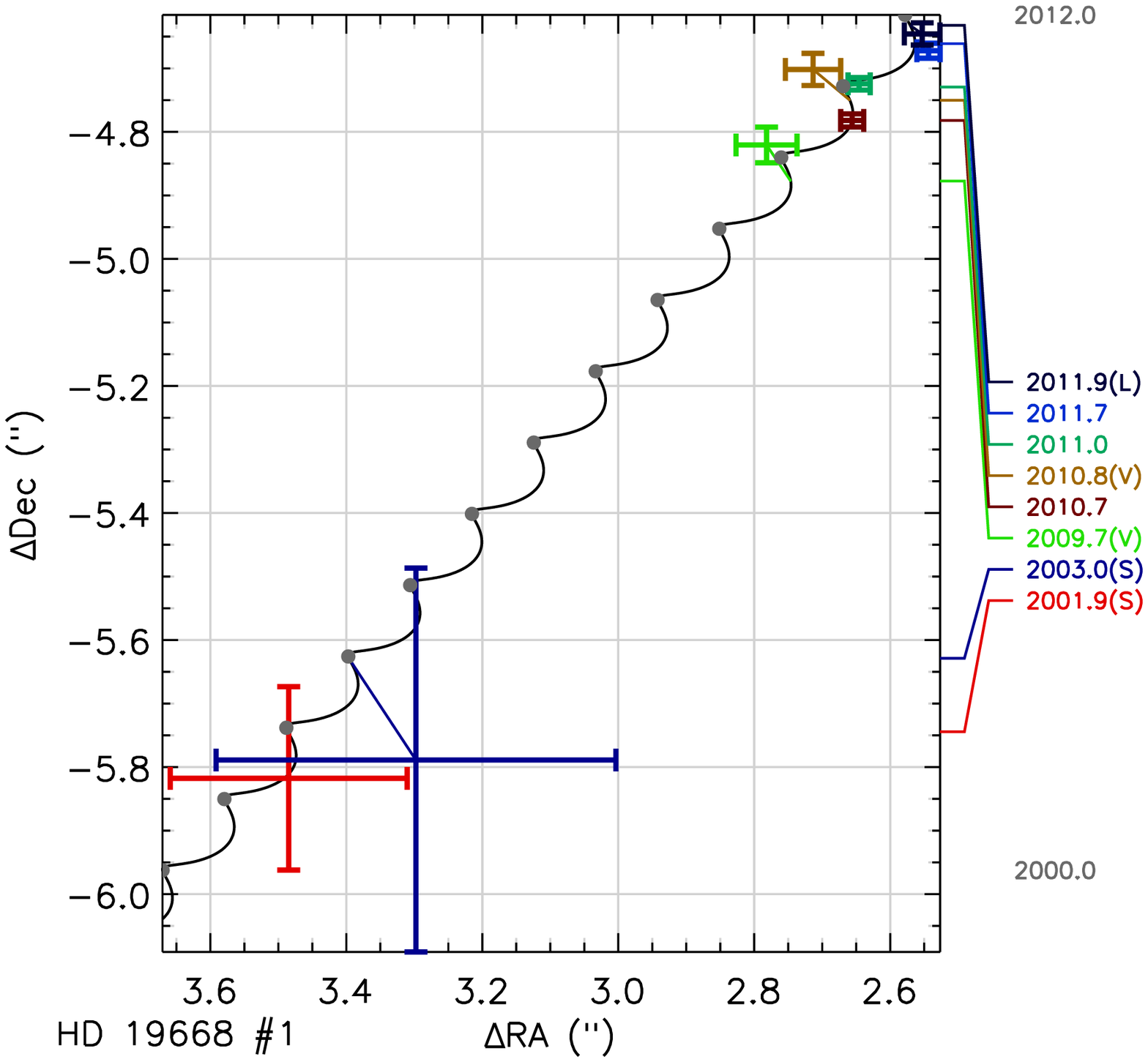}
}
\vskip -0.2in
\centerline{
\includegraphics[width=2.0in]{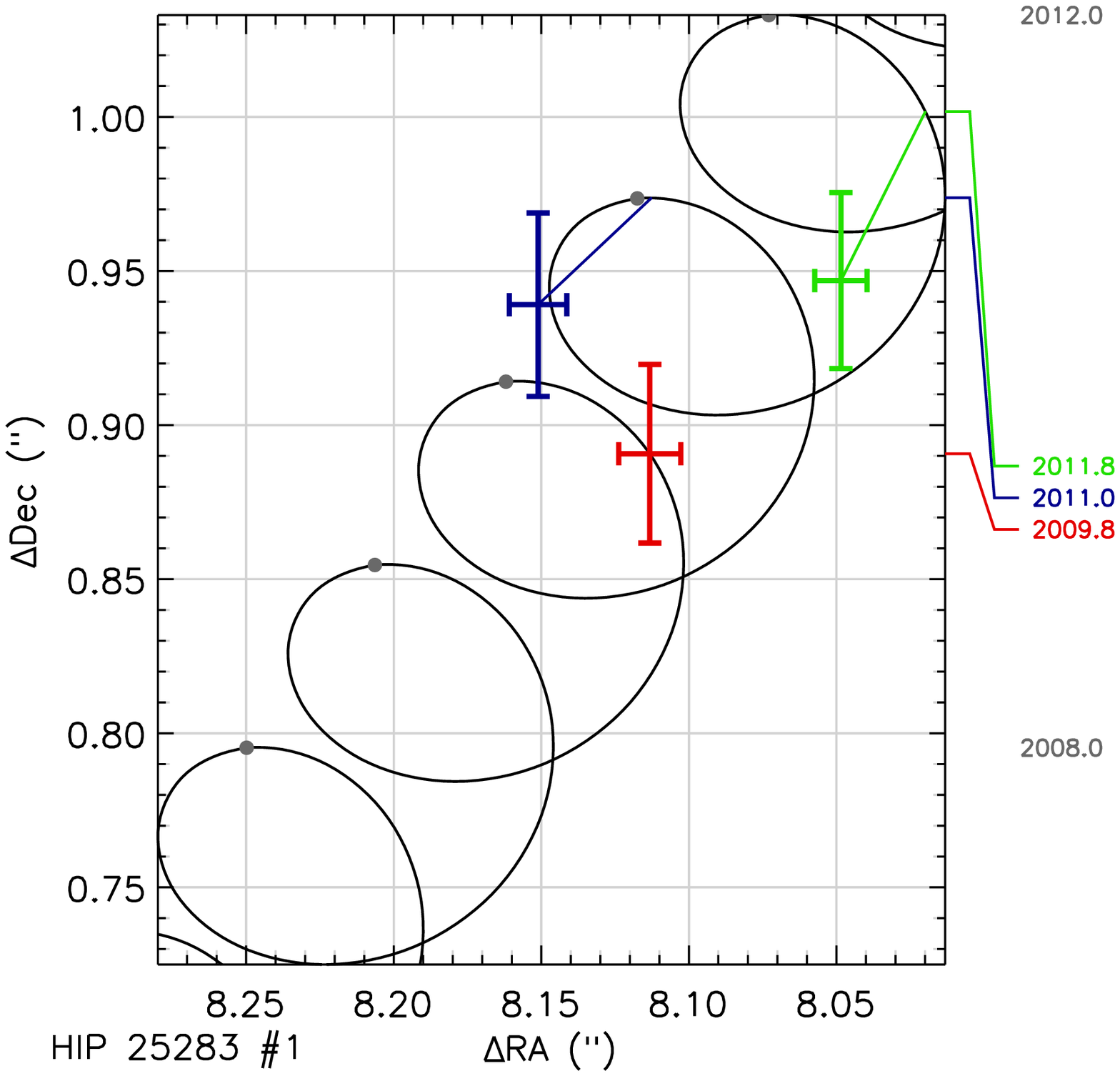}
\hskip -0.3in
\includegraphics[width=2.0in]{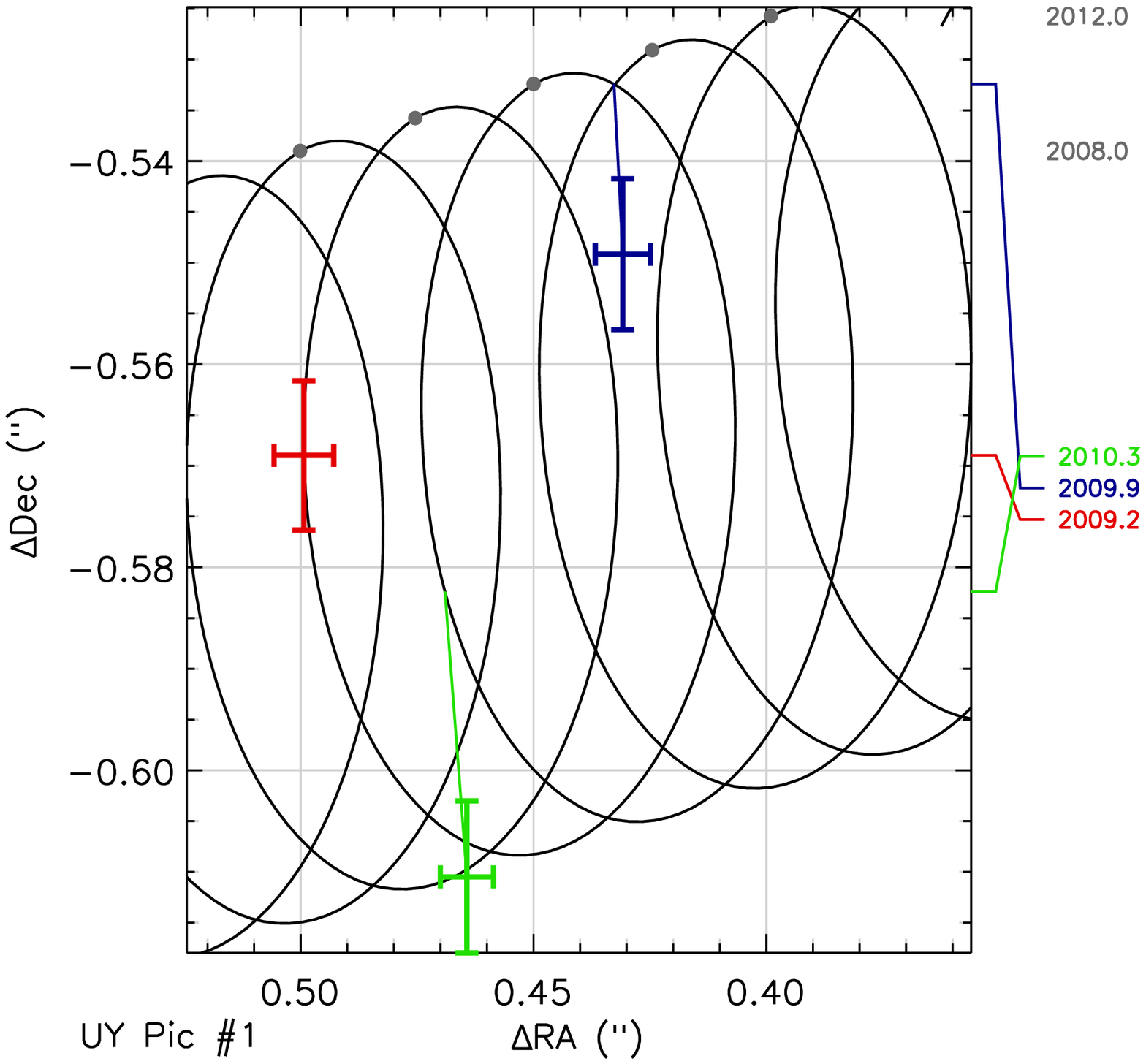}
\hskip -0.3in
\includegraphics[width=2.0in]{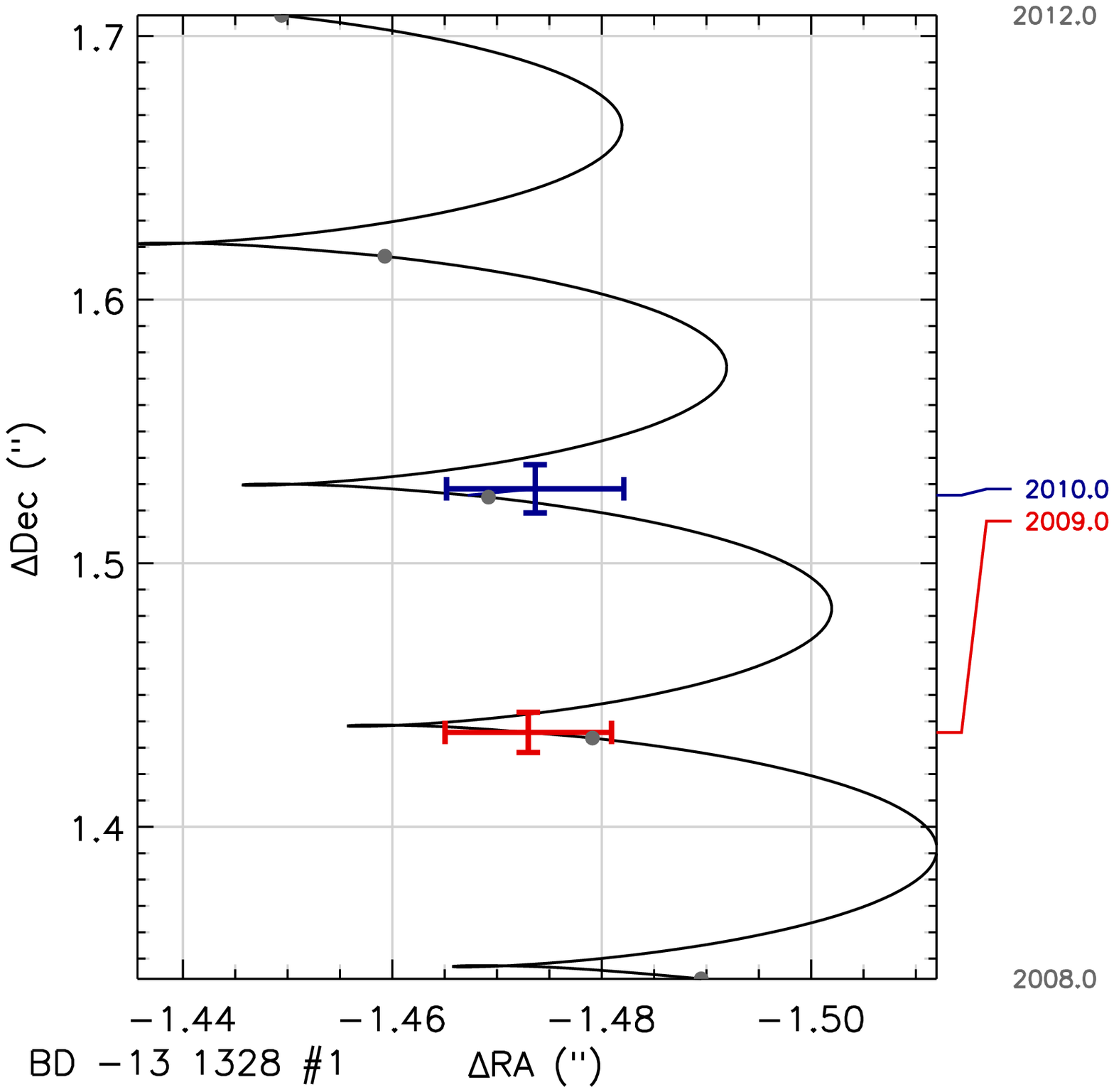}
\hskip -0.3in
\includegraphics[width=2.0in]{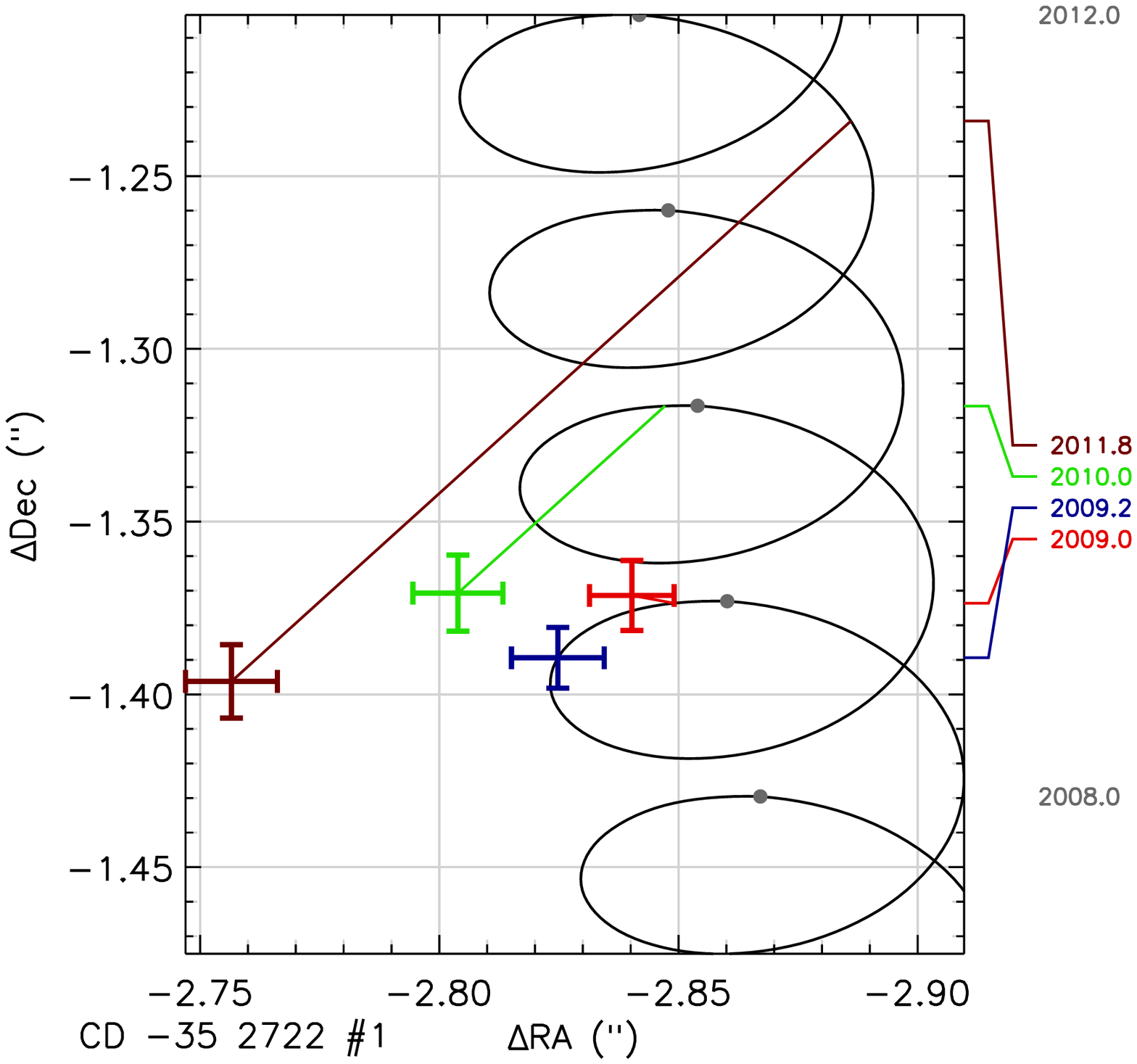}
}
\vskip -0.2in
\centerline{
\includegraphics[width=2.0in]{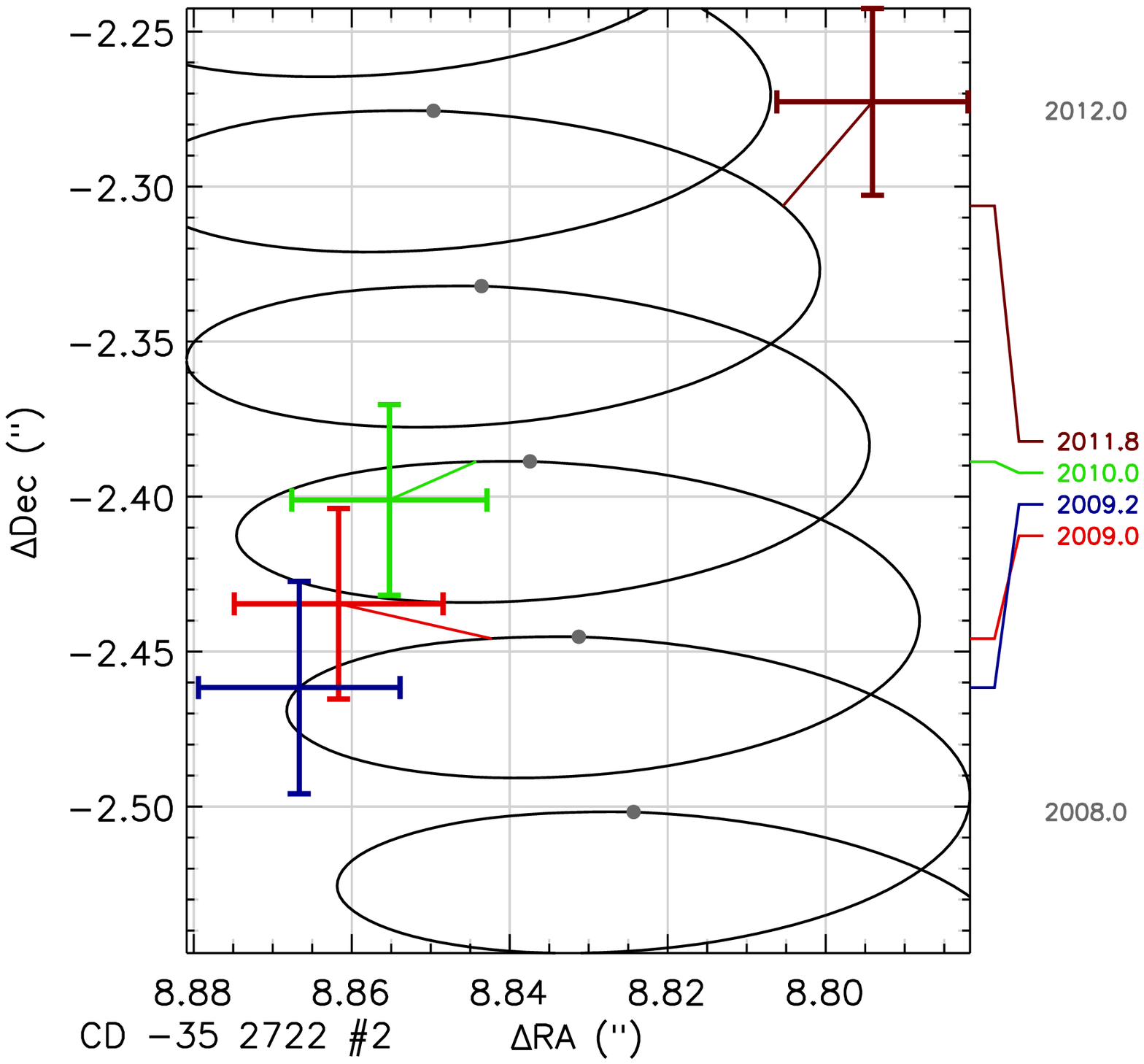}
\hskip -0.3in
\includegraphics[width=2.0in]{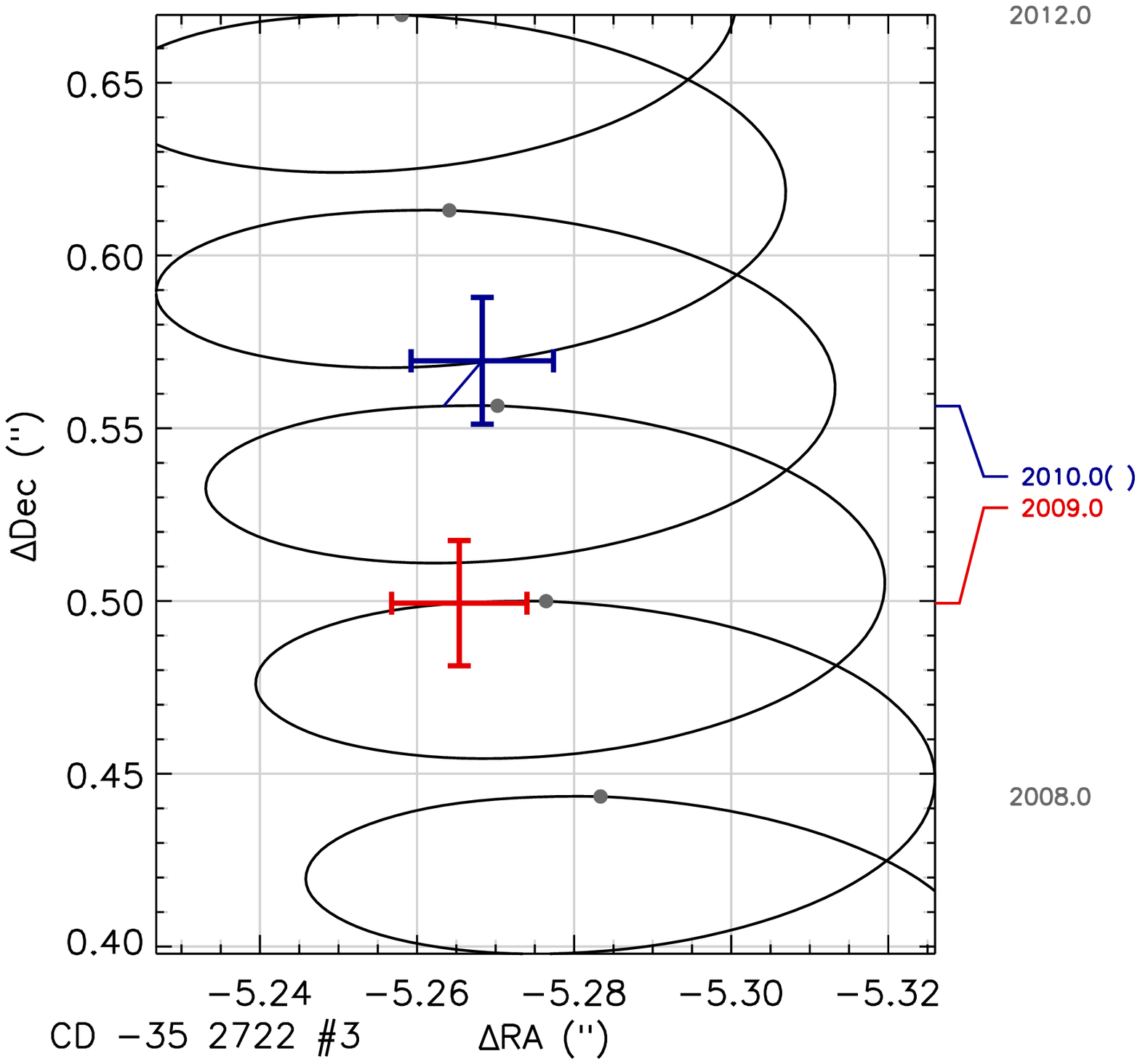}
\hskip -0.3in
\includegraphics[width=2.0in]{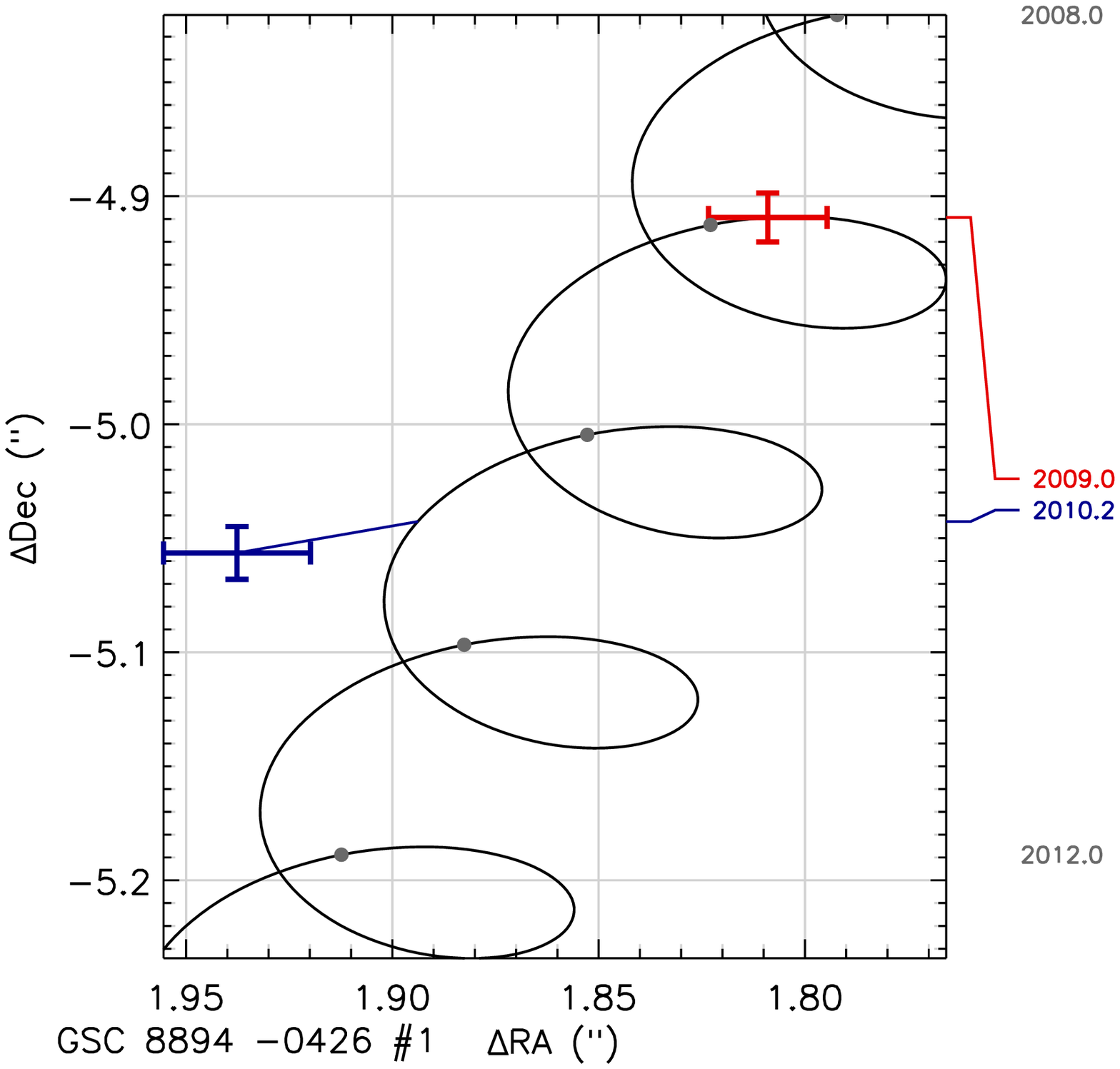}
\hskip -0.3in
\includegraphics[width=2.0in]{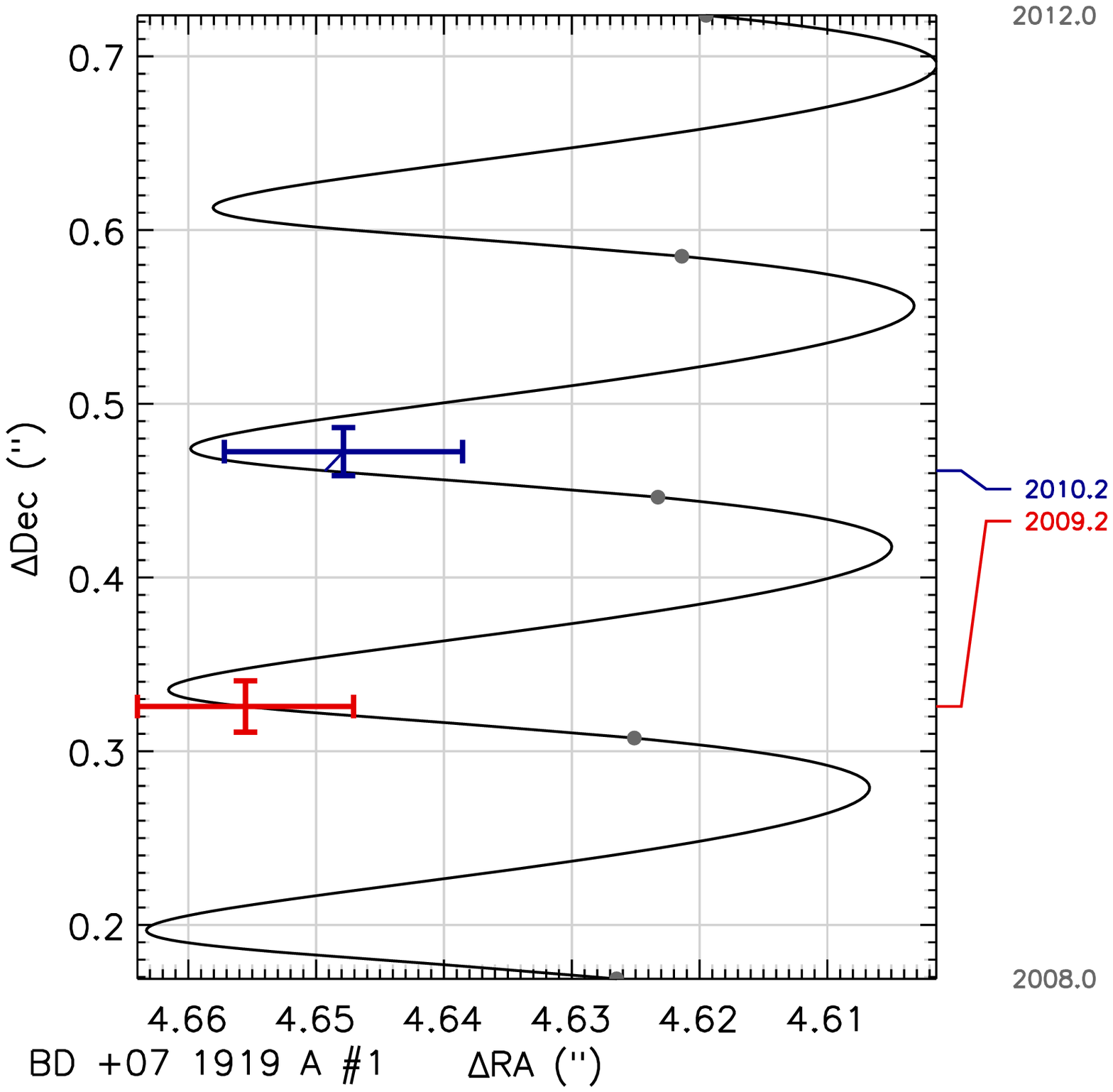}
}
\vskip -0.2in
\centerline{
\includegraphics[width=2.0in]{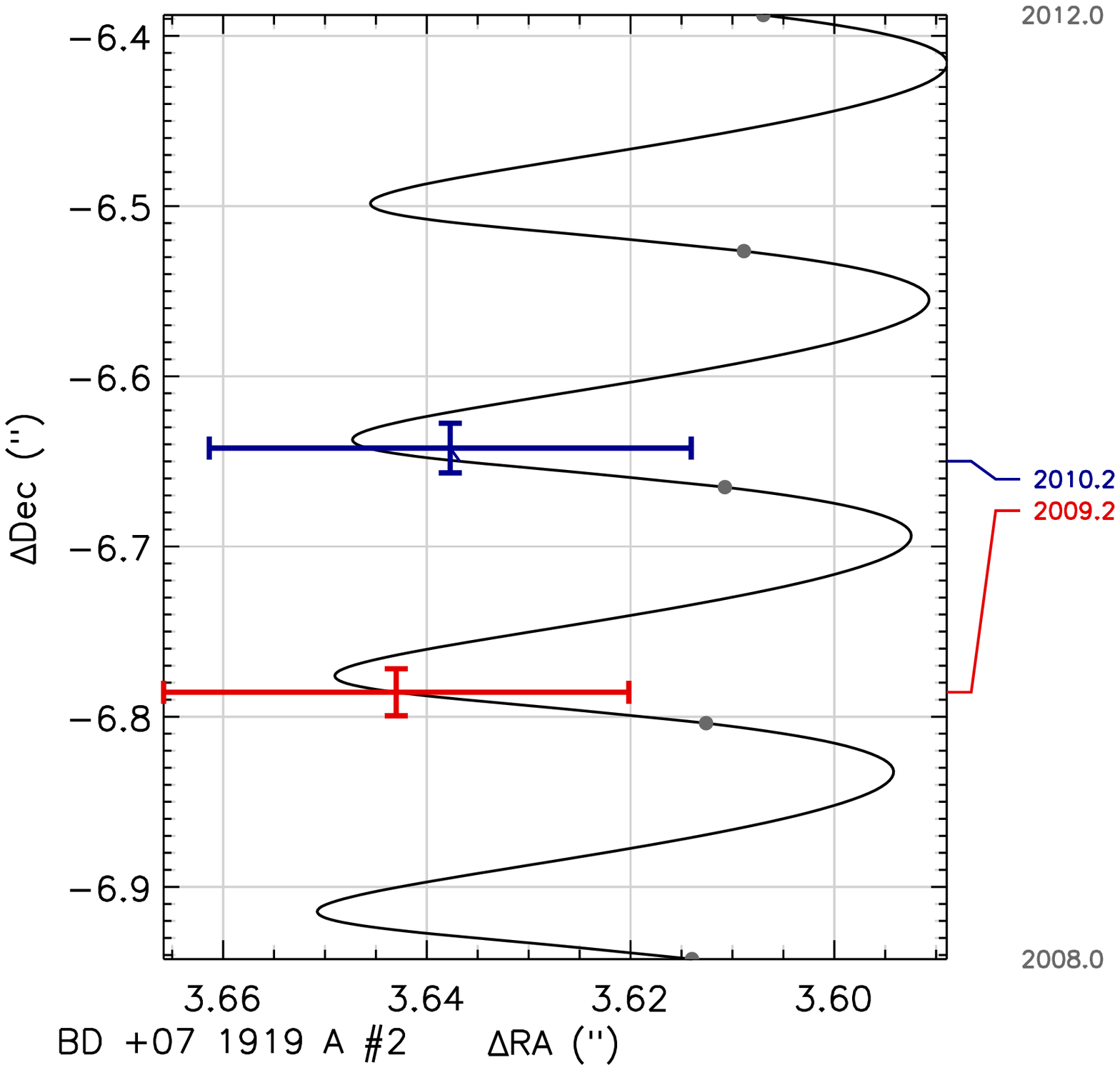}
\hskip -0.3in
\includegraphics[width=2.0in]{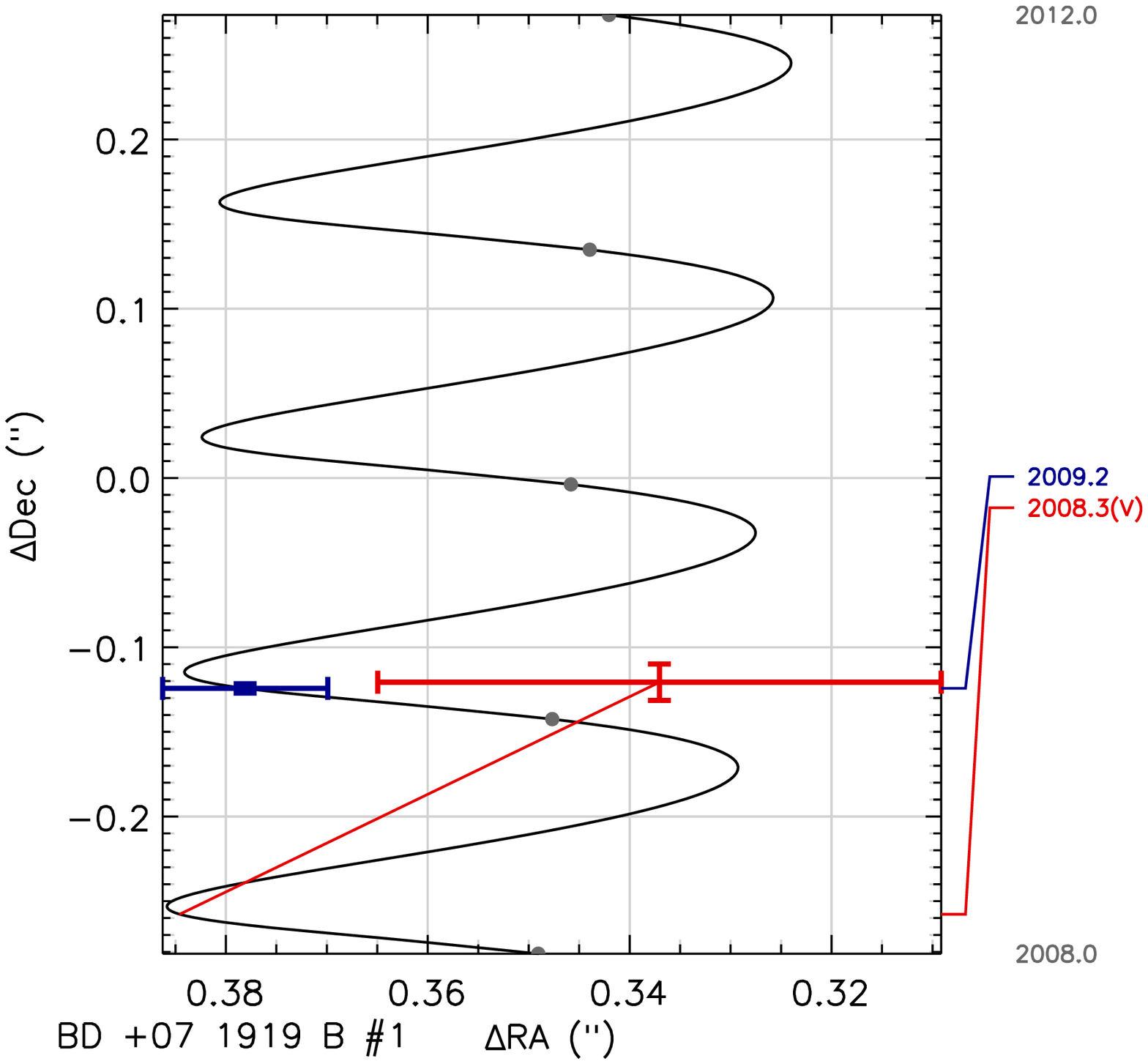}
\hskip -0.3in
\includegraphics[width=2.0in]{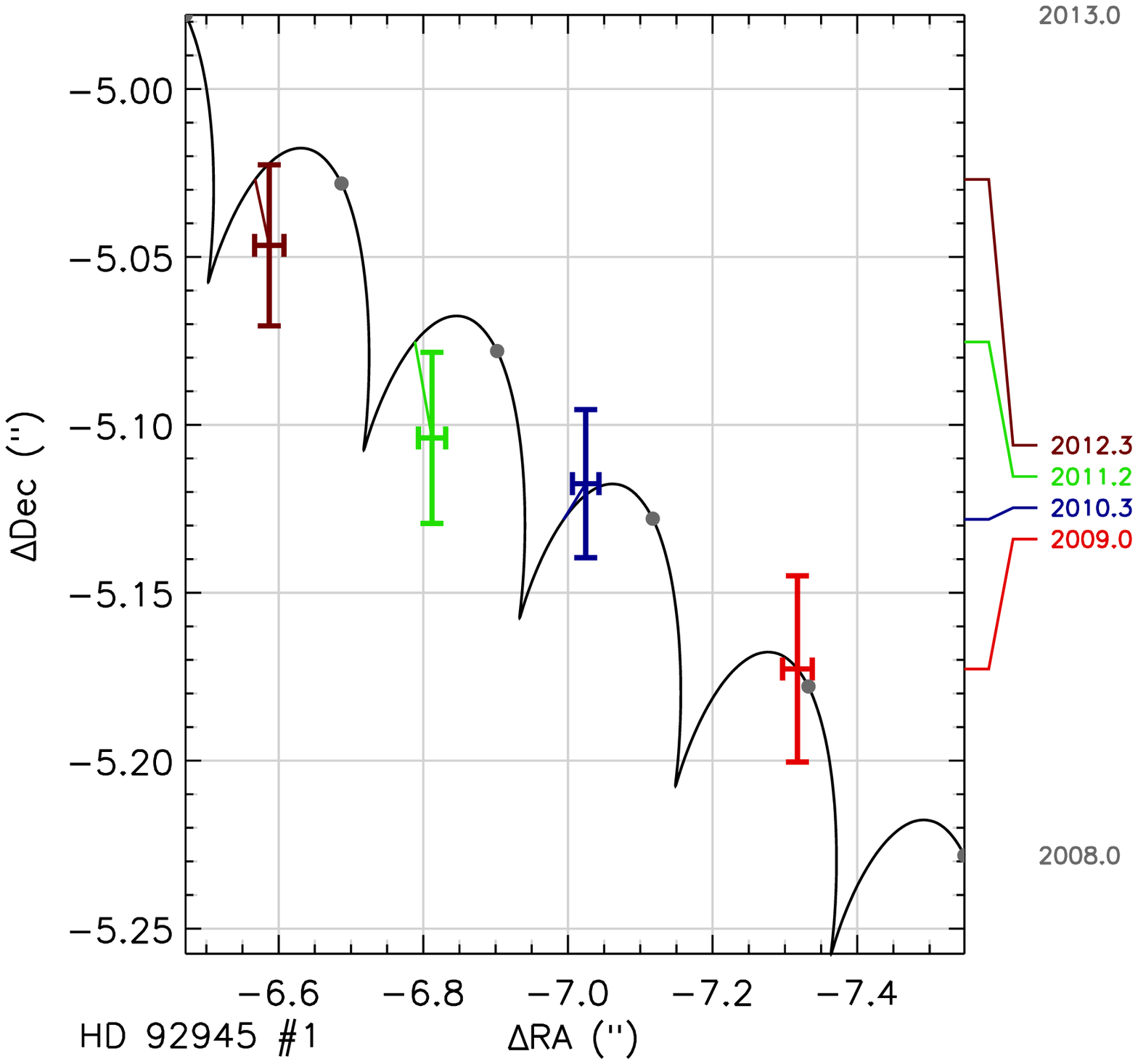}
\hskip -0.3in
\includegraphics[width=2.0in]{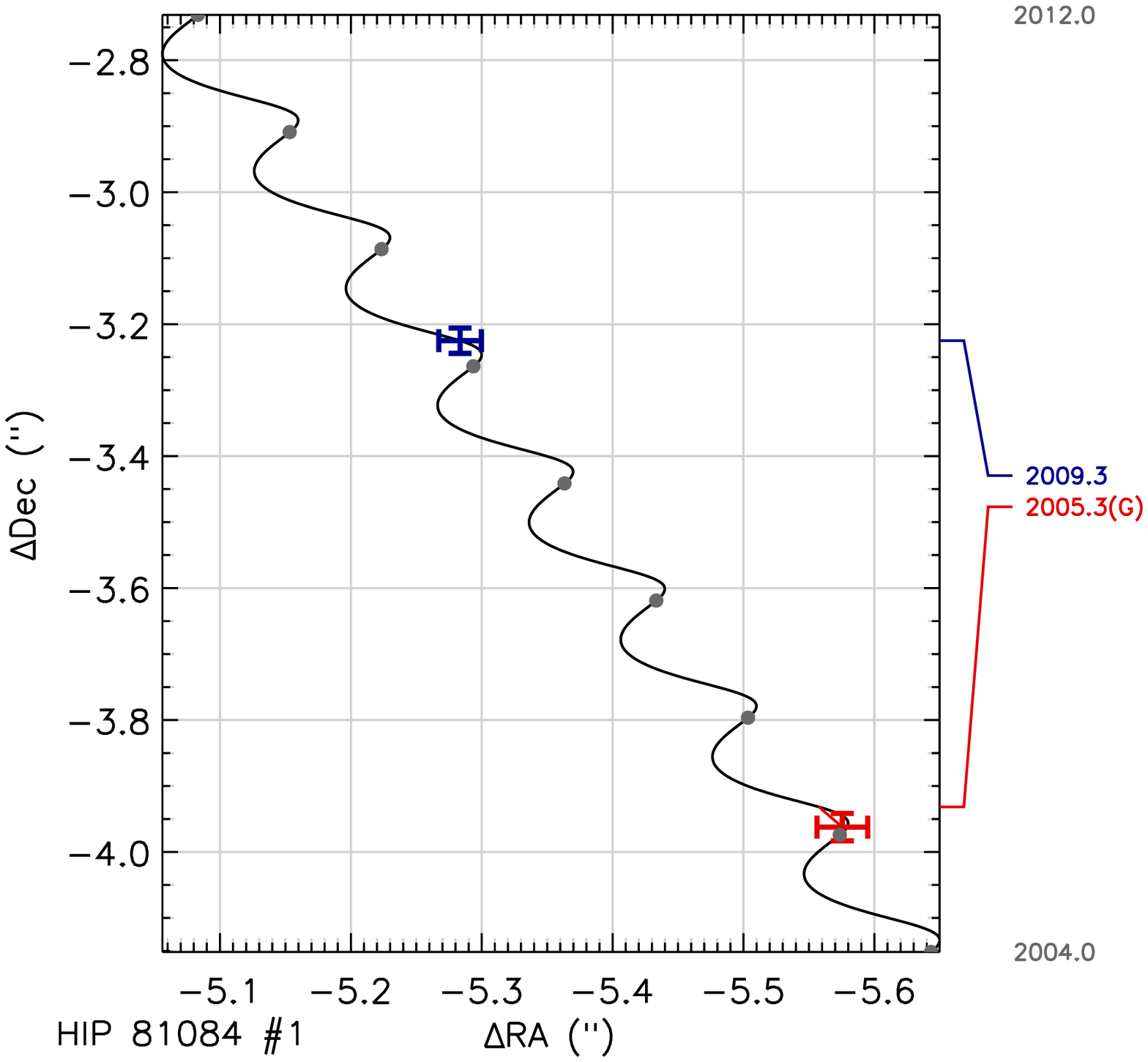}
}
\vskip -0.2in
\centerline{
\includegraphics[width=2.0in]{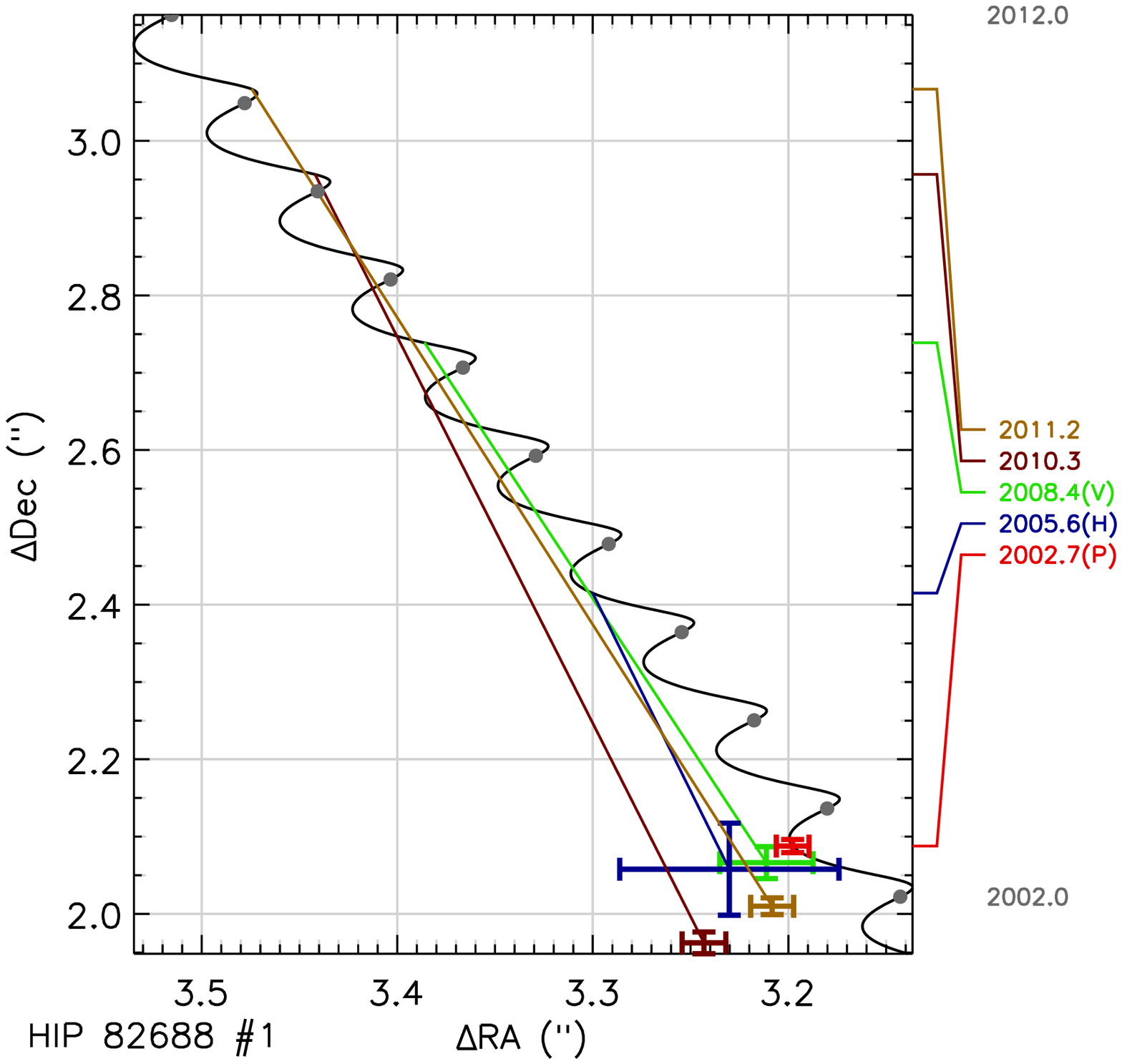}
\hskip -0.3in
\includegraphics[width=2.0in]{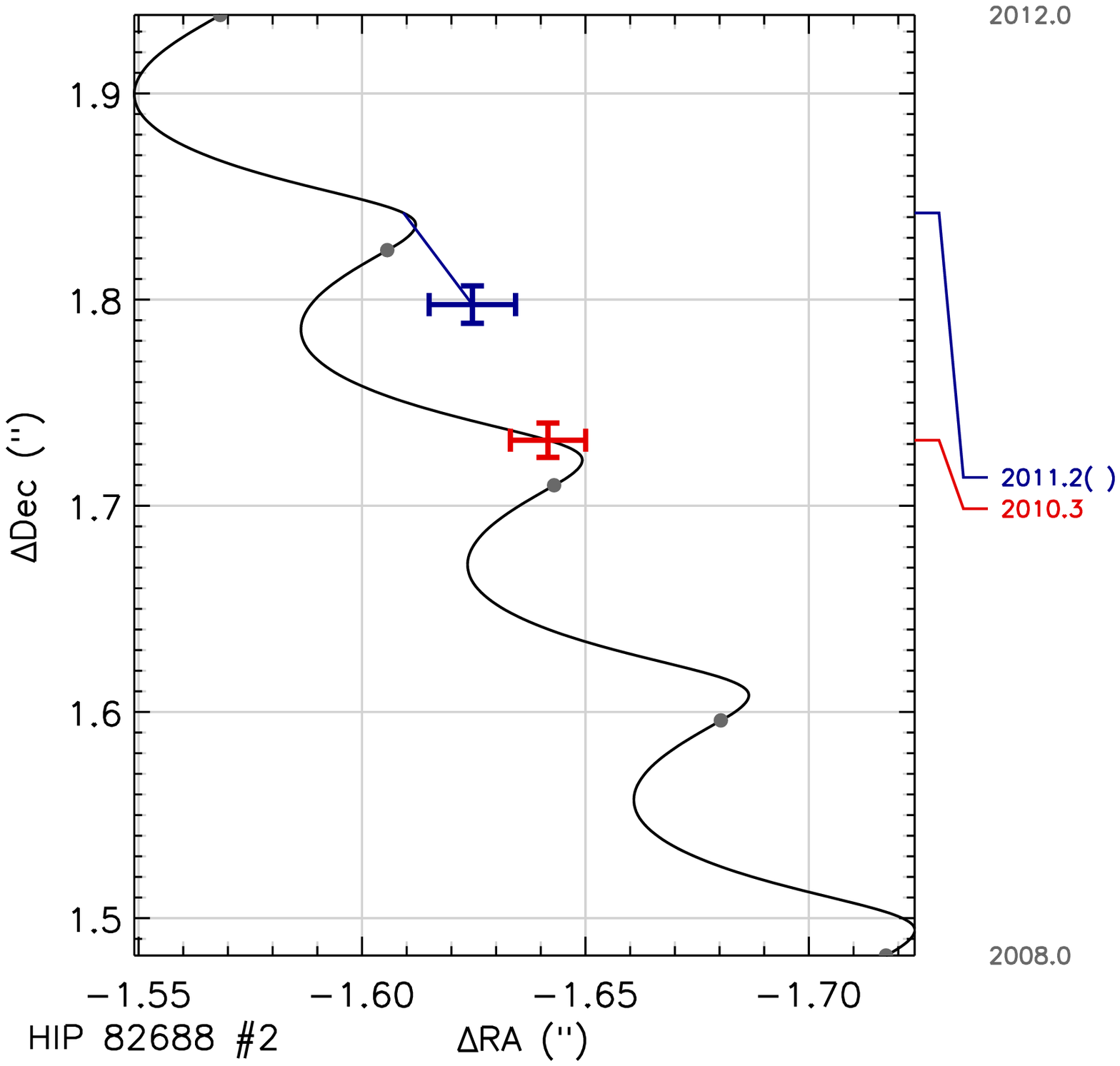}
\hskip -0.3in
\includegraphics[width=2.0in]{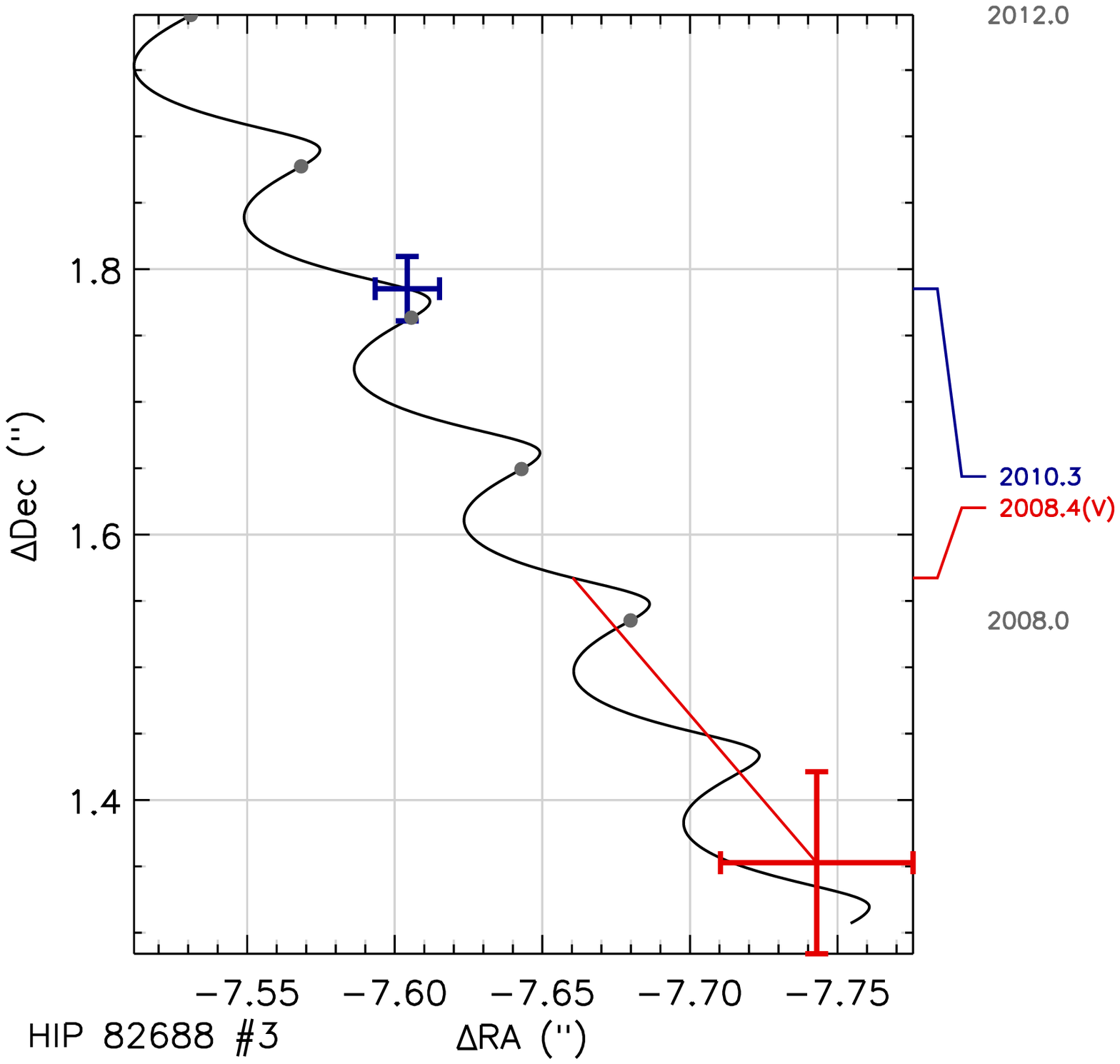}
\hskip -0.3in
}
\vskip -0.2in
\centerline{
}
\caption{On-Sky plots for AB Dor MG objects. \label{fig:ABDor_skyplots}}
\end{figure}

From the proper motions and parallaxes of our
MG sample stars and pinning to the NICI first epoch position,
we can calculate the expected 
motion relative to the primary star for each candidate companion, 
assuming that the candidate is a background object.
On-sky plots presenting background ephemerides and the 
actual on-sky motion of each candidate companion relative to the
primary are presented in Figures~\ref{fig:TWHya_skyplots} to \ref{fig:ABDor_skyplots}.
We compute the $\chi^2$ value for the expected background track
position relative to the actual sky
position for each candidate (see Nielsen et al. 2013).  
$\chi^2$ values are shown in Table~\ref{table_comp2a}.  Candidates
with reduced $\chi^2$ values
close to 1 are confirmed to be background objects.
In total, 81 candidate companions were tested for common proper motion 
with either archival or 2nd epoch Gemini NICI data.
Of these candidates, 77 were background objects;
however, four co-moving brown dwarf or stellar 
companions (discussed in more detail in 
Section 7) were detected for the first time in the moving group sample: PZ Tel B
\citep{Bil10}, CD -35 2722B \citep{Wah11}, HD 12894B (this work)
and BD+07 1919C (this work).  We also retrieve the known 
stellar companion to HD 82688 \citep{Met09}, as well as the 
brown dwarf companions AB Pic B \citep{Cha05b}
and HR 7329B \citep{Low00, Gue01}.  

A number of stars 
(HD 139084 B, V343 Nor, CD-54 7336, CD-31 16041, HD 159911, GJ 560 A, 
and TYC 7443-1102-1) were near the Galactic Bulge and often possessed extremely dense
starfields ($>$20 objects in the NICI images).  As we expect almost all of these candidates to be background
objects, we assigned these stars lower priority for second epoch NICI
followup and consequently they were not observed before the end of the
NICI Campaign.  Astrometry for candidates observed at only one epoch and thus
unconfirmed as background or common proper motion is 
presented in Table \ref{tab:1epoch}.


\subsection{New Stellar Binaries}

\begin{figure}
\centerline{
\includegraphics[width=2.0in]{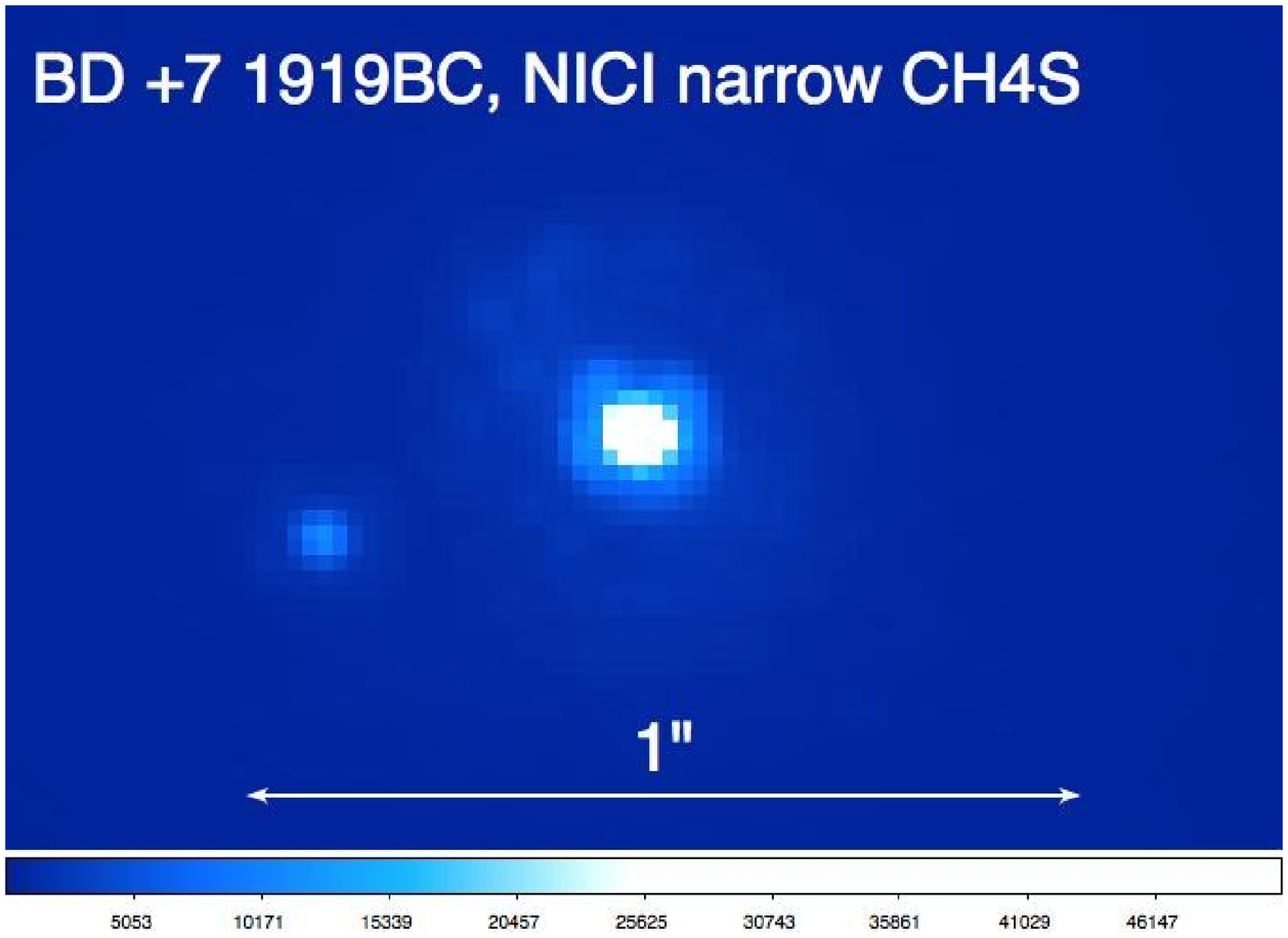}
\includegraphics[width=2.0in]{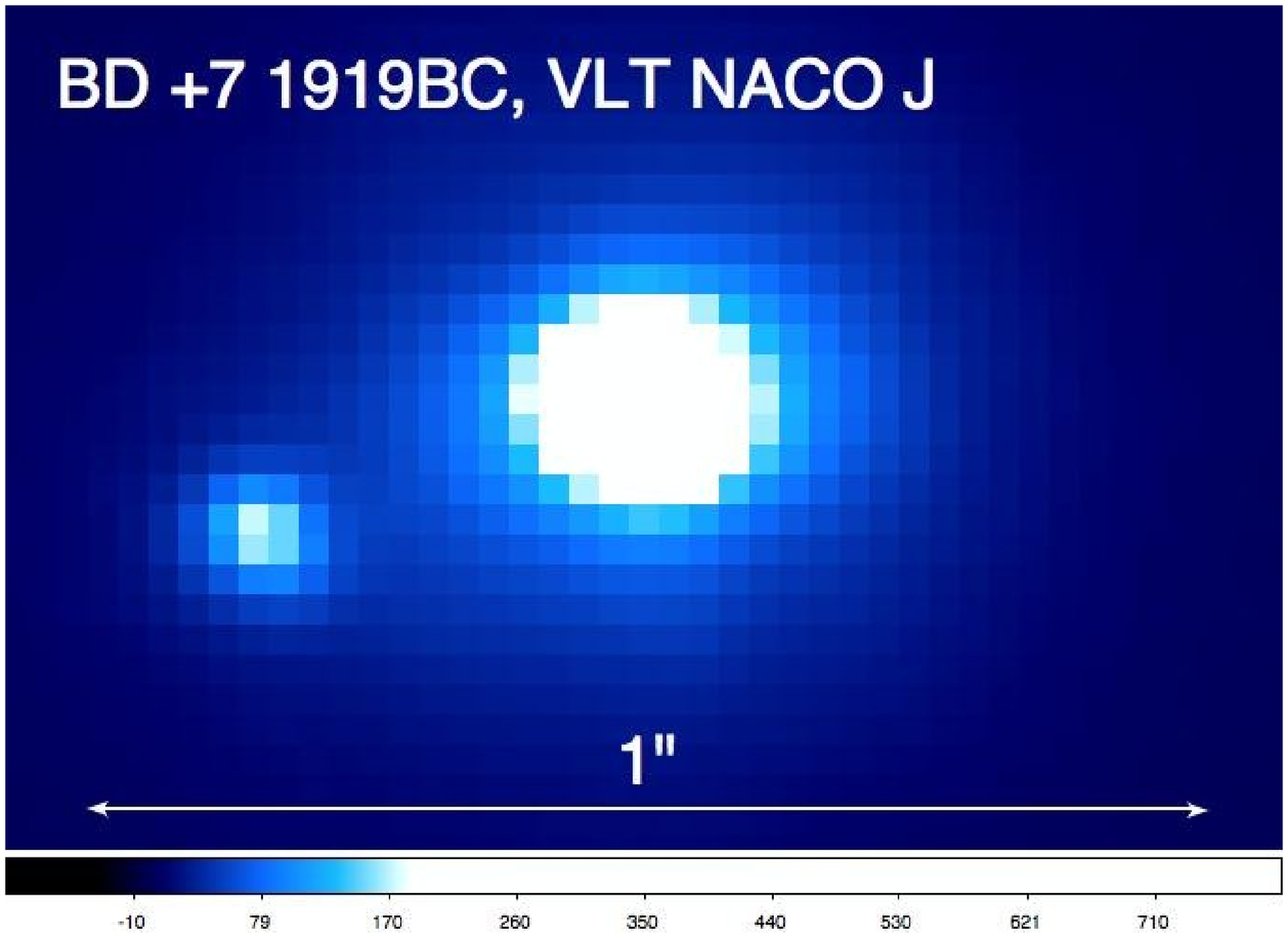}
}
\centerline{
\includegraphics[width=2.0in]{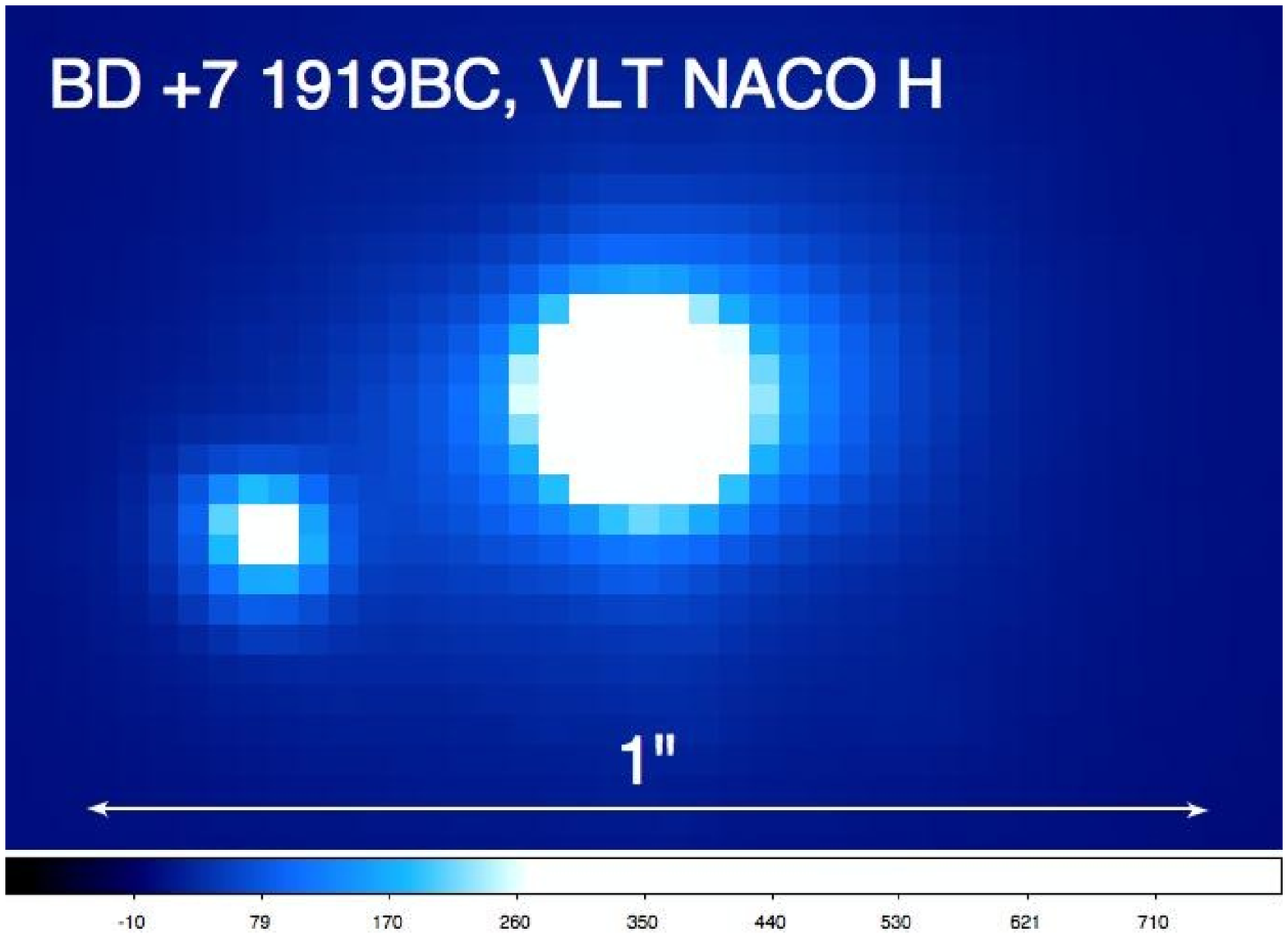}
\includegraphics[width=2.0in]{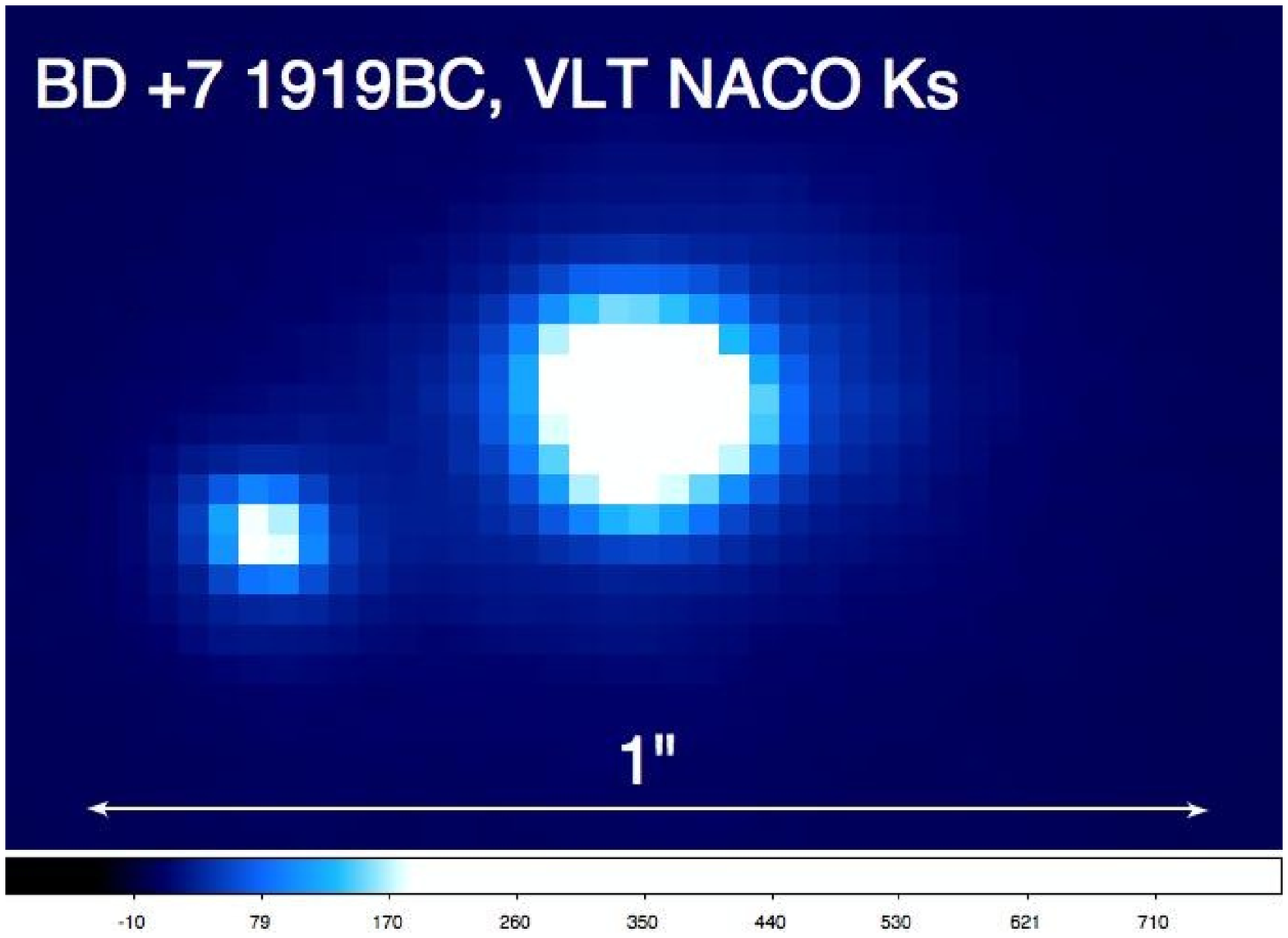}
}
\vskip 0.2in
\centerline{
\includegraphics[width=2.0in]{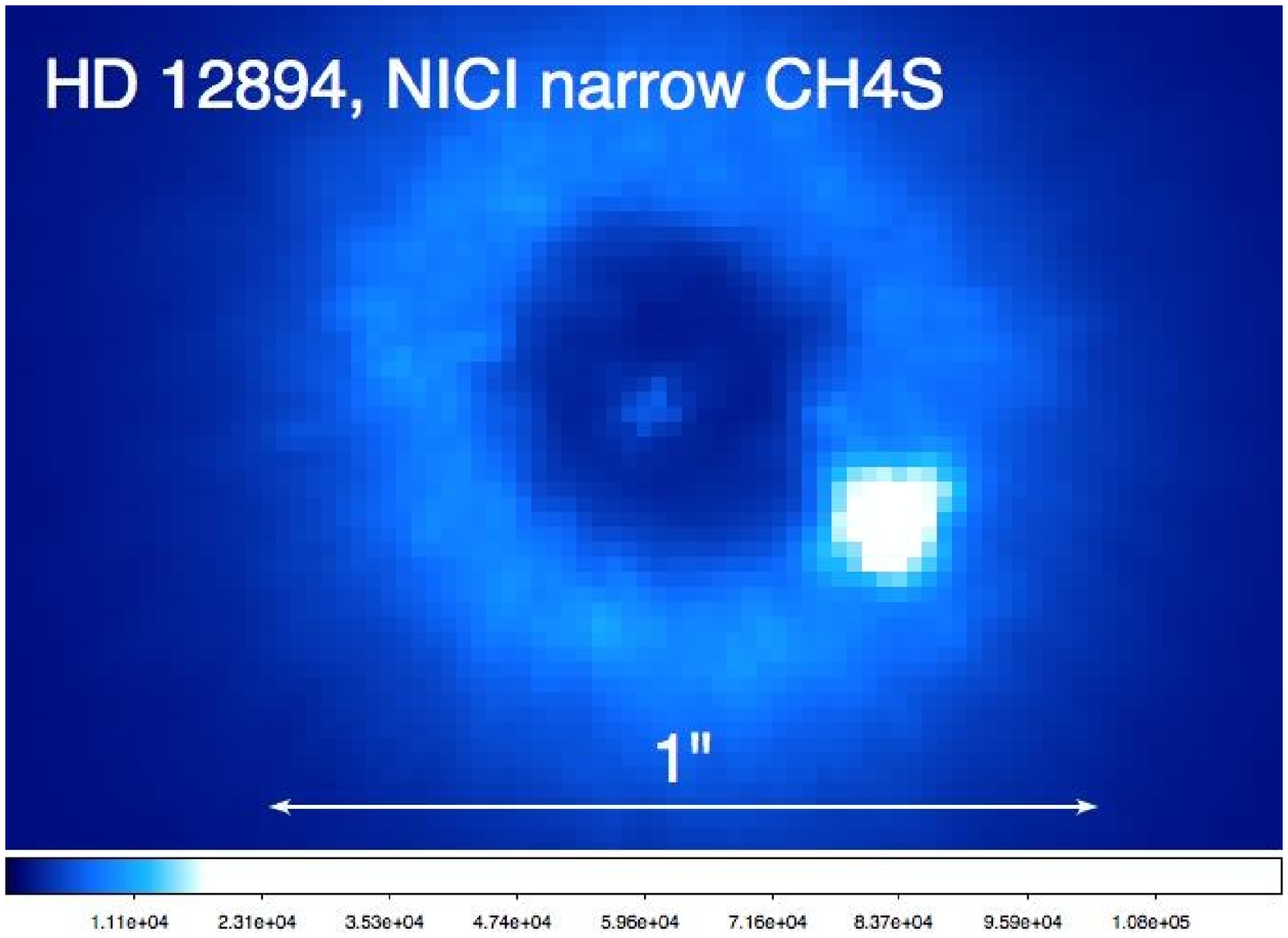}
\includegraphics[width=2.0in]{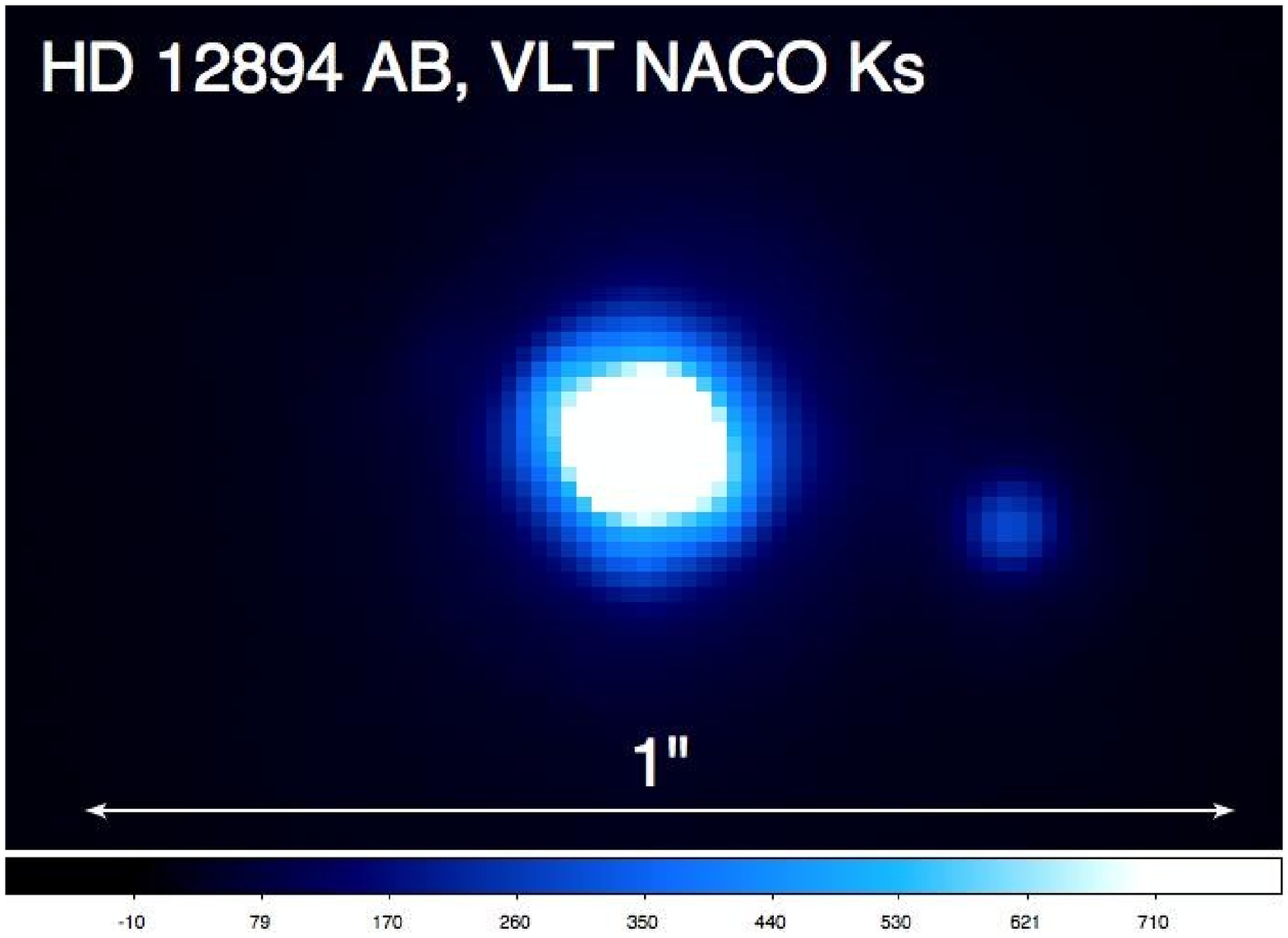}
\includegraphics[width=2.0in]{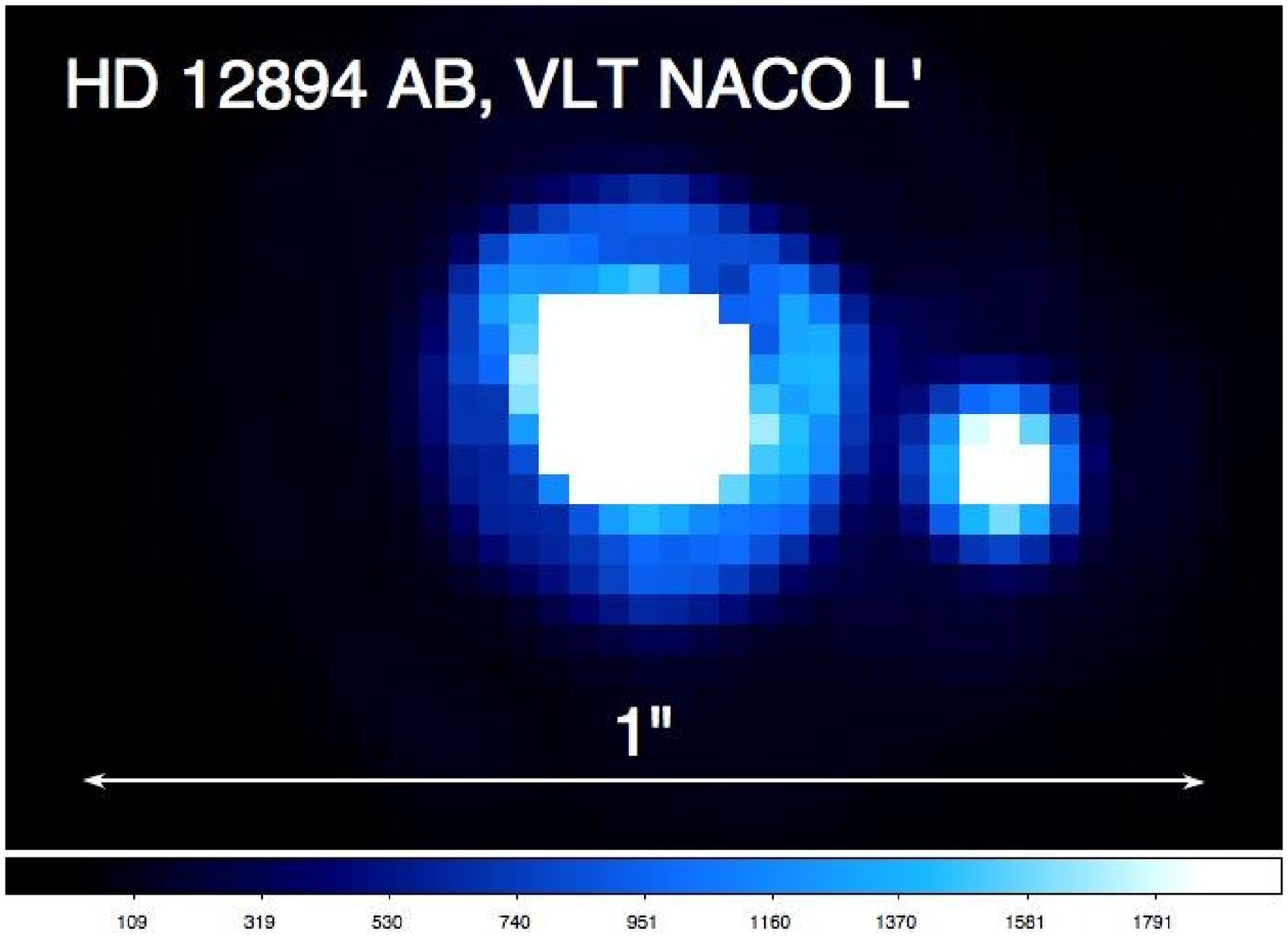}
}
\caption{Gemini NICI and archival VLT NACO images of two newly discovered 
low-mass stellar companions, HD 12894B (0.46$\pm$ 0.08 M$_{\odot}$) 
and BD +07 1919C (0.20$\pm$0.03 M$_{\odot}$).  North is up and East is 
left in all images.
   \label{fig:binaryimages}}
\end{figure}

\begin{figure}
\centerline{
\includegraphics[width=3in]{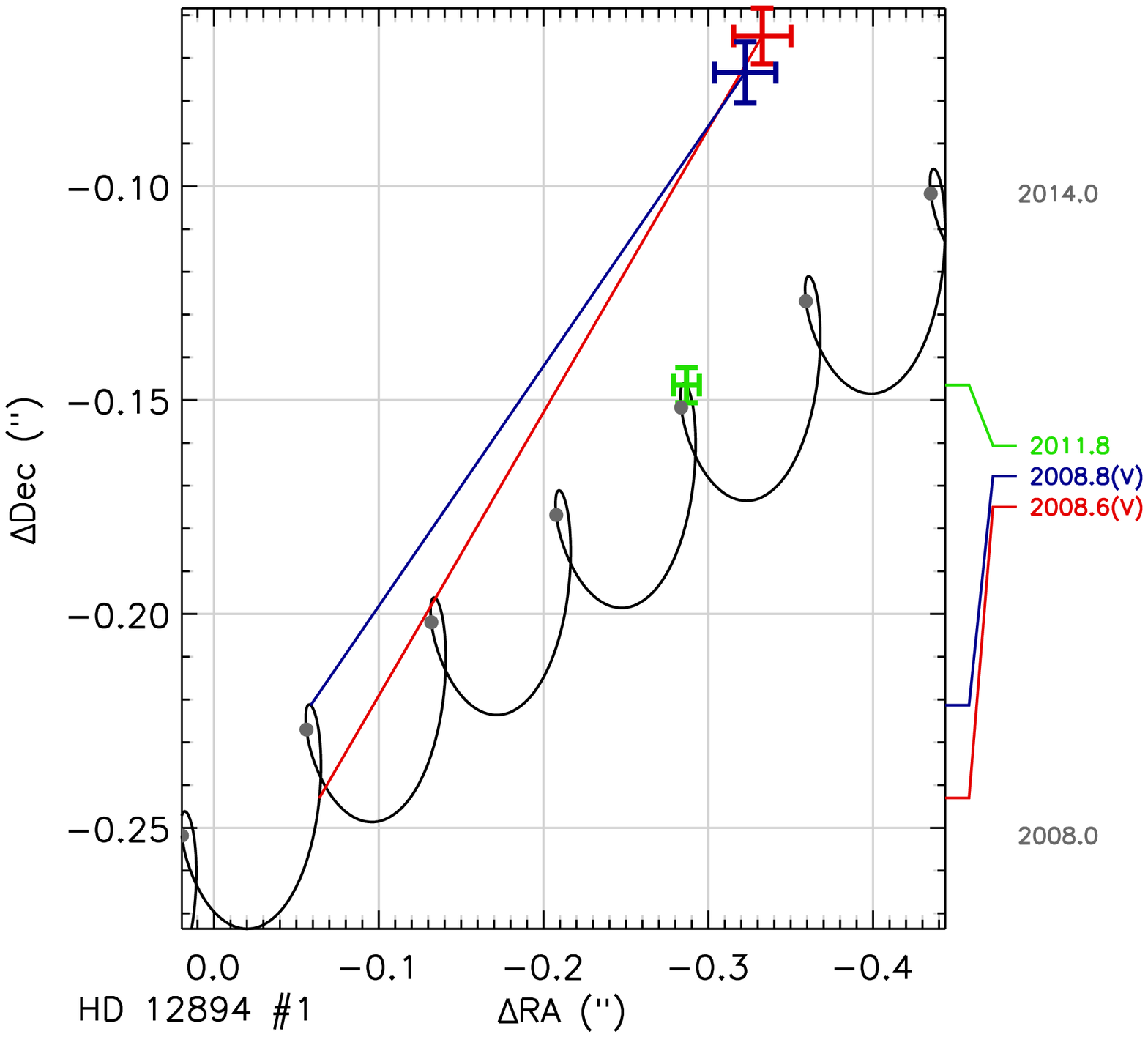}
\includegraphics[width=3in]{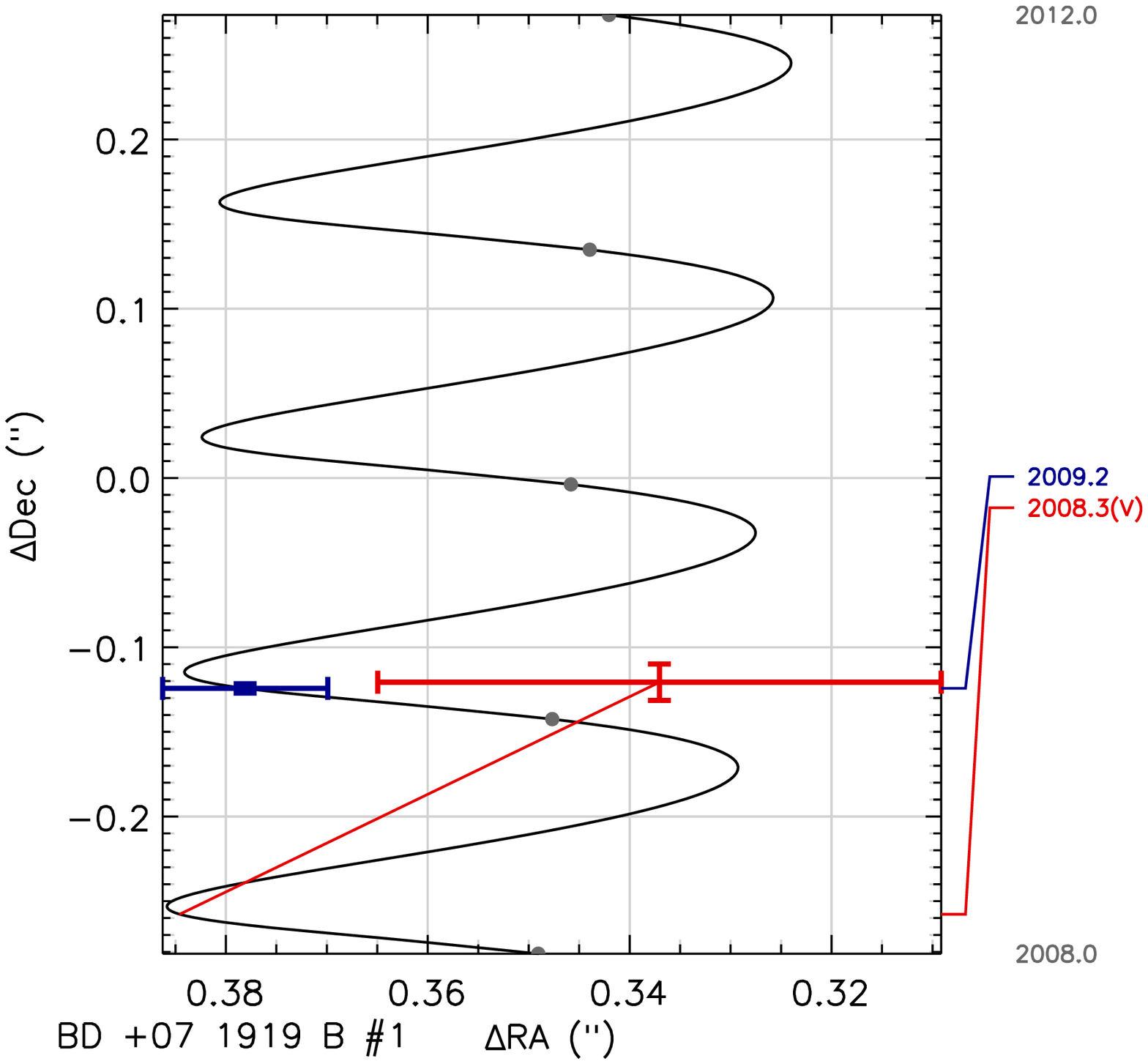}
}
\caption{Sky plots for newly discovered stellar binaries, HD
  12894B (left) and BD +07 1919C (right).  Both new binaries share 
common proper motion with their parent star and 
HD 12894B shows clear orbital motion.  
\label{fig:binaryskyplots}}
\end{figure}

In the course of the survey, we discovered two new low mass stellar 
companions, HD 12894B and BD+07 1919C (Figure~\ref{fig:binaryimages}).
NICI and archival datasets analyzed are tabulated in Table~\ref{tab:binarydata}.
Both companions have been confirmed to be common proper motion 
with their primary using VLT NACO archival data.  Sky plots are shown in Figure~\ref{fig:binaryskyplots}
and astrometry is presented in Tables~\ref{tab:HD12894} and \ref{tab:BD+07}.
Archival images were sky-subtracted and flat-fielded.  Bad pixels
identified from a dark image were removed.  Images at different 
dither positions were registered and stacked.  
Astrometry was derived from both NICI and NACO archival datasets
using the star and companion centroids measured from the 
final reduced stacked images.    

Since the NICI datasets for these binaries either had the starspot
saturated or were taken in the narrow methane filters, we calculated 
broadband photometry from the VLT NACO archival images.
Both companions sit on the wings of the primary PSF.  For these datasets, the PSF shape 
was generally azimuthally symmetric.  To obtain photometry, 
we thus subtracted out a PSF radial profile generated from the azimuthal
median of the star image, excluding the position angle range within 
$\pm$20 degrees of the detected companion.
Aperture photometry was performed using 2, 3, 4, 5, and 6-pixel
apertures.  All apertures produced consistent results; we adopt the 
results using the 4-pixel aperture here.  To estimate photometric
errors, photometry was calculated both for individual reduced frames
and the final reduced image.  We adopt the rms of the values from the individual
reduced frames as the photometric error.  Our photometry
is presented in Tables~\ref{tab:HD12894} and \ref{tab:BD+07}.

We estimate companion masses based on the models of \citet[]{Bar98}.
We adopt Monte Carlo methods to account for the photometric
uncertainties as well as the range of possible distances and
ages for these binaries.  We simulate an ensemble of 10$^{6}$
realizations of the system, drawing from Gaussian distributions
in age, parallax, and photometry with 1$\sigma$ widths taken from the 
measured uncertainties on these parameters.
For each realization, we then interpolate with age and single-band 
absolute magnitude to estimate the mass of the companion 
from the models of \citet[]{Bar98}.  The adopted mass is then 
the peak of the output distribution of simulated realizations, 
with error bars drawn from the 68\% confidence limits of the output distribution.
Results using {\it J}, {\it H}, and {\it K}$_{s}$ single band absolute magnitudes
yielded consistent results; {\it K}$_{s}$ band results are presented in 
Tables~\ref{tab:HD12894} and \ref{tab:BD+07}.  No estimate was
made for the {\it L'} band observations of HD 12894, as we could not 
find an apparent {\it L'} magnitude for HD 12894 in the literature.
We find best mass estimates of 0.46$\pm$ 0.08 M$_{\odot}$
for HD 12894B and 0.20$\pm$0.03 M$_{\odot}$ for BD +07 1919C.

These relatively massive (0.2--0.5 M$_{\odot}$) companions have, unsurprisingly, shown
some orbital motion between the archival and NICI epochs.
Thus, these orbits may yield dynamical mass measurements 
on a 10-20 year timescale.  
To determine the necessary timescales 
to measure these orbits, we estimate their semimajor axes and 
periods.  Assuming a uniform eccentricity distribution
between 0 $<$ e $<$ 1 and random viewing angles, 
\citet[]{Dup11} compute a median correction factor between
projected separation and semimajor axis of 1.10$^{+0.91}
_{-0.36}$ (68.3\% confidence limits).  Using this correction
factor, we derive a semimajor axis of 16.9$^{+14.1}_{-5.6}$ AU
for HD 12894AB and a semimajor axis of 13.8$^{+11.5}_{-4.8}$ AU
for BD +07 1919BC (neglecting the presence or influence of A, 
which lies several arcsec and $>$200 AU away).  To convert from semi-major axis to period
requires an estimate of the total system mass. 
We estimate the primary masses using the same Monte Carlo method
as described above for the secondary masses, giving a mass 
of 1.10$\pm$0.06 M$_{\odot}$ for HD 12894 and 0.70$\pm$0.05 M$_{\odot}$
and 0.66$\pm$0.05 M$_{\odot}$ for BD+07 1919B and C respectively.
Combining with the previously estimated companion masses, 
we estimate periods of 56$^{+70}_{-28}$ yr 
for HD 12894AB and 55$^{+69}_{-29}$ yr for BD+07 1919BC.  
Further orbital monitoring will thus be necessary to better
constrain the semi-major axes and periods of these orbits.
 
\subsection{PZ Tel  -- No debris disk}

In \citet[][]{Bil10}, we reported the detection of a 36$\pm$6 M$_{Jup}$
companion to the young solar analogue PZ Tel, a member of the 
$\beta$ Pic moving group.  Due to the considerable on-sky motion of PZ
Tel B, we were able to constrain the eccentricity of the PZ
Tel B orbit to $>$0.6 through Monte Carlo orbital simulations with
just two epochs of NICI astrometry.  Recently, this result has been confirmed
by \citet[]{Mug12}.  

PZ Tel had previously been reported to have 
70 $\mu$m excess emission and hence a debris disk \citep[][]{Reb08}.  
The existence of a debris disk is hard to
reconcile with the highly eccentric orbit of the brown dwarf
companion, which would likely disrupt the outer debris disk as it
moves through it.  However, recent analysis of both {\it Spitzer} 24 and 70 $\mu$m data 
as well as {\it Herschel} 70, 100, and 160 $\mu$m data yield no
detection of excess in any band
at the location of PZ Tel AB (G. Bryden, private communication).  There is 
a very red source $\sim$25$\arcsec$ north of PZ Tel AB which is likely
extragalactic.  The centroiding algorithm used by \citet[][]{Reb08} allows for the
centroid position to move from the target position in order to 
account for pointing errors and as a result likely mis-identified the
extragalactic source as PZ Tel. (L. Rebull, private communication). 
Thus, PZ Tel does not possess a debris disk.
 
\subsection{AB Pic B -- Typical L0.5 colors with NICI}

\begin{figure}
\includegraphics[width=6in]{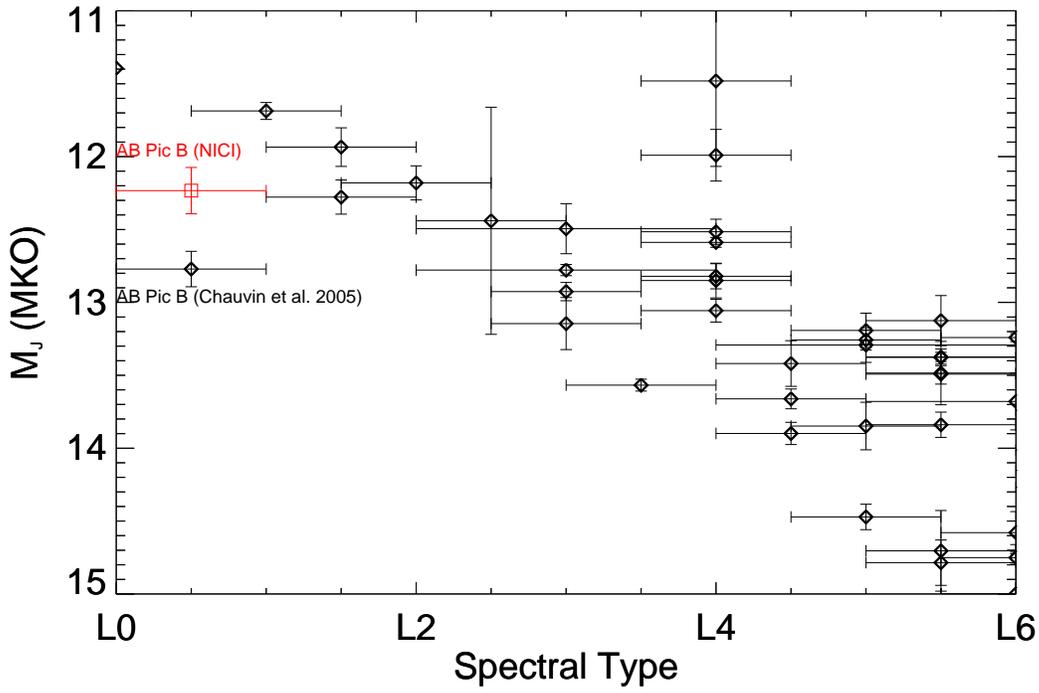}
\caption{Infrared spectral type vs. absolute J magnitude for AB Pic B
  and other L0-L6 field dwarfs.  Photometry for AB Pic B from
  \citet{Cha05b} and the present study are plotted for
  comparison.  We find a comparatively brighter J
  magnitude for AB Pic B compared to the photometry
from \citet{Cha05b}, which shifts AB Pic B from being anomalously faint
into the brightness sequence expected for its
spectral type.  Data for the comparison objects are taken from \citet{Dup12}.
}
\label{fig:ABPic}
\end{figure}

\citet{Cha05b} reported the discovery of a faint
companion to the Tuc-Hor association star AB Pic, with an estimated
mass of 13-14 M$_{Jup}$.  Compared to other objects of its spectral
type, AB Pic B's published $J$-band absolute magnitude 
is anomalously faint for its spectral type of L0.5
\citep{All13, Dup12}.  The published
colors of this object are also quite red for its spectral type.
During the NICI Campaign, we acquired new $J$ and $K_{S}$ photometry
for AB Pic B, presented here in Table~\ref{tab:ABPic}.  
While we measure a similarly red $J - K_{S}$=1.78$\pm$0.17 mag 
\citep[vs.~2.04$\pm$0.13~mag~from][]{Cha05b}, we find a considerably brighter
$J$ magnitude of 7.97$\pm$0.14 mag for b 
\citep[vs.~8.6$\pm$0.1~mag~from][]{Cha05b}.  In Figure~\ref{fig:ABPic}, we plot
spectral type vs. $J$ magnitude for AB Pic B and a number of comparison
objects.  The \citet{Cha05b} photometry places AB Pic B 
fainter than expected for its spectral type.
Assuming the measured difference in 
photometry is not due to true variability, our brighter $J$-band magnitude 
places AB Pic B firmly into the expected position for its spectral type.


\section{Statistical Analyses of the NICI MG Survey}

Here we present limits on the frequency of wide giant extrasolar
planets based on two different statistical analyses of our achieved sensitivities
for the MG sample.  Two of our sample stars have confirmed
planetary or planet-brown dwarf boundary companions, specifically
$\beta$ Pic and AB Pic.  The bona fide planet around $\beta$ Pic was not detected 
in our first epoch Campaign data while AB Pic B was clearly detected.  
Including two stars with known $<$20 M$_{Jup}$ companions
poses issues for determining an unbiased estimate of planet 
frequency from our survey.  Specifically, it is unclear how much 
we bias our estimate of planet frequency towards higher values by including 
a priori known companions.
Thus, for the purposes of this analysis, we exclude these two
stars from the sample.  In Section~\ref{Bayesdet}, we consider the effect of adding
these two stars and their companions.

\subsection{Monte Carlo Constraints on Planet Fraction}

Following the method of \citet[][]{Nie08} and \citet[][]{Nie10}, we use
Monte Carlo methods to constrain our sensitivity to planets 
around each target star and combine these results to place 
constraints on planet fraction across our entire moving 
group sample.  First,  we simulate 10000 planets with a given semimajor 
axis and mass, as well as randomly selected orbital 
parameters and eccentricity drawn from the eccentricity
distribution of radial velocity planets \citep[][]{Nie10}.  The ensemble of 
simulated planets in {mass, semimajor axis} variables are
then converted to equivalent contrasts and projected separations
using the COND models of \citet{Bar03} and the simulated orbital parameters.
This simulation was repeated at masses of 0.5 - 16.9 M$_{Jup}$, in steps of
0.164 M$_{Jup}$, and at semi-major axes of 0 - 4200 AU, with step size
varying as a function of distance 
(0.286 AU out to 20 AU, 5.333 AU from 20 - 100 AU, 7.333 AU from
100-210 AU, 10 AU from 210 - 500 AU, 20 AU from 500 - 1000 AU, 
40 AU from 1000-2000 AU, and 100 AU from 2000 to 4200 AU).
The converted ensemble is then compared with the attained 
ASDI and ADI contrast curves for the star to derive the percentage of simulated 
planets detected at the particular combination of semimajor axis
and mass.  In cases where a candidate companion was observed in only a single
epoch and thus not confirmed as background or common proper motion, 
we cut off the contrast curve at the separation of the unconfirmed
candidate companion or utilized a shallower contrast curve from an
earlier epoch where the unconfirmed candidate was not detected.
A number of stars near the Galactic bulge have been dropped from this
analysis due to numerous unconfirmed candidate companions,
specifically: CD -54 7336, CD -31 16041, HD 159911, V343 Nor, 
and HD 139084B. In total, 73 stars were used for this analysis.    
For the ASDI contrast curve comparison, the fluxes of 
simulated planets are modified to simulate the effect of
ASDI self-subtraction using the SpeX Prism Library of ultracool
dwarfs to partition flux between the on- and off-methane
absorption images (see Nielsen et al. 2013 and Nielsen \& Close 2010).
This contrast curve comparison procedure is then repeated along a grid of semimajor 
axes and masses.

After calculating the detection probability grid for each 
star in the sample, we use these values to place constraints 
on the planet frequency over the entire sample as a function of
semi-major axis and mass.   For a given bin in \{semi-major axis, mass\}, 
the number of planets we expect to detect is given by:
\begin{equation}
N(a,m) = \sum_{i=1}^{N_{obs}} f_{p}(a, m) P_{i}(a, m) \label{eqn:mcprob}
\end{equation}
where $P_{i}(a,m)$ is the fraction of planets with semimajor axis and mass
($a,m$) we could detect given the achieved contrast for star $i$
(i.e. the quantity calculated in our Monte Carlo 
simulations) and $f_{p}(a, m)$ is the fraction of stars that
have such a planet to detect, hereafter referred to as ``planet fraction''.

According to 
radial velocity studies, higher mass stars
may preferentially host giant planets compared to lower mass stars 
\citep{Joh07, Joh10}.  To account for this variation, we introduce a 
mass correction to adjust the probability that a given star hosts a
planet based on that star's mass:
\begin{equation}
C_{1.0}(M_{*}) = \frac{F_{p}(M_{*})}{F_{p}(1.0~M_{\odot})}
\end{equation}
where $F_{p}(M_{*})$ is the relative probability of hosting giant 
planets as a function of mass, based on the linear fit of 
planet frequency as a function of mass for RV planets 
from \citet{Joh10}.   The mass-corrected version of 
Equation~\ref{eqn:mcprob} is then:
\begin{equation}
N(a,m) = \sum_{i=1}^{N_{obs}} f_{p}(a, m) P_{i}(a, m) C_{1.0}(M_{*,i}) 
\end{equation}
We normalize this correction at 1 $M_{\odot}$ since our 
sample is composed primarily of FGK stars.
To estimate the mass of each of our sample stars,  
we interpolated from the models of \citet[][]{Sie00}.
First, we converted V and V-K to M$_{bol}$ and T$_{eff}$ 
using the lookup table developed for pre-main-sequence stars in 
\citet[][]{Ken95}.  Then we used the \citet[][]{Sie00} 
solar metallicity tracks for 0.1 - 7 M$_{\odot}$ stars
to find the stellar mass which best reproduces the observed M$_{bol}$ 
and T$_{eff}$. 

In the zero-detection case, we use Poisson statistics to 
set an upper limit on the planet fraction for our entire 
ensemble.  Assuming that planet fraction at a given 
semi-major axis and mass is the same for all survey stars, 
we remove $f_{p}$ from the sum.  The 95\% confidence 
level upper limit on planet fraction, $f_{p,95\%}$ is then:
\begin{equation}
f_{p,95\%}(a, m) \leq \frac{3}{\sum_{i=1}^{N_{obs}} P_{i}(a, m) C_{1.0}(M_{*,i})}
\end{equation}
where 3 is the Poisson expectation value to set a
95\% confidence upper limit on planet fraction
in the null result case.  

Many of our sample stars have binary companions, which may
disrupt the formation of planets in that system.  To account for
the effect of binary companions, we have followed the approach
detailed in Nielsen et al. (2013) and define an ``exclusion zone''
around each of the binaries in our sample in which we do not 
expect planets to form and thus where we do not simulate planets.
Binaries in our sample are listed in Table~\ref{tab:binaries}.

We also account for nonuniform position angle coverage of our observations at large angular
separations. The NICI detector is square, with the focal plane mask and target star placed offset
from the center. As a result, while we image 360$^{\circ}$ in position angle at small separations, at larger
separations ($\gtrsim$6.3$\arcsec$) our coverage declines as some position angles are off the edge of the detector.
In our Monte Carlo simulations we account for this effect by
generating a uniform random variable between 0 and 1 for each simulated
planet. If that random variable is greater than the fractional angular coverage at the projected
separation of the simulated planet, then that planet is considered undetectable even if it is brighter
than the contrast curve. This parameter is similar to the position angle of nodes (rotation of the
orbit on the plane of the sky), which follows a uniform distribution. When multiple contrast curves
are available for a single target star, this random variable is also preserved across all epochs so that
the same set of simulated planets are compared to each contrast curve for the same star.

\begin{figure}
\includegraphics[width=6in]{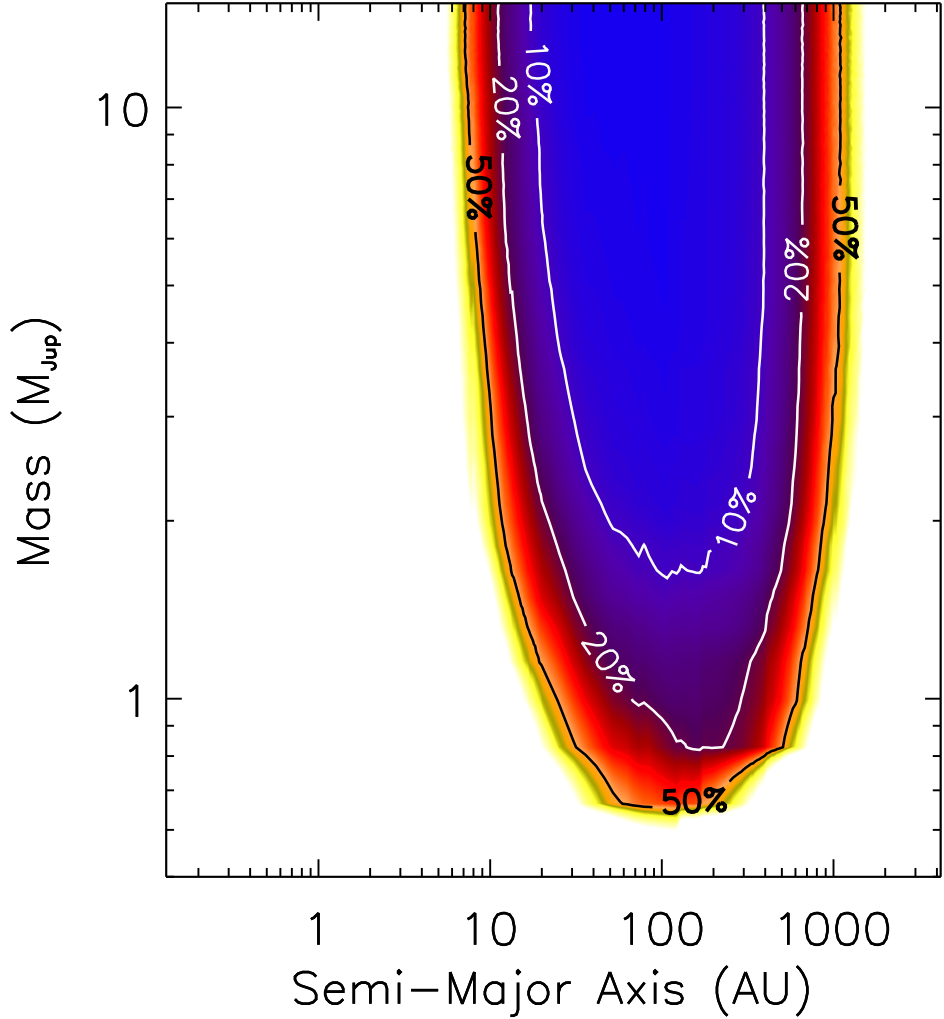}
\caption{The 95\% confidence upper limit on planet fraction 
as a function of semimajor axis and planet mass 
for our entire MG survey sample, using the models of \citet{Bar02, Bar03}
and the Monte Carlo method described in \citet{Nie08} and \citet{Nie10}.
We also utilize here a mass correction to adjust the probability that a given star hosts a
planet based on that star's mass, drawn from the linear fit of 
planet frequency as a function of mass for RV planets from \citet{Joh10}.
In general, we find that giant planets are rare at wide separations
in our sample: we expect less than 10$\%$ of stars to possess a 
2 M$_{Jup}$ planet at separations of 49 to 290 AU, at a 95$\%$ 
confidence level.  Note that this analysis does not assume a particular distribution 
of planets as a function of mass and semi-major axis.
 \label{fig:MG_CT}
}
\end{figure}

Figure~\ref{fig:MG_CT} gives the upper limit on planet fraction $f_{p,95\%}$ 
as a function of semimajor axis and planet mass 
for our entire moving group sample, using the models of \citet{Bar02}
to convert between achieved survey contrast and predicted detectable
planet mass.  Upper limits on planet fraction as a function of semi-major axis
for this analysis (i.e. single mass cuts from Figure~\ref{fig:MG_CT})
are presented in Table~\ref{table_pf}.  
Giant planets are rare at wide separations;
for instance we expect less than 10$\%$ of stars to possess a 
2 M$_{Jup}$ planet at semi-major axes of 49 to 290 AU.
Note that this analysis does not assume a particular distribution 
of planets as a function of mass and semi-major axis.

\subsection{A Bayesian Analysis of the NICI MG Survey}

Bayesian methods provide a powerful complement to frequentist
Monte-Carlo methods for interpreting large-scale direct imaging surveys
for exoplanets \citep[see~e.g.~][]{Nie08, Nie10, Bon12}.
Frequentist Monte-Carlo methods produce useful star-by-star
constraints but sometimes have difficulties interpreting positive detections.
In contrast, Bayesian methods produce less useful star-by-star
constraints but can seamlessly handle both null and positive
detections, as well as data analysis using 
multiple parameter models.  Here we apply
the Bayesian statistical analysis method pioneered by \citet{All07} to the NICI MG sample.
Our goal is to estimate the frequency of planets 
based on the observational constraints produced by our survey, 
given the limits of our sample size and sensitivity.

\subsubsection{Bayes' Theorem}

Bayes' theorem can be simply derived from the basic rules of
probability and provides a powerful means to analyze and interpret
data \citep[e.g.~][]{Siv06}.
At the end of our experiment, the quantity we would like to determine
is:
\begin{equation}
Prob(model | data, I)
\end{equation}
which is the probability that a given model is correct
given the data in hand as well as any other prior information $I$.
This quantity is the posterior probability distribution 
function (henceforth posterior PDF).  The power of Bayes'
theorem is it allows us to relate the posterior PDF to other, 
more easily calculated quantities:
\begin{equation}
Prob(model | data, I) \propto Prob(data | model, I) \times Prob(model| I) \label{eqn:Bayes}
\end{equation}

The quantity $Prob(data | model, I)$ is known as the likelihood function -- it is 
the probability of obtaining the data on hand given a specific
model and additional prior information.  The quantity $Prob(model | I)$
is known as the prior probability (or simply the prior) 
and includes any additional prior information
we know about the problem.  Thus, by formulating reasonable likelihood
functions and priors for a direct imaging planet detection survey, 
we can derive the 
posterior PDF and constrain models for the underlying planet
population.  \footnote[1]{By presenting Bayes' theorem as a proportionality, we have 
omitted a possible term of interest.  The value $Prob(data|I)$ which
we have omitted from the denominator of Equation~\ref{eqn:Bayes} is known as
the Bayes factor or the evidence.  The Bayes factor allows for a full normalization of
the probability and can be used to compare the likelihoods of 
competing models.  For our current parameter estimation case, it is 
not necessary to calculate the Bayes factor.}

\subsubsection{Description of Method}

\subsubsubsection{Calculating the posterior PDF for one bin in observable space}

We adapt the method established in \citet{All07} and 
\citet{Kra11} for studying stellar binarity in the context of a direct imaging survey
of exoplanets.  \citet{All07} model the distributions of 
substellar and stellar binary mass ratios and semi-major axis as a power
law in mass ratio and a Gaussian in semi-major axis (henceforth $a$).
For exoplanet companions to stars, we adopt instead the form of the 
power law distributions derived for RV planets by \citet{Cum08}, 
and consider only the planet mass (henceforth $m$) rather than 
the mass ratio adopted for binaries:
\begin{equation}
\frac{dN}{dm} \propto m^{\alpha}
\end{equation}
\begin{equation}
\frac{dN}{da} \propto a^{\beta}
\end{equation}
Following the procedure of \citet{Nie08} and \citet{Nie10}, we
extend the semi-major axis power-law out to a limiting cutoff value, since
earlier studies already rule out a significant population of giant planets
at very wide separations\citep{Nie10}.   While such planets do exist
\citep[e.g.~][]{Laf08, Ire11}, they are much less common than 
planets detected via radial velocity at closer separations
\citep{Fis05, Cum08, Nie10}.

Thus, we are left with 4 parameters to our models: the two 
power-law indices $\alpha$ and $\beta$, the outer cutoff of the semi-major axis
distribution (henceforth $a_{max}$), and $F$, the fraction of stars with planets.  
We define $F$ such that the planet fraction over a 
given range of semimajor axes and masses is:
\begin{equation}
F = C_{0} \int_{m_{min}}^{m_{max}} \int_{a_{min}}^{a_{max}} \!m^{\alpha} a^{\beta} \, \mathrm{d} m~\mathrm{d} a 
\label{eqn:F}
\end{equation}
where C$_{0}$ is a normalization constant (and thus a function of F, $\alpha$,
$\beta$, and a$_{max}$).

The probability to find a planet around a star in a given
\{semi-major axis, mass\} bin, for a particular set of values for 
$\alpha$, $\beta$, $a_{max}$, and $F$, is then the planet fraction
within that bin:
\begin{equation}
R(a, m | \alpha, \beta, a_{max}, F ) =
F_{bin}=C_{0}(F,\alpha,\beta,a_{max}) m^{\alpha} a^{\beta},~~~~~~a \leq a_{max} 
\end{equation}
\begin{equation}
R(a, m | \alpha, \beta, a_{max}, F ) = F_{bin}=0,~~~~~~ a > a_{max}
\end{equation}
where C$_{0}$ can be determined from Equation~\ref{eqn:F}.

To compute the likelihood, we calculate how many planets 
we expect to detect with this model in each \{semimajor axis, mass\} bin and 
compare with the actual number of planets (generally 0) detected in each bin,
accounting for projection effects between semi-major axis and projected
separation.  For a given \{semimajor axis, mass\} bin and set of model
parameters, the number of planets predicted will be:
\begin{equation}
N_{pred} (a, m) = N_{obs} R(a, m | \alpha, \beta, a_{max}, F )
\end{equation}  
where N$_{obs}$ is the number of times this \{semimajor axis, mass\} bin 
was observed in our survey (derived from the contrast curves and
stellar properties of each survey star).    For instance, if we
observe 50 stars in our survey and 30 of the observed stars have 
contrasts deep enough to image a 10 M$_{Jup}$ planet at a semimajor
axis of 10 AU, then $N_{obs}$(10 AU, 10 $M_{Jup})$ = 30.
We then wish to compare $N_{pred} $ with $N_{det}$, the 
number of planets detected for a given \{semimajor axis, mass\} bin.  
To compare data and model, we need to adopt a likelihood estimator.
Since we expect to detect only small numbers of planets, our
survey can be treated as a counting experiment.  Thus, we adopt
Poisson statistics to calculate the likelihood:
\begin{equation}
likelihood = prob(N_{det} | \alpha, \beta, a_{max}, F ) = \frac{N_{pred}^{N_{det}} exp(-N_{det})}{N_{det}!}
\end{equation}
To derive the posterior PDF for this bin, we must multiply 
the likelihood by any prior probability distribution for our
parameters. For now, we adopt the simple uniform priors for $\alpha$, 
$\beta$, F, and a$_{max}$:
\begin{equation}
prob(\alpha|I) = prob(\beta|I) = prob(F|I) = prob(a_{max}|I) = 1
\end{equation}
Multiplying the likelihood and prior then yields the posterior PDF for
this \{semimajor axis, mass\} bin. 

\subsubsubsection{Generalization across observable space}

In the last section,  we showed how to calculate the posterior PDF for one 
\{semimajor axis, mass\} bin.  This can be generalized across all \{semimajor axis, mass\} bins
for the survey fairly easily.  We generalize $N_{obs}$ and $N_{det}$ 
into 2d arrays for each \{projected separation, mass\} bin observed, which we will henceforth call the window function
and detection array, respectively.  

To build the window function, we use the contrast curve for each survey star
to define the ranges in separation and mass where planets can be detected 
and the ranges where the contrast is insufficient to do so.
Often stars were observed in both ADI and ASDI modes; in these cases, 
we adopt the best contrast value from the available curves at each given separation.   
When ASDI contrast curves are used, they are
corrected for spectral self-subtraction, assuming no methane
absorption (i.e. the most conservative contrast case).
In cases where a candidate companion was observed in only a single
epoch and thus not confirmed as background or common proper motion, 
we cut off the contrast curve at the separation of the unconfirmed
candidate companion or utilized a shallower contrast curve from an
earlier epoch where the unconfirmed candidate was not detected.
A number of stars near the Galactic bulge have been dropped from this
analysis due to numerous unconfirmed candidate companions,
specifically: CD -54 7336, CD -31 16041, HD 159911, V343 Nor, 
and HD 139084B.   In total, 73 stars were used for this analysis.
Bins where a planet can be detected are assigned a value of 1 and
bins where no planet can be detected are assigned a value of 0.
We account for nonuniform position angle coverage of our contrast curves
by multiplying the window function for each star by the fractional
coverage at each separation.
We then convert the window function expressed in projected angular 
separation and contrast 
to projected physical separation and estimated mass using the known distance
and age of each star and either the DUSTY models of \citet{Bar02}
or the COND models of \citet{Bar03}.  
The detection array is set up in a similar manner --- as a
simple array with the number of objects detected in each
\{separation,mass\} bin.  
As exoplanets cool with age, dust should condense from their atmospheres, 
producing a transition from red, dusty spectra (DUSTY)
to bluer spectra characterized by methane absorption (COND). 
However, no directly imaged planet to date has yet to show methane
absorption in the near-IR, so we choose here to
present results using both of these models.  

We then calculate the posterior PDF for each \{separation, mass\} point.
This calculation is accomplished using a 
small scale Monte-Carlo simulation.
At each physical separation point, we simulate 10$^6$ planetary
orbits, drawing eccentricity, orbital phase, and other orbital elements randomly.
We solve for the semi-major implied for each simulated
orbit, then produce a histogram of the result with 
a 5 AU binsize.  The posterior PDF is calculated for each 
semi-major axis bin in this histogram and then weighted according to the 
number of simulated orbits falling into that bin to 
produce the posterior PDF at each given \{separation, mass\} point.
We calculate the posterior in this manner at each 
\{separation, mass\} point and then multiply the posterior PDFs 
across all these points to get the 
full posterior PDF across observable space for this set of model
parameters.  This process is repeated for all sets of model parameters
of interest to derive the full posterior pdf as a function of the four
model parameters.  
 
\subsubsection{Results with no planet detections \label{Bayesnodet}}

To determine what section of parameter space can be ruled out 
by our exoplanet non-detection around MG stars, we ran the  
Bayesian analysis with all four parameters allowed to vary.
Since our contrast curves are only 95\% complete, 
we systematically estimate a slightly low planet fraction, but this
effect is likely minor.  Additionally, we have
  adopted hot-start models here, which predict considerably brighter
planets at these young ages compared to cold-start models
\citep[for~instance~][]{Spi12}.  Thus, we predict systematically
more stringent upper limits on planet fractions than would be found 
with cold start models.
For window functions and the detection function, we
considered a linear grid in separation (in AU) running from 10.5 to 1015.5
AU, with points every 5 AU and a linear grid in mass (in Jupiter
masses) running from 0.2 to 19.2 $M_{Jup}$, with points every 1 
$M_{Jup}$, thus fully covering the mass range of possible planets 
as well as low mass brown dwarfs 
which could plausibly form via core accretion \citep{Sch11}.
The grids for $\alpha$ and $\beta$ were centered on
the values $\alpha$=-1.16 and $\beta$=-0.61, 
derived from radial velocity planet distributions \citep{Cum08}
and converted from the logarithmic units used in \citet{Cum08}
to linear units here.  We allowed $\alpha$ to run from -2.09 to 
-0.16 in increments of 0.066, $\beta$ to run from -1.54 to 0.39, in increments
of 0.066, $F$ to run from 0.005 to 0.972 in increments of 0.033, and 
semi-major axis cutoff $a_{max}$ to run from 12.5 AU to 152.5 AU in increments 
of 5 AU.  We choose to investigate this range of semi-major axis 
cutoff values as a value of $a_{max} <$ 10 AU is ruled out by 
radial velocity studies \citep{Cum08, Fis05} and a value of 
$a_{max} >$ 150 AU is ruled out by previous directly imaging
studies \citep{Nie10}.   Planet fraction $F$, and hence also the normalization 
constant $C_0$,  are calculated over the range 10 -- 150 AU.  

\begin{figure}
\begin{tabular}{cc}
\includegraphics[width=3.2in]{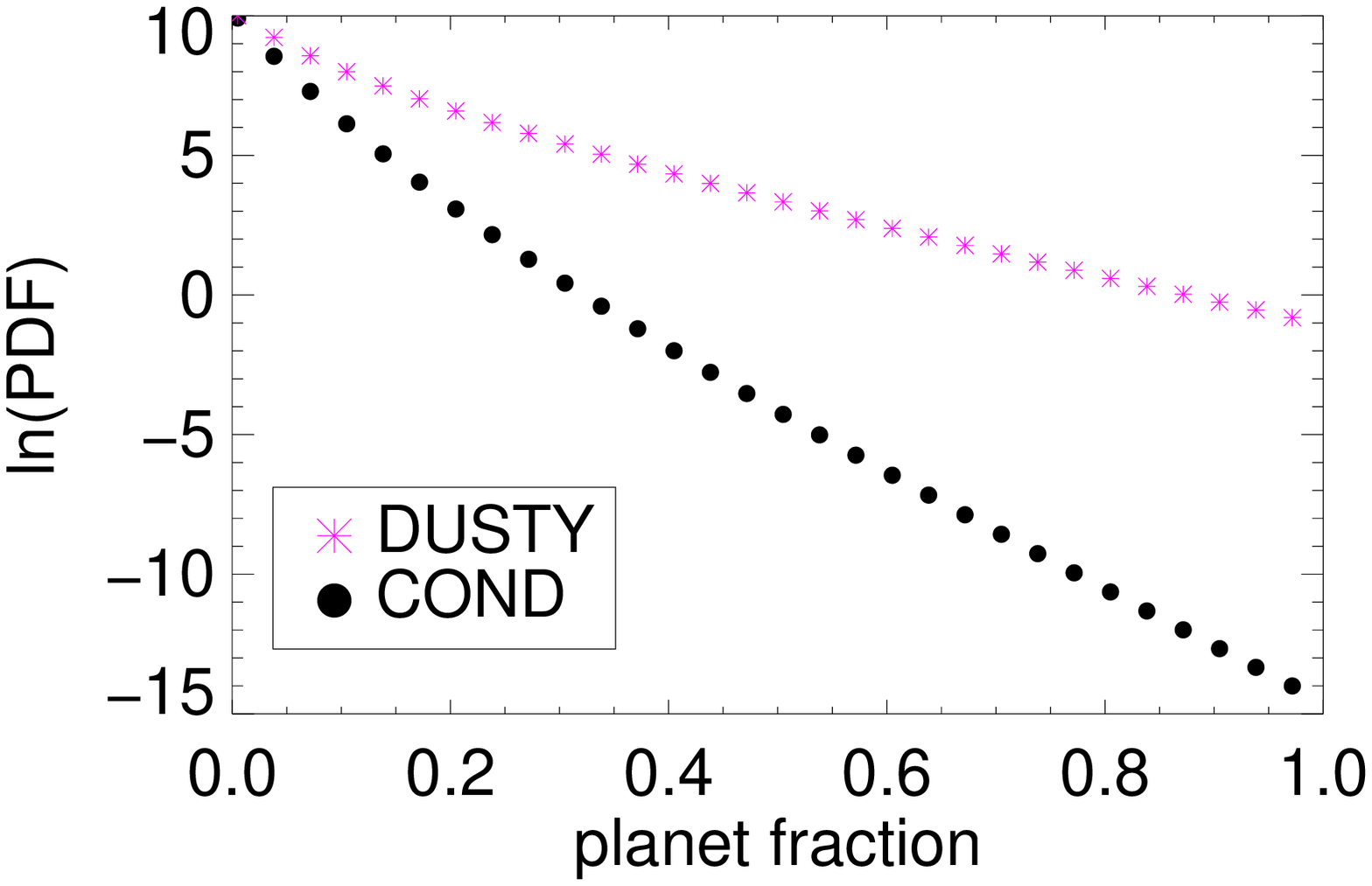} &
\includegraphics[width=3.2in]{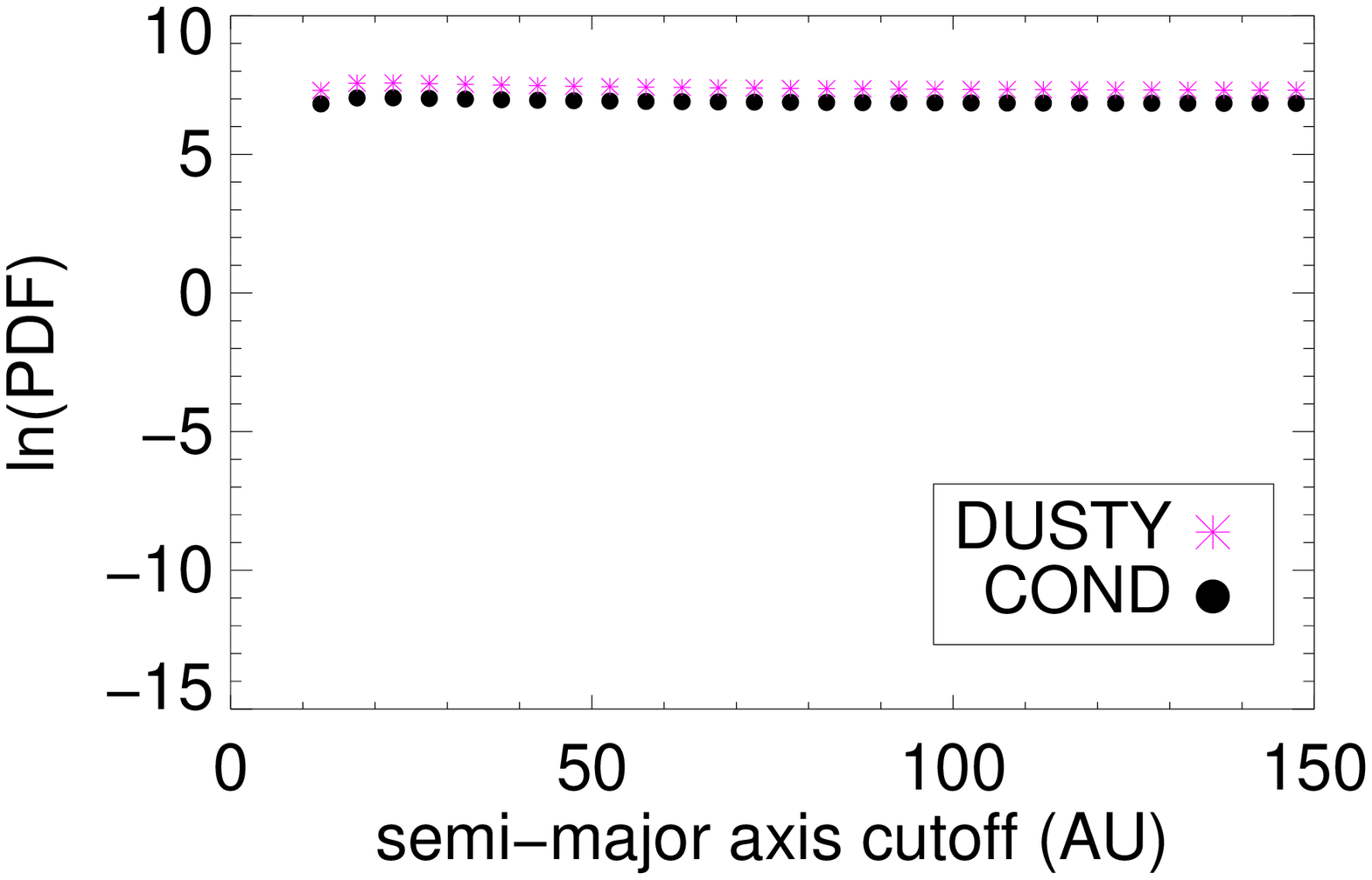} \\
\includegraphics[width=3.2in]{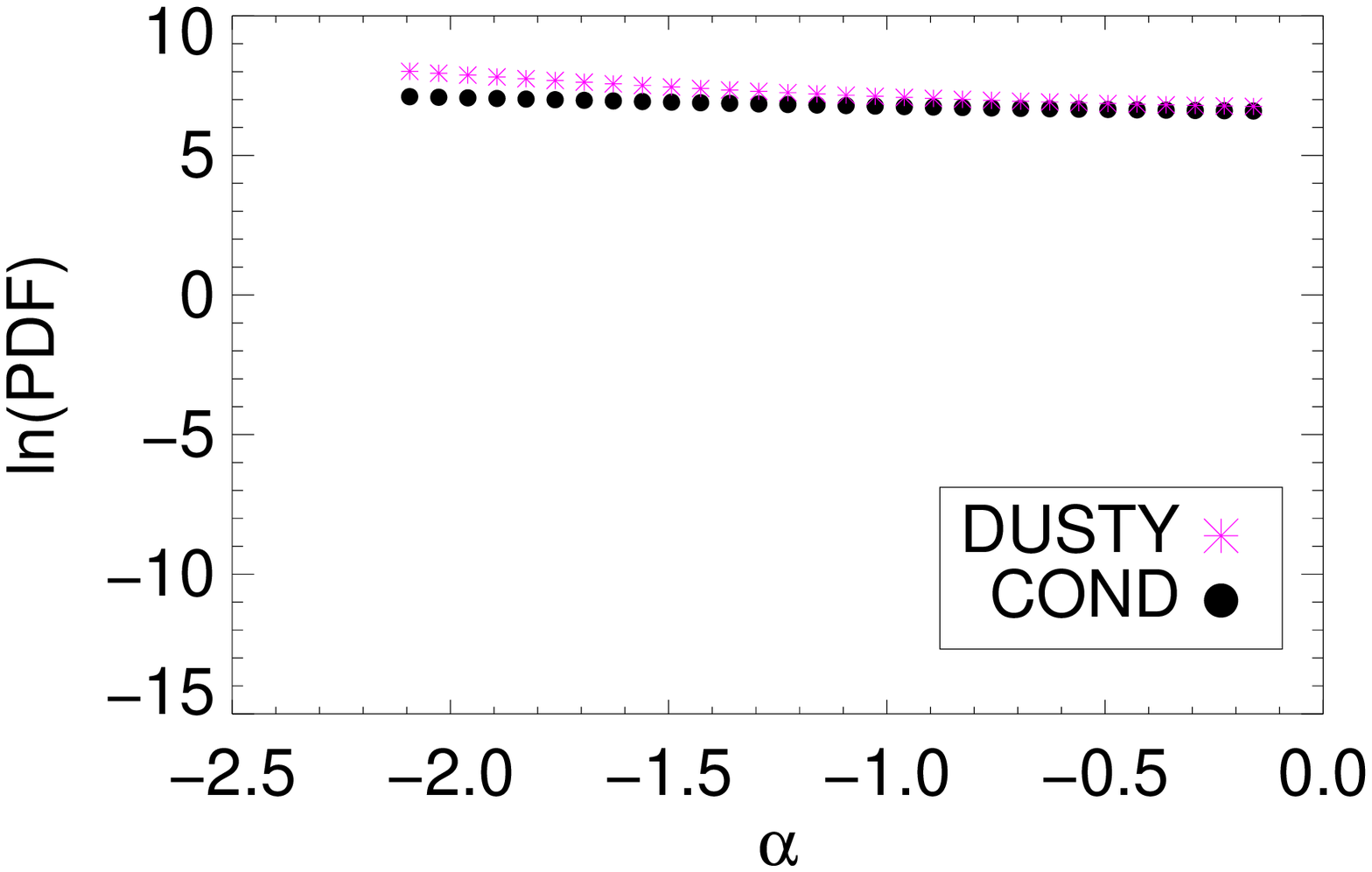} &
\includegraphics[width=3.2in]{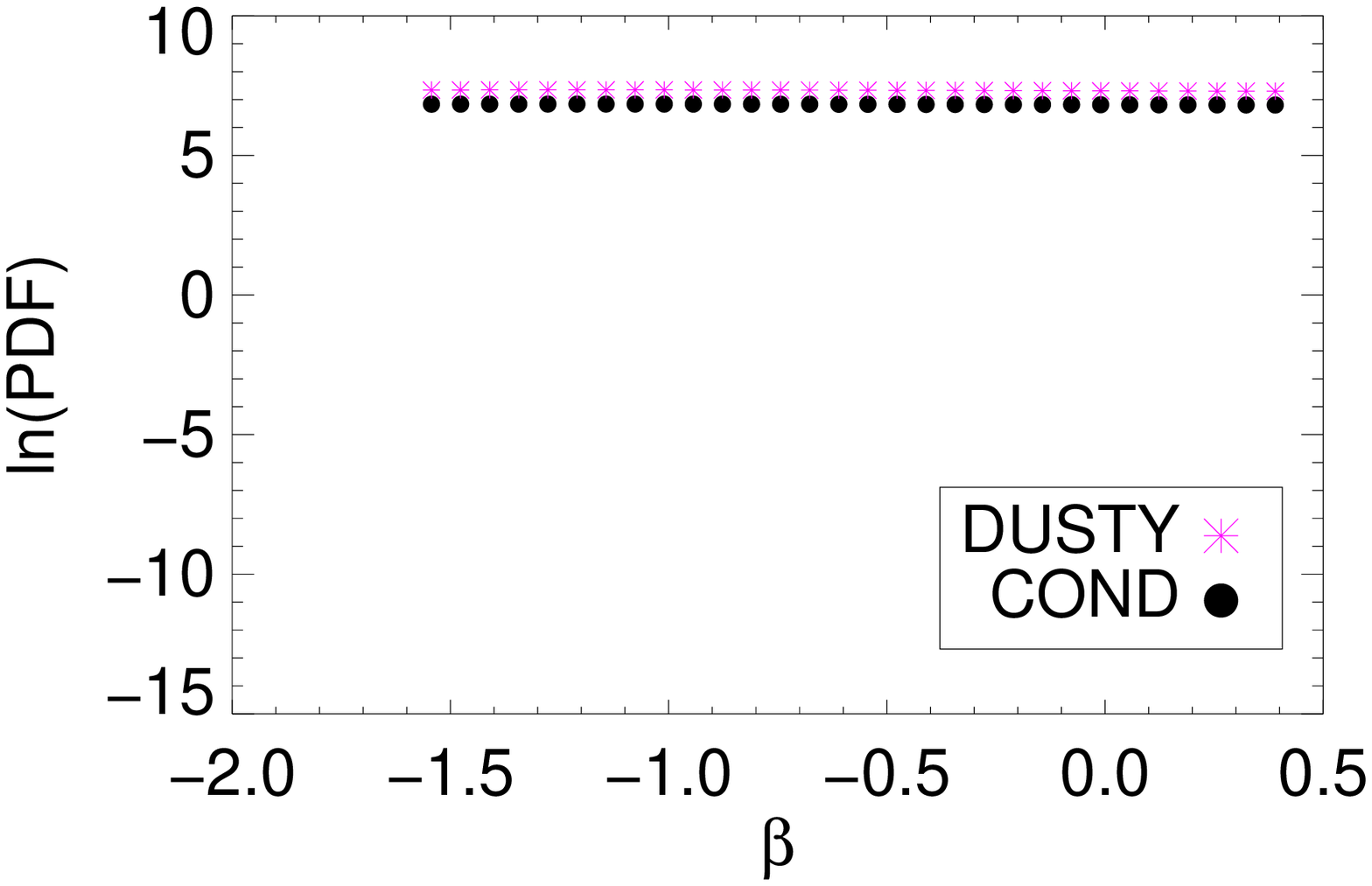} \\
\end{tabular}
\caption{1-d marginalized posterior PDFs including 4 free parameters,
  using the DUSTY models of \citet{Bar02} (magenta asterisks) and 
  the COND models of \citet{Bar03} (black filled circles).  Posterior probability 
  is plotted in natural logarithmic units.  The same plot range in
  ln(PDF) is used for each 1-d marginalized posterior PDF in order to
  clearly illustrate that, except for the planet frequency $F$,  these marginalized
  posteriors remain unconstrained (i.e. no
  clear peak or trailing off to 0).  Thus, we have only put confidence
  intervals on the planet fraction for semi-major axes between 10-150 AU.
  Using the DUSTY models, at a 95.4$\%$ confidence level, planet fraction must be $\leq$18$\%$
  marginalized over the ranges alpha=[-2.1, -0.2], 
  beta=[-1.5, 0.4], and cutoff=[10 AU, 150 AU].
  Using the COND models, at a 95.4$\%$ confidence level, planet fraction must be $\leq$6$\%$, 
  marginalized over the ranges alpha=[-2.1, 0.2], 
  beta=[-1.5, 0.4], and cutoff=[10 AU, 150 AU].
\label{fig:1d_4param_DUSTY}
}
\end{figure}

\begin{figure}
\begin{tabular}{cc}
\includegraphics[width=3.0in]{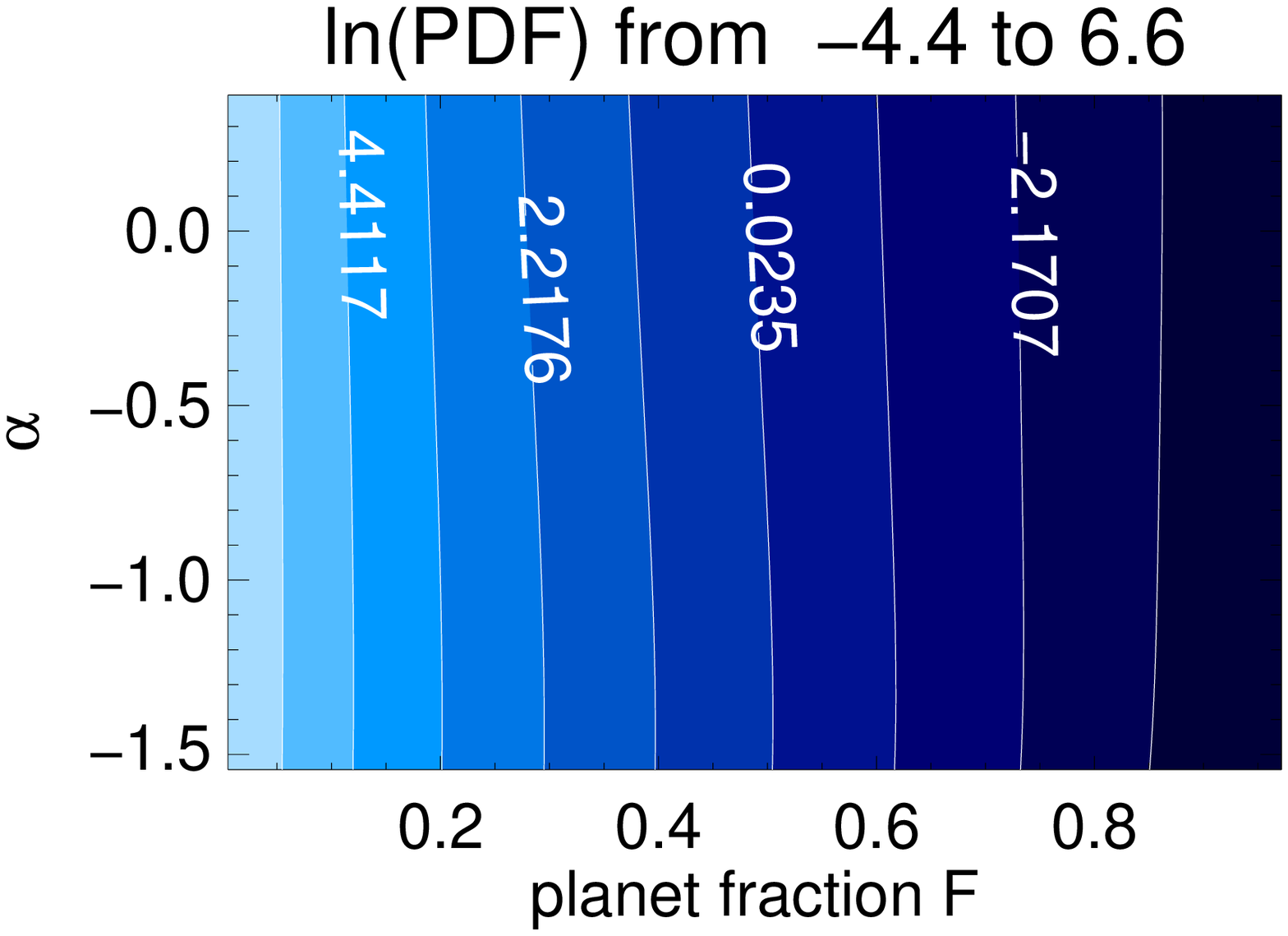} &
\includegraphics[width=3.0in]{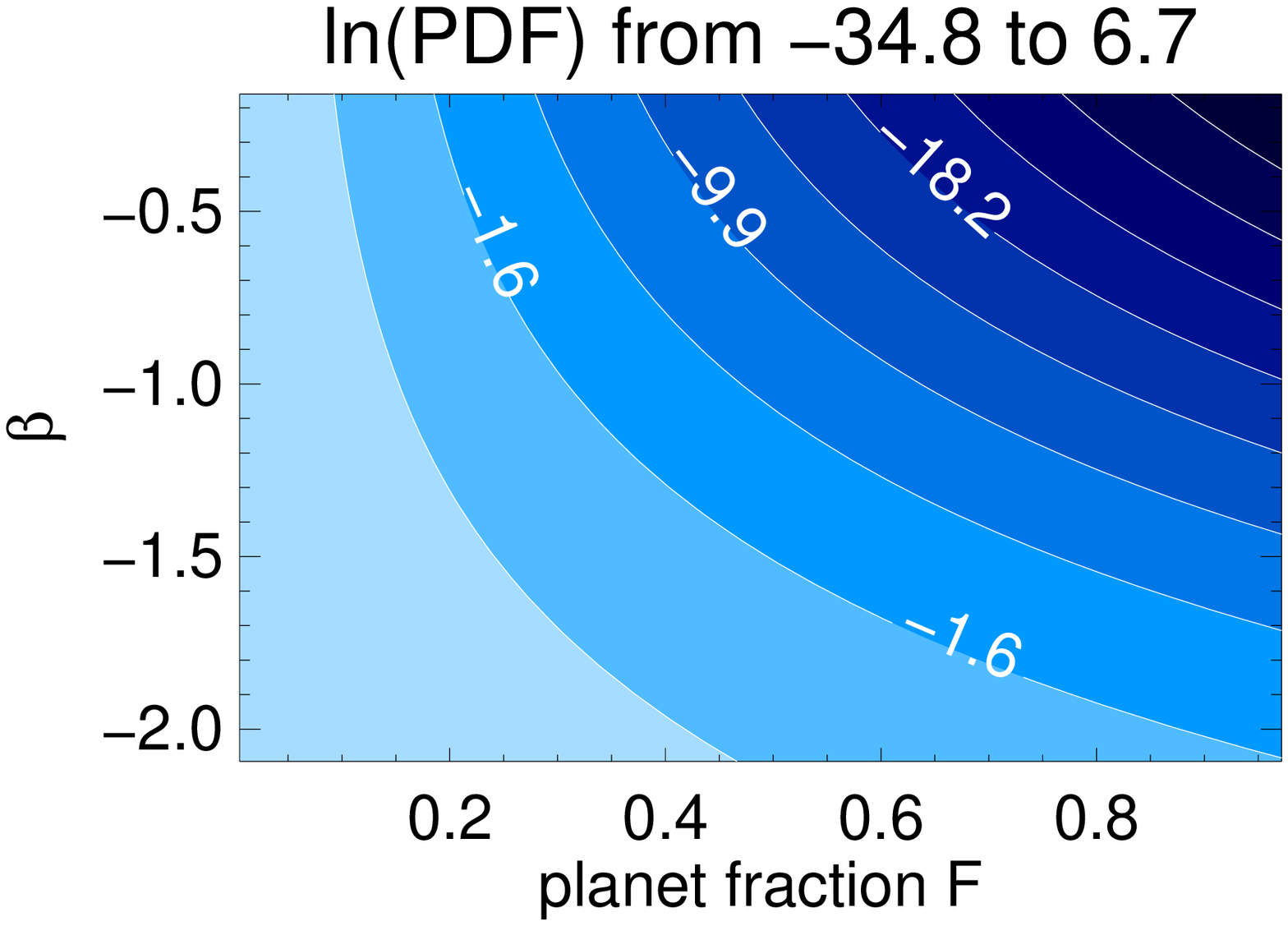} \\
\includegraphics[width=3.0in]{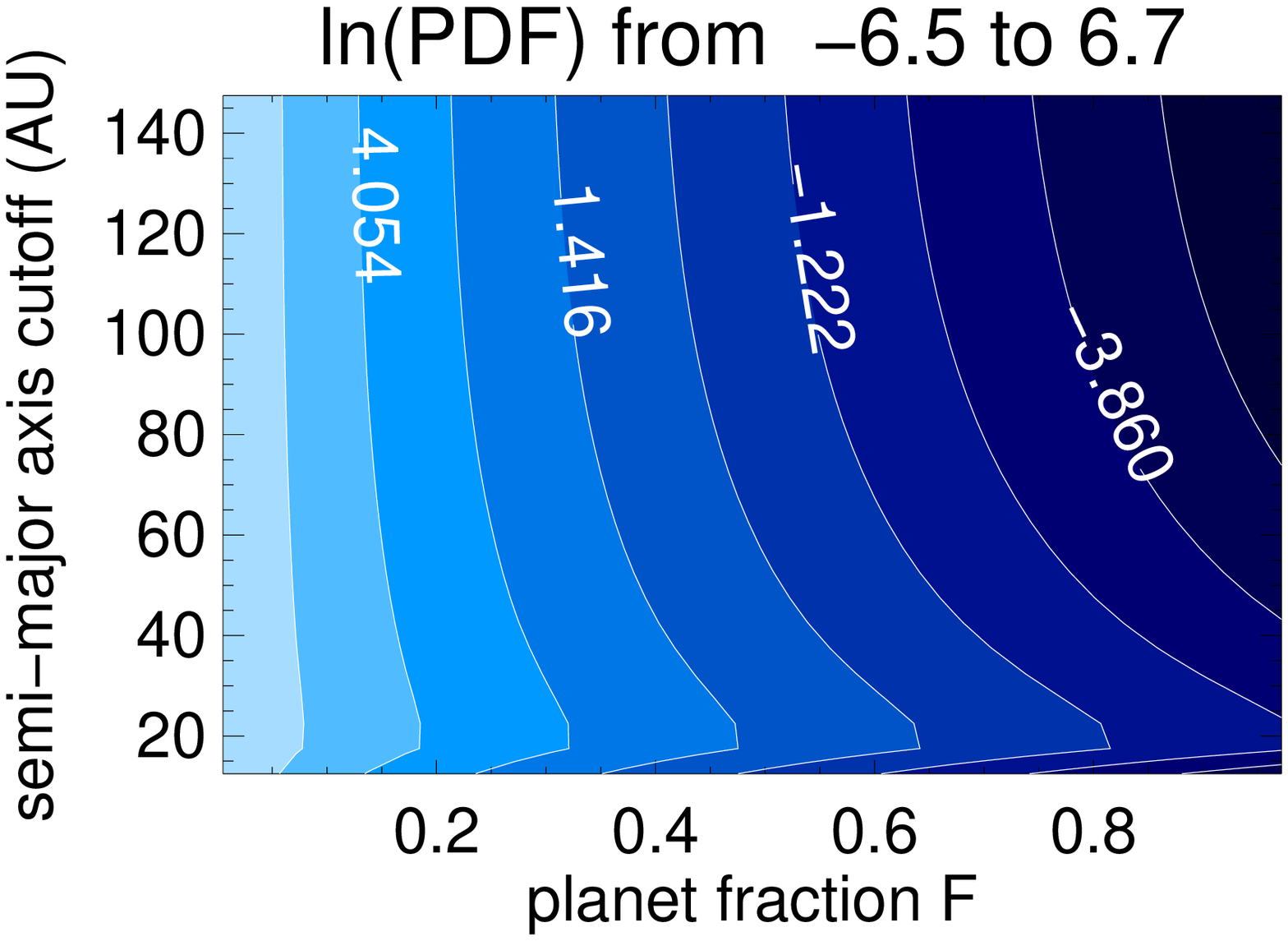} &
\includegraphics[width=3.0in]{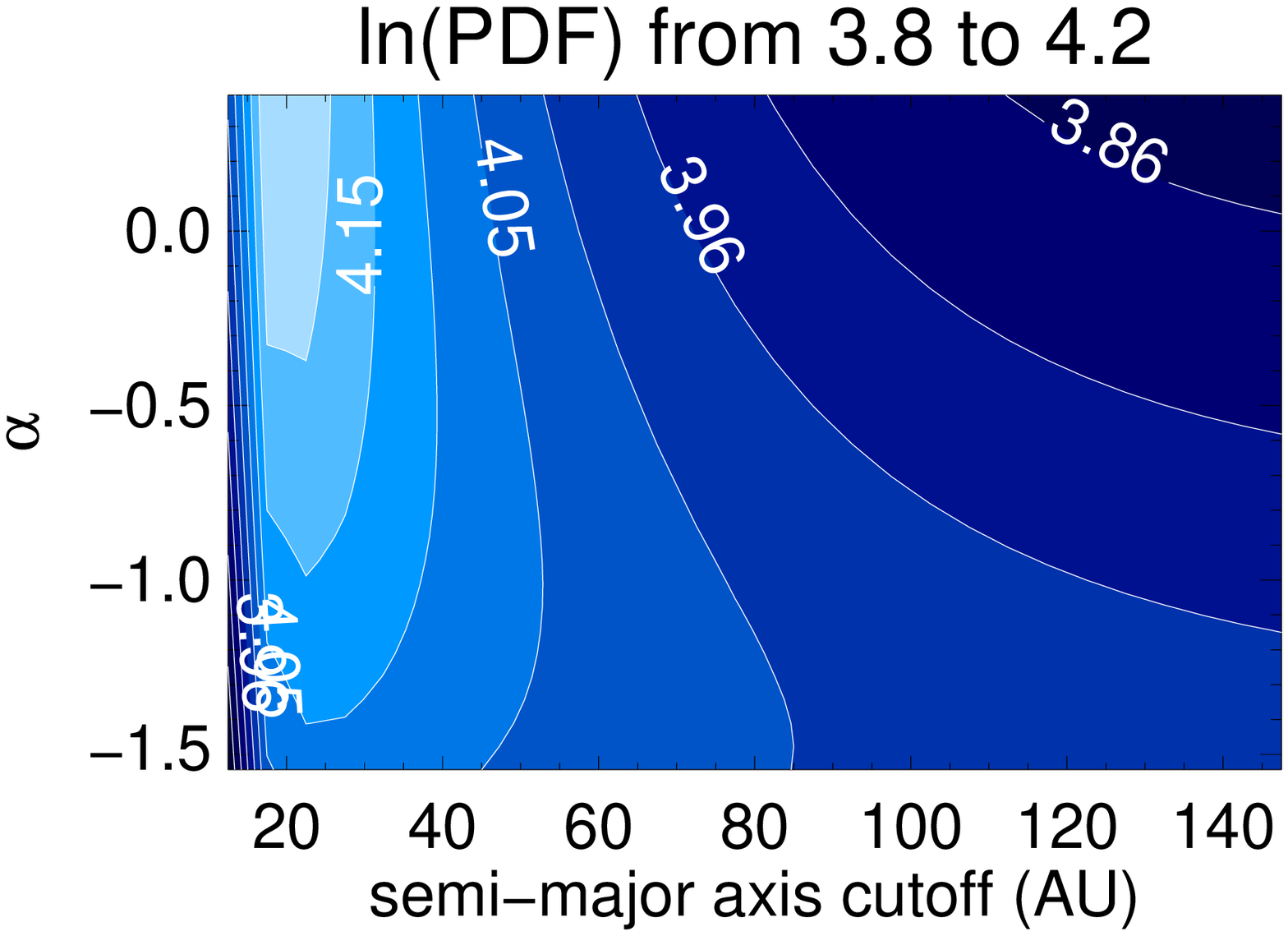} \\
\includegraphics[width=3.0in]{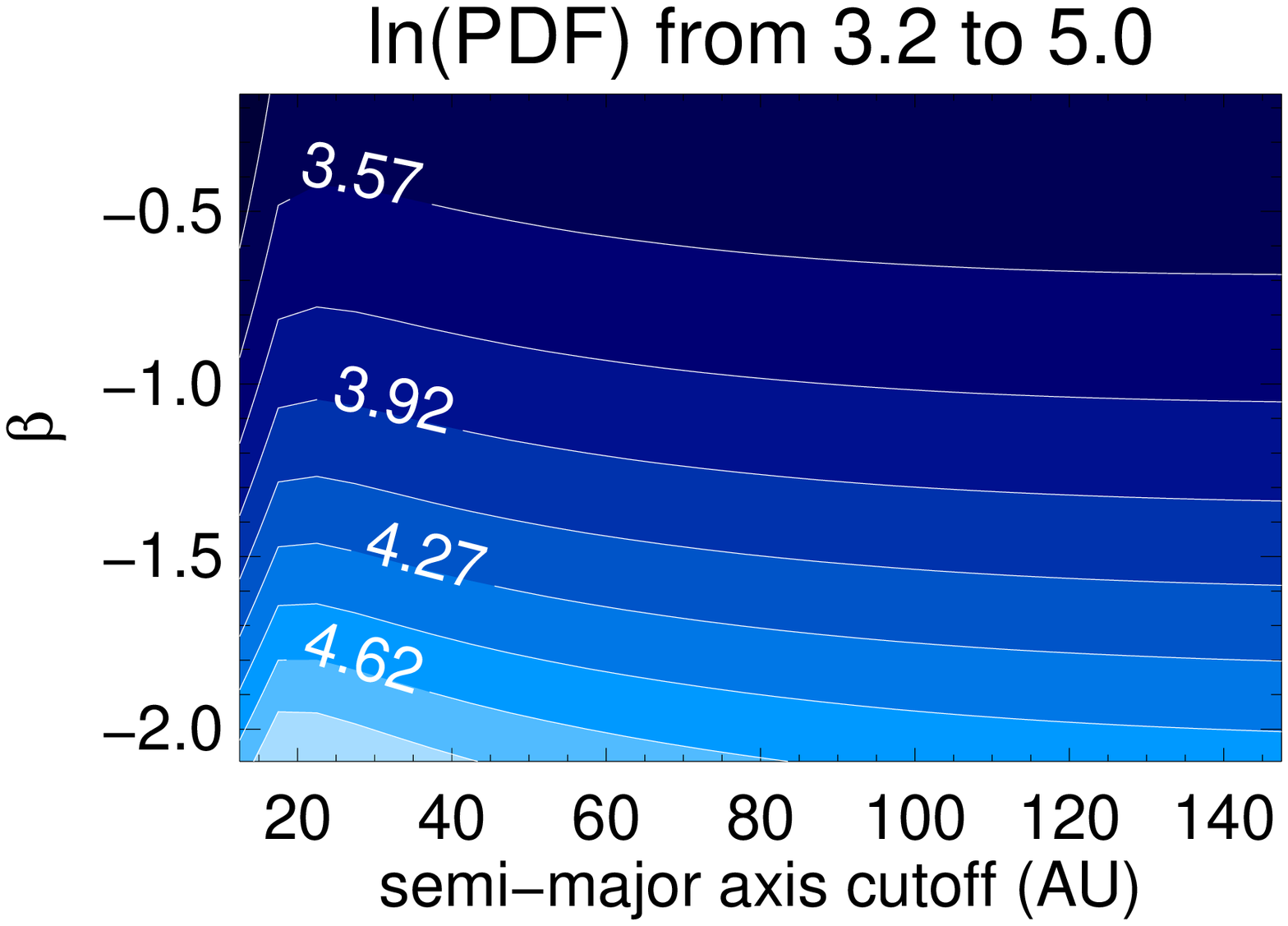} &
\includegraphics[width=3.0in]{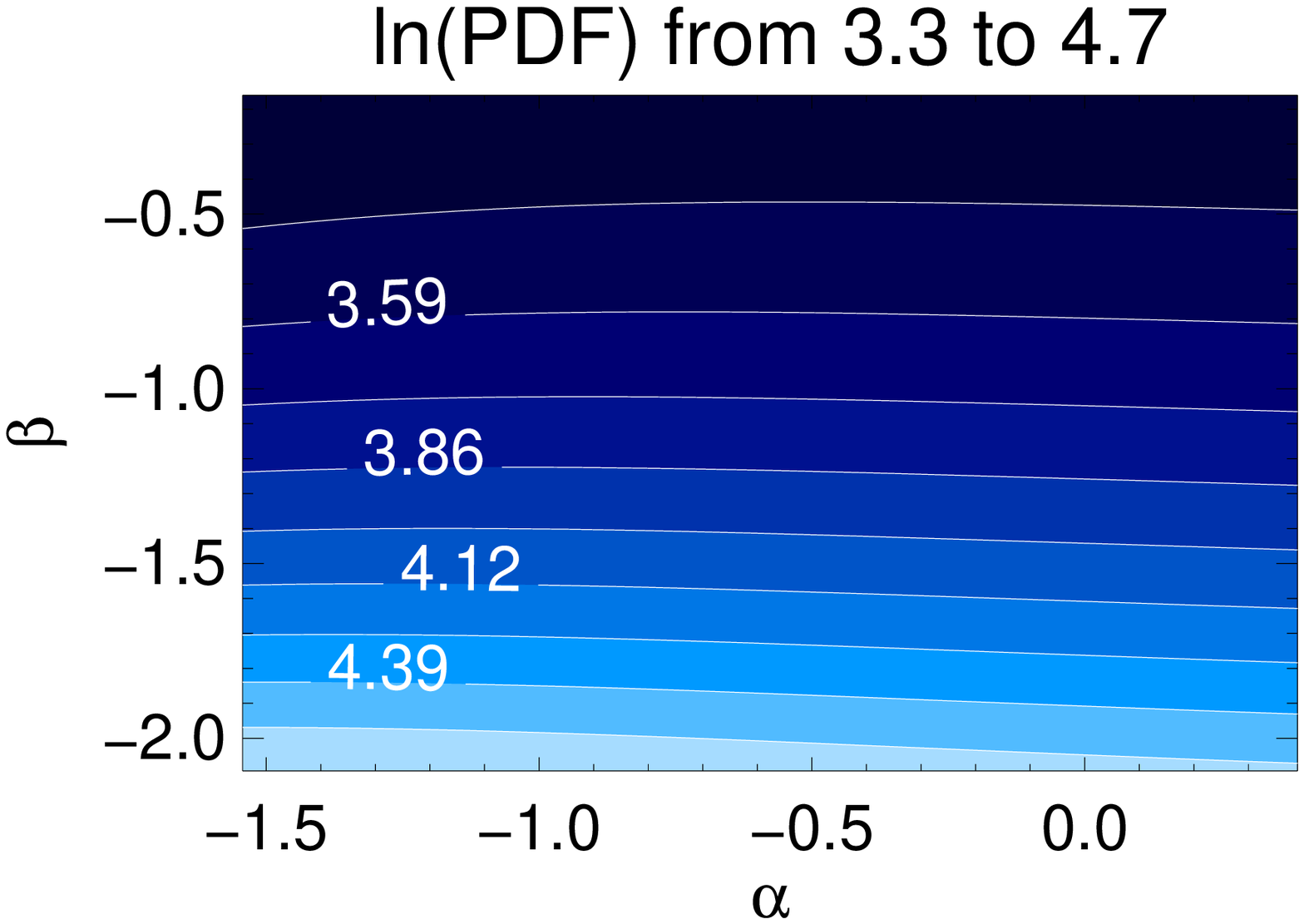} \\
\end{tabular}
\caption{2-d marginalized posterior PDFs with 4
  free parameters, with contours displayed as ln(PDF)
and using the DUSTY models of \citet{Bar02}.  Posterior probability 
  is plotted in logarithmic units, with 10 contour levels
equally spaced in logarithmic space from the minimum
value of each posterior PDF to the maximum value.
As different posterior PDFs traverse very different 
probability ranges, we have included the range of
ln(PDF) values plotted in the title for each subplot.
Posterior PDFs which cover a greater probability range
are more constrained.  The ratio of these units for two different contour levels
yields the relative likelihood of parameter combinations
along those respective contours.  Darker regions 
indicate parameter combinations with lower likelihood.
We have not normalized these marginalized posterior PDFs, as 
they remain unconstrained in the null detection case.
\label{fig:2d_4param_DUSTY}}
\end{figure}

\begin{figure}
\begin{tabular}{cc}
\includegraphics[width=3.0in]{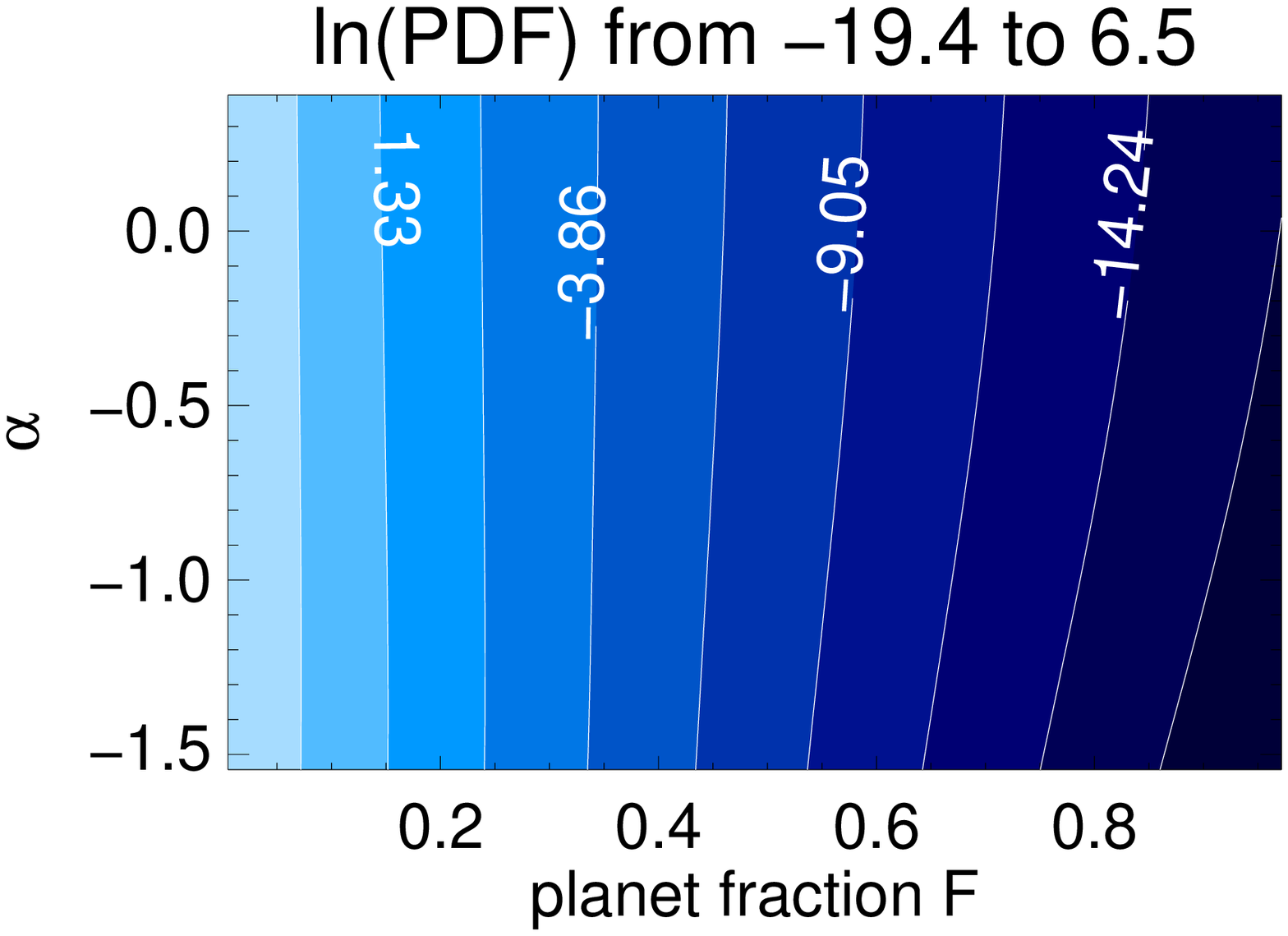} &
\includegraphics[width=3.0in]{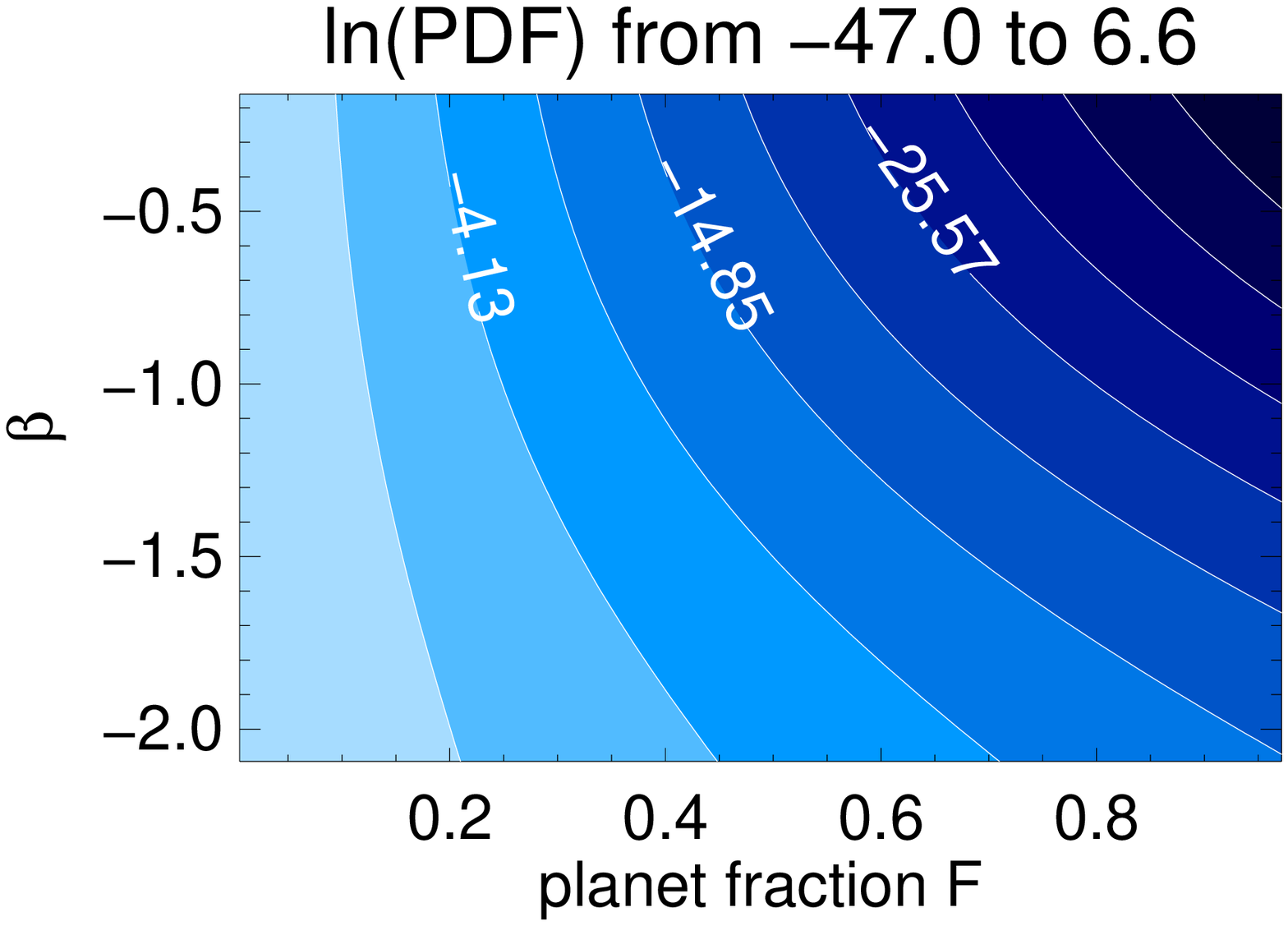} \\
\includegraphics[width=3.0in]{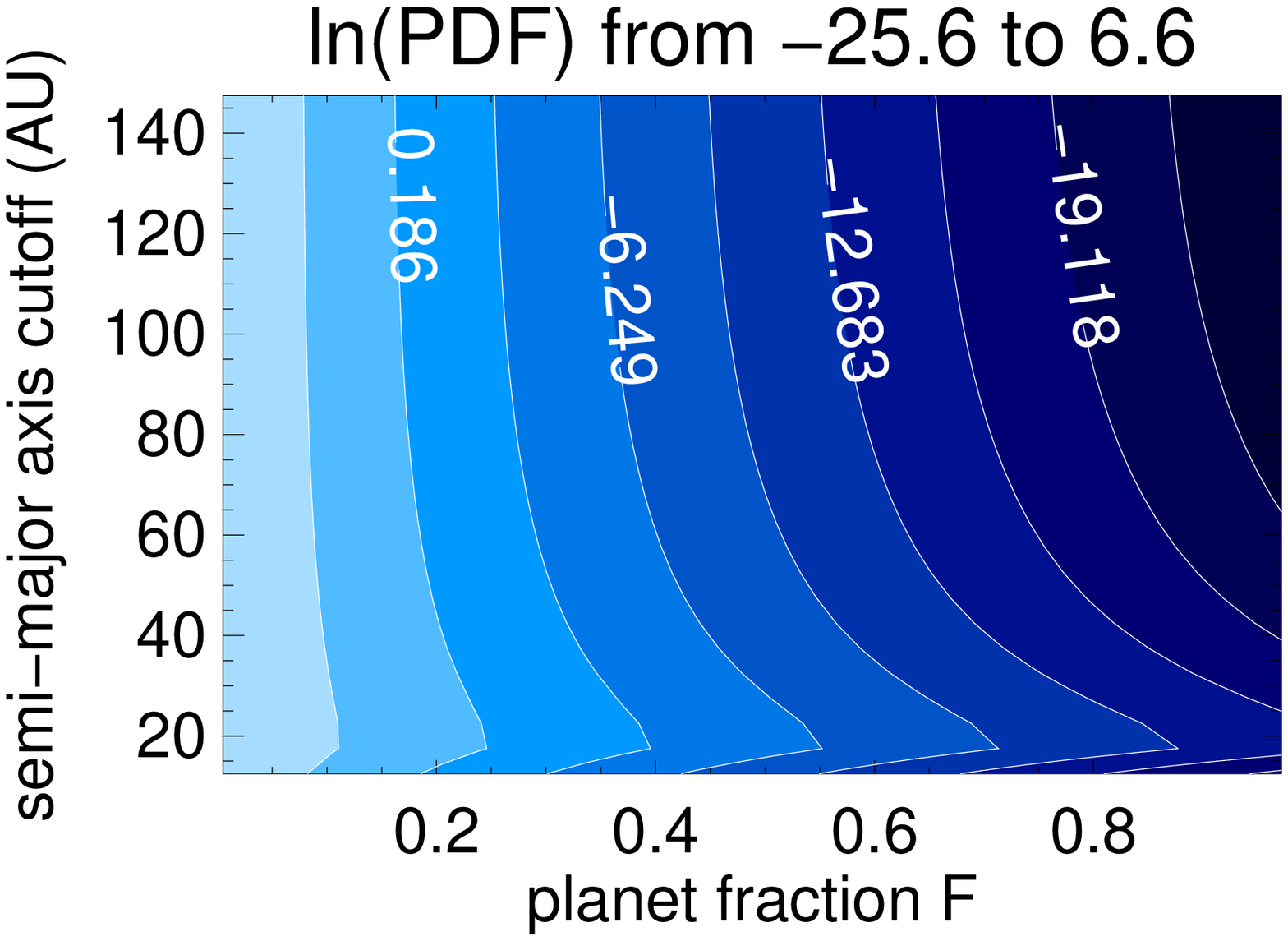} &
\includegraphics[width=3.0in]{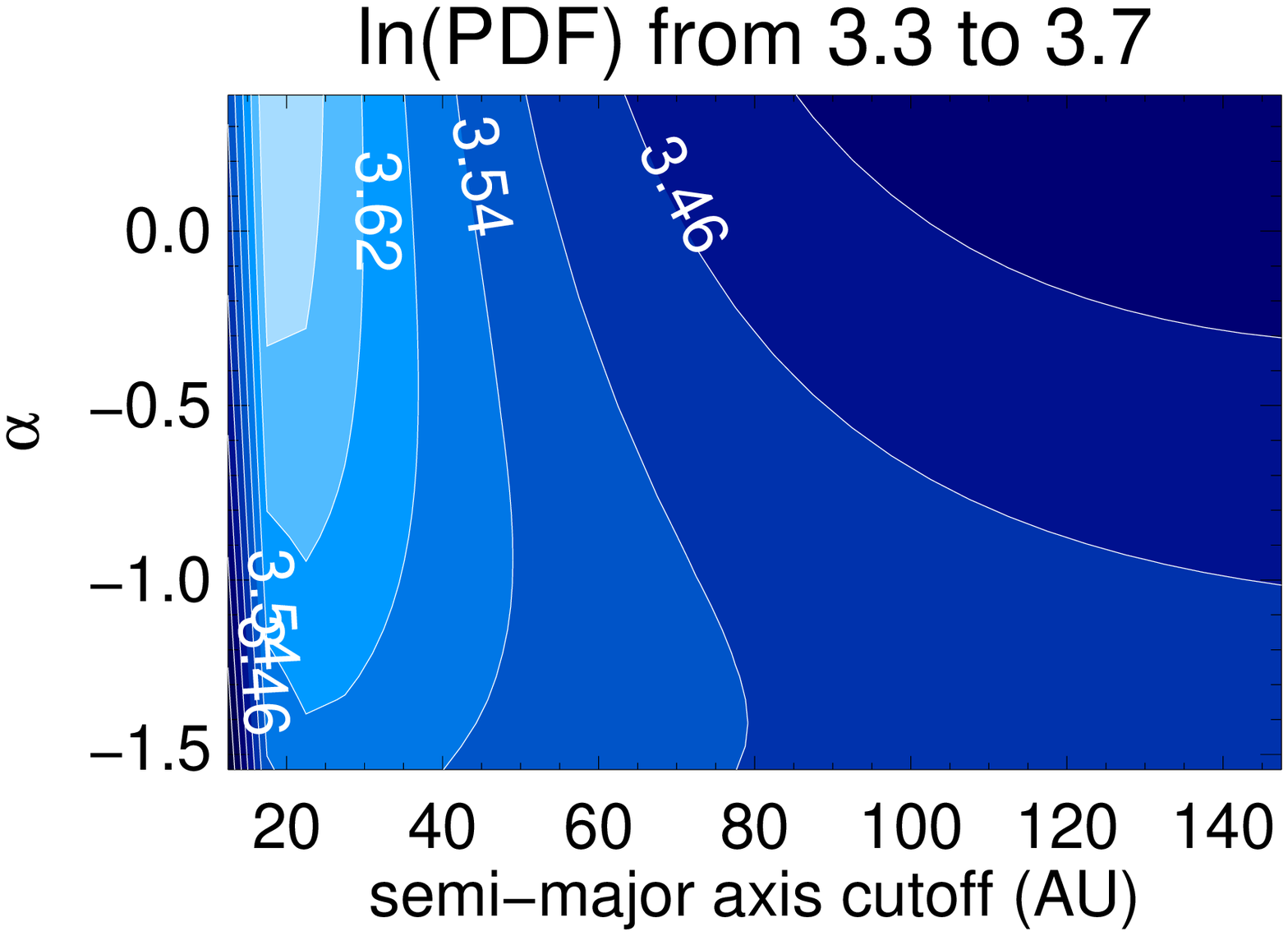} \\
\includegraphics[width=3.0in]{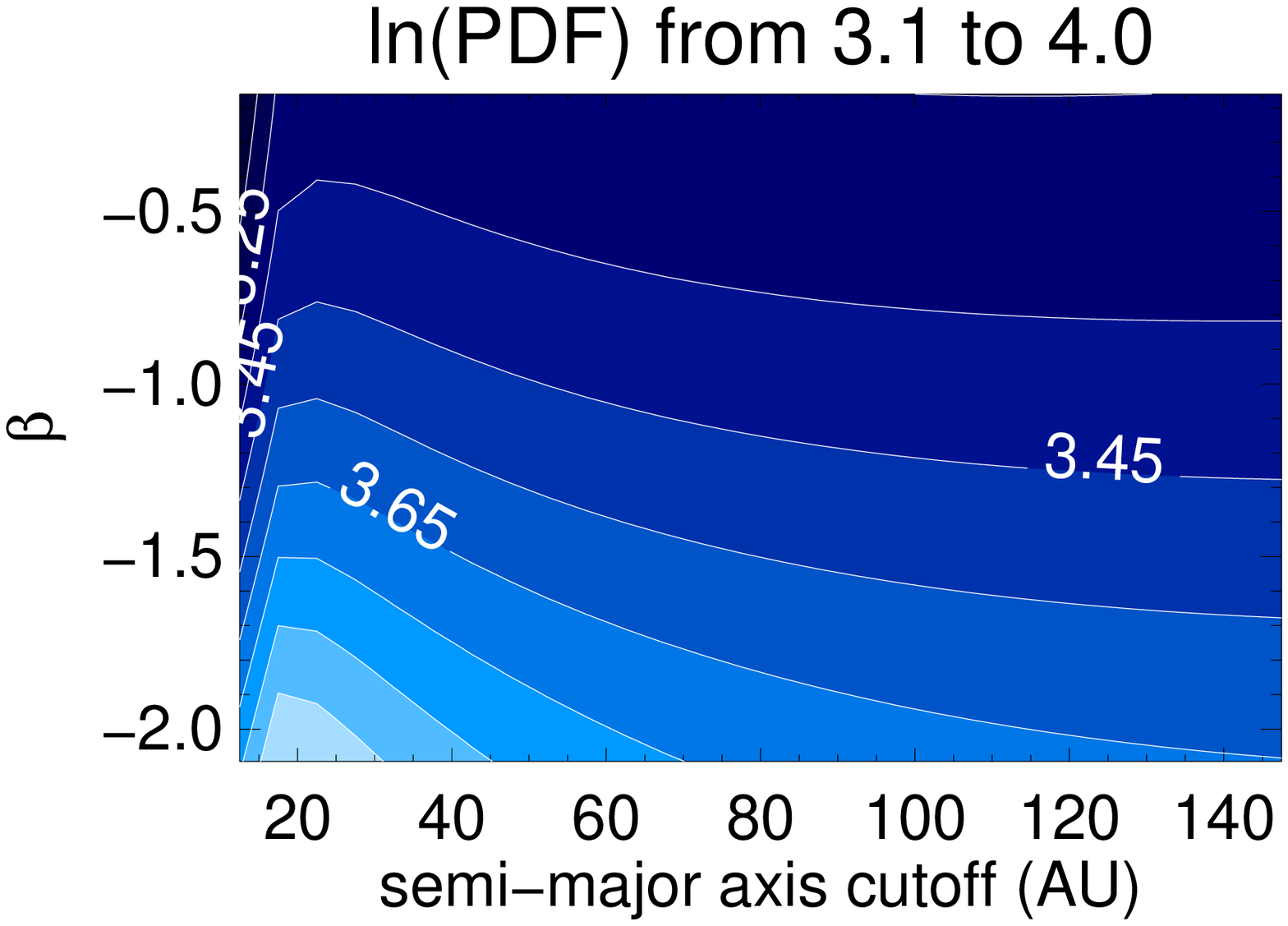} &
\includegraphics[width=3.0in]{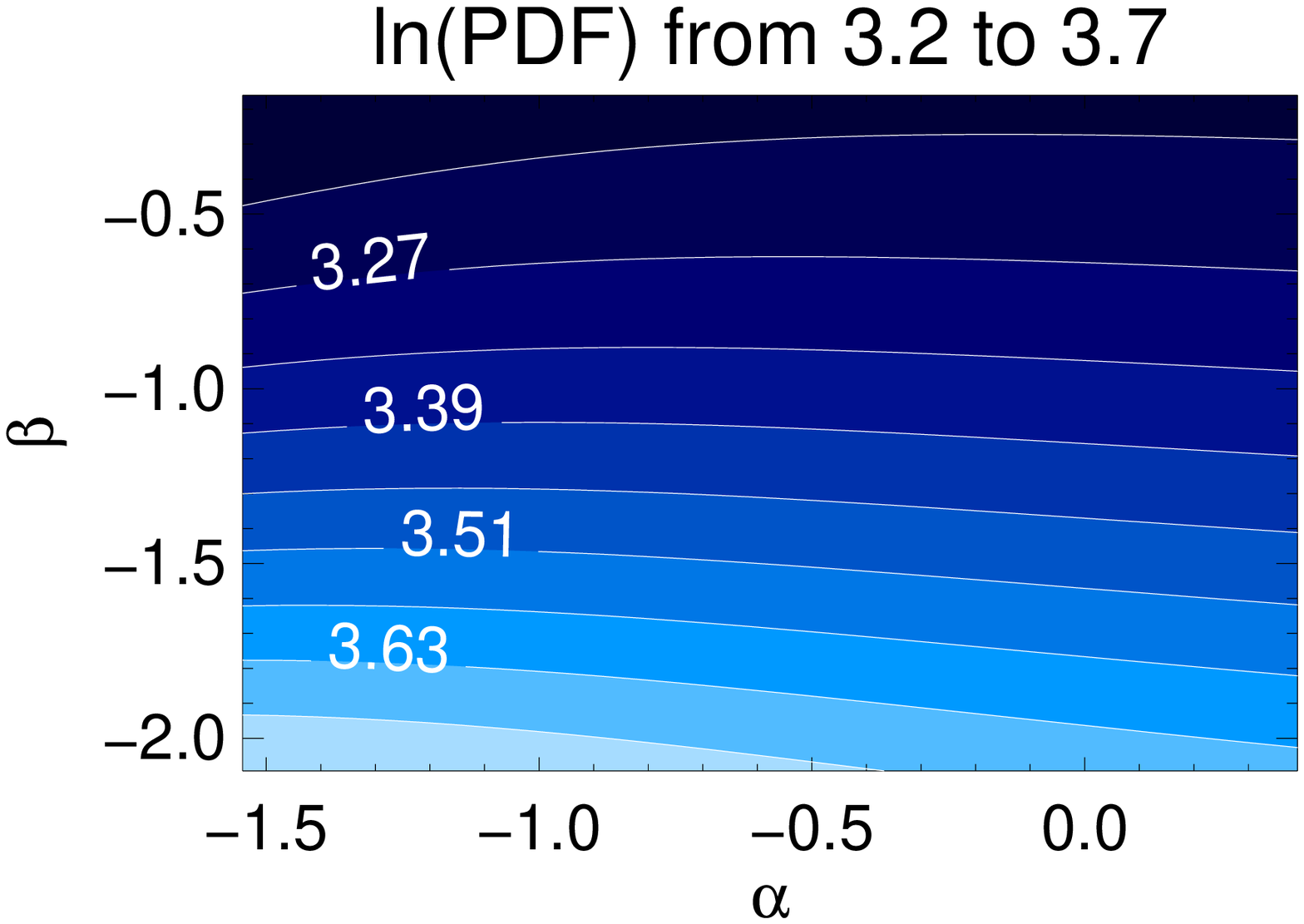} \\
\end{tabular}
\caption{2-d marginalized posterior PDFs with 4
  free parameters, with contours displayed as ln(PDF)
and using the COND models of \citet{Bar03}.  
Posterior probability is plotted in logarithmic units, with 10 contour levels
equally spaced in logarithmic space from the minimum
value of each posterior PDF to the maximum value.
As different posterior PDFs traverse very different 
probability ranges, we have included the range of
ln(PDF) values plotted in the title for each subplot.
Posterior PDFs which cover a greater probability range
are more constrained. 
The ratio of these units for two different contour levels
yields the relative likelihood of parameter combinations
along those respective contours.  Darker regions 
indicate parameter combinations with lower likelihood.
We have not normalized these marginalized posterior PDFs, as 
they remain unconstrained in the null detection case.
\label{fig:2d_4param_COND}}
\end{figure}

Obviously, the complete 4-dimensional posterior 
PDF cannot be fully visualized, so to present the results, we have
calculated 1-d and 2-d marginalized posterior PDFs by integrating 
over some of the parameters.  1-d marginalized posterior PDFs are presented 
in Figure~\ref{fig:1d_4param_DUSTY} for both DUSTY \citep{Bar02} and 
COND models \citep{Bar03}.   The 2-d marginalized posterior PDFs 
are presented in Figure~\ref{fig:2d_4param_DUSTY} for the DUSTY models
and in Figure~\ref{fig:2d_4param_COND} for the COND models.  All PDFs are plotted in 
logarithmic units.

Non-detection of planets with such a large sample and deep contrasts 
places the strongest constraints to date on the planet fraction $F$
for directly imaged exoplanets.   We derive upper limits on planet
fraction $F$ by normalizing our 1-d marginalized 
posterior PDF for $F$.  Upper limits on planet fraction $F$ are 
tabulated in Table~\ref{table_pf_Bayesian}.
For the DUSTY models, a semi-major axis range of 10-150 AU, and companion masses of
1-20 M$_{Jup}$, our 95.4$\%$ confidence limit on F is $\leq$18$\%$, and at a 99.7$\%$ confidence level, 
$F\leq44\%$.  For the same parameter ranges and the COND models, 
at a 95.4$\%$ confidence level, $F\leq$6$\%$, and at a 99.7$\%$ confidence level, 
$F\leq$14$\%$.  This is consistent with the results from our Monte Carlo simulations
as well (see Table~\ref{table_pf}) and is valid for a wide range of possible planet distributions.
Our results strongly constrain the frequency of planets within
semi-major axes of 50 AU as well.
For the DUSTY models, a semi-major axis range of 10-50 AU, and companion masses of
1-20 M$_{Jup}$, at a 95.4$\%$ confidence level, $F\leq$21$\%$, and at a 99.7$\%$ confidence level, 
$F\leq51\%$.  For the same parameter ranges and the COND models, 
at a 95.4$\%$ confidence level, $F\leq$7$\%$, and at a 99.7$\%$ confidence level, 
$F\leq$17$\%$.  The similar constraints obtained for 10-50 AU as for
10-150 AU suggests that the 50-150 AU semi-major axis range is quite
devoid of planets.

Other than for $F$, however, our 
marginalized posterior PDFs remain unconstrained (i.e. no
clear peak or trailing off to 0) and do not cover a wide range in 
ln(PDF) .  While the marginalized 1-d posterior PDF for 
planet fraction $F$ varies by over 10 orders of magnitude
(Figure~\ref{fig:1d_4param_DUSTY}), the marginalized 1-d
posteriors for the other 3 parameters 
vary by $<$1.5 orders of magnitude (a factor of 4.5 at most). 
Thus, we do not place confidence intervals on parameters other 
than the planet frequency $F$.

While we can place strong limits on $F$ by
marginalizing over the other three parameters,
determining the best-fit power law parameters for directly
imaged planet populations must be deferred until there is a
statistically significant population of such objects to fit.
The choice of a power-law model for directly imaged planet
distributions is based on fits to the properties of radial velocity
planet \citep{Cum08}; it is not known yet whether this is the best
model to describe directly imaged planet distributions.

\subsubsection{Results with the AB Pic and $\beta$ Pic detections \label{Bayesdet}}

Our Bayesian approach can seamlessly handle both planet detections
and non-detections.  Here we rerun the Bayesian analysis described
above, this time adding in AB Pic and $\beta$ Pic, the two stars 
with already known planetary or low mass brown dwarf ($<$20 M$_{Jup}$) companions in our sample.
We adopt a mass estimate of 8 M$_{Jup}$ 
\citep[i.e.~the~middle~of~the~range~found~by~][]{Bon13} and a projected separation
of 8.5 AU \citep{Cha12} for $\beta$ Pic b. 
For AB Pic B, we adopt a mass estimate of 13.5 M$_{Jup}$ \citep{Bon10} and a 
projected separation of 275 AU \citep{Cha05b}.

\begin{figure}
\begin{tabular}{p{2.2in}p{2.2in}p{2.2in}}
\Large $\beta$ Pic b only & \Large  AB Pic B only & \Large  $\beta$ Pic b and AB Pic B \\
\end{tabular}
\vskip -0.2in
\centerline{
\includegraphics[width=2.5in]{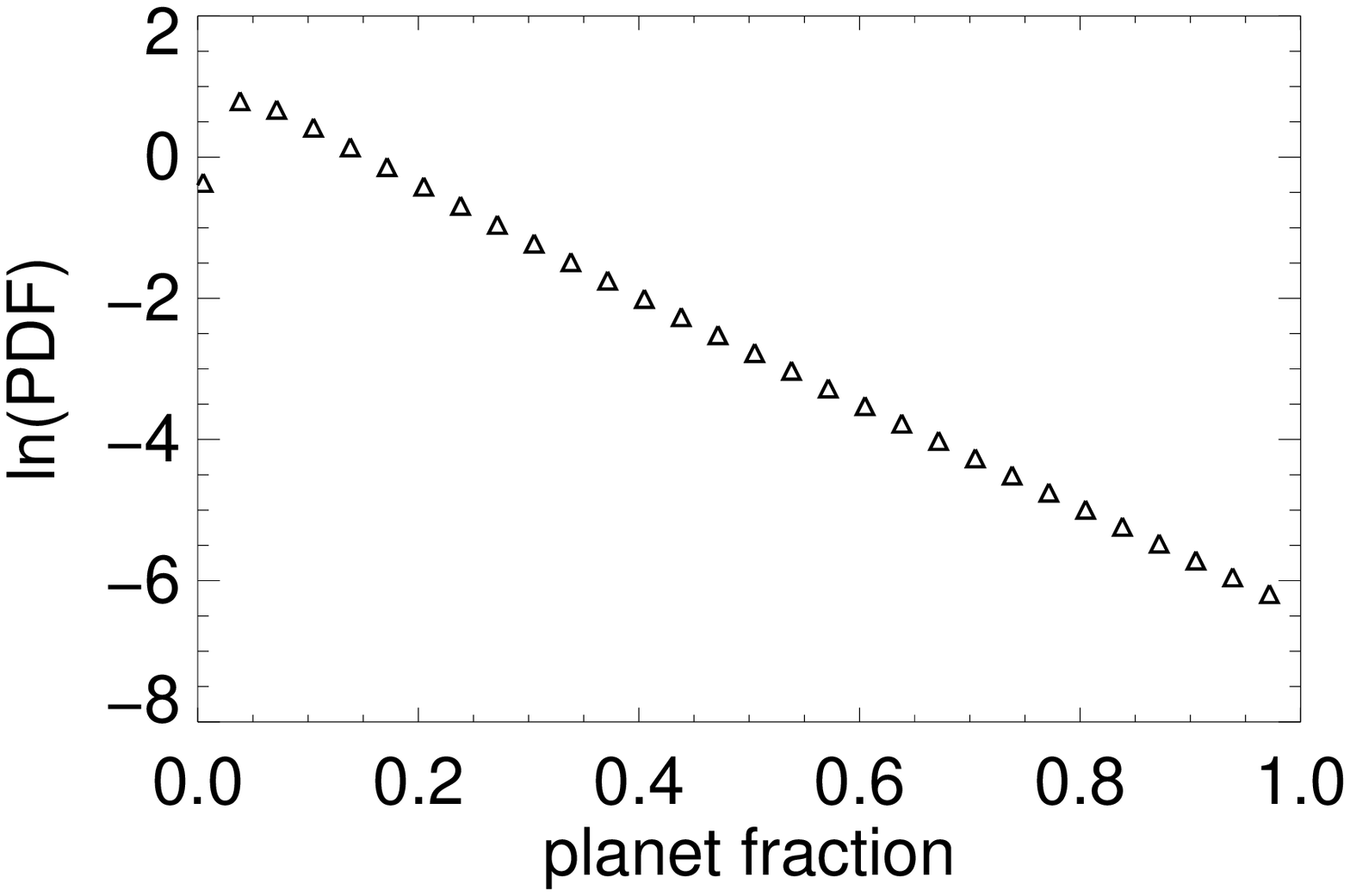} 
\hskip -0.2in
\includegraphics[width=2.5in]{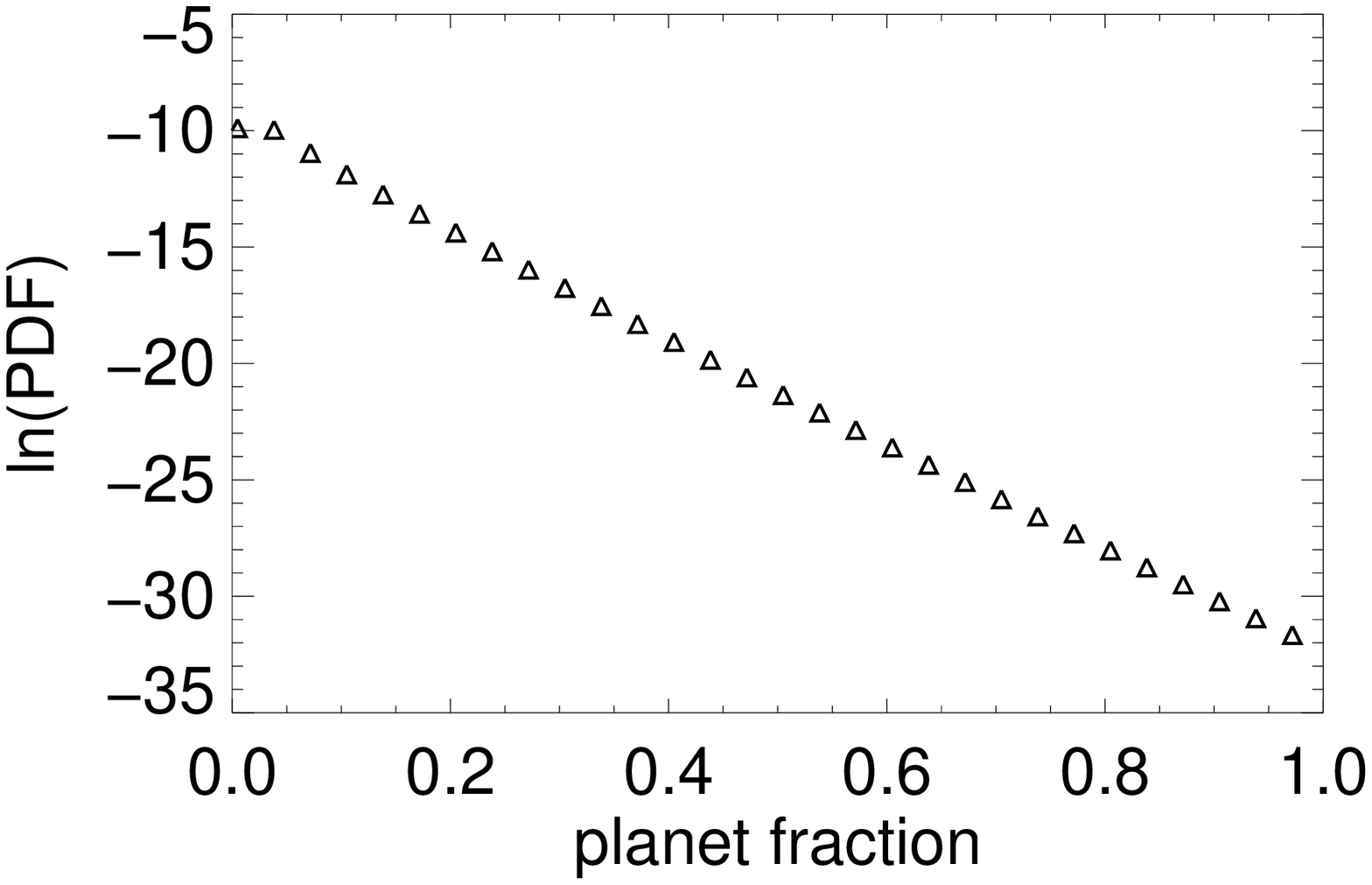} 
\hskip -0.2in
\includegraphics[width=2.5in]{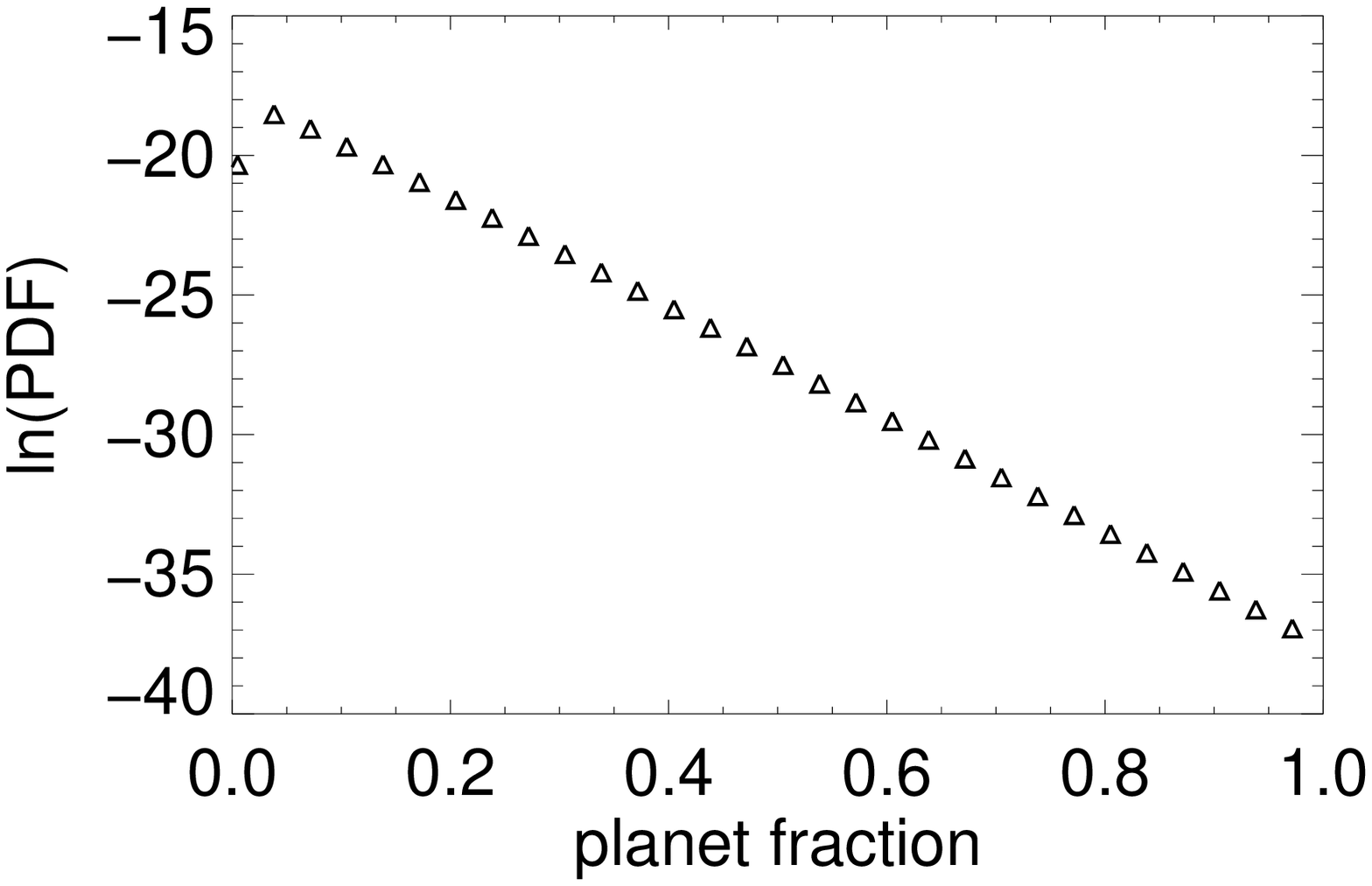} 
}
\vskip -0.2in
\centerline{
\includegraphics[width=2.5in]{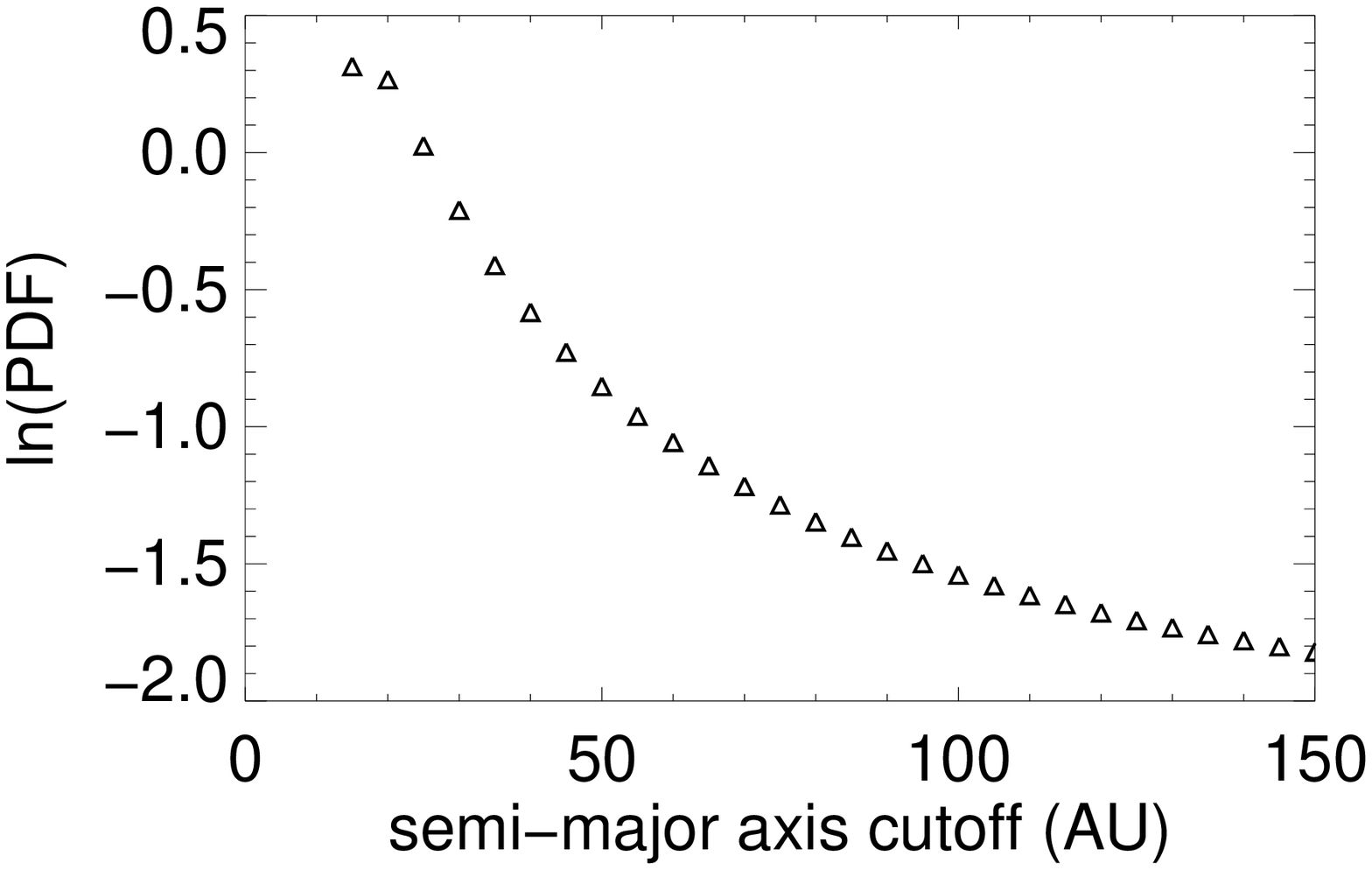} 
\hskip -0.2in
\includegraphics[width=2.5in]{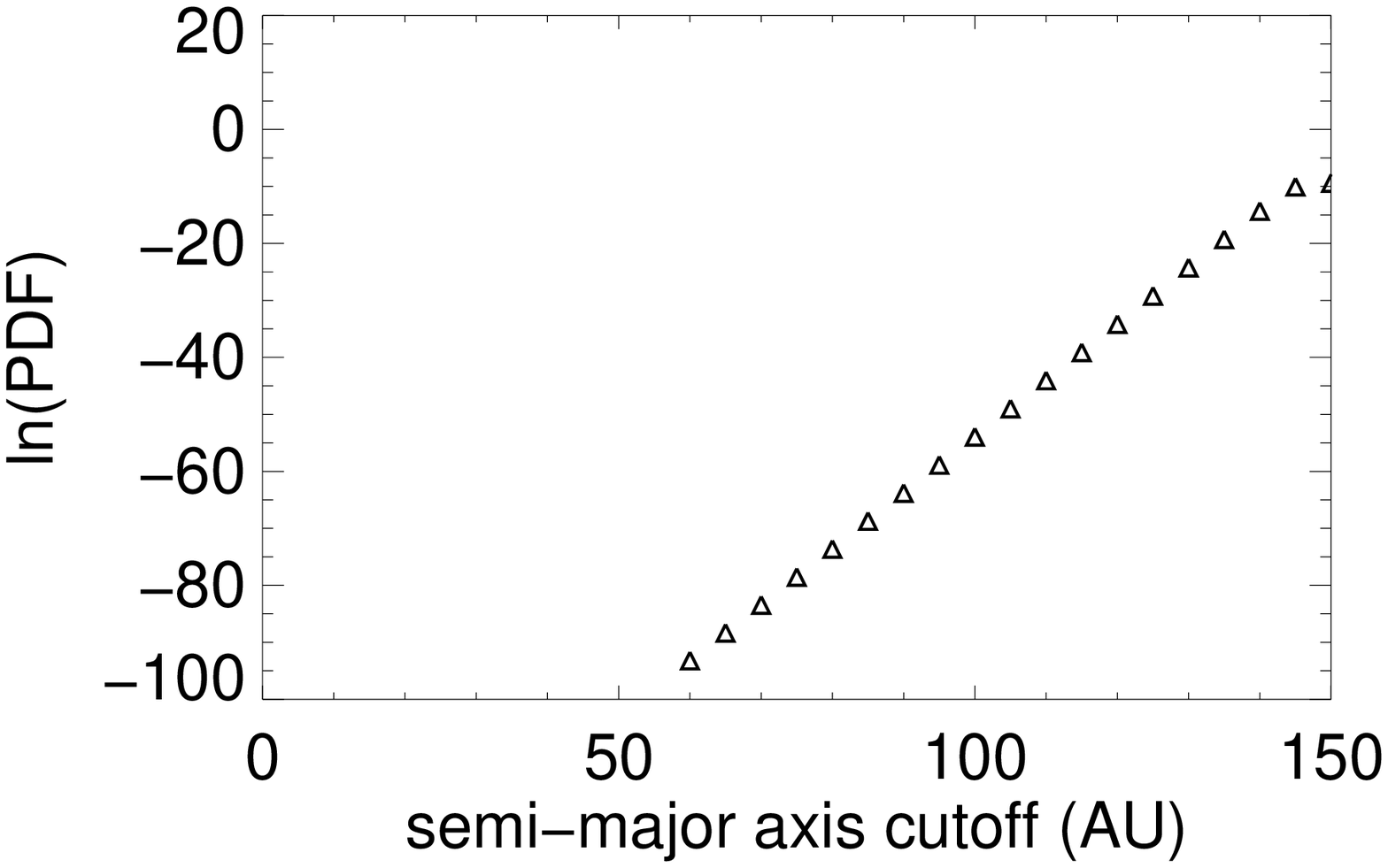} 
\hskip -0.2in
\includegraphics[width=2.5in]{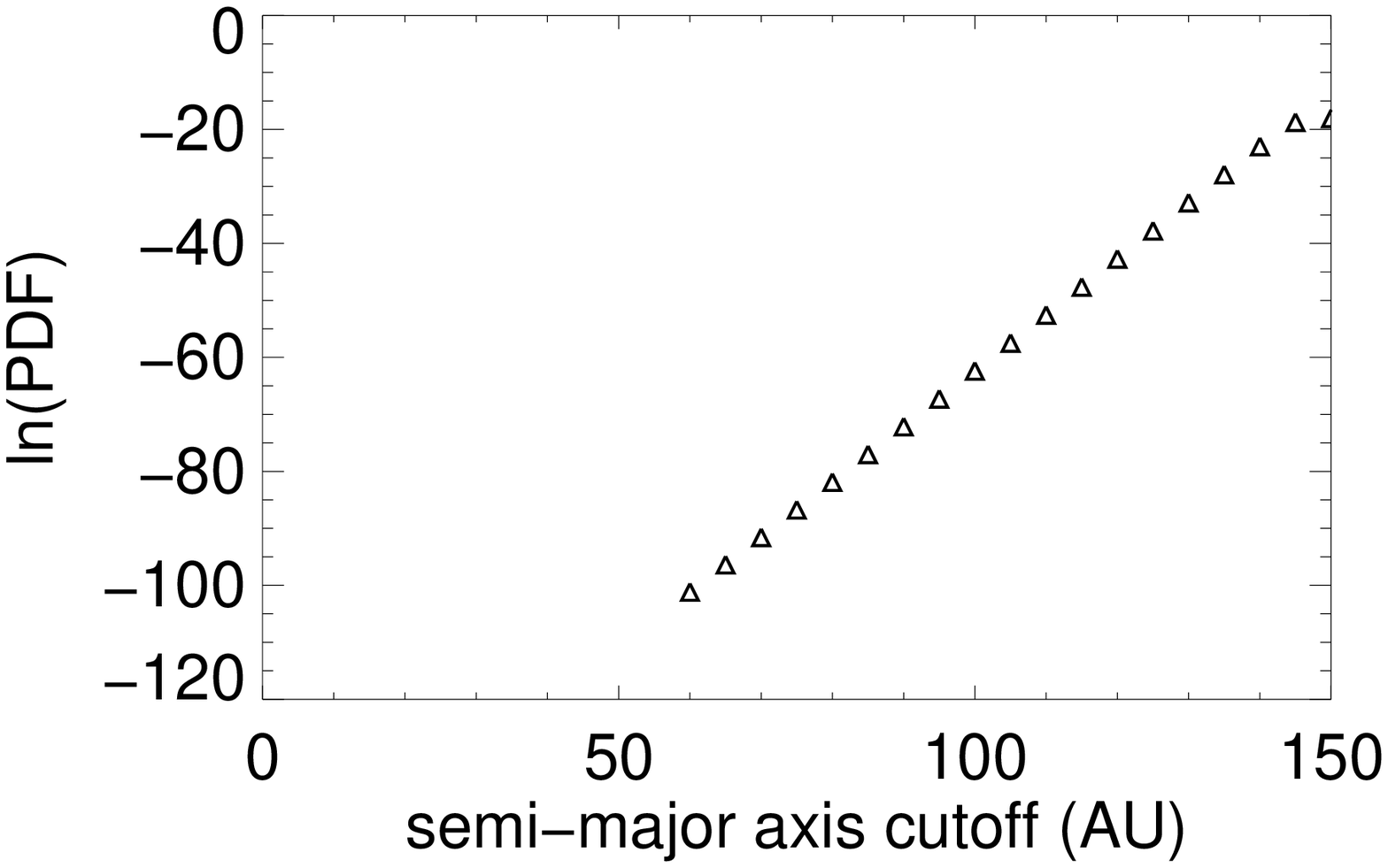} 
}
\vskip -0.2in
\centerline{
\includegraphics[width=2.5in]{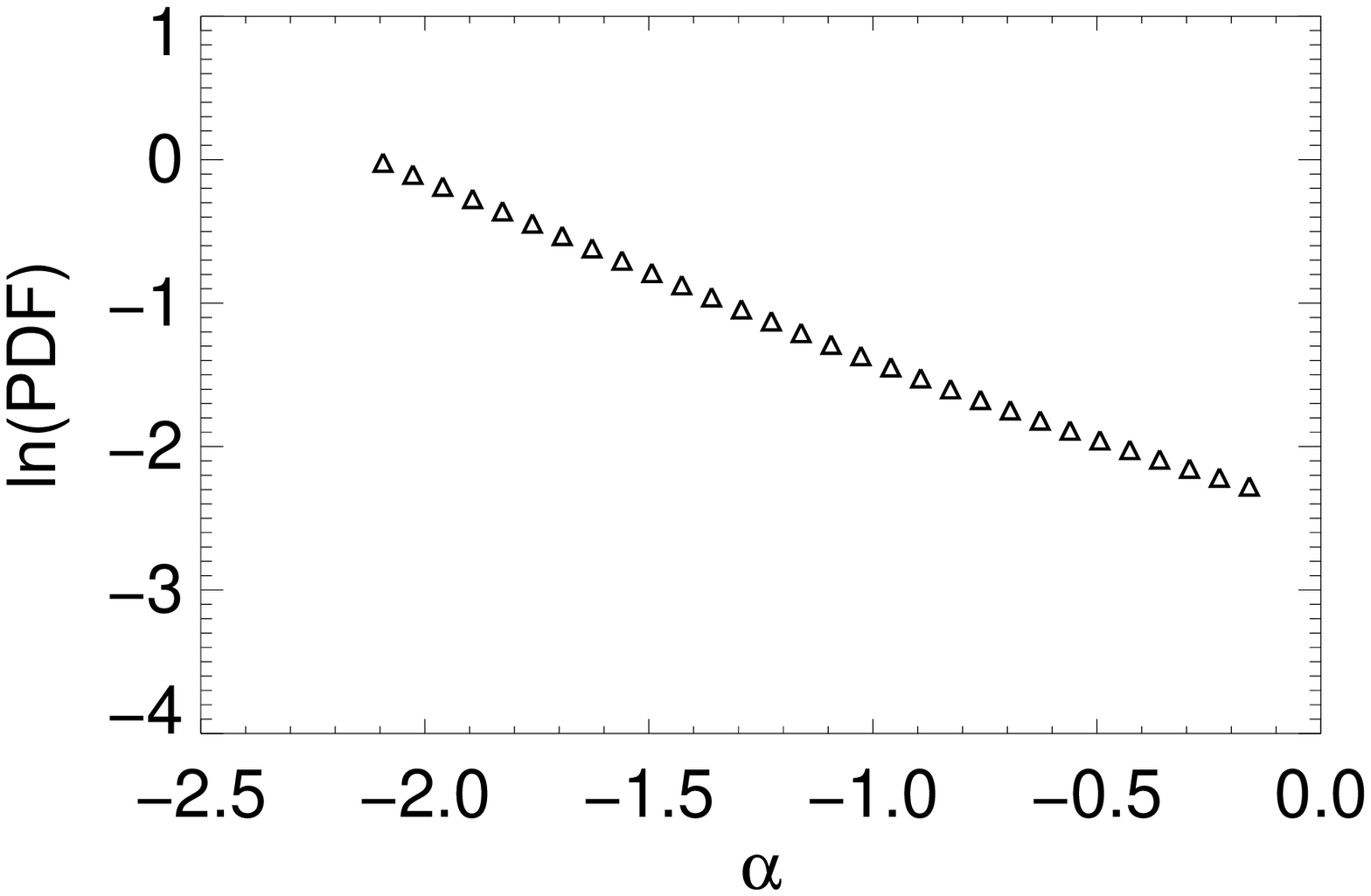} 
\hskip -0.2in
\includegraphics[width=2.5in]{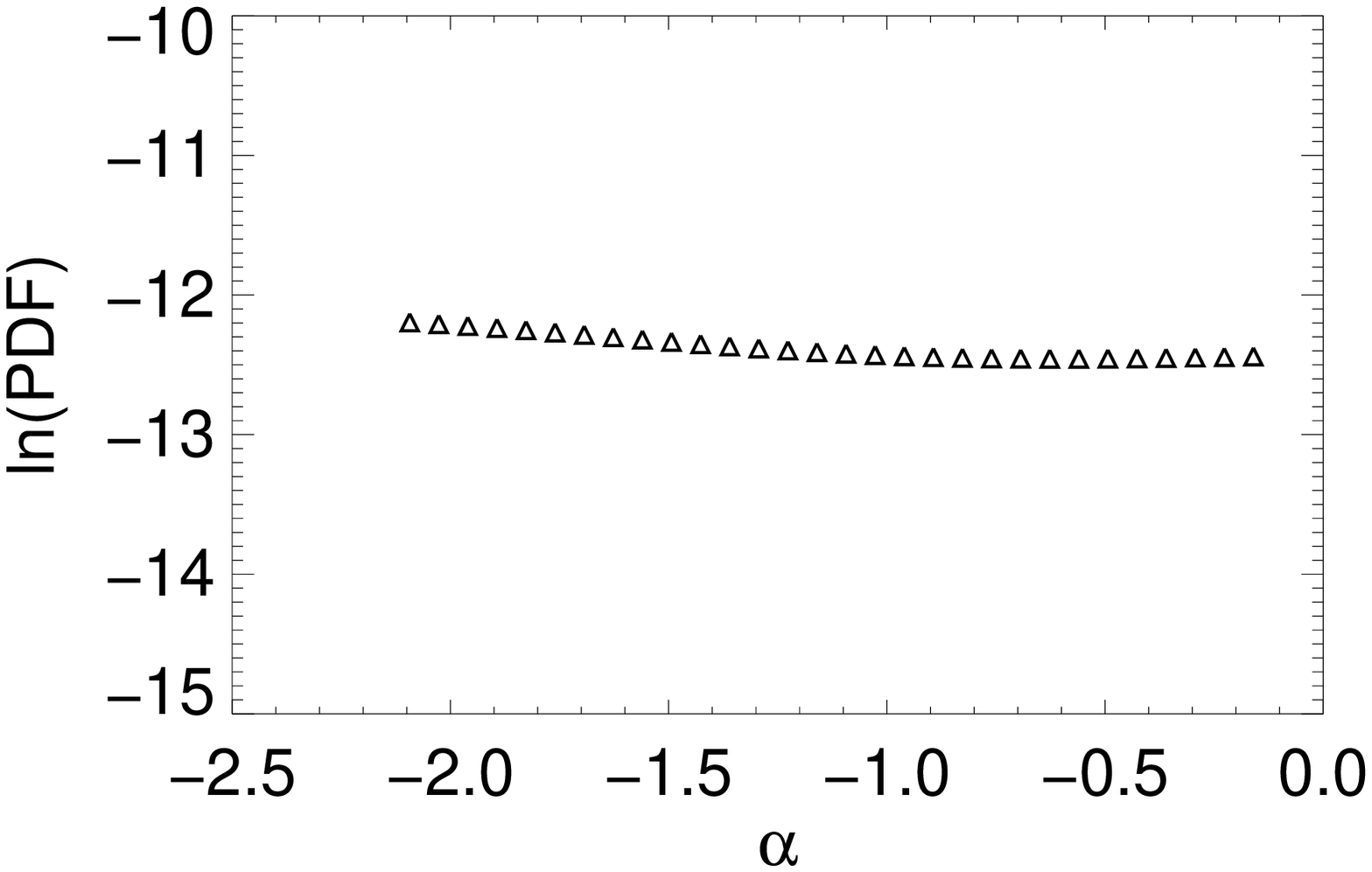} 
\hskip -0.2in
\includegraphics[width=2.5in]{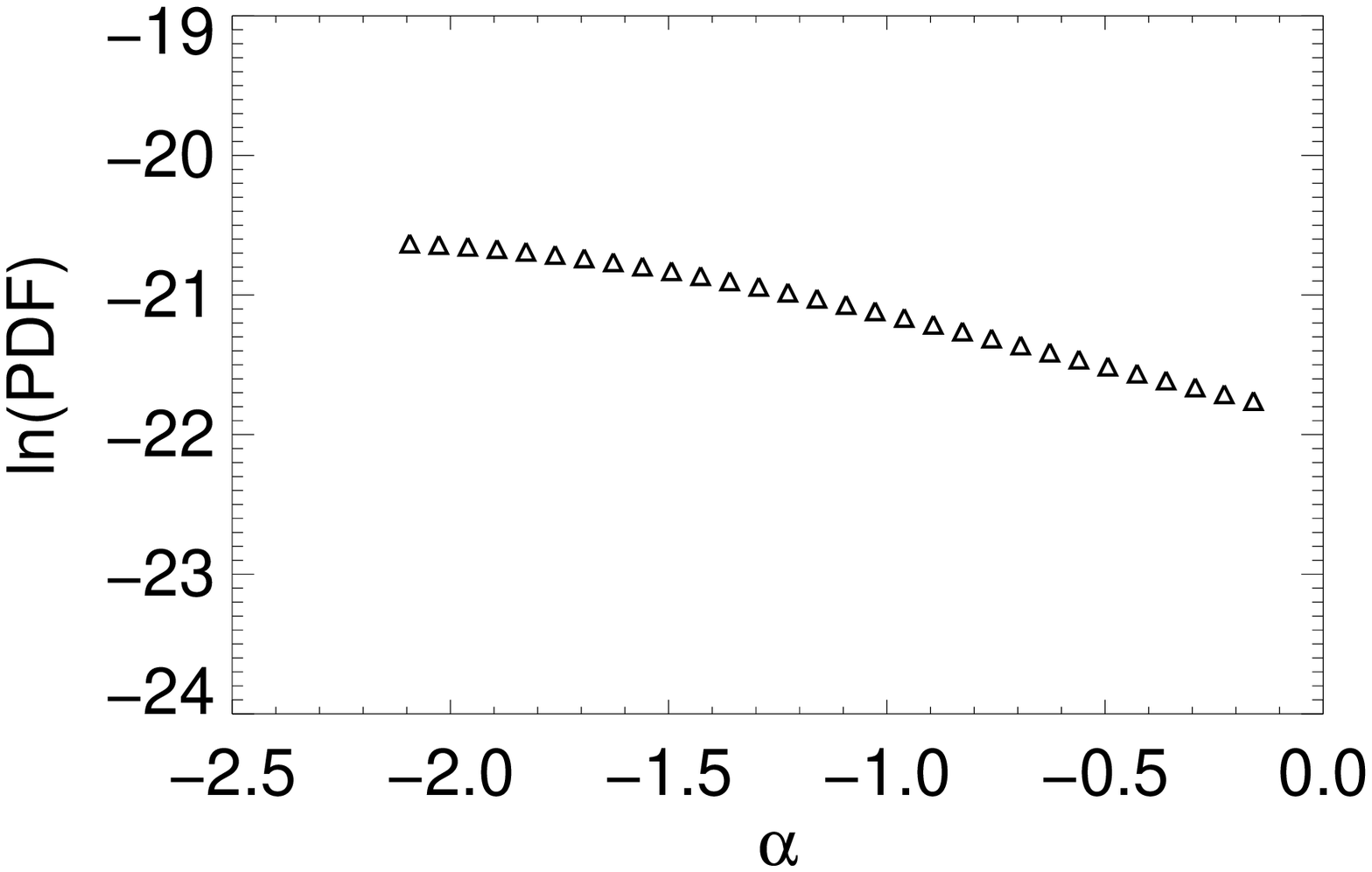} 
}
\vskip -0.2in
\centerline{
\includegraphics[width=2.5in]{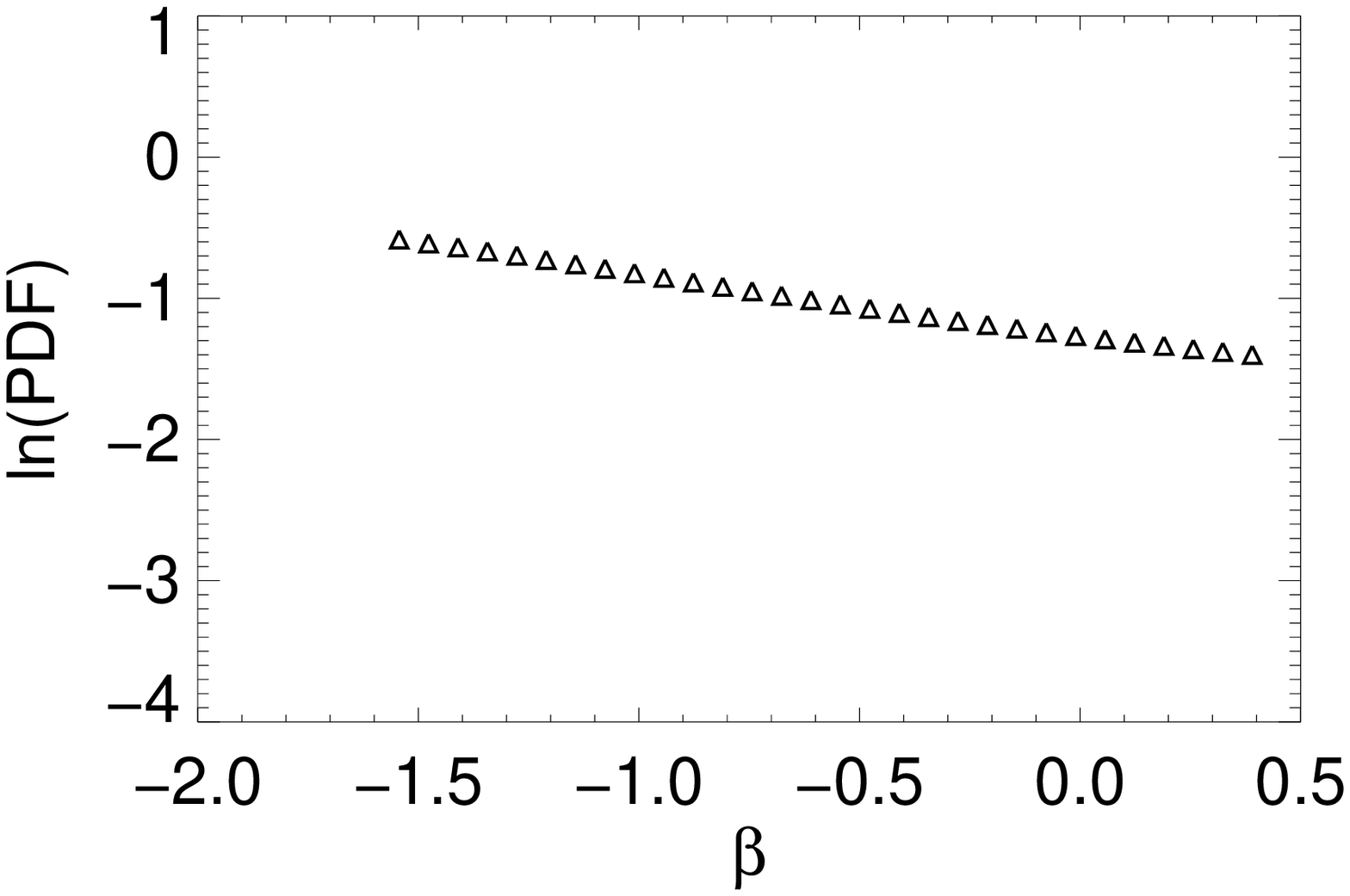} 
\hskip -0.2in
\includegraphics[width=2.5in]{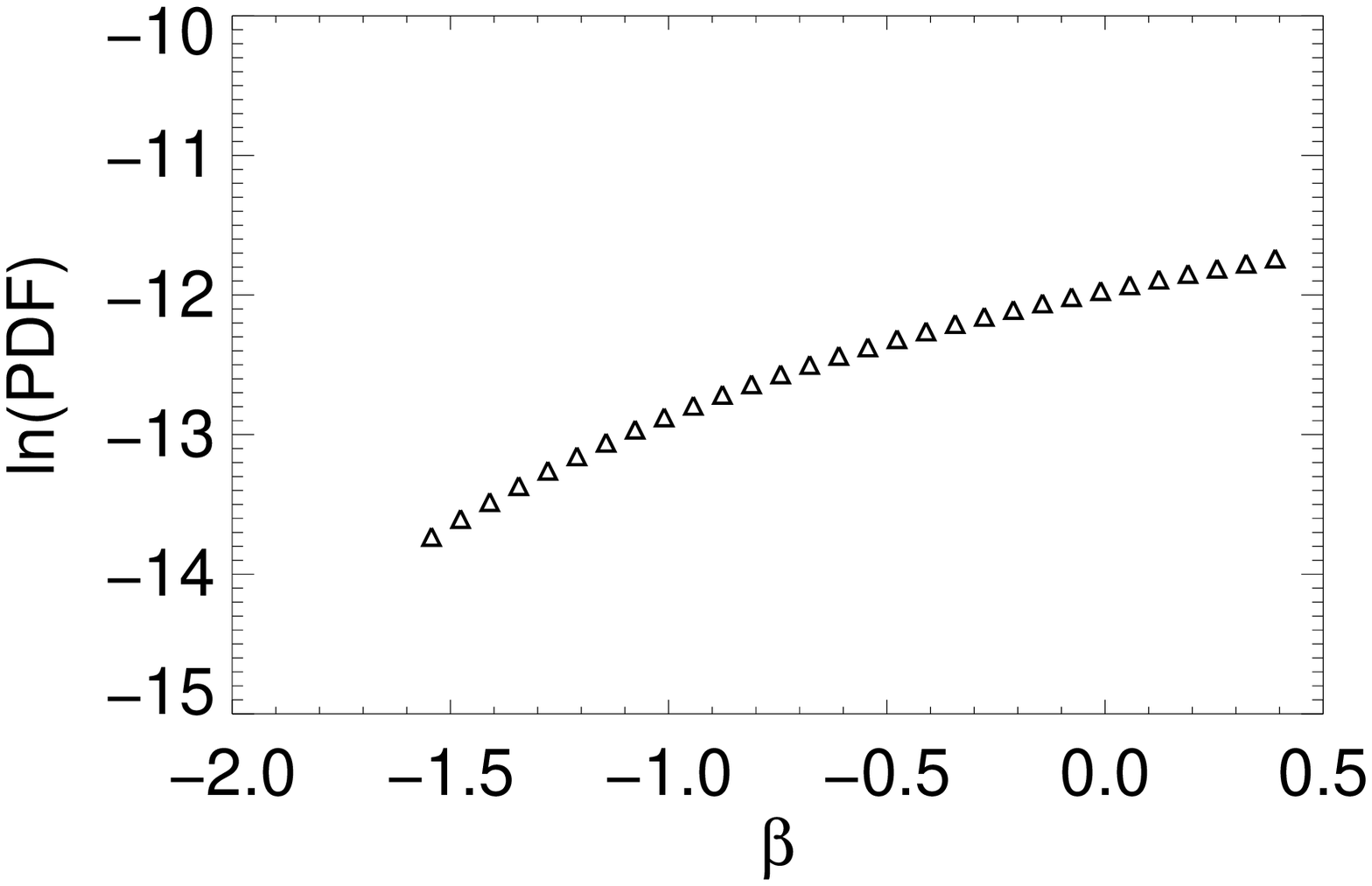} 
\hskip -0.2in
\includegraphics[width=2.5in]{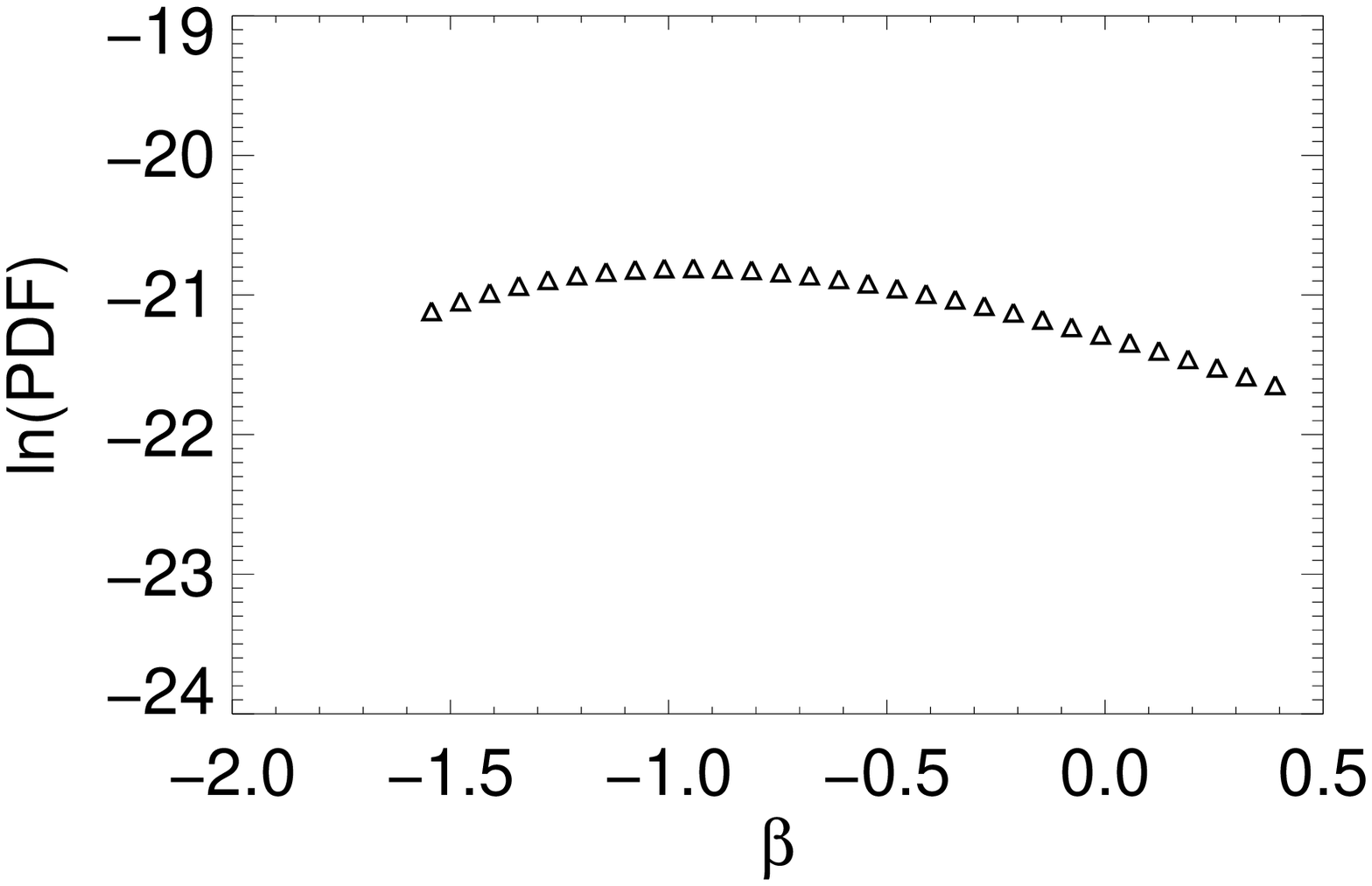} 
}
\caption{Comparison of 1-d marginalized posterior pdfs including 
{\bf (left)} only the $\beta$ Pic b detection, {\bf (center)} only the
AB Pic B detection, and {\bf (right)} both detections.  
Posterior probability is plotted in logarithmic units. 
For AB Pic B, in the case of small ($<$50 AU) cutoffs, ln(PDF) trended to
negative infinity and is not plotted here. 
\label{fig:1d_det}
}
\end{figure}

The Bayesian analysis described above was rerun with the 73 original
stars, and including 1) only the $\beta$ Pic b detection, 2) only the AB
Pic B detection, and 3) both detections.  Results are presented in
Fig.~\ref{fig:1d_det}.   Including the companions
affects the shape of the marginalized PDF for planet fraction $F$
and also provide a significant constraint on $a_{max}$, the 
semi-major axis cutoff.  The marginalized PDFs for planet fraction
$F$ with $\beta$ Pic b only and for both detections
now show a peak at $\sim$4$\%$, as a clear detection 
of a close companion
rules out a zero value for planet fraction $F$ in the 10-150 AU range.
For the $\beta$ Pic b only case, planet fraction
$F$=0.04$_{-0.04}^{+0.35}$,  with 95.4$\%$ confidence level error bars, 
so, as expected, detection of a single object does not highly
constrain $F$.
The AB Pic B detection is at much larger separation than the
$\beta$ Pic b detection, so
it provides much less of a constraint on planet fraction in this
range, as there is only a very slight chance that this companion 
has a semi-major axis $<$ 150 AU (i.e. a highly eccentric orbit).
The marginalized PDF for planet fraction $F$ in this case
shows some flattening at small values of $F$ but no clear peak.
As in the non-detection case, $\alpha$
and $\beta$ do not cover enough range in ln(PDF) to yield
useful constraints.  Given that the two companions
were detected in very different separation regimes, they provide
contradictory constraints on cutoff, with the AB Pic B detection
strongly ruling out $a_{max}$ $<$ 100 AU and the $\beta$ Pic b
weakly ruling $a_{max}$ $>$ 100 AU.  
While it is informative to investigate how single detections with 
varying estimated masses and separations affect the shape of the
posterior PDF, it is dangerous to draw conclusions based on 
such a small sample of detections.  Detection of a larger cohort 
of similar companions is necessary to put consistent
constraints on the properties of such objects 

\section{Discussion}

While a number of giant planets and giant planet candidates have now been imaged at separations
$>$20 AU, large-scale surveys illustrate that such planets are 
comparatively rare around main-sequence solar analogues and low mass stars.  
Of the ensemble of directly-imaged planets known to date, 
most have been discovered around A stars
\citep[HR~8799bcde,~Fomalhaut~b,~$\beta$~Pic~b,~WD~0806-661;][]{Mar08,
  Mar10, Kal08, Lag09, Lag10,Luh11,Qua13,Ram13b}, with a few also 
discovered around very young solar analogues \citep[1RXS J160929.1$-$210524b,
 LkCa 15b, GSC 06214$-$00210b,][]{Laf08, Kra12, Ire11}.  Only 
one planet has been directly-imaged to date around a main-sequence
solar analogue \citep[GJ~504b;][]{Kuz13} .  Two companions right
at the deuterium burning limit have recently been reported around M stars
\citep[][]{Bow13, Del13}, but no companion with estimated mass $<$10
M$_{Jup}$ has yet been imaged around a low mass star.

The small number of detected planets
 is not due to a lack of stars surveyed.  Our NICI survey of 80 stars is the
largest single sample of MG stars observed.  Significant
numbers of MG stars have also been observed as part of the Gemini
Deep Planet survey \citep{Laf07b}, International Deep Planet
Survey \citep{Vig12}, SDI survey \citep{Bil07}, Deep Imaging Survey 
of Young, Nearby Austral Stars \citep{Cha10}, 
A Survey of Young, Nearby, and Dusty Stars \citep{Ram13a}, 
NACO Large Program (Vigan et al. 2013), 
and SEEDS (Brandt et al. 2013).  
Based on a sample of 118 stars
compiled from the surveys of \citet{Mas05}, \citet{Bil07}, and \citet{Laf07b},   
\citet{Nie10} found that planets more massive than 4 M$_{Jup}$
are found around $<$20$\%$ of FGKM stars
in orbits between 22 and 507 AU, at 95$\%$ confidence.
\citet{Cha10} find a qualitatively similar result
based on a sample of 88 stars (51 of which are members of 
young moving groups), constraining the fraction 
of stars with giant planets to $<$10$\%$ at semi-major 
axes $>$40 AU for a planet distribution extended from radial velocity 
power laws.  With considerably higher contrasts
and better inner working angles (0.3'' vs. typically 0.5-0.7''), 
our work here directly extends these results 
to lower masses and smaller separations.
Our results are qualitatively similar to those of \citet{Nie10}
but at a considerably higher confidence level.
As discussed in Section~\ref{Bayesnodet}, we confirm and extend the result of \citet{Nie10}:
$>$5 M$_{Jup}$ companions to FGKM stars are rare at separations 
$>$10 AU.

\citet{Joh07} and \citet{Joh10} find that for RV planets, host star mass and 
planet mass are related.  Higher mass stars seem to preferentially
host more high mass RV planets ($>$1 M$_{Jup}$) than lower mass stars,
attributable to the fact that more massive stars also likely
possessed more massive primordial circumstellar disks.
Indeed, \citet{Joh07,Joh10} find that $>$1 M$_{Jup}$ radial velocity
planets are quite rare around M stars at semi-major axes
$<$5 AU.  The fact that the majority of directly-imaged planets to date
have been found around higher mass stars
 qualitatively suggests a similar conclusion may hold for the 
wide planet population probed by direct imaging.  

We examine here whether
the statistics from direct imaging surveys to date supports this assertion.
The first constraints on directly imaged planet fraction
for high mass AB stars have only recently been published. 
\citet{Jan11} found that $<$30$\%$
of massive stars have giant planet ($>$1 M$_{Jup}$) or brown 
dwarf companions that formed via gravitational instability 
with mass $<$100 M$_{Jup}$ within 300 AU at the 99$\%$ confidence level
for a sample of 18 high-mass stars in the solar neighborhood, 
however this work does not place limits on core-accretion planets
around these hosts.
For a 42-star sample, \citet{Vig12} found that the fraction 
of A stars with 1 massive planet (3-14 M$_{Jup}$) from 5-300 AU was
5.9-18.8$\%$ at the 68$\%$ confidence level
(assuming power law distributions for mass and semi-major 
axis appropriate for core-accretion planets), however
the age determination for their survey stars may be 
overly optimistic (Nielsen et al. 2013).
Our current sample is comprised of 70$\%$ stars with
spectral type of K or later and contains 33 M stars, and thus
can be directly compared to the samples of 
\citet{Vig12}  to test whether planet fraction indeed falls with
stellar mass.  Using the DUSTY models, 
we limit the planet fraction $F$ of our sample to $\leq$3.5$\%$
at the 68$\%$ confidence level result for 1--20 M$_{Jup}$ companions
at semi-major axes of 10--150 AU; this is lower than the 5.9-18.8$\%$ 
planet fraction at a 68$\%$ confidence level found by \citet{Vig12}.
Thus, the current set of direct imaging surveys may hint
that directly imaged giant planets are less
common around lower mass GKM stars compared to AB stars.


\section{Conclusions}

As part of the Gemini NICI Planet-Finding Campaign, we imaged 
80 members of nearby young moving groups, 
with ages from 10--200 Myr and within 100 pc.  
In ASDI mode, we attain median contrasts of $\Delta$(mag)=12.4, 13.9, 
and 14.5 mag at 0.5$\arcsec$, 1$\arcsec$, and 
2$\arcsec$ respectively in the narrow band methane 
filters ($\lambda$=1.58 $\mu$m), with a typical standard deviation of 0.9 mag.
In ADI mode, we attain median contrasts of $\Delta$(mag)=10.4,
13.2, and 15.1 mag at 0.5$\arcsec$, 1$\arcsec$, and 
2$\arcsec$ respectively in $H$ band.  We achieve median minimum detectable masses 
of 11, 5, and 3 M$_{Jup}$ at 0.5$\arcsec$, 1$\arcsec$, and 
2$\arcsec$ using the DUSTY models \citep{Bar02}.

Candidate companions within 400 AU from the star and not in dense stellar fields
that could not be confirmed or rejected as common-proper
motion companions using archival data were
reobserved with NICI. 
A total of 77 candidate companions were detected and eliminated
as background contaminants. 
Four comoving brown 
dwarf or substellar companions were 
discovered in the moving group sample: 
PZ Tel B \citep{Bil10}, CD -35 2722B \citep{Wah11}, HD 12894B (this work)
and BD+07 1919C (this work).  PZ Tel B and CD-35 2722B are both 
30-40 M$_{Jup}$ brown dwarf companions, while HD 12894B and 
BD+07 1919C are stellar companions with estimated masses of 
0.46$\pm$0.08 M$_{\odot}$ and 0.20$\pm$0.03 M$_{\odot}$ respectively.
We also retrieved the substellar companions AB Pic B \citep{Cha05b}
and HR 7329 B \citep{Low00} as well as the known stellar 
companion to HD 82688 \citep{Met09}. 
To compare to previous published surveys, we have
adopted hot start models in our statistical analysis, which predict considerably brighter
planets at these young ages compared to cold start models
\citep[for~instance,~][]{Spi12}.  Thus, earlier surveys as well as 
our own predict systematically
more stringent upper limits on planet fraction than would be found 
with cold start models.  Nonetheless, 
our constraints on planet fraction are consistent with and
more stringent than previous work.
From a Bayesian analysis for a wide range of parameters and power-law
models of planet distributions, we restrict the frequency
of 1--20 M$_{Jup}$ companions at semi-major
axes from 10--150 AU to $<$18\% at a 95.4$\%$ confidence level
using DUSTY models \citep{Bar02} and to $<$6\% at a 95.4$\%$ 
confidence level using COND models \citep{Bar03}.  

\acknowledgements

We thank Geoff Bryden and Luisa Rebull for clarification on the
Spitzer detection of PZ Tel.  We thank the referee for useful suggestions which helped strengthen
this work.  B.A.B. was supported by Hubble Fellowship grant HST-HF-01204.01-A
awarded by the Space Telescope Science Institute, which is operated by AURA for NASA, under
contract NAS 5-26555. This work was supported in part by NSF grants AST-0713881 and AST-
0709484 awarded to M. Liu, NASA Origins grant NNX11 AC31G awarded to M. Liu, and NSF grant
AAG-1109114 awarded to L. Close. The Gemini Observatory is operated by the Association of Universities
for Research in Astronomy, Inc., under a cooperative agreement with the NSF on behalf of
the Gemini partnership: the National Science Foundation (United States), the Science and Technology
Facilities Council (United Kingdom), the National Research Council (Canada), CONICYT
(Chile), the Australian Research Council (Australia), CNPq (Brazil), and CONICET (Argentina).
Based on observations made with the European Southern Observatory telescopes obtained from
the ESO/ST-ECF Science Archive Facility.  This publication makes use
of data products from the Two Micron All Sky Survey, which is a joint project of the University of Massachusetts and the
Infrared Processing and Analysis Center/California Institute of Technology, funded by the National
Aeronautics and Space Administration and the National Science Foundation. This research has
made use of the SIMBAD database, operated at CDS, Strasbourg, France. This research has made
use of the VizieR catalogue access tool, CDS, Strasbourg, France. The Digitized Sky Survey was
produced at the Space Telescope Science Institute under U.S. Government grant NAG W-2166.
The images of these surveys are based on photographic data obtained using the Oschin Schmidt
Telescope on Palomar Mountain and the UK Schmidt Telescope. The plates were processed into
the present compressed digital form with the permission of these institutions.

\clearpage

\begin{deluxetable}{lcccccccc}
\rotate
\tablecolumns{10}
\tablewidth{0pc}
\tabletypesize{\small}
\tablecaption{Properties of MG Survey Stars}
\tablehead{
\colhead{Target} & \colhead{RA} & \colhead{DEC} & \colhead{Distance (pc)}\tablenotemark{1} 
& \colhead{SpT}  & \colhead{$V$} & \colhead{$H$} & 
\colhead{$Ks$} & \colhead{Comments}}
\startdata
TW Hya, 10 Myr \\ \hline
         TWA 6   &         10:18:28.8   &          -31:50:02   & (77)\tablenotemark{2}  & M0  &     12.00  &      8.18  &      8.04  & \\
       TWA 7   &         10:42:30.1   &          -33:40:16   & (28)\tablenotemark{3}  & M2  &     11.06  &      7.12  &      6.90  & \\
         TW Hya   &         11:01:52.0   &          -34:42:17   &53.7  & K6  &     11.34  &      7.56  &      7.30  & \\
          TWA 14   &         11:13:26.5   &    -45:23:43.0   & 95.9\tablenotemark{4}  &  M0\tablenotemark{2}  & \nodata     &      8.73  &      8.49  & \\
       TWA 13A   &         11:21:17.2   &          -34:46:46   & 55.6\tablenotemark{4}  & M1  &      11.46 \tablenotemark{5}  &      7.73  &      7.49  & \\
         TWA 13B   &         11:21:17.4   &         -34:46:50   & 59.7\tablenotemark{4}  & M1  &     11.96\tablenotemark{5}  &      7.68  &      7.46  & \\
        TWA 8A   &         11:32:41.2   &          -26:51:56   & 47.0\tablenotemark{6}  & M3  &     12.23\tablenotemark{5}  &      7.66  &      7.43  & \\
\enddata
\tablenotetext{1}{Unless otherwise specified, distances are drawn from
  the Hipparcos survey, spectral types are drawn from \citet{Tor06}, 
V magnitudes are drawn from the Tycho survey \citep{Hog00} and H and Ks
  magnitudes are drawn from the 2MASS survey.  Distances presented 
in parenthesis are kinematic and calculated by the divergence method.}
\tablenotetext{2}{\citet{Zuc04}}
\tablenotetext{3}{\citet{Tor08}}
\tablenotetext{4}{\citet{Wei13}}
\tablenotetext{5}{\citet{Tor06}}
\tablenotetext{6}{\citet{Mal13}, from Riedel et al. in prep}
\tablenotetext{7}{\citet{Mes10}}
\tablenotetext{8}{\citet{Bar06}}
\tablenotetext{9}{\citet{Moo06}}
\tablenotetext{10}{\citet{Mal13}}
\tablenotetext{11}{\citet{Lep09}}
\tablenotetext{12}{\citet{Ria06}}
\tablenotetext{13}{\citet{Kis11}}
\tablenotetext{t}{\citet{Whi07}}
\tablenotetext{15}{\citet{Koe10}}
\tablenotetext{16}{\citet{Lop06}}
\tablenotetext{17}{\citet{Zuc04b}}
\tablenotetext{18}{\citet{Hog00, Tor06}}
\tablenotetext{19}{\citet{Zuc04}}
\tablenotetext{20}{\citet{Mon01}}
\tablenotetext{21}{\citet{Gra06}}
\tablenotetext{22}{\citet{Lan09}}
\end{deluxetable}

\begin{deluxetable}{lccccc}
\tabletypesize{\small}
\tablecaption{NICI Campaign Observations of Moving Group Targets}
\tablewidth{0pt}
\tablehead{ \colhead{Target}  &\colhead{Date (UT)} & \colhead{Obs Mode} & \colhead{Images}    & \colhead{Exp. Time (s)}     & \colhead{Rotation (deg)} }
\startdata
TW Hya, 10 Myr \\ \hline
TWA 6 & 2009-01-17 & ADI &      45 &    2718 &   153.0 \\
TWA 7 & 2009-02-11 & ASDI &     45  &  2701  &  119.3 \\
TWA 7 & 2009-02-11 & ADI &      20 &    1208 &     14.2 \\
TWA 7 & 2010-02-28 & ADI &      20 &    1208 &    12.0 \\
TW Hya & 2009-01-18 & ADI &      20 &    1208 &     7.0 \\
TW Hya & 2009-01-18 & ASDI &     55 &    3302 &   100.7 \\
TW Hya & 2009-02-12 & ADI &      20 &    1208 &     7.1 \\
\enddata
\label{tab:observations}
\end{deluxetable}

\begin{deluxetable}{lrrrrrrrrrrrrrrrrrr}
\tabletypesize{\scriptsize}
\rotate
\tablecaption{95\% completeness $CH_{4}$-band ASDI contrasts ($\Delta$mag)}
\tablewidth{0pt}
\tablehead{ \colhead{Target} &\colhead{0.36$''$} & \colhead{0.5$''$} & \colhead{0.75$''$} & \colhead{1$''$}&\colhead{1.5$''$} & \colhead{2$''$} & \colhead{3$''$} & \colhead{4$''$}&\colhead{5$''$}&\colhead{7$''$} & \colhead{Cov.} & \colhead{9$''$} & \colhead{Cov.}&\colhead{12$''$} & \colhead{Cov.} & \colhead{14.8$''$} & \colhead{Cov.}}
\startdata
\\ \hline
All Stars \\ \hline
Median Contrast & 11.0 & 12.4 & 13.4 & 13.9 & 14.3 & 14.5 & 14.6 & 14.5 & 14.4 & - & - & - & - & - & - & - & - \\ 
\\ \hline
TW Hya, 10 Myr \\ \hline
TWA 7 & 11.1 & 12.2 & 13.2 & 13.6 & 14.1 & 14.3 & 14.3 & 14.4 & 14.4 & 14.3 & 1.00 & 13.8 & 1.00 & 12.9 & 0.73 & 12.6 & 0.27 \\
TW Hya & 10.5 & 11.6 & 12.5 & 12.8 & 13.4 & 13.5 & 13.7 & 13.7 & 13.7 & 13.2 & 1.00 & 11.8 & 1.00 & 10.6 & 0.69 & 11.9 & 0.20 \\
TWA 13A & 10.7 & 12.1 & 12.7 & 13.0 & 13.4 & 13.4 & 13.3 & 13.4 & 13.1 & 13.1 & 0.95 & 13.1 & 0.65 & 12.7 & 0.32 & 12.1 & 0.07 \\
\enddata
\label{tab:contrastsASDI}
\tablecomments{Beyond 6.3$"$ only a fraction of each final image has data. This coverage fraction is given beside 
the contrast at these large separations. All the contrasts are 95\% completeness contrasts, except at separations beyond 6.3$"$ where the 
nominal 5$\sigma$ contrast curve is used. A constant is added to the
5$\sigma$ curve so that both curves match at 6.3$"$.  Median contrasts are only given for regions of 100\% coverage.}
\end{deluxetable}

\begin{deluxetable}{lrrrrrrrrrrrrrrrrrr}
\tabletypesize{\scriptsize}
\rotate
\tablecaption{95\% completeness $H$-band ADI contrasts ($\Delta$mag)}
\tablewidth{0pt}
\tablehead{ \colhead{Target} &\colhead{0.36$''$} & \colhead{0.5$''$} & \colhead{0.75$''$} & \colhead{1$''$}&\colhead{1.5$''$} & \colhead{2$''$} & \colhead{3$''$} & \colhead{4$''$}&\colhead{5$''$}&\colhead{7$''$} & \colhead{Cov.} & \colhead{9$''$} & \colhead{Cov.}&\colhead{12$''$} & \colhead{Cov.} & \colhead{14.8$''$} & \colhead{Cov.}}
\startdata
\\ \hline
All Stars \\ \hline
Median Contrast & 9.1 & 10.4 & 11.6 & 13.2 & 14.4 & 15.1 & 15.4 & 15.5 & 15.4 & - & - & - & - & - & - & - & - \\ 
\\ \hline
TW Hya, 10 Myr \\ \hline
TWA 6 &  8.9 & 10.0 & 11.4 & 12.5 & 13.6 & 13.9 & 13.8 & 13.6 & 13.5 & 12.7 & 1.00 & 11.2 & 1.00 & 10.5 & 0.73 & 11.0 & 0.38 \\
TWA 7 &  7.8 &  9.1 & 10.6 & 12.3 & 14.3 & 15.2 & 15.3 & 15.7 & 15.5 & 15.4 & 0.90 & 15.0 & 0.67 & 13.6 & 0.29 & 13.7 & 0.05 \\
TW Hya &  7.4 &  9.3 & 11.7 & 13.0 & 14.4 & 14.9 & 14.9 & 15.0 & 14.9 & 15.1 & 0.88 & 14.7 & 0.63 & 14.0 & 0.26 & 13.4 & 0.04 \\
\enddata
\tablecomments{Beyond 6.3$"$ only a fraction of each final image has data. This coverage fraction is given beside 
the contrast at these large separations. All the contrasts are 95\% completeness contrasts, except at separations beyond 6.3$"$ where the 
nominal 5$\sigma$ contrast curve is used. A constant is added to the
5$\sigma$ curve so that both curves match at 6.3$"$.  Median contrasts are only given for regions of 100\% coverage.}
\label{tab:contrastsADI}
\end{deluxetable}

\begin{deluxetable}{lrrrrrrrrr}
\tabletypesize{\scriptsize}
\tablecaption{95\% completeness $H$-band limiting magnitudes}
\tablewidth{0pt}
\tablehead{ \colhead{Target} &\colhead{0.3$''$} & \colhead{0.5$''$} & \colhead{0.75$''$} & \colhead{1$''$}&\colhead{1.5$''$} & \colhead{2$''$} & \colhead{3$''$} & \colhead{4$''$}& \colhead{5$''$}}
\startdata
\\ \hline
All Stars\\ \hline
Median Limiting Magnitude & 16.4 & 17.7 & 18.9 & 20.1 & 21.6 & 22.0 & 22.5 & 22.5 & 22.5 \\
\\ \hline
TW Hya, 10 Myr \\ \hline
TWA 6 & 17.1 & 18.2 & 19.6 & 20.7 & 21.8 & 22.1 & 22.0 & 21.8 & 21.7 \\
TWA 7 & 15.5 & 16.9 & 18.3 & 20.1 & 22.0 & 22.9 & 23.1 & 23.4 & 23.3 \\
TW Hya & 15.0 & 16.8 & 19.2 & 20.6 & 22.0 & 22.4 & 22.5 & 22.6 & 22.5 \\
\enddata
\label{tab:appmagADI}
\end{deluxetable}

\begin{deluxetable}{lrrrrrrrrr}
\tabletypesize{\scriptsize}
\tablecaption{95\% completeness minimum detectable masses (M$_{Jup}$) in ADI mode, DUSTY models}
\tablewidth{0pt}
\tablehead{ \colhead{Target} &\colhead{0.3$''$} & \colhead{0.5$''$} & \colhead{0.75$''$} & \colhead{1$''$}&\colhead{1.5$''$} & \colhead{2$''$} & \colhead{3$''$} & \colhead{4$''$}& \colhead{5$''$}}
\startdata
\\ \hline
All Stars\\ \hline
Median Detectable Mass & 12.3 & 11.1 & 6.8 & 6.0 & 4.4 & 3.7 & 3.2 & 3.1 & 3.1 \\
\\ \hline
TW Hya, 10 Myr \\ \hline
TWA 6 &  7.7 &  6.4 &  5.1 &  4.2 &  3.5 &  3.3 &  3.3 &  3.5 &  3.5 \\
TWA 7 &  6.9 &  5.5 &  4.3 &  3.1 &  1.9 &  1.4 &  1.3 &  1.1 &  1.2 \\
TW Hya & 10.9 &  7.1 &  4.7 &  3.7 &  2.8 &  2.6 &  2.5 &  2.4 &  2.5 \\
\enddata
\label{tab:minmassADI_DUSTY}
\end{deluxetable}

\begin{deluxetable}{lrrrrrrrrr}
\tabletypesize{\scriptsize}
\tablecaption{95\% completeness minimum detectable masses (M$_{Jup}$) in ADI mode, COND models}
\tablewidth{0pt}
\tablehead{ \colhead{Target} &\colhead{0.3$''$} & \colhead{0.5$''$} & \colhead{0.75$''$} & \colhead{1$''$}&\colhead{1.5$''$} & \colhead{2$''$} & \colhead{3$''$} & \colhead{4$''$}& \colhead{5$''$}}
\startdata
\\ \hline
All Stars \\ \hline
Median Detectable Mass & 10.8 & 6.1 & 3.6 & 2.0 & 1.4 & 1.1 & 1.1 & 1.1 & 1.1 \\
\\ \hline
TW Hya, 10 Myr \\ \hline
TWA 6 &  6.1 &  4.2 &  2.6 &  1.8 &  1.3 &  1.1 &  1.2 &  1.3 &  1.3 \\
TWA 7 &  4.9 &  3.0 &  1.9 &  1.0 &  1.0 &  1.0 &  1.0 &  1.0 &  1.0 \\
TW Hya &  9.7 &  5.1 &  2.2 &  1.4 &  1.0 &  1.0 &  1.0 &  1.0 &  1.0 \\
\enddata
\label{tab:minmassADI_COND}
\end{deluxetable}

\begin{deluxetable}{ccccccccccc}
\tabletypesize{\scriptsize}
\tablecaption{Properties of Candidate Companions\label{table_comp2a}}
\tablewidth{0pt}
\tablehead{
\colhead{Name} & \colhead{\#} & \colhead{Sep} & 
\colhead{PA} & 
\colhead{$\Delta H$} & \colhead{$\Delta T$} & 
\colhead{Epochs} & \colhead{$\chi^2_{\nu}$(BG)} & 
\colhead{$\chi^2_{\nu}$(CPM)} & \colhead{dof} & \colhead{Comp?} \\
\colhead{} & \colhead{} & \colhead{($"$)} & \colhead{(deg)} & 
\colhead{(mag)} & \colhead{(years)} & \colhead{} & \colhead{} & 
\colhead{} & \colhead{} & \colhead{}
}
\startdata
HIP 2729 & 1 & 9.986 & 30.9 & 10.2 & 8.33 & 3 & 6.61 & 31.72 & 4 & BG  \\ 
HD 12894 & 1 & 0.322 & 242.9 & 3.0 & 3.22 & 3 & 322.08 & 110.70 & 4 & CPM  \\ 
CD -58 553 & 1 & 6.924 & 85.3 & 10.8 & 4.01 & 4 & 0.06 & 83.96 & 6 & BG  \\ 
HD 19668 & 1 & 5.470 & 151.0 & 9.8 & 9.93 & 8 & 0.49 & 30.57 & 14 & BG  \\ 
LP 776-25 & 1 & 2.570 & 87.8 & 7.5  & 3.13 & 2 & 0.16 & 67.91 & 2 & BG  \\ 
LP 776-25 & 2 & 6.340 & 264.7 & 12.9 & 3.13 & 2 & 0.17 & 14.56 & 2 & BG  \\ 
LP 776-25 & 3 & 7.010 & 270.1 & 12.8 & 3.13 & 2 & 1.28 & 14.48 & 2 & BG  \\ 
GJ 182 & 1 & 4.870 & 224.7 & 12.5 & 4.25 & 2 & 2.21 & 249.10 & 2 & BG  \\
\enddata
\tablecomments{Summary of the candidate companions detected around 
each target star.  For each candidate we list the separation and 
position angle at the reference epoch, the contrast between 
candidate and host star, the time baseline for our astrometric 
data, the number of epochs, the reduced chi-square statistic for 
the companion to be background (BG) or common proper motion (CPM), 
the number of degrees of freedom, and the final determination of 
each companion: background or common proper motion.  Astrometry 
for individual epochs is in Table~\ref{table_comp2b}.
}
\tablenotetext{a}{\citet{Wah11}}
\tablenotetext{b}{known common proper motion substellar companion,
  discovered by \citet{Cha05b}}
\tablenotetext{c}{see Lowrance et al. 2005 for previous background verification}
\tablenotetext{d}{spectroscopically verified to be background by Jura et al. (1998) }
\tablenotetext{e}{\citet{Met09}}
\tablenotetext{f}{\citet{Bil10}}
\tablenotetext{g}{\citet{Low00,Low05}}
\end{deluxetable}

\begin{deluxetable}{cccccccccccccc}
\tabletypesize{\scriptsize}
\tablecaption{Astrometry of Candidate Companions\label{table_comp2b}}
\tablewidth{0pt}
\tablehead{
 & & & \multicolumn{4}{c}{Measured Position} & & 
\multicolumn{4}{c}{Background Position} & 
& \\
\cline{4-7}
\cline{9-12}
\colhead{Name} & \colhead{\#} & \colhead{Epoch} & \colhead{Sep} & 
\colhead{$\sigma_{Sep}$} & \colhead{PA} & \colhead{$\sigma_{PA}$} & 
\colhead{} & \colhead{Sep} & \colhead{$\sigma_{Sep}$} & 
\colhead{PA} & \colhead{$\sigma_{PA}$} & 
\colhead{Inst.} & \colhead{Comp?} \\
\colhead{} & \colhead{} & \colhead{} & \colhead{($"$)} & 
\colhead{} & \colhead{(deg)} & \colhead{} & \colhead{} & 
\colhead{($"$)} & \colhead{} & \colhead{(deg)} & \colhead{} & 
\colhead{} & \colhead{} \\
}
\startdata
HIP 2729 & 1 & 2011.89 & 9.986 & 0.009 & 30.9 & 0.2 &   & \ldots & \ldots & \ldots & \ldots & N & BG \\
 &   & 2003.56 & 10.174 & 0.024 & 35.3 & 0.5 &   & 9.990 & 0.012 & 35.671 & 0.191 & V & BG \\
  &   & 2005.78 & 10.044 & 0.288 & 33.6 & 1.0 &   & 10.001 & 0.011 & 34.415 & 0.191 & S & BG \\
HD 12894 & 1 & 2011.84 & 0.322 & 0.009 & 242.9 & 0.2 &   & \ldots & \ldots & \ldots & \ldots & N & CPM \\
  &   & 2008.62 & 0.339 & 0.020 & 259.0 & 1.0 &   & 0.251 & 0.006 & 194.735 & 1.455 & V & CPM \\
  &   & 2008.85 & 0.331 & 0.020 & 257.2 & 1.0 &   & 0.229 & 0.006 & 194.856 & 1.596 & V & CPM \\
CD -58 553 & 1 & 2011.70 & 6.924 & 0.009 & 85.3 & 0.2 &   & \ldots & \ldots & \ldots & \ldots & N & BG \\
  &   & 2007.69 & 7.265 & 0.040 & 85.2 & 1.0 &   & 7.266 & 0.016 & 85.734 & 0.194 & V & BG \\
\enddata
\tablecomments{Astrometry for each candidate companion detected around 
our target stars from NICI and archival observations.  At each 
epoch we give the measured separation, position angle, and 
uncertainties as well as the predicted separation and position 
angle for a background object based on the proper motion and 
parallax of the primary and the candidate position at the reference 
epoch, which is the first epoch listed for each candidate.  
Astrometry is taken from Gemini-South NICI (N), VLT NACO (V), 
Keck NIRC2 (K), VLT ISAAC (I), ESO 3.6m (E), HST NICMOS (H),
HST ACS (A), Gemini-North NIRI (G), 2MASS (M), NTT SOFI (S), LBT (L), 
Palomar (P), and DSS (D).
}
\tablenotetext{a}{Mouillet et al. 1997}
\end{deluxetable}


\begin{deluxetable}{lrccccc}
\tabletypesize{\scriptsize}
\tablecaption{Archival Datasets for New Stellar Binaries}
\tablewidth{0pt}
\tablehead{ \colhead{Target}  &\colhead{Date (UT) } &\colhead{Telescope} &
  \colhead{Filter} & \colhead{Mode} & \colhead{Length (s)} & \colhead{Notes}}
\startdata
BD+07 1919  & 2008-04-02 & VLT NACO & $J$ & direct imaging & 450 & \\
& 2008-04-02 & VLT NACO & $H$ & direct imaging & 450 & \\
& 2008-04-02 & VLT NACO & $K$s & direct imaging & 150 & \\
 & 2009-03-12 & Gemini NICI & $CH_4$ & ASDI  &   2701 & B saturated \\
 & & & & & & except in short exps \\
 & 2009-03-12 & Gemini NICI & $H$ & ADI &  1208 &   B saturated \\
 & & & & & &  in all exposures \\
 & 2010-02-28 & Gemini NICI & $H$ & ADI &  1389 &   B saturated \\
 & & & & & & in all exposures \\
HD 12894 & 2008-08-14 & VLT NACO & $L'$ & direct imaging & 1230 & \\
 & 2008-11-07 & VLT NACO & $K$s & direct imaging & 480 & \\
 & 2011-11-03 & Gemini NICI & $H$ & ADI & 2372 &  B saturated \\
 & & & & & & except in short exps \\
           & & & & & & A under mask  \\ 
\enddata
\label{tab:binarydata}
\end{deluxetable}


\begin{deluxetable}{lcc}
\tabletypesize{\footnotesize}
\tablecaption{Properties of the HD 12894 AB System\label{tab:HD12894}}
\tablewidth{0pt}
\tablehead{
\colhead{} & \colhead{Primary} & \colhead{Secondary}}

\startdata

Parallax  (mas)    & \multicolumn{2}{c}{20.9$\pm$0.5 \tablenotemark{a}} \\
Age  (Myr)         & \multicolumn{2}{c}{30$\pm$10} \\
Proper Motion ($\mu_{\alpha}$, $\mu_{\delta}$) (mas/yr) & \multicolumn{2}{c}{(75.7$\pm$0.4, $-$25.0$\pm$0.5)\tablenotemark{a}}  \\

Separation: 2008.62, VLT NACO   & \multicolumn{2}{c}{0.33$\pm$0.02$\arcsec$ (15.7$\pm$1.0~AU)} \\ 
Position Angle: 2008.62, VLT NACO  & \multicolumn{2}{c}{258.9$\pm$1.0$^{\circ}$} \\
Separation: 2008.85, VLT NACO   & \multicolumn{2}{c}{0.32$\pm$0.02$\arcsec$ (15.4$\pm$1.0~AU)} \\ 
Position Angle: 2008.85, VLT NACO  & \multicolumn{2}{c}{257.5$\pm$1.0$^{\circ}$} \\
Separation: 2011.84, Gemini NICI   & \multicolumn{2}{c}{0.323$\pm$0.009$\arcsec$  (15.4$\pm$0.5~AU)} \\
\enddata
\tablenotetext{a}{\citet[][]{vLe07}}
\tablenotetext{b}{initial combined $Ks$ mag from 2MASS}
\end{deluxetable}

\begin{deluxetable}{lccc}
\tabletypesize{\footnotesize}
\tablecaption{Properties of the BD+07 1919 ABC System\label{tab:BD+07}}
\tablewidth{0pt}
\tablehead{
\colhead{} & \colhead{Primary} & \colhead{Secondary} & \colhead{Tertiary}}

\startdata

Parallax  (mas)    & \multicolumn{3}{c}{28.52$\pm$2.65 \tablenotemark{a}} \\
Age (Myr)        & \multicolumn{3}{c}{70$\pm$10} \\
Proper Motion ($\mu_{\alpha}$, $\mu_{\delta}$) (mas/yr) & \multicolumn{3}{c}{(1.79$\pm$3.75, $-$138.68$\pm$3.25)\tablenotemark{a}}  \\

BC Separation: 2008.25, VLT NACO   & \multicolumn{3}{c}{0.358$\pm$0.02$\arcsec$ (12.5$\pm$1.4~AU)} \\ 
BC Position Angle: 2008.25, VLT NACO  & \multicolumn{3}{c}{109.7$\pm$1.0$^{\circ}$} \\
BC Separation: 2009.19, Gemini NICI   & \multicolumn{3}{c}{0.398$\pm$0.009$\arcsec$  (14.0$\pm$1.3~AU)} \\ 
BC Position Angle: 2009.19, Gemini NICI  & \multicolumn{3}{c}{108.2$\pm$0.2$^{\circ}$} \\


$\Delta J$(mag)  & \nodata & \nodata & 2.45$\pm$0.08\tablenotemark{b} \\
\enddata
\tablenotetext{a}{\citet[][]{vLe07}}
\tablenotetext{b}{$\Delta$(mag) relative to B component}
\tablenotetext{c}{2MASS}
\tablenotetext{d}{initial combined mag from 2MASS}
\end{deluxetable}

\begin{deluxetable}{lcc}
\tabletypesize{\footnotesize}
\tablecaption{Photometry for AB Pic AB \label{tab:ABPic}}
\tablewidth{0pt}
\tablehead{
\colhead{} & \colhead{Primary}\tablenotemark{a} & \colhead{Secondary} }

\startdata
$\Delta J$(mag) & \nodata & 7.97$\pm$0.14 \\
$\Delta Ks$(mag) & \nodata & 6.79$\pm$0.10 \\
$J$ (mag)                       & 7.58$\pm$0.03 & 15.55$\pm$0.14 \\
$Ks$ (mag)                       & 6.98$\pm$0.03 & 13.77$\pm$0.10 \\
$J$ - $Ks$ (mag)              & 0.60$\pm$0.04 & 1.78$\pm$0.17 \\
$M_{J}$ (mag)                   & 4.26$\pm$0.07 & 12.23$\pm$0.16 \\
$M_{K_s}$ (mag)                   & 3.66$\pm$0.07 & 10.45$\pm$0.12 \\
\enddata
\tablenotetext{a}{Primary photometry from 2MASS.}
\end{deluxetable}

\begin{deluxetable}{lccc}
\tabletypesize{\scriptsize}
\tablecaption{Stellar and Brown Dwarf Binaries to Sample Stars\label{tab:binaries}}
\tablewidth{0pt}
\tablehead{ \colhead{Primary Star} &\colhead{Projected Separation (AU)} &
\colhead{Sp. Type or Est. Mass} &
\colhead{Reference} }
\startdata
\hline
BD +7 1919A & 211 & K5 & \citet{CCDM} \\
BD +7 1919B & 14 & $\sim$0.2 M$_{\odot}$   & discovered in this work \\  
HD 12894 & 15.5 & $\sim$0.5 M$_{\odot}$   & discovered in this work \\
PZ Tel & 16.4 & $\sim$40 M$_{Jup}$ & \citet{Bil10} \\
CD-35 2722 & 67 & $\sim$30 M$_{Jup}$ & \citet{Wah11} \\
HR 7329 & 200 &  M7-M8 & \citet{Low00, Gue01}\\
AB Pic &  250  & L0.5 & \citet{Cha05b} \\
TYC 7443-1102-1 & 1520 & M0 & \citet{Lep09} \\
CD-53 544 & 936 & K6 & companion is AF Hor, \citet{WDS} \\
AF Hor & 936 & M2 & companion is CD -53 544, \citet{WDS} \\
GJ 2036 A & 129 & M3 & companion is GJ 2036 B, \citet{CCDM} \\
GJ 2036 B & 129 & M3 & companion is GJ 2036 A, \citet{CCDM} \\
HIP 26369 &  621 & K6 & companion is UY Pic, \citet{CCDM} \\
UY Pic & 621 & K0 & companion is HIP 26369, \citet{CCDM} \\
HD 45270 & 376 & \nodata & \citet{CCDM}\\
DX Leo & 1150 & M4.5 & \citet{CCDM} \\
HIP 82688 & 181 & 0.31 M$_{\odot}$ & \citet{Met09} \\
HR 4796 A & 560, 2900 & M2.5, --- &  \citet{CCDM} \\
HR 4796 B & 560, 2900 & A0, --- & \citet{CCDM} \\
GJ 560 A & 260 & K5 & \citet{CCDM} \\
TWA 8A & 611 & M5 & \citet{WDS} \\
TWA 9A & 271  & M1 & comp. is TWA 9A, \citet{CCDM} \\
TWA 9B &  271  & K5 & comp. is TWA 9B, \citet{CCDM} \\
HIP 10679 & 375 & F5 & \citet{CCDM} \\
BD-21 1074A  & 398 & M3 & \citet{WDS} \\ 
V343 Nor &  411 & M4 & \citet{WDS} \\
HD 164249B & 309  & F6 & \citet{WDS} \\
HD 172555A & 2052 & K5 & \citet{WDS} \\
GJ 799A & 39  &  M4 & \citet{WDS} \\ 
HIP 112312A & 839 & M5 & companion to TX PsA, \citet{CCDM} \\
TX PsA & 839 & M4 & companion  to HIP 112312A, \citet{CCDM} \\
\hline
\enddata
\tablenotetext{1}{Semi-major axis}
\label{tab:binaries}
\end{deluxetable}

\begin{deluxetable}{ccccc}
\tablecaption{Upper Limits on Planet Fraction from the Monte Carlo Analysis\label{table_pf}}
\tablewidth{0pt}
\tablehead{
 & \multicolumn{4}{c}{Planet Fraction Upper Limit} \\
\colhead{Mass (M${_{Jup}}$)}  & \colhead{$\leq$5\%}  & \colhead{$\leq$10\%}  & \colhead{$\leq$20\%}  &
\colhead{50\%}  \\
}
\startdata
1& \ldots & \ldots & 67 -- 290 AU & 24 -- 610 AU \\
2& \ldots & 49 -- 290 AU & 22 -- 550 AU & 12 -- 880 AU \\
4& \ldots & 25 -- 380 AU & 14 -- 640 AU & 9.3 -- 1100 AU \\
10& \ldots & 19 -- 390 AU & 12 -- 660 AU & 7.4 -- 1100 AU \\
\enddata
\tablenotetext{a}{
This table gives the range in semi-major axis at which a
given upper limit on planet fraction is reached for a given planet mass.
For example, between semi-major axis of 67 and 290 AU fewer than 10\%
of MG stars can have giant planets more massive than 1 M$_{Jup}$.}
\end{deluxetable}

\begin{deluxetable}{ccc}
\tablecaption{Upper Limits on Planet Fraction from the Bayesian Analysis\label{table_pf_Bayesian}}
\tablewidth{0pt}
\tablehead{
\colhead{Model}  & \colhead{95.4\%
  Confidence Limit}  & \colhead{99.7\% Confidence Limit} }
\startdata
  \multicolumn{3}{c}{Planet Fraction Upper Limit, 1--20 M$_{Jup}$, 10--150 AU} \\
DUSTY & $\leq$18$\%$ & $\leq44\%$ \\
COND & $\leq$6$\%$ & $\leq14\%$ \\\hline
  \multicolumn{3}{c}{Planet Fraction Upper Limit, 1--20 M$_{Jup}$,  10--50 AU} \\
DUSTY & $\leq$21$\%$ & $\leq51\%$ \\
COND & $\leq7\%$ & $\leq17\%$ \\
\enddata
\end{deluxetable}

\begin{figure}
\includegraphics[width=6in]{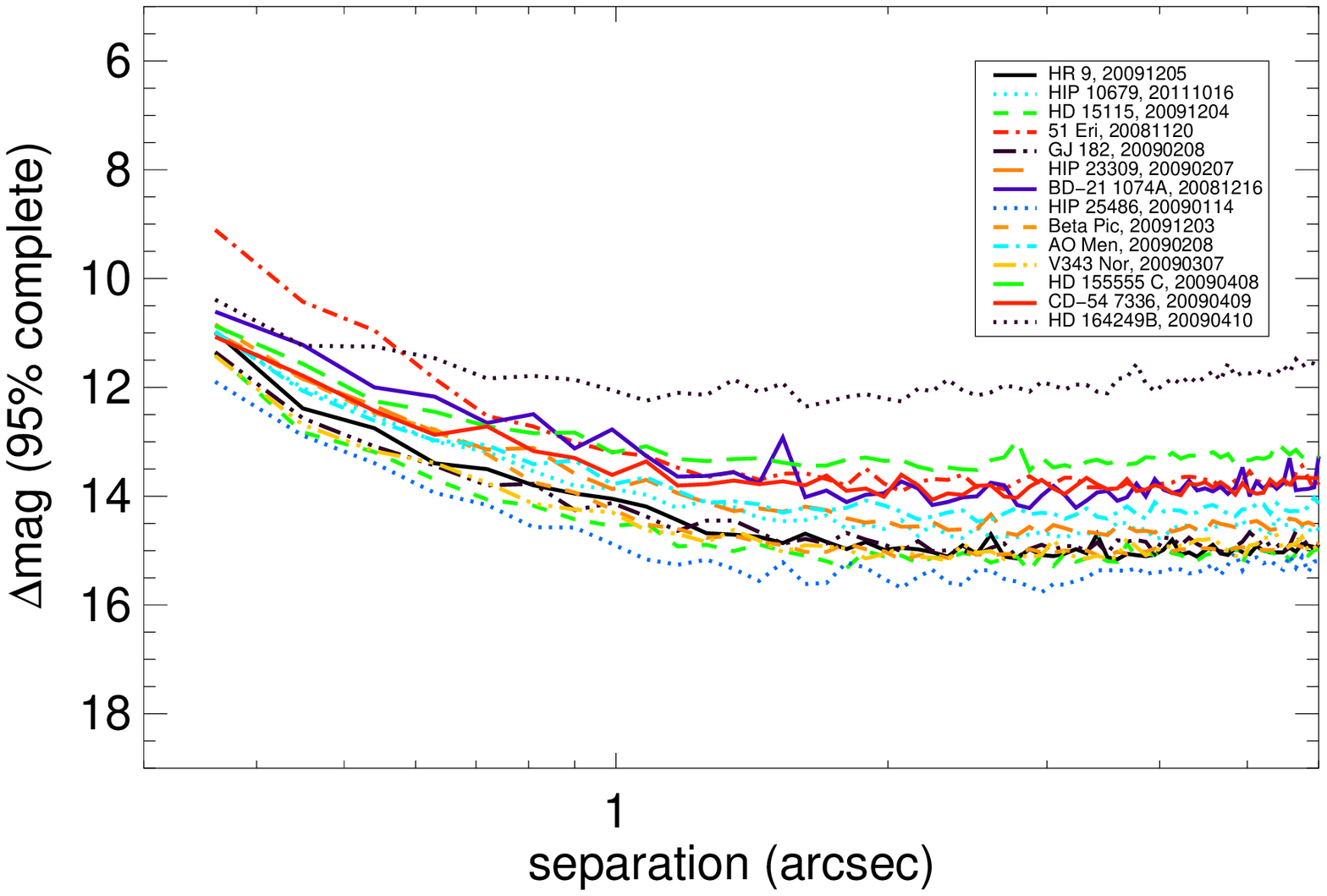}
\includegraphics[width=6in]{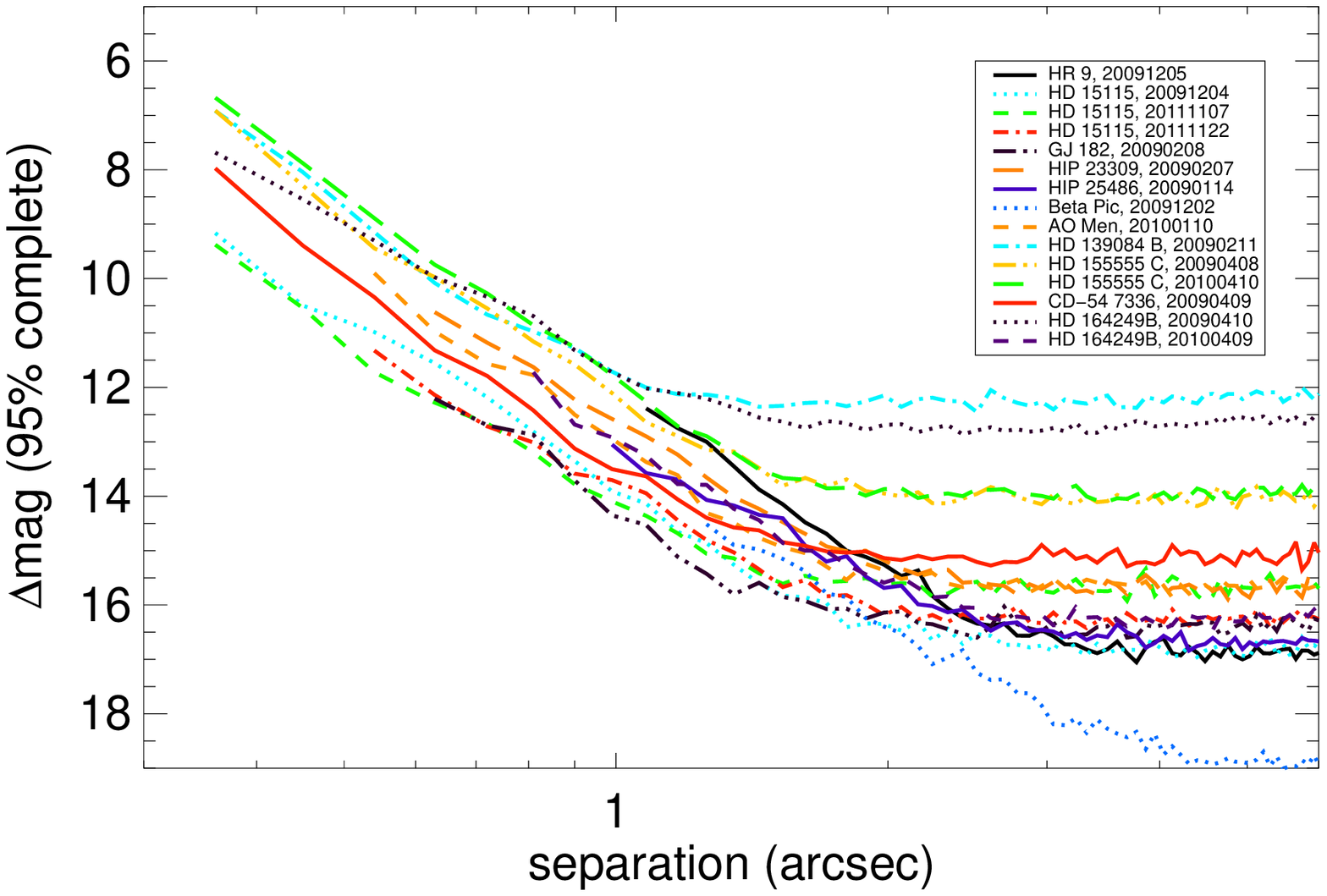}
\caption{95\% completeness contrast curves for $\beta$ Pic moving group stars (1st
  half), with plot range from 0.3'' (coronagraph inner working angle)
  to 6''.  Top: ASDI contrasts.  Bottom: ADI contrasts.}
\end{figure}

\begin{figure}
\includegraphics[width=6in]{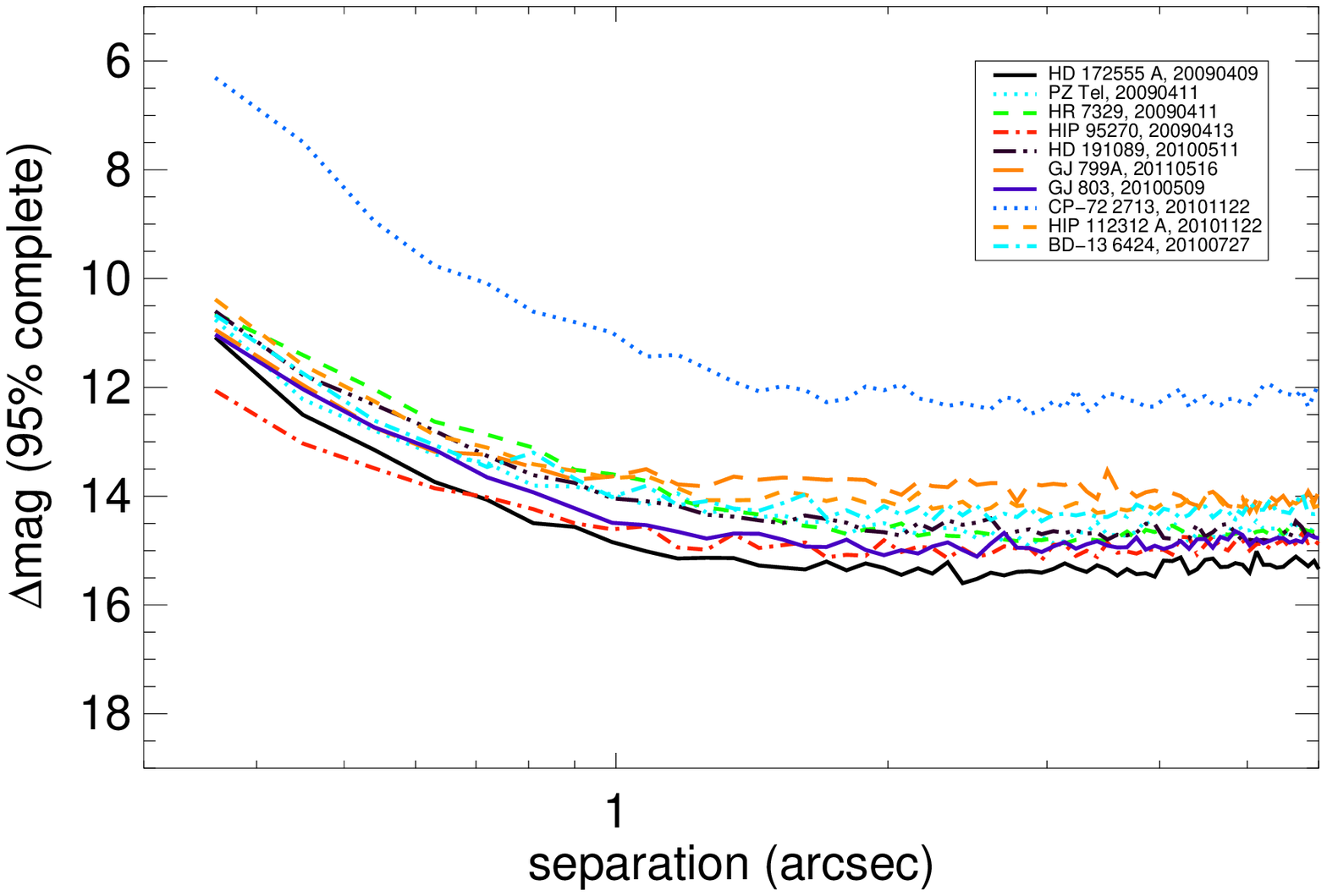}
\includegraphics[width=6in]{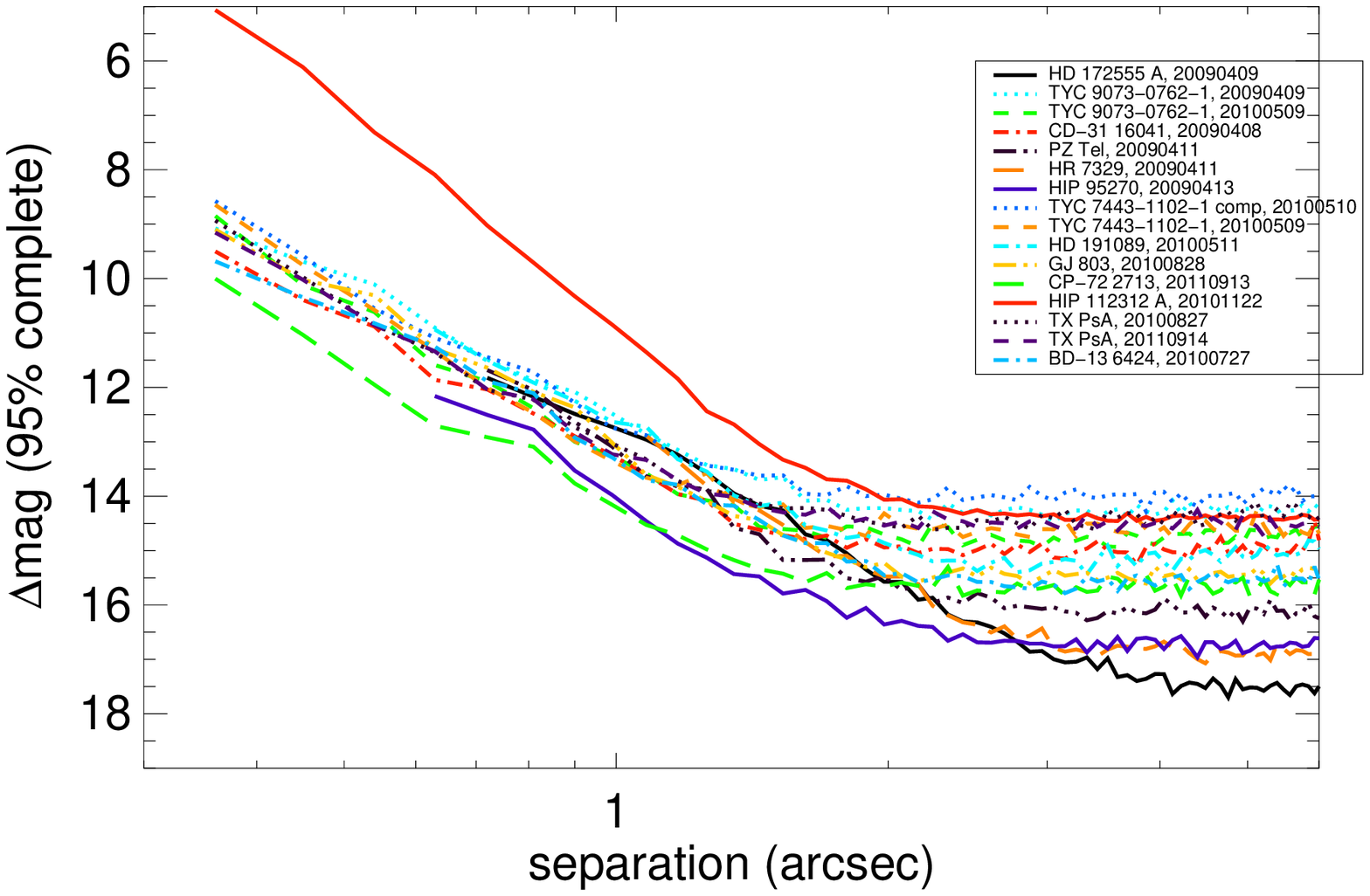}
\caption{95\% completeness contrast curves for $\beta$ moving group (2nd
  half), with plot range from 0.3'' (coronagraph inner working angle) to 6''.  Top:
   ASDI contrasts.  Bottom:  ADI contrasts.}
\end{figure}

\begin{figure}
\includegraphics[width=6in]{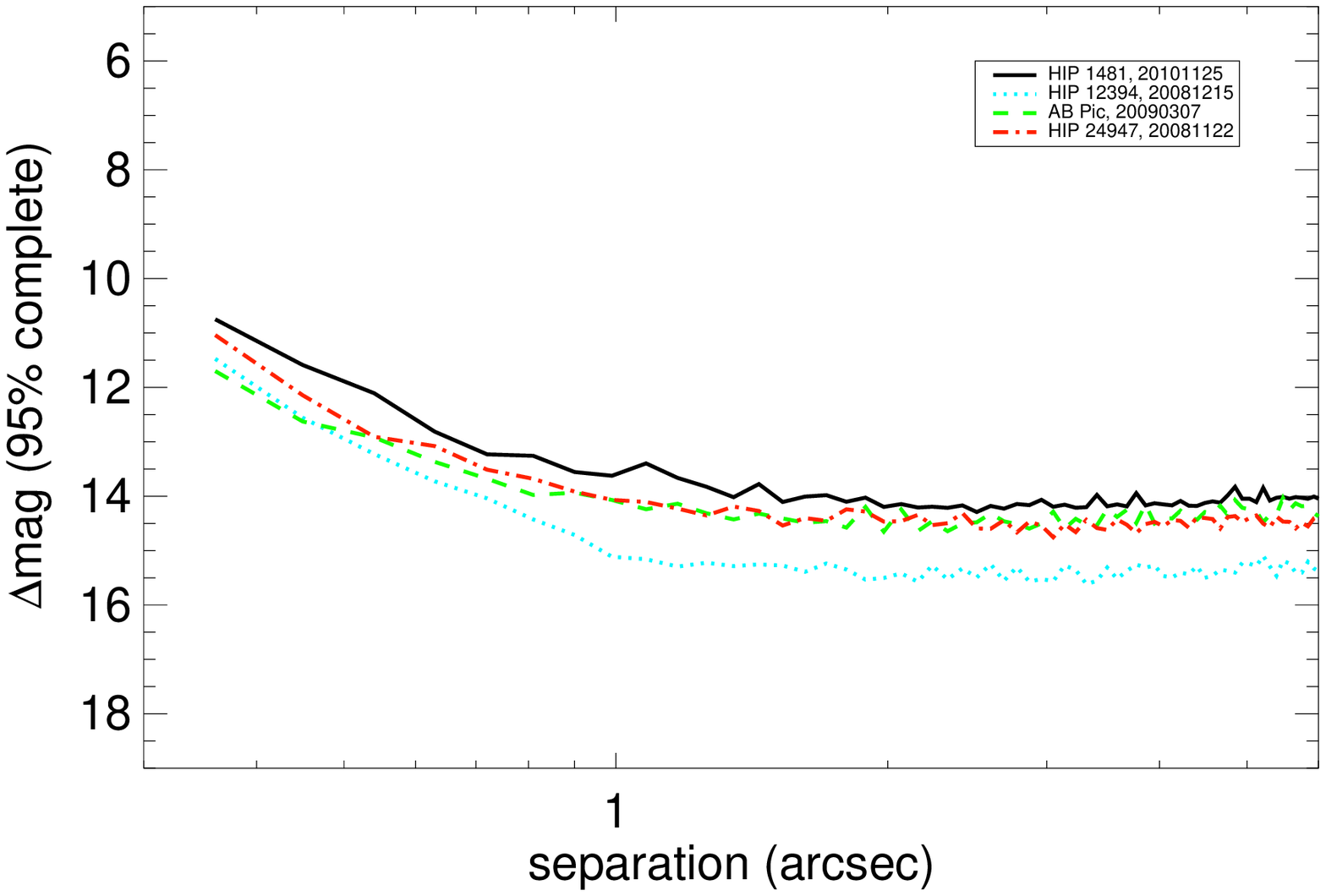}
\includegraphics[width=6in]{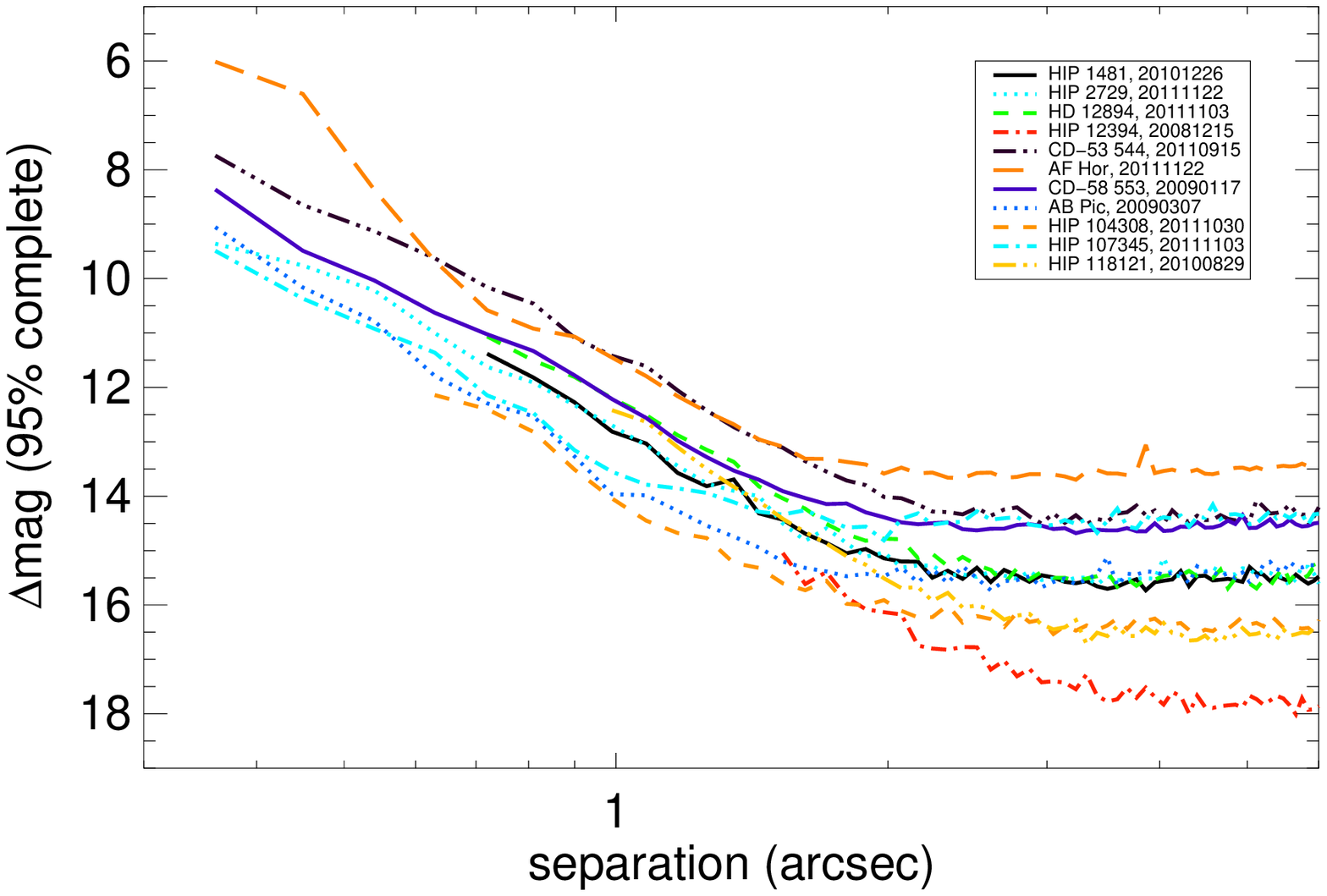}
\caption{95\% completeness contrast curves for Tucana-Horologium
  association stars, with plot range from 0.3'' (coronagraph inner working angle) to 6''.  Top:
   ASDI contrasts.  Bottom:  ADI contrasts.}
\end{figure}

\begin{figure}
\includegraphics[width=6in]{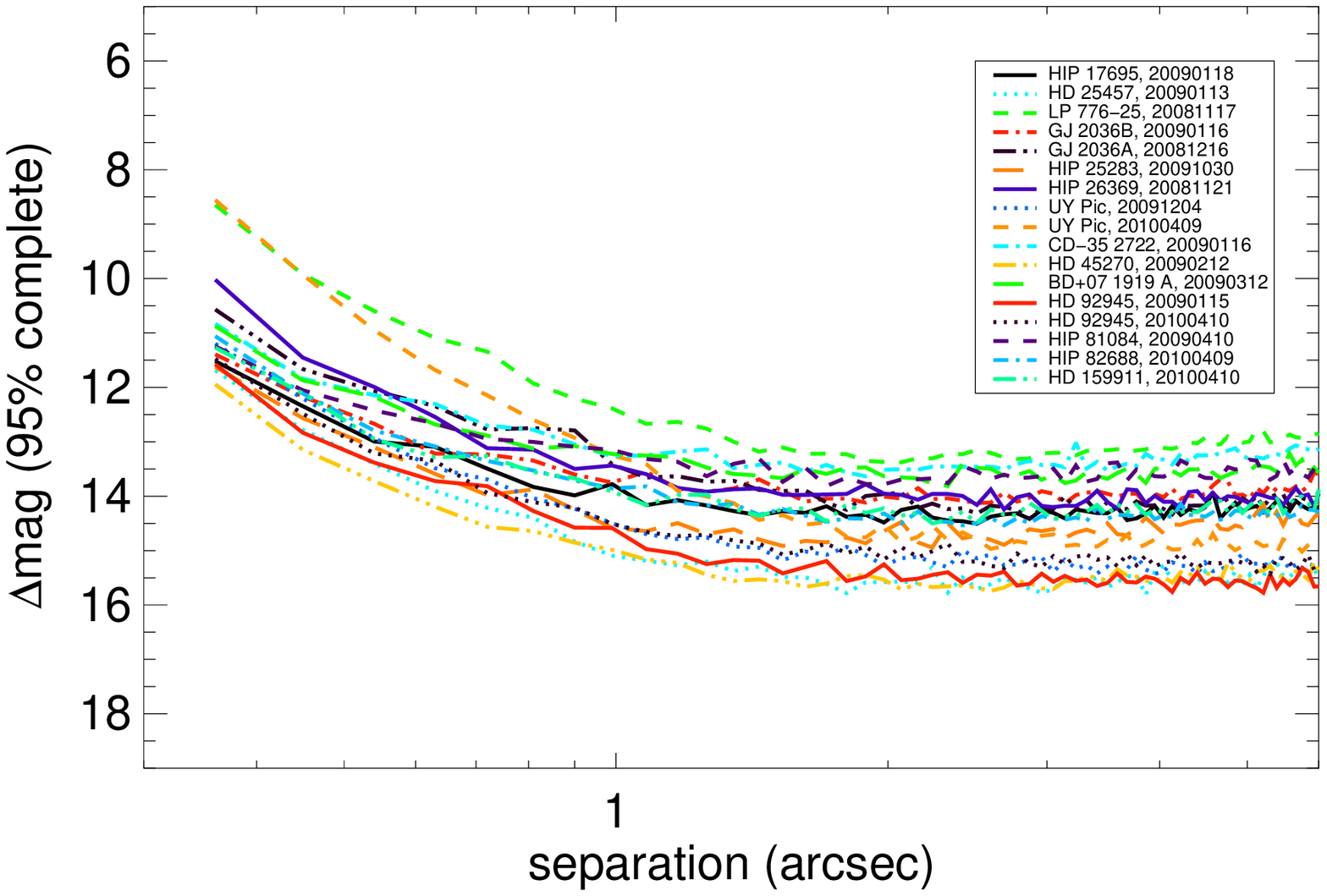}
\includegraphics[width=6in]{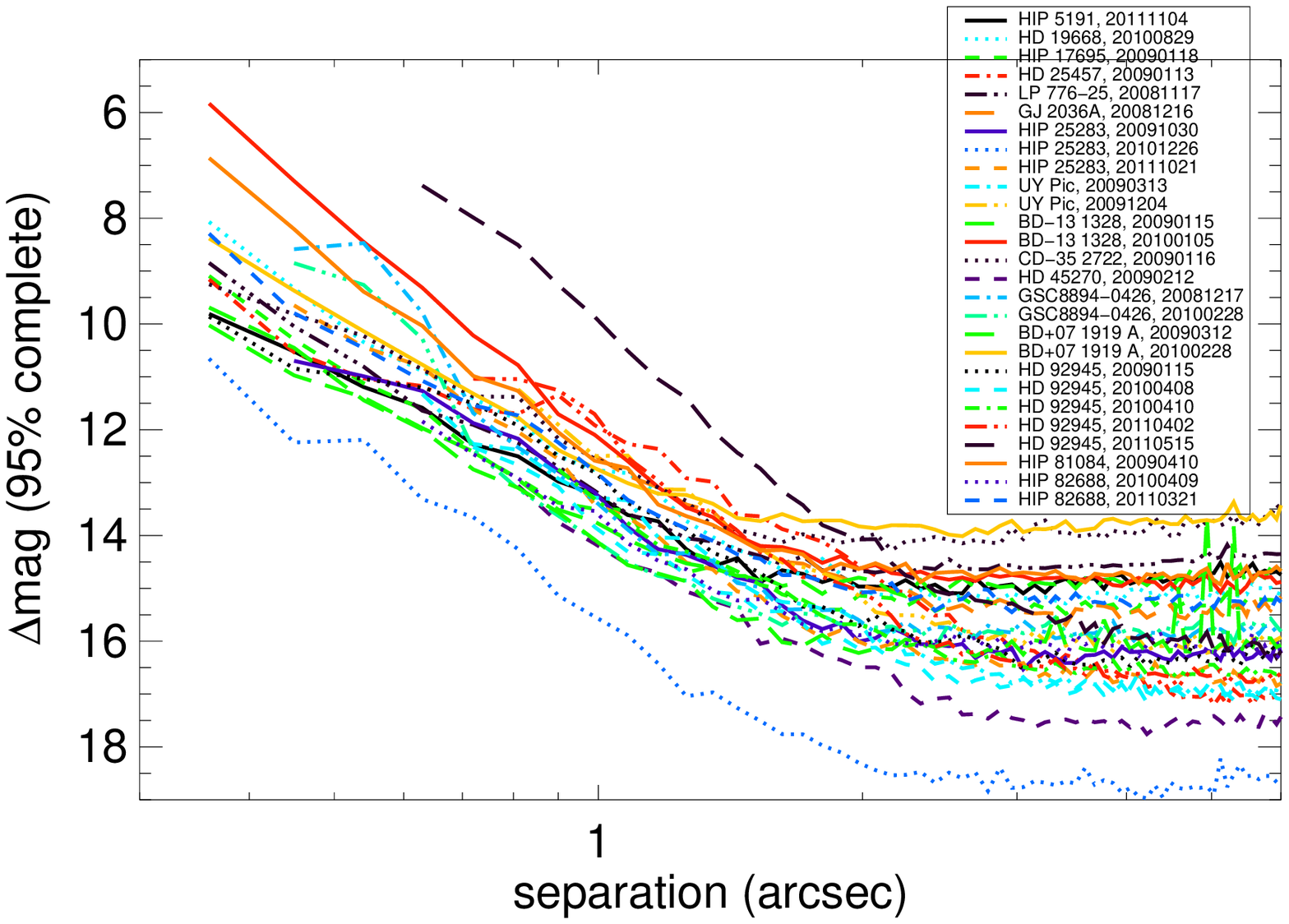}
\caption{95\% completeness contrast curves for AB Dor moving group 
  stars, with plot range from 0.3'' (coronagraph inner working angle) to 6''.  Top:
   ASDI contrasts.  Bottom:  ADI contrasts.}
\end{figure}

\clearpage

\clearpage

\clearpage

\clearpage

\clearpage

\clearpage
\clearpage
\appendix

\section{Candidate companions observed at only one epoch}

By the end of NICI Campaign observations, we were unable to 
obtain followup observations for a number of candidate companions 
at projected separations $>$400 AU or in dense stellar fields ($>$20
objects in the field) located in the Galactic bulge or disk.
Here we list target stars with candidate
companions for which we only have a single epoch of data.  Since
we cannot classify these as either CPM or background, we provide the
properties of these candidates here and note the changes we
make to the contrast curves in the Action column.  Contrast curves are
edited either by reverting to a less sensitive contrast curve or
restricting the contrast curve to within a given separation.
For cases where we have
only a single epoch and there are
candidates inside the 100\% coverage region for
position angle, we drop the star from our analysis.

\begin{deluxetable}{cccccccc}
\tabletypesize{\scriptsize}
\tablecaption{Candidate Companions with One Epoch of
Data\label{tab:1epoch}}
\tablewidth{0pt}
\tablehead{
\colhead{Star} & \colhead{\#} & \colhead{Sep} & \colhead{Sep} & \colhead{PA}
& \colhead{$\Delta H$} & \colhead{Epoch} & \colhead{Action} \\
\colhead{} & \colhead{} & \colhead{(``)} & \colhead{(AU)} & \colhead{(deg)}
& \colhead{(mag)} & \colhead{} & \colhead{} \\
}
\startdata
AO Men & 1 & 4.74 & 183 & 108.1 & 15.6 & 2009.1041 &  Revert to 2009.10 ASDI \\
    & 2 & 6.87 & 264 & 32.1 & 15.6 & 2009.1041 &     \\
    & 3 & 6.92 & 266 & 102.4 & 14.7 & 2009.1041 &     \\
    & 4 & 7.86 & 303 & 65.4 & 15.2 & 2009.1041 &     \\
CD-54 7336 & 1 & 2.92 & 193 & 252.5 & 8.1 & 2009.2684 & Drop \\
    & 2 & 4.26 & 281 & 228.6 & 15.0 & 2009.2684 &     \\
    & 3 & 4.32 & 285 & 27.7 & 10.6 & 2009.2684 &     \\
    & 4 & 4.74 & 313 & 207.1 & 13.2 & 2009.2684 &     \\
    & 5 & 5.39 & 356 & 264.5 & 14.1 & 2009.2684 &     \\
    & 6 & 6.28 & 415 & 132.0 & 11.6 & 2009.2684 &     \\
    & 7 & 6.36 & 420 & 100.6 & 5.7 & 2009.2684 &     \\
    & 8 & 6.55 & 433 & 332.6 & 14.6 & 2009.2684 &     \\
    & 9 & 6.95 & 459 & 296.6 & 13.3 & 2009.2684 &     \\
    & 10 & 7.04 & 465 & 167.6 & 11.5 & 2009.2684 &     \\
    & 11 & 7.15 & 472 & 91.6 & 11.7 & 2009.2684 &     \\
    & 12 & 7.68 & 507 & 105.5 & 11.1 & 2009.2684 &     \\
    & 13 & 8.17 & 539 & 284.1 & 13.4 & 2009.2684 &     \\
    & 14 & 8.24 & 544 & 346.0 & 13.2 & 2009.2684 &     \\
    & 15 & 8.67 & 572 & 100.1 & 14.7 & 2009.2684 &     \\
CD-31 16041 & 1 & 1.94 & 99 & 65.3 & 11.7 & 2009.2657 & Drop \\
    & 2 & 2.59 & 132 & 87.2 & 12.9 & 2009.2657 &     \\
    & 3 & 3.12 & 159 & 123.5 & 14.2 & 2009.2657 &     \\
    & 4 & 3.47 & 177 & 176.1 & 11.1 & 2009.2657 &     \\
    & 5 & 3.60 & 184 & 156.5 & 11.8 & 2009.2657 &     \\
    & 6 & 3.62 & 185 & 12.0 & 14.8 & 2009.2657 &     \\
    & 7 & 3.63 & 185 & 174.3 & 14.6 & 2009.2657 &     \\
    & 8 & 3.69 & 188 & 160.7 & 15.1 & 2009.2657 &     \\
    & 9 & 3.69 & 188 & 16.7 & 10.9 & 2009.2657 &     \\
    & 10 & 3.89 & 199 & 111.4 & 14.4 & 2009.2657 &     \\
    & 11 & 4.45 & 227 & 188.1 & 12.1 & 2009.2657 &     \\
    & 12 & 4.50 & 229 & 71.1 & 14.0 & 2009.2657 &     \\
    & 13 & 4.85 & 247 & 115.4 & 14.8 & 2009.2657 &     \\
    & 14 & 5.01 & 256 & 65.0 & 10.7 & 2009.2657 &     \\
    & 15 & 5.09 & 260 & 77.6 & 12.8 & 2009.2657 &     \\
    & 16 & 5.13 & 262 & 69.1 & 14.9 & 2009.2657 &     \\
    & 17 & 5.61 & 286 & 273.7 & 13.6 & 2009.2657 &     \\
    & 18 & 5.62 & 287 & 201.2 & 12.2 & 2009.2657 &     \\
    & 19 & 5.77 & 294 & 77.6 & 14.7 & 2009.2657 &     \\
    & 20 & 5.85 & 298 & 345.2 & 14.2 & 2009.2657 &     \\
    & 21 & 6.01 & 307 & 282.4 & 14.0 & 2009.2657 &     \\
    & 22 & 6.50 & 332 & 16.0 & 13.1 & 2009.2657 &     \\
    & 23 & 6.79 & 346 & 235.9 & 15.0 & 2009.2657 &     \\
    & 24 & 6.87 & 350 & 177.0 & 12.1 & 2009.2657 &     \\
    & 25 & 7.12 & 363 & 249.0 & 13.1 & 2009.2657 &     \\
    & 26 & 7.92 & 404 & 90.4 & 14.3 & 2009.2657 &     \\
    & 27 & 8.09 & 413 & 326.3 & 13.7 & 2009.2657 &     \\
    & 28 & 8.35 & 426 & 72.7 & 14.6 & 2009.2657 &     \\
    & 29 & 8.42 & 430 & 3.8 & 14.5 & 2009.2657 &     \\
    & 30 & 8.98 & 458 & 340.4 & 13.2 & 2009.2657 &     \\
    & 31 & 9.12 & 465 & 76.2 & 11.4 & 2009.2657 &     \\
    & 32 & 9.17 & 468 & 37.4 & 14.8 & 2009.2657 &     \\
    & 33 & 9.32 & 475 & 74.1 & 11.2 & 2009.2657 &     \\
    & 34 & 10.05 & 512 & 328.5 & 13.2 & 2009.2657 &     \\
HD 159911 & 1 & 3.79 & 171 & 118.3 & 13.8 & 2010.2712 &  Drop \\
    & 2 & 3.99 & 179 & 79.1 & 10.7 & 2010.2712 &     \\
    & 3 & 4.86 & 219 & 288.3 & 14.0 & 2010.2712 &     \\
    & 4 & 4.95 & 223 & 315.9 & 13.3 & 2010.2712 &     \\
    & 5 & 5.24 & 236 & 159.3 & 13.1 & 2010.2712 &     \\
    & 6 & 5.35 & 241 & 315.1 & 11.2 & 2010.2712 &     \\
    & 7 & 5.74 & 258 & 343.1 & 10.5 & 2010.2712 &     \\
    & 8 & 5.75 & 259 & 357.5 & 12.5 & 2010.2712 &     \\
    & 9 & 5.79 & 261 & 259.3 & 13.2 & 2010.2712 &     \\
    & 10 & 6.24 & 281 & 165.8 & 12.9 & 2010.2712 &     \\
    & 11 & 6.27 & 282 & 248.5 & 14.8 & 2010.2712 &     \\
    & 12 & 7.14 & 321 & 287.0 & 11.3 & 2010.2712 &     \\
    & 13 & 7.23 & 326 & 248.6 & 12.8 & 2010.2712 &     \\
    & 14 & 7.67 & 345 & 18.6 & 8.0 & 2010.2712 &     \\
TYC 7443-1102-1A & 1 & 7.94 & 458 & 155.4 & 12.3 & 2010.3507 & Sep $<$ 7.94" \\
    & 2 & 8.89 & 513 & 355.0 & 13.6 & 2010.3507 &     \\
TYC 7443-1102-1B & 1 & 8.85 & 511 & 12.7 & 12.8 & 2010.3534 & Sep $<$ 8.85"  \\
GJ 560 A & 1 & 5.71 & 94 & 230.7 & 15.9 & 2009.1891 &  Sep $<$ 5.88" \\
    & 2 & 5.88 & 96 & 185.6 & 15.8 & 2009.1891 &     \\
    & 3 & 6.15 & 101 & 27.5 & 16.3 & 2009.1891 &     \\
    & 4 & 6.30 & 103 & 52.6 & 16.1 & 2009.1891 &     \\
    & 5 & 6.86 & 113 & 269.0 & 14.6 & 2009.1891 &     \\
    & 6 & 7.34 & 120 & 284.4 & 15.0 & 2009.1891 &     \\
HD 139664 & 1 & 4.49 & 79 & 241.6 & 16.2 & 2010.3507 &  Revert to 2009.10 ASDI \\
    & 2 & 5.19 & 91 & 311.6 & 17.2 & 2010.3507 &     \\
    & 3 & 6.08 & 106 & 335.6 & 15.8 & 2010.3507 &     \\
    & 4 & 6.45 & 113 & 310.0 & 16.8 & 2010.3507 &     \\
    & 5 & 7.04 & 123 & 16.6 & 15.8 & 2010.3507 &     \\
    & 6 & 8.48 & 148 & 68.1 & 15.7 & 2010.3507 &     \\
    & 7 & 9.13 & 160 & 301.2 & 15.2 & 2010.3507 &     \\
    & 8 & 10.55 & 185 & 37.7 & 16.7 & 2010.3507 &     \\
HD 92945 & 1 & 8.43 & 182 & 253.4 & 15.9 & 2009.0383 & Sep $<$ 8.43" \\
TYC 9073-762-1 & 1 & 8.01 & 433 & 27.9 & 16.7 & 2010.3507 & No Change \\
    & 2 & 9.31 & 503 & -78.2 & 15.4 & 2010.3507 &     \\
GSC 8894-0426 & 1 & 9.93 & 228 & -118.2 & 9.6 & 2010.1589 & Sep $<$ 9.93" \\
HD 164249B & 1 & 4.60 & 216 & 136.3 & 12.6 & 2009.2712 & Sep $<$ 6.27", Revert to 2009.27 ASDI \\
    & 2 & 6.27 & 295 & 80.9 & 7.0 & 2009.2712 &     \\
    & 3 & 7.72 & 363 & 58.1 & 12.9 & 2009.2712 &     \\
    & 4 & 8.27 & 389 & 36.5 & 10.2 & 2009.2712 &     \\
HR 4796 B & 1 & 8.86 & 595 & 103.4 & 11.8 & 2009.1151 & Sep $<$ 8.86"  \\
V343 Nor & 1 & 1.32 & 53 & 317.2 & 14.6 & 2009.1781 & Drop  \\
    & 2 & 1.77 & 70 & 170.6 & 12.8 & 2009.1781 &     \\
    & 3 & 1.81 & 72 & 15.6 & 13.9 & 2009.1781 &     \\
    & 4 & 1.91 & 76 & 330.1 & 14.0 & 2009.1781 &     \\
    & 5 & 2.08 & 83 & 251.9 & 14.8 & 2009.1781 &     \\
    & 6 & 2.16 & 86 & 108.9 & 15.5 & 2009.1781 &     \\
    & 7 & 2.30 & 91 & 198.7 & 9.1 & 2009.1781 &     \\
    & 8 & 2.41 & 96 & 188.1 & 12.3 & 2009.1781 &     \\
    & 9 & 2.49 & 99 & 195.0 & 13.8 & 2009.1781 &     \\
    & 10 & 2.61 & 104 & 251.9 & 16.1 & 2009.1781 &     \\
    & 11 & 2.67 & 106 & 71.2 & 13.7 & 2009.1781 &     \\
    & 12 & 2.69 & 107 & 324.0 & 15.9 & 2009.1781 &     \\
    & 13 & 2.93 & 117 & 131.0 & 13.5 & 2009.1781 &     \\
    & 14 & 2.93 & 117 & 45.2 & 13.8 & 2009.1781 &     \\
    & 15 & 3.04 & 121 & 86.6 & 14.0 & 2009.1781 &     \\
    & 16 & 3.04 & 121 & 350.1 & 8.8 & 2009.1781 &     \\
    & 17 & 3.07 & 122 & 169.2 & 15.8 & 2009.1781 &     \\
    & 18 & 3.09 & 123 & 91.7 & 15.7 & 2009.1781 &     \\
    & 19 & 3.27 & 130 & 51.2 & 15.0 & 2009.1781 &     \\
    & 20 & 3.33 & 132 & 27.0 & 15.8 & 2009.1781 &     \\
    & 21 & 3.34 & 133 & 206.2 & 12.5 & 2009.1781 &     \\
    & 22 & 3.38 & 135 & 86.7 & 15.8 & 2009.1781 &     \\
    & 23 & 3.67 & 146 & 11.6 & 13.6 & 2009.1781 &     \\
    & 24 & 3.76 & 150 & 33.3 & 13.9 & 2009.1781 &     \\
    & 25 & 3.77 & 150 & 257.4 & 10.2 & 2009.1781 &     \\
    & 26 & 3.78 & 150 & 331.6 & 14.1 & 2009.1781 &     \\
    & 27 & 3.83 & 152 & 18.6 & 13.7 & 2009.1781 &     \\
    & 28 & 3.91 & 156 & 293.9 & 11.8 & 2009.1781 &     \\
    & 29 & 3.96 & 157 & 54.8 & 13.2 & 2009.1781 &     \\
    & 30 & 4.14 & 165 & 110.3 & 12.0 & 2009.1781 &     \\
    & 31 & 4.22 & 168 & 11.5 & 13.9 & 2009.1781 &     \\
    & 32 & 4.25 & 169 & 80.0 & 7.8 & 2009.1781 &     \\
    & 33 & 4.28 & 170 & 328.0 & 14.6 & 2009.1781 &     \\
    & 34 & 4.30 & 171 & 93.4 & 15.1 & 2009.1781 &     \\
    & 35 & 4.41 & 175 & 95.8 & 14.9 & 2009.1781 &     \\
    & 36 & 4.45 & 177 & 148.9 & 12.3 & 2009.1781 &     \\
    & 37 & 4.48 & 178 & 237.3 & 14.3 & 2009.1781 &     \\
    & 38 & 4.52 & 180 & 356.1 & 15.3 & 2009.1781 &     \\
    & 39 & 4.58 & 182 & 205.1 & 11.2 & 2009.1781 &     \\
    & 40 & 4.62 & 184 & 133.4 & 15.5 & 2009.1781 &     \\
    & 41 & 4.67 & 186 & 334.7 & 15.2 & 2009.1781 &     \\
    & 42 & 4.69 & 187 & 27.4 & 14.6 & 2009.1781 &     \\
    & 43 & 4.73 & 188 & 196.3 & 12.3 & 2009.1781 &     \\
    & 44 & 4.74 & 189 & 212.0 & 10.7 & 2009.1781 &     \\
    & 45 & 4.77 & 190 & 102.6 & 11.5 & 2009.1781 &     \\
    & 46 & 4.80 & 191 & 136.5 & 13.1 & 2009.1781 &     \\
    & 47 & 4.82 & 192 & 181.7 & 14.8 & 2009.1781 &     \\
    & 48 & 4.86 & 193 & 26.1 & 15.2 & 2009.1781 &     \\
    & 49 & 4.88 & 194 & 309.0 & 14.7 & 2009.1781 &     \\
    & 50 & 4.91 & 195 & 319.6 & 14.3 & 2009.1781 &     \\
    & 51 & 4.93 & 196 & 272.5 & 12.4 & 2009.1781 &     \\
    & 52 & 4.93 & 196 & 314.7 & 12.8 & 2009.1781 &     \\
    & 53 & 4.94 & 196 & 230.5 & 14.3 & 2009.1781 &     \\
    & 54 & 5.01 & 199 & 359.8 & 13.9 & 2009.1781 &     \\
    & 55 & 5.18 & 206 & 237.7 & 14.9 & 2009.1781 &     \\
    & 56 & 5.23 & 208 & 291.5 & 13.7 & 2009.1781 &     \\
    & 57 & 5.24 & 208 & 308.9 & 15.1 & 2009.1781 &     \\
    & 58 & 5.39 & 214 & 155.6 & 15.3 & 2009.1781 &     \\
    & 59 & 5.39 & 215 & 295.5 & 11.3 & 2009.1781 &     \\
    & 60 & 5.42 & 216 & 210.9 & 15.0 & 2009.1781 &     \\
    & 61 & 5.45 & 217 & 229.3 & 15.6 & 2009.1781 &     \\
    & 62 & 5.46 & 217 & 239.6 & 15.4 & 2009.1781 &     \\
    & 63 & 5.48 & 218 & 127.5 & 14.2 & 2009.1781 &     \\
    & 64 & 5.53 & 220 & 335.4 & 8.8 & 2009.1781 &     \\
    & 65 & 5.55 & 221 & 196.6 & 15.3 & 2009.1781 &     \\
    & 66 & 5.60 & 223 & 313.8 & 12.6 & 2009.1781 &     \\
    & 67 & 5.63 & 224 & 354.5 & 14.2 & 2009.1781 &     \\
    & 68 & 5.65 & 225 & 164.5 & 14.7 & 2009.1781 &     \\
    & 69 & 5.74 & 228 & 165.9 & 15.0 & 2009.1781 &     \\
    & 70 & 5.76 & 229 & 331.1 & 14.0 & 2009.1781 &     \\
    & 71 & 5.83 & 232 & 201.5 & 11.9 & 2009.1781 &     \\
    & 72 & 5.84 & 233 & 232.7 & 15.7 & 2009.1781 &     \\
    & 73 & 5.87 & 234 & 81.4 & 13.8 & 2009.1781 &     \\
    & 74 & 5.94 & 236 & 94.0 & 13.9 & 2009.1781 &     \\
    & 75 & 5.95 & 237 & 191.1 & 13.5 & 2009.1781 &     \\
    & 76 & 5.96 & 237 & 306.7 & 12.9 & 2009.1781 &     \\
    & 77 & 6.07 & 241 & 149.2 & 12.1 & 2009.1781 &     \\
    & 78 & 6.08 & 242 & 312.6 & 14.6 & 2009.1781 &     \\
    & 79 & 6.09 & 242 & 336.5 & 15.4 & 2009.1781 &     \\
    & 80 & 6.10 & 243 & 60.6 & 12.0 & 2009.1781 &     \\
    & 81 & 6.17 & 246 & 102.9 & 15.7 & 2009.1781 &     \\
    & 82 & 6.21 & 247 & 107.5 & 13.5 & 2009.1781 &     \\
    & 83 & 6.23 & 248 & 330.8 & 14.3 & 2009.1781 &     \\
    & 84 & 6.25 & 249 & 209.4 & 14.1 & 2009.1781 &     \\
    & 85 & 6.26 & 249 & 288.2 & 13.7 & 2009.1781 &     \\
    & 86 & 6.36 & 253 & 116.5 & 15.7 & 2009.1781 &     \\
    & 87 & 6.42 & 255 & 59.6 & 10.5 & 2009.1781 &     \\
    & 88 & 6.47 & 258 & 202.9 & 14.1 & 2009.1781 &     \\
    & 89 & 6.53 & 260 & 202.1 & 14.6 & 2009.1781 &     \\
    & 90 & 6.53 & 260 & 3.9 & 15.3 & 2009.1781 &     \\
    & 91 & 6.55 & 261 & 226.6 & 15.1 & 2009.1781 &     \\
    & 92 & 6.57 & 261 & 325.7 & 15.2 & 2009.1781 &     \\
    & 93 & 6.63 & 264 & 105.2 & 15.4 & 2009.1781 &     \\
    & 94 & 6.67 & 265 & 248.3 & 13.5 & 2009.1781 &     \\
    & 95 & 6.67 & 266 & 40.6 & 13.6 & 2009.1781 &     \\
    & 96 & 6.68 & 266 & 121.5 & 15.3 & 2009.1781 &     \\
    & 97 & 6.71 & 267 & 70.8 & 9.0 & 2009.1781 &     \\
    & 98 & 6.71 & 267 & 9.2 & 14.5 & 2009.1781 &     \\
    & 99 & 6.75 & 269 & 230.7 & 13.9 & 2009.1781 &     \\
    & 100 & 6.77 & 269 & 185.6 & 10.5 & 2009.1781 &     \\
    & 101 & 6.80 & 271 & 85.0 & 15.7 & 2009.1781 &     \\
    & 102 & 6.85 & 273 & 353.4 & 13.8 & 2009.1781 &     \\
    & 103 & 6.86 & 273 & 292.1 & 15.3 & 2009.1781 &     \\
    & 104 & 6.89 & 274 & 279.8 & 13.7 & 2009.1781 &     \\
    & 105 & 6.98 & 278 & 84.8 & 15.3 & 2009.1781 &     \\
    & 106 & 7.14 & 284 & 182.5 & 14.4 & 2009.1781 &     \\
    & 107 & 7.17 & 285 & 148.1 & 13.5 & 2009.1781 &     \\
    & 108 & 7.25 & 289 & 37.6 & 12.9 & 2009.1781 &     \\
    & 109 & 7.26 & 289 & 269.4 & 14.1 & 2009.1781 &     \\
    & 110 & 7.28 & 290 & 300.6 & 14.2 & 2009.1781 &     \\
    & 111 & 7.35 & 292 & 89.6 & 12.8 & 2009.1781 &     \\
    & 112 & 7.41 & 295 & 92.8 & 13.4 & 2009.1781 &     \\
    & 113 & 7.46 & 297 & 257.8 & 13.9 & 2009.1781 &     \\
    & 114 & 7.48 & 298 & 80.9 & 12.5 & 2009.1781 &     \\
    & 115 & 7.63 & 304 & 297.5 & 9.1 & 2009.1781 &     \\
    & 116 & 7.74 & 308 & 201.9 & 13.6 & 2009.1781 &     \\
    & 117 & 7.77 & 309 & 112.1 & 12.2 & 2009.1781 &     \\
    & 118 & 7.85 & 312 & 204.0 & 12.3 & 2009.1781 &     \\
    & 119 & 7.96 & 317 & 24.3 & 13.8 & 2009.1781 &     \\
HD 139084 B & 1 & 0.92 & 37 & 206.6 & 8.1 & 2009.1123 & Drop \\
    & 2 & 1.19 & 48 & 347.8 & 9.5 & 2009.1123 &     \\
    & 3 & 1.38 & 55 & 119.6 & 11.2 & 2009.1123 &     \\
    & 4 & 1.51 & 60 & 32.4 & 12.3 & 2009.1123 &     \\
    & 5 & 1.78 & 71 & 351.3 & 12.1 & 2009.1123 &     \\
    & 6 & 1.84 & 73 & 241.4 & 11.1 & 2009.1123 &     \\
    & 7 & 1.96 & 78 & 355.3 & 11.6 & 2009.1123 &     \\
    & 8 & 2.11 & 84 & 258.7 & 9.4 & 2009.1123 &     \\
    & 9 & 2.22 & 88 & 14.0 & 11.3 & 2009.1123 &     \\
    & 10 & 2.23 & 89 & 267.2 & 11.4 & 2009.1123 &     \\
    & 11 & 2.26 & 90 & 226.3 & 7.3 & 2009.1123 &     \\
    & 12 & 2.39 & 95 & 57.3 & 9.3 & 2009.1123 &     \\
    & 13 & 2.43 & 97 & 241.7 & 11.7 & 2009.1123 &     \\
    & 14 & 2.45 & 97 & 302.4 & 9.8 & 2009.1123 &     \\
    & 15 & 2.73 & 109 & 290.8 & 11.9 & 2009.1123 &     \\
    & 16 & 2.79 & 111 & 326.0 & 10.0 & 2009.1123 &     \\
    & 17 & 2.82 & 112 & 39.6 & 6.2 & 2009.1123 &     \\
    & 18 & 2.85 & 113 & 26.6 & 11.7 & 2009.1123 &     \\
    & 19 & 2.86 & 114 & 129.4 & 12.0 & 2009.1123 &     \\
    & 20 & 2.88 & 115 & 303.3 & 11.9 & 2009.1123 &     \\
    & 21 & 3.02 & 120 & 183.7 & 9.8 & 2009.1123 &     \\
    & 22 & 3.45 & 137 & 335.7 & 11.8 & 2009.1123 &     \\
    & 23 & 3.60 & 143 & 188.6 & 9.7 & 2009.1123 &     \\
    & 24 & 3.62 & 144 & 83.4 & 11.4 & 2009.1123 &     \\
    & 25 & 3.69 & 147 & 198.4 & 10.7 & 2009.1123 &     \\
    & 26 & 3.72 & 148 & 192.6 & 12.2 & 2009.1123 &     \\
    & 27 & 3.77 & 150 & 139.6 & 11.6 & 2009.1123 &     \\
    & 28 & 3.87 & 154 & 23.2 & 10.3 & 2009.1123 &     \\
    & 29 & 3.98 & 158 & 8.8 & 11.0 & 2009.1123 &     \\
    & 30 & 4.06 & 162 & 210.9 & 11.6 & 2009.1123 &     \\
    & 31 & 4.17 & 166 & 182.7 & 12.3 & 2009.1123 &     \\
    & 32 & 4.22 & 168 & 132.6 & 10.6 & 2009.1123 &     \\
    & 33 & 4.30 & 171 & 60.4 & 9.6 & 2009.1123 &     \\
    & 34 & 4.46 & 178 & 6.0 & 10.6 & 2009.1123 &     \\
    & 35 & 4.48 & 178 & 158.4 & 11.1 & 2009.1123 &     \\
    & 36 & 4.56 & 182 & 181.4 & 11.0 & 2009.1123 &     \\
    & 37 & 4.65 & 185 & 134.4 & 10.5 & 2009.1123 &     \\
    & 38 & 4.66 & 186 & 194.7 & 11.5 & 2009.1123 &     \\
    & 39 & 4.80 & 191 & 268.3 & 9.2 & 2009.1123 &     \\
    & 40 & 4.85 & 193 & 310.9 & 11.2 & 2009.1123 &     \\
    & 41 & 4.88 & 194 & 154.9 & 9.0 & 2009.1123 &     \\
    & 42 & 4.91 & 195 & 164.1 & 9.3 & 2009.1123 &     \\
    & 43 & 5.03 & 200 & 203.5 & 10.5 & 2009.1123 &     \\
    & 44 & 5.04 & 201 & 130.7 & 5.0 & 2009.1123 &     \\
    & 45 & 5.08 & 202 & 248.9 & 9.8 & 2009.1123 &     \\
    & 46 & 5.13 & 204 & 289.6 & 9.3 & 2009.1123 &     \\
    & 47 & 5.16 & 205 & 310.6 & 11.2 & 2009.1123 &     \\
    & 48 & 5.27 & 210 & 210.1 & 11.7 & 2009.1123 &     \\
    & 49 & 5.36 & 213 & 22.4 & 10.1 & 2009.1123 &     \\
    & 50 & 5.38 & 214 & 326.3 & 10.4 & 2009.1123 &     \\
    & 51 & 5.38 & 214 & 25.1 & 9.1 & 2009.1123 &     \\
    & 52 & 5.40 & 215 & 158.1 & 11.9 & 2009.1123 &     \\
    & 53 & 5.42 & 216 & 147.3 & 10.8 & 2009.1123 &     \\
    & 54 & 5.47 & 218 & 82.3 & 10.5 & 2009.1123 &     \\
    & 55 & 5.50 & 219 & 151.9 & 9.3 & 2009.1123 &     \\
    & 56 & 5.56 & 221 & 105.7 & 9.8 & 2009.1123 &     \\
    & 57 & 5.58 & 222 & 345.7 & 9.9 & 2009.1123 &     \\
    & 58 & 5.58 & 222 & 228.0 & 5.9 & 2009.1123 &     \\
    & 59 & 5.65 & 225 & 333.8 & 8.8 & 2009.1123 &     \\
    & 60 & 5.66 & 225 & 27.4 & 10.4 & 2009.1123 &     \\
    & 61 & 5.67 & 226 & 260.1 & 8.5 & 2009.1123 &     \\
    & 62 & 5.70 & 227 & 182.5 & 12.0 & 2009.1123 &     \\
    & 63 & 5.72 & 228 & 340.9 & 11.6 & 2009.1123 &     \\
    & 64 & 5.74 & 228 & 277.4 & 12.0 & 2009.1123 &     \\
    & 65 & 5.76 & 229 & 237.6 & 11.4 & 2009.1123 &     \\
    & 66 & 5.83 & 232 & 43.4 & 10.8 & 2009.1123 &     \\
    & 67 & 5.90 & 235 & 77.5 & 9.7 & 2009.1123 &     \\
    & 68 & 6.00 & 239 & 269.7 & 9.8 & 2009.1123 &     \\
    & 69 & 6.02 & 240 & 60.5 & 8.3 & 2009.1123 &     \\
    & 70 & 6.05 & 241 & 140.8 & 11.2 & 2009.1123 &     \\
    & 71 & 6.10 & 243 & 168.5 & 7.8 & 2009.1123 &     \\
    & 72 & 6.13 & 244 & 328.2 & 11.1 & 2009.1123 &     \\
\enddata
\tablecomments{Target stars with candidate
companions for which we only have a single epoch of data.  Since
we cannot classify these as either CPM or background, we provide the
properties of these candidates here and note the changes we
make to the contrast curves in the Action column.  Contrast curves are
edited either by reverting to a less sensitive contrast curve or
restricting the contrast curve to within a given separation.
For cases where we have
only a single epoch and there are
candidates inside the 100\% coverage region for
position angle, we drop the star from our analysis.}
\end{deluxetable}

\end{document}